\documentclass[%
 reprint,
showpacs,
 amsmath,amssymb,
 aps,
rmp
]{revtex4-1}

\usepackage{graphicx}
\usepackage{dcolumn}
\usepackage{bm}
\usepackage[british,UKenglish,USenglish,american]{babel}

\usepackage[table]{xcolor}
\newcommand*{\nuc}[2]{\hbox{$^{#1}$#2}}
\usepackage{xr}
\begin{document}


\title{Evolution of shell structure in exotic nuclei}

\author{Takaharu Otsuka}
\affiliation{
Department of Physics and Center for Nuclear Study, University of Tokyo, Hongo, Bunkyo-ku, 
Tokyo 113-0033, Japan
}
\affiliation{
RIKEN Nishina Center, 2-1 Hirosawa, Wako, Saitama 351-0198, Japan
}
\affiliation{
Instituut voor Kern- en Stralingsfysica, Katholieke Universiteit Leuven, B-3001 Leuven, Belgium
}
\affiliation{
National Superconducting Cyclotron Laboratory,
Michigan State University, East Lansing, Michigan 48824, USA
}
\affiliation{
Department of Physics and Astronomy, 
Michigan State University, East Lansing, Michigan 48824, USA
}

\author{Alexandra Gade}
\affiliation{
National Superconducting Cyclotron Laboratory,
Michigan State University, East Lansing, Michigan 48824, USA
}
\affiliation{
Department of Physics and Astronomy, 
Michigan State University, East Lansing, Michigan 48824, USA
}

\author{Olivier Sorlin}
\affiliation{
Grand Acc\'el\'erateur National d'Ions Lourds (GANIL),
CEA/DSM-CNRS/IN2P3, BP 55027, F-14076 Caen Cedex 5, France
}

\author{Toshio Suzuki}
\affiliation{
Department of Physics and Graduate School of Integrated Basic Sciences, 
College of Humanities and Sciences,
Nihon University, Sakurajosui, Setagaya-ku, Tokyo 156-8550, Japan
}
\affiliation{
National Astronomical Observatory of Japan, Mitaka, Tokyo 181-8588, Japan
}

\author{Yutaka Utsuno}
\affiliation{
Advanced Science Research Center, Japan Atomic Energy Agency, 
Tokai, Ibaraki 319-1195, Japan
}
\affiliation{
Center for Nuclear Study, University of Tokyo, Hongo, Bunkyo-ku, 
Tokyo 113-0033, Japan
}

\date{\today}

\begin{abstract}
The atomic nucleus is a quantum many-body system whose constituent nucleons (protons and neutrons) are subject to complex nucleon-nucleon interactions that include spin- and isospin-dependent components.
For stable nuclei, already several decades ago, emerging seemingly regular patterns 
in some observables could be described successfully within a shell-model picture that results in particularly stable nuclei at certain magic fillings of the shells with protons and/or neutrons: N,Z = 8, 20, 28, 50, 82, 126. However, in short-lived, so-called exotic nuclei or rare isotopes,  characterized by a large N/Z asymmetry and located far away from the valley of beta stability on the nuclear chart,  these magic numbers, viewed through observables, were shown to change. 
These changes in the regime of exotic nuclei offer an unprecedented view at the roles of the various components of the nuclear force  
when theoretical descriptions are confronted 
with experimental data on exotic nuclei where certain 
effects are enhanced.  
This article reviews the driving forces 
behind shell evolution from a theoretical point of view and connects this to experimental signatures.
\end{abstract}

\maketitle

\tableofcontents

\section{Introduction}
\label{sec:introduction}

The atomic nucleus is composed of protons and neutrons (collectively called {\it nucleons}) bound into one entity 
by nuclear forces.   Its properties have been studied extensively for over a century
since its discovery by E. Rutherford in 1911 \cite{rutherford_1911}, 
providing a rather comprehensive picture of stable nuclei, {\it i.e.}, nuclei
with infinite or almost infinite lifetimes that are characterized by a balanced ratio of the number of neutrons 
($N$) and protons ($Z$), e.g., $N/Z \sim 1 - 1.5$.  
Matter found on the earth is essentially made up of stable nuclei, including long-lived 
primordial isotopes like $^{235}$U.  
Almost all matter in the visible universe is comprised of atomic nuclei. 

While the overall picture had thus been conceived for stable nuclei, 
the landscape of atomic nuclei has been significantly expanded in recent years.   
This is associated with a major shift in the frontiers of nuclear physics
from stable to exotic (or unstable) nuclei.  Here, exotic nuclei imply atomic nuclei
with an unbalanced $N/Z$ ratio as compared to stable ones, thus losing  
binding energy due to a large difference in $Z$ and $N$  
\cite{weizsacker_1935,bethe_1936}.   Relatively smaller binding energies 
mean that $\beta$-decay channels open up, proceeding towards more $N/Z$ balanced
systems and resulting   
in finite (often short, sub-second) lifetimes.  

Such extreme $N/Z$ ratios impact not only lifetimes of exotic nuclei but also 
their quantum many-body structure relative to that of stable nuclei.  This is
the main subject of this review article, with a particular emphasis on the
variations of the nuclear shell structure. 

\begin{figure}[bt]      
\begin{center}
\includegraphics[width=8.5cm,clip=]{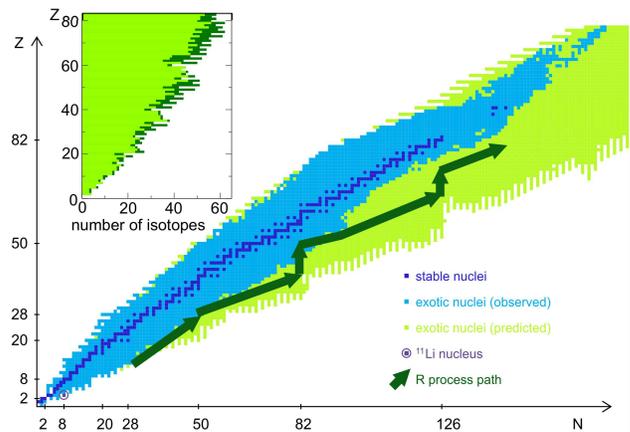}  
\end{center}
\caption{  
The nuclear chart as a function of neutron and proton number, $N$ and $Z$. 
Each nucleus is represented by a box specified by $Z$ and $N$.
Blue squares indicate stable nuclei. Exotic nuclei experimentally observed as of the year 2012 
are shown by light-brown squares, while light-green squares denote those predicted 
by a theoretical model \cite{koura2005}. The $^{11}$Li nucleus is highlighted in purple. 
A possible path of the $r$ process is indicated schematically by the green arrows. Inset: Number of bound neutron-rich exotic nuclei as a function of $Z$ based on Ref. \cite{koura2005}. The light and dark blue parts count nuclei with two-neutron separation energy 
$S_{2n}  > $ and $<$ 2 MeV, respectively.
Adapted from \citet{otsuka_nobel,otsuka_nuppec}.
}
\label{NC_all}
\end{figure}

Figure~\ref{NC_all} shows a nuclear chart (or Segr\`e chart), where an individual nucleus is specified by two  
coordinates: $Z$ and $N$.  
In Fig.~\ref{NC_all}, 
stable nuclei (blue squares) stretch along a ``line'', called the $\beta$-stability line. Exotic nuclei are
widely distributed as indicated by light brown or light green squares. Their existence limit on the neutron-rich 
(proton-rich) side is called the neutron (proton) dripline.  
Although a certain number of exotic nuclei have been familiar to nuclear physics
since the field's early days,
systematic studies of them have begun in the 1980's. One of the examples is the
measurement of the matter radius of $^{11}$Li  \cite{Tanihata1985}, marking a
visible milestone in the development of experiments with radioactive ion (or
rare isotope) beams with the discovery of the neutron halo.   
Many other experiments have been conducted in the last decades, re-drawing the
nuclear landscape.  

The nucleus $^{11}$Li is known for its extraordinarily large matter radius due
to the formation of a neutron halo, inherent to the last two loosely-bound neutrons \cite{hansen_1987}.  
The neutron halo is a characteristic phenomenon at and near the dripline that led us to 
change the canonical assumption that the nucleon density is almost constant
inside the nucleus and that the nuclear radius is proportional to $A^{1/3}$ where $A=Z+N$ is the mass number.
While 
$^{11}$Li  is located only four units away from the $\beta$-stability line on the nuclear  chart, 
the distance between the $\beta$-stability line and the neutron dripline
increases with $Z$ (see Fig. \ref{NC_all}).  
The nuclei shown in Fig.~\ref{NC_all} are all bound. The inset of Fig.\ref{NC_all} counts the number of bound neutron-rich exotic nuclei. 
It starts with just a few for $Z \sim 1$, but grows rapidly up to more than fifty for $Z$=82. 
Weakly-bound nuclei near the dripline are shown in dark blue, where neutron halo
or phenomena connected to the continuum can be expected.  One notices, however,
that the majority of isotopes are still well bound. Partly because such
well-bound exotic nuclei are so plentiful, but also because they span a
remarkable range of $N/Z$, we can ask ourselves whether the structure of those many nuclei 
is just like that of the stable ones. If not, an intriguing question arises:
what changes can be expected in extremely $N/Z$ asymmetric nuclei and why? 

We also note that the $r$-process, which creates heavy elements in explosive
scenarios such as neutron star mergers or supernovae in a series of neutron capture reactions and
decays, actually proceeds through extremely neutron-rich exotic nuclei (as shown
schematically in Fig.~\ref{NC_all}). Thus, for understanding how the
elements in the universe are formed, the study of the properties of exotic
nuclei is essential. 

The advent of radioactive ion beam facilities worldwide, together with
constantly improved experimental techniques, has enabled a more thorough
verification/discovery of the structure changes in exotic nuclei and ultimately
allowed reaching the nuclear driplines for some isotopes. 
 
Atomic nuclei show shell structure expressed in terms of the single-particle
orbits of protons and neutrons, similar to electrons in an atom. Such shell structure was proposed
originally by \citet{Mayer1949,Haxel1949}, and has provided a firm footing for
various studies on the structure of stable nuclei.  It has been found in  
recent years that the shell structure changes as a function of $Z$ and $N$ in
exotic nuclei, and this change is often referred to as {\it shell evolution}.   
While there has been enormous progress in the physics of exotic nuclei, we
rather concentrate in this article on the shell evolution, partly because this
subject alone is exhaustive and also because shell evolution is linked to a
large variety of observables, phenomena and features of current interest in the field.  A primer on nuclear shell
structure will be  presented in Sec.~\ref{sec:primer}. 

In Sec.~\ref{sec:monopole}, we review the definition of the monopole component of the $NN$ interaction in a pedagogical way.  
Although the monopole interaction has been discussed since  \cite{Bansal}, open
questions remain. The effective single-particle energies (ESPEs) are then derived from the monopole
interaction and are shown to be consistent with earlier derivations (see e.g. \cite{Baranger1970}).
The variation of the ESPEs as a function of $N$ or $Z$ is shown to be a
robust mechanism behind shell evolution.

In Sec.~\ref{sec:SE}, we discuss the major 
sources of the monopole interaction. In addition to the  central force,  the
tensor force is considered and the unique features of its monopole
interaction are reviewed.   
The treatment of the tensor force in other theories is summarized.  
The monopole effects of the two-body spin-orbit force is discussed in Sec.~\ref{subsec:2-body LS}.

Several features of nuclear forces related to the shell evolution are presented in Sec.~\ref{sec:force}, starting with the renormalization property of the tensor force 
followed by some properties obtained by a spin-tensor decomposition.   
The monopole effect of the three-nucleon force is discussed.  Finally, in
Sec.~\ref{sec:force} we also present a brief
overview of ab-initio approaches.     

Examples of structural changes are discussed in Sec.~\ref{sec:actual} before a summary is
given in Sec.~\ref{sec:summary}.  



\section{Nuclear shell structure: a primer}
\label{sec:primer}

\begin{figure} [tb]     
\includegraphics[width=8 cm] {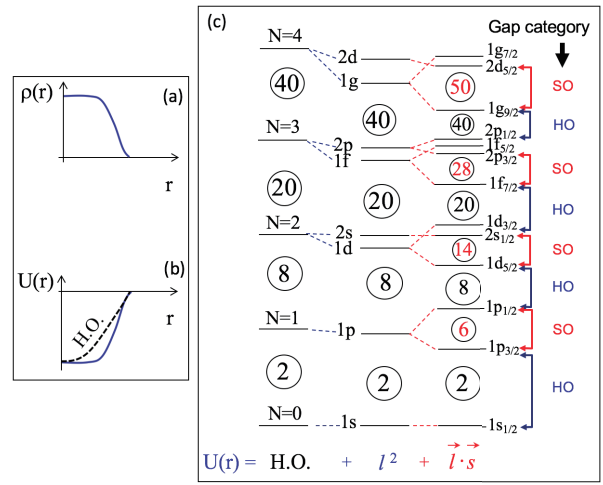} 
\caption {(a) Nucleon density distribution $\rho(r)$ and (b) mean potential $U(r)$ are shown as a function of the distance from the center of the nucleus, r. 
(c) Single-particle energies for a Harmonic Oscillator (HO) potential well, with
an added $\ell^2$ term and a spin-orbit interaction (SO) $\vec{\ell} \cdot
\vec{s}$.  Shell-gap categories are shown by HO and SO. The $N$ label refers
here to the oscillator shell $N=2(n-1)+\ell$, with $(n-1)$ being the number of
nodes of the radial wave function and $\ell$ the orbital angular
momentum. Figure based on \cite{Ragnarsson1995}. }   
\label{fig:HO-LS}
\end{figure}

In this section, we briefly describe the nuclear shell structure, starting with the nucleon distributions in nuclei. 
Extensive precision electron scattering experiments carried out on stable targets starting in the 60's 
combined with other experiments showed that 
the nucleon density $\rho(r)$ is essentially constant well inside the nucleus 
with smooth but rapid damping at the surface
as shown in Fig.~\ref{fig:HO-LS} (a): the paradigm of density saturation.  
The mean potential for a nucleon inside the nucleus represents 
the mean effects of the nucleon-nucleon ($NN$) interaction, or the nuclear
force, as generated by the other nucleons.  
The $NN$ interaction between free nucleons is strongly repulsive at short distances (below 0.7 fm), becomes 
attractive at medium range ($\approx$1.0 fm), and practically vanishes at large distances (beyond 2 fm). 
In the nuclear interior, the nuclear density is $\sim$ 0.17 nucleons/fm$^{3}$.
For the description of nucleons 
confined in the nucleus, an effective $NN$ interaction,
that incorporates various renormalization effects, such as in-medium effects, short-range correlation effects, {\it etc.}, is used. 
Those nucleons interact mainly with their immediate neighbors, which leads
to a saturation of the binding energy.  
Combining those properties of the density and the nuclear force, a nucleon well inside the nucleus
is subject to the same mean effect independent of its location.  In other words,  
the mean potential has a flat bottom.  The potential becomes gradually shallower 
towards the surface, as shown in Fig.~\ref{fig:HO-LS} (b).
Such a mean potential can further be modeled by a Harmonic Oscillator (HO) potential also shown in 
Fig.~\ref{fig:HO-LS} (b).   For that, the nucleons  move on the orbits which are the eigenstates of this 
HO potential, and the energies are given in terms of the oscillator quanta $\mathcal{N}$  as shown in the column ``H.O.''.
In order to resolve systematic discrepancies with experiment, 
Mayer and Jensen included the spin-orbit (SO) coupling ($\vec{\ell} \cdot \vec{s}$)
where $\vec{s}$ denotes the nuclear spin \cite{Mayer1949,Haxel1949}.  
This ($\vec{\ell} \cdot \vec{s}$) term with the proper strength produces the spin-orbit splitting, where the orbit with the total angular momentum $j_>=\ell+1/2$ becomes lower than the one with $j_<=\ell-1/2$.
The resultant single-particle levels are shown in the right column in Fig.~\ref{fig:HO-LS} (c).

Without the SO coupling, the single-particle states are classified by the $\mathcal{N}$ and $\ell$ quantum numbers 
as shown in the center column of Fig.~\ref{fig:HO-LS} (c).  The single-particle states are grouped 
according to $\mathcal{N}$, forming {\it shells}.  Shells are separated by {\it shell gaps}.  The number of protons or neutrons  
below a certain gap defines a {\it magic number}.  The magic number is related to the stability of the nucleus:
for instance, up to 20 protons can be put into the shells formed by the 2$s$, 1$d$, 1$p$ and 1$s$ orbits, 
whereas the 21$^{st}$ proton must occupy either the 1$f$ or 2$p$ orbit at higher
energy ({\it i.e.}, leading to a smaller 
binding energy).  Beyond the magic number 20, the SO coupling splits the 1$f$ 
orbit into 1$f_{7/2}$ and 1$f_{5/2}$ sufficiently strong and creates a magic
number at 28, as shown in 
Fig.~\ref{fig:HO-LS} (c).  
The 1$f_{7/2}$ orbit is bordered in this figure by two magic numbers 20 and 28:
the former has HO origin, whereas the latter has SO origin.  
Other shells and magic numbers are shown in the same figure. While $N$=40 is a
sub-shell gap, all magic nuclei above $N$=40 are of the SO origin.  
The major magic numbers, which correspond to large shell gaps, are 
2, 8, 20, 28, 50, 82, 126.
This shell structure and the corresponding magic numbers turned out to be extremely successful in the description of the 
nuclei.

\begin{figure} [tb]     
\includegraphics[width=8.5 cm] {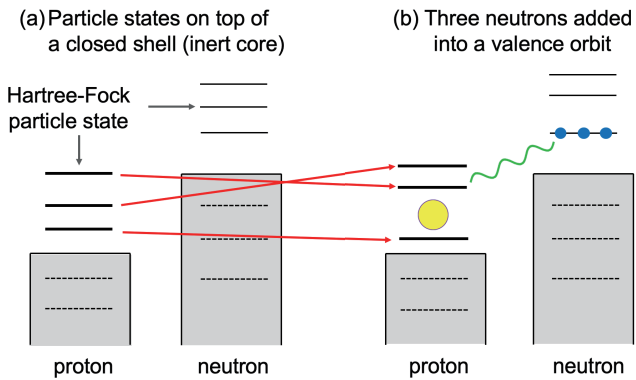}  
\caption {
Schematic illustrations of (a) closed shell and single-particle states in a Hartree-Fock picture and (b) singe-particle states with additional neutrons in a valence orbit. 
The circle indicates a subshell gap. The green wavy line denotes the monopole interaction. } 
\label{fig:HF}
\end{figure}

We note that the above argument is based only on a few robust properties: density
saturation, the short range of the 
nuclear force, and the existence of spin-orbit splitting.   
This independent-particle model, where nucleons are confined by a potential
without interacting with each other, can formally be refined through the
Hartree-Fock (HF) method, based on effective $NN$ interactions.
Figure~\ref{fig:HF} (a) shows this schematically: a HF calculation for $Z$ and
$N$ being magic numbers is supposed to produce the corresponding HF ground
state, which is a closed shell.  For this ground state, single-particle energies
for particle (and hole) states are obtained within the HF framework,
yielding Mayer-Jensen's shell structure (Fig.~\ref{fig:HO-LS} (c)).   
We now add nucleons to orbits above the closed shell, called valence orbits.  
Figure~\ref{fig:HF} (b) shows, still schematically, that the single-particle energies are shifted due to those added nucleons, mediated by the monopole interaction (indicated by the green wavy line in the figure), which is a component of the nuclear force.  
The monopole interaction shifts single-particle energies effectively without mixing different orbits, and its effect  depends only on the occupation numbers of individual orbits (Sec.~\ref{sec:monopole} for details).  
Such energy shifts represent shell evolution and   
manifest themselves systematically in a variety of observables measured for exotic nuclei. They are also one of the main subjects of this review article. 

\begin{figure} [tb]     
\includegraphics[width=8.5 cm] {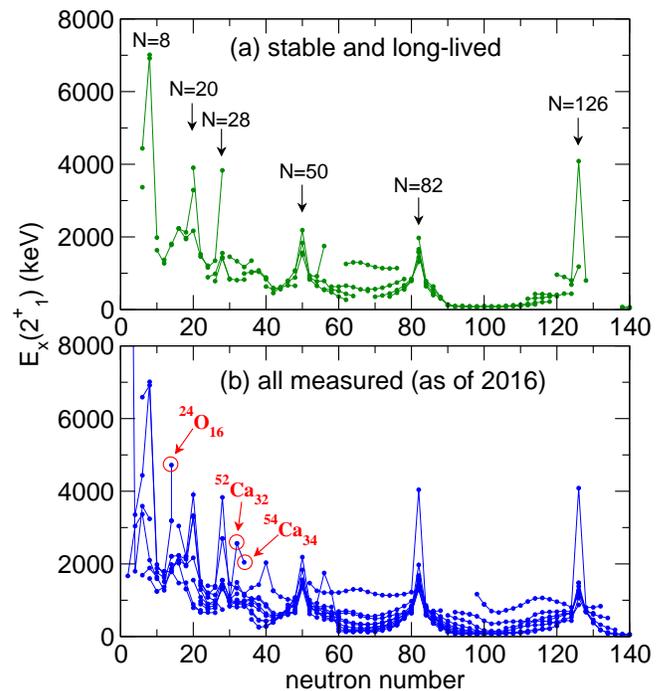}  
\caption { 
Systematics of the first 2$^+$ levels,  (a) for stable and long-lived nuclei,
and (b) all nuclei included 
measured up to 2016. Created from \cite{Pritychenko2016}. }
\label{fig:ex2+}
\end{figure}

Figure~\ref{fig:HF} (b) shows a small energy gap between two proton orbits (yellow circle).  Such energy gaps can appear as $Z$ and/or $N$ changes.   If such gaps become large enough, they may result in new magic numbers.  Alternatively, some of conventional magic numbers may disappear.   
We shall see how the shell structure changes or evolves over the Segr\`e chart. 

We stress that the single-particle orbits shown in Fig.~\ref{fig:HF} are
obtained for a spherical closed shell, {\it i.e.}, a spherical HF ground state.
This is the picture for most of the discussions in this article.  For the
majority of nuclei, however, their shape is non-spherical (deformed).
Nuclear deformation has been studied extensively since \cite{rainwater1950},
\cite{bohr1952}, \cite{bohr_mottelson1953}, as one of the major subjects of
nuclear physics \cite{BM2}.  The deformation can be described in terms of
various correlations of nucleons in the single-particle orbits.
Besides this, the HF solution itself can be deformed in some cases, where the
mean field is not isotropic and the HF ground state is not spherical.   
With the onset of deformation, deformed shell gaps can develop and lead to deformed magic numbers.
This is related to nuclear shape coexistence, {\it i.e.}, the appearance of states with different shapes at similar energies  
(see reviews \cite{heyde85}, \cite{wood92}, \cite{heyde2011},
\cite{wood2016}). Such a situation can be found in many nuclei \cite{heyde85},
\cite{wood92}, \cite{heyde2011}.  

Coming back to spherical magic numbers, experimental hints of their appearance/disappearance are visible, for example, in the excitation energy of the first 2$^+$ state.   
In the ground state of a magic nucleus, protons and neutrons fill
single-particle orbits up to a magic number and the corresponding large energy gap, and hence nucleons must be excited
across those gaps to form excited states.   Thus, the excitation energy becomes 
large, similarly to the relevant energy gaps.   Because the first 2$^+$ state is
the lowest excited state in many nuclei with even numbers of $Z$ and $N$, high
values of the lowest 2$^+$ level may indicate the occurrence of magic numbers.       
Figure~\ref{fig:ex2+} (a) shows the 2$^+_1$ energies obtained for stable and long-lived (half life $>$ 30 days) nuclei as a function of $N$ for many isotopic chains.  Higher 2$^+_1$ levels point remarkably well to Mayer-Jensen's magic numbers.   Panel (b) plots the 2$^+_1$ levels for all nuclei, including exotic ones, as of the year 2016.
Now, additional elevated 2$^+_1$ energies stand out, {\it e.g.} at $N$=16 ($^{24}$O),
32 ($^{52}$Ca), and 34 ($^{54}$Ca) as well as at $N=40$.
We note that 2$^+_1$ energies are impacted by a variety of correlations, such as
pairing, for example, but for the extreme values they can be attributed to magic
numbers. It should be remarked, however, that, while they provide useful
first indicators for magic numbers, they are not a decisive fingerprint. 

\begin{figure}[tb!]     
\begin{center}
\includegraphics[width=7cm]{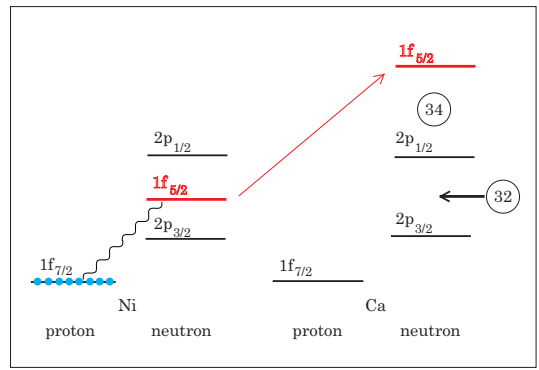}  
\end{center}
\caption{Schematic illustration of shell evolution from Ni to Ca for neutron orbits.
Light blue circles represent protons.  The wavy line implies the interaction between the proton 
$1f_{7/2}$ and the neutron $1f_{5/2}$ orbit.  The numbers in circles indicate (semi-)magic numbers. 
From \citet{Otsuka2016}.
}
\label{fig:SENiCa}
\end{figure}

Figure~\ref{fig:SENiCa} indicates schematically how shell closures may appear
for $N$=32 and 34 from the neutron single-particle levels for the Ca and Ni isotopes.
The figure shows the relevant single-particle level scheme of the
Ni isotopes, which is consistent with Fig.~\ref{fig:HO-LS} (c), representing
the situation in stable nuclei. This is confronted with the single-particle
levels of the Ca isotopes, where additional sub-shell closures at neutron number
32 and 34 are shown, resulting in an ordering of the neutron orbitals in
$^{52,54}$Ca that is different from Fig.~\ref{fig:HO-LS} (c).  We shall discuss
throughout this article why and how such shell evolution occurs.  

Thus, the magic numbers and shell structure are not immutable and undergo
change. 
As we look back several decades ago,     
the concept of rigid magic numbers was questioned already in the 1970's by
the observation of anomalies in experimental masses, nuclear radii and spectroscopy of nuclei far from stability, around $N=20$ since \cite{Thibault1975} with \cite{Huber1978, Detraz1979, Guillemaud-Mueller1984}. 
A much weakened effect of the $N=20$ gap, combined with the emergence of
deformed intruder states, was seen in various observables and interpreted to
signal a change in the shell structure.  
We note that another earlier observation questioning the conventional
understanding was marked by the discovery of the abnormal ground state of $^{11}$Be by  
\cite{Wilkinson1959} followed by a theoretical analysis 
\cite{Talmi1960}.   

Over the years, the local disappearance of many of the previously well-established shell gaps has been 
pointed out far from stability, leading to a revised picture of 
the magic numbers and shell structure in general.  One of the goals of the present article is to summarize the presently available understanding, to extract basic underlying mechanisms of the shell evolution, and 
to overview various nuclear phenomena related to them.  
Such outcomes allow us to anticipate new physics in hitherto unexplored regions of the Segr\`e chart. 

\begin{figure} [tb]     
\includegraphics[width=8.5 cm] {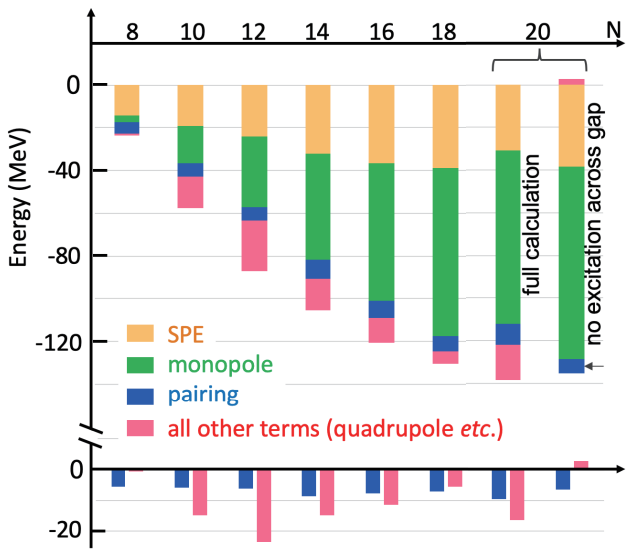}  
\caption { 
Contributions of the monopole, pairing and all other terms of the $NN$ interaction 
to the ground-state energy of Mg isotopes with even $N$.
The contribution from the single-particle energy (SPE) is included also.  The other terms are dominated by a  quadrupole interaction.  The upper panel shows the cumulative contributions, while the lower one shows the variations of pairing and the other terms.
For $N$=20, the full calculation, including the excitations to the $pf$ shell, is shown in comparison to the calculation without such excitations.  For latter, the small arrow indicates the ground-state energy because the "all other terms" gives a repulsive contribution (a positive value).  See the text for details.} 
\label{fig:Mg_decomp}
\end{figure}

In this article, the theoretical description of the structure of those nuclei is
given mainly within the 
shell-model framework, which is known as Configuration Interaction method in
other disciplines. Protons (neutrons) in valence orbits are called valence
protons (neutrons).  The shell-model description of atomic nuclei is made
in terms of such valence orbits on top of the inert core ({\it i.e.} closed shell). Effects of states outside this scheme are expected to be included in effective $NN$ interaction and effective operators, obtained in phenomenological, microscopic or hybrid manners.

The valence nucleons interact each other through the effective $NN$ interaction, and various configurations of valence nucleons are mixed in a shell-model eigenstate.  Many-body correlations are important for the resulting eigenstates 
and can often be related to certain parts of the effective $NN$ interaction.  
We decompose the effective $NN$ interaction into three parts: 
(a) monopole as discussed above, (b) pairing and (c) all other terms.
The pairing part corresponds to proton-proton and neutron-neutron interactions coupled to
the angular momentum $J=0$, which is an extended version of the usual BCS-type pairing. 
All other terms account for the remaining parts.  

Although the remaining parts include various types of interactions, 
dominant effects on the energy and structure of the ground and the low-lying states, which are of the current interest, are due to the quadrupole interaction.  
Such quadrupole interactions can be modeled, to a good extent, by the coupling between quadrupole-moment operators.  The quadrupole interaction has been studied extensively over decades, for instance, in \cite{nilsson1955,elliott1958a,bes1969,Dufour1996,kaneko2011}, while SU(3) picture of the quadrupole moment \cite{elliott1958a} has been generalized, for instance,  in \cite{zuker2015}.  

Figure~\ref{fig:Mg_decomp} shows the ground-state expectation values of
these parts for the Mg isotopes with even $N$=8-20, calculated with a shell model in
the $sd$ shell for $^{20-30}$Mg, and in the $sd$-$pf$ shell for $^{32}$Mg, where 
$pf$-shell configurations become important as mentioned above.  
The USDA interaction 
\cite{Brown2006a} is used for $^{20-30}$Mg.  The SDPF-M interaction
\cite{Utsuno1999} is taken for $^{32}$Mg, while the calculation without the excitation from the $sd$ to the $pf$ shells is also shown for comparison.  The expectation value of the SPE contribution  increases in magnitude up to $N$=18, since more neutrons occupy well-bound orbits  (negative energies).
The magnitude of the ``monopole" contribution increases up to $N$=20.  
The ``pairing'' contribution and that of the ``other terms'' are also shown separately in the bottom part of the figure.   
This part indicates clearly that the contribution of the pairing interaction does not change much.  In contrast, the contribution of the  
``other terms", dominated by the quadrupole interaction as mentioned already, varies sharply
with its maximum (in magnitude) at $N$=12 ($^{24}$Mg).  This is consistent with a large quadrupole moment of $^{24}$Mg.  The contribution of the ``other terms'' decreases up to $N$=18, as the quadrupole deformation weakens.  
These trends resemble the ones shown in Fig. 4 of \citep{heyde2011} across a shell for heavy nuclei. 
At $N$=20, an intruder state composed of many particle-hole excitations is energetically favored over normal configurations ({\it i.e.} no particle-hole excitations across the $N$=20 magic gap) and becomes the ground state.
To this state, the contribution of the "other terms" becomes large, only a
little smaller than for $^{24}$Mg, pointing to a strongly deformed ground state.
The ``monopole" and ``pairing" contributions increase as well from $N$=18 to 20,
but the ``SPE" contribution is reduced due to particle-hole excitations across the $N=20$
gap.   
For comparison, results of a calculation are shown without cross-shell excitations, resulting in almost no quadrupole correlations and a higher energy.
Thus, the Mg isotopes show varying deformation and the phenomenon of shape
coexistence at $N$=20 \citep{heyde2011}.  The features shown here apply to  many
isotopic chains across the nuclear chart. 
The three parts, ``monopole", ``pairing'' and "other terms", exhibit sizable contributions with notable variations.   
In this article, we will highlight the important role of the monopole interaction in describing structural changes mainly from the shell-model viewpoint.

The Hartree-Fock calculation discussed in Fig.~\ref{fig:HF} corresponds to a
spherical ground state. Considering the strong quadrupole and higher multipole
interactions, deformed ground states may occur and can be described through
deformed HF configurations.  A  non-spherical mean potential is obtained, and
the HF ground state becomes the intrinsic state of a rotational band
\cite{Ring-Schuck}. There exits extensive literature on the deformed HF
description of shape coexistence, see, e.g., \cite{wood92}, \cite{Reinhard1999}
and \cite{heyde2011}.  

Before closing this section, we comment on the comparisons of the shell structure of 
atomic nuclei to the shell structure of other many-body systems.
First, as the nuclear potential is generated by its constituents, shell
structure changes from nucleus to nucleus, leading to shell evolution.
We mention here that shell structure appears in other mesoscopic
systems such as metallic clusters as described, for instance, in \cite{Sugano},
\cite{Knight84}, 
and \cite{Clemenger85}, where the correspondence       
to the classical motion and geometrical symmetries is important.

It has been argued that the damping of the nucleon density in the radial
direction may be more gradual in neutron-rich exotic nuclei than in stable
nuclei, causing a reduced spin-orbit splittings and single-particle levels
distributed more evenly \cite{Dobaczewski1994}, sometimes referred to as ``shell
quenching''. This hypothetical 
phenomenon is predicted to be found at or near to the dripline, confined to weakly-bound
systems, and remains a challenge for future experiments.   
The present article addresses the shell evolution driven by the combination of
characteristic features of nuclear forces and extreme neutron/proton ratios
of the nucleus. 


\section{Monopole interaction and empirical analysis based on it}
\label{sec:monopole}

The shell structure can be specified by a set of single-particle energies of valence (or active) orbits 
on top of a closed shell (or inert core).     
As more neutrons or protons are added to a nucleus, the single-particle energies of 
those valence orbits may change due to the interaction between valence nucleons.  
This implies some changes of shell structure, called {\it shell evolution} as introduced in 
Sec.~\ref{sec:introduction}.   
The shell evolution is generated by the monopole part of the nucleon-nucleon ($NN$) interaction, 
which will be abbreviated hereafter as the {\it monopole interaction}.  
The $NN$ interaction here means an effective one for nucleons in nuclei.  Although there can be a variety of such interactions from fitted to microscopically-derived ones including hybrid versions, we shall discuss their general properties.
In this section, we first introduce the definition of the monopole interaction, and discuss how it acts. 
The monopole interaction has been discussed in the past, for instance, by 
Bansal and French \cite{Bansal} and by Poves and Zuker \cite{Poves1981}.  
We introduce the monopole interaction in a different way, as an average of correlation energies 
of two nucleons in an open-shell nucleus, without referring to closed-shell energies. 
The final outcome of this formulation turns out to be basically consistent with those earlier works.   

The effective single-particle energy will then be defined for open-shell nuclei 
in a close connection to the monopole interaction there,  
in a possibly more transparent and straightforward way than the simple interpolation between
the beginning and end of a given shell.     

We will then move forward to 
the evolution of the shell structure by defining effective single-particle energies with this monopole interaction.  
We also present applications of the monopole interaction to some examples taken from
actual nuclei.  At this point, we stress that the monopole interaction is a part 
of the $NN$ interaction, and that the rest of the interaction produces various dynamical correlations 
and must be taken into account for 
an actual description of the nuclear structure.  Nevertheless, as the monopole interaction generates unique and crucial effects, it deserves special efforts and attention.

\subsection{Monopole interaction \label{subsec:mono}}

We start with single-particle orbits.  For each orbit, the total angular momentum is specified by 
$\vec{j} = \vec{\ell} + \vec{s}$ with its orbital angular momentum $\vec{\ell}$ and spin $\vec{s}$. 
The single-particle orbits are labelled by the magnitudes of their $\vec{j}$'s, 
referred to as $j$, $j'$, ...  hereafter.  
They are combined with the corresponding magnetic quantum numbers, $m$, $m'$, ... as $(j, m)$, $(j', m')$, ... 
The symbol $j$, $j'$, .... are put in a fixed order, and may carry implicitly such a sequential ordering 
as well as other quantum numbers like the node of the radial wave function $n$.  
Having these single-particle orbits on top of the inert core ({\it i.e.}, closed shell), 
we denote the single-particle energies (SPEs) of those orbits 
as $\epsilon^0_j$, $\epsilon^0_{j'}$, ....  
As usual, this SPE $\epsilon^0_j$ stands for the sum of 
the kinetic energy of a nucleon on this orbit $j$ and the total effects of nuclear forces on this nucleon 
from all nucleons in the inert core.   
 
We shall begin with the simpler case by assuming that there is  
only one kind of nucleons, {\it e.g.}, neutrons.   
The Hamiltonian is expressed then as 
\begin{equation}
\hat{H_{n}} \,  =  \, \sum_{j} \,  \epsilon^0_j  \, \hat{n}_{j} + \hat{v}_{nn} \,  ,
\label{eq:Hn}
\end{equation}
where $\hat{n}_{j}$ denotes the number operator for the orbit $j$ and $\hat{v}_{nn}$ stands for
the neutron-neutron effective interaction. 

\begin{figure*}[bt]     
\includegraphics[width=14cm]{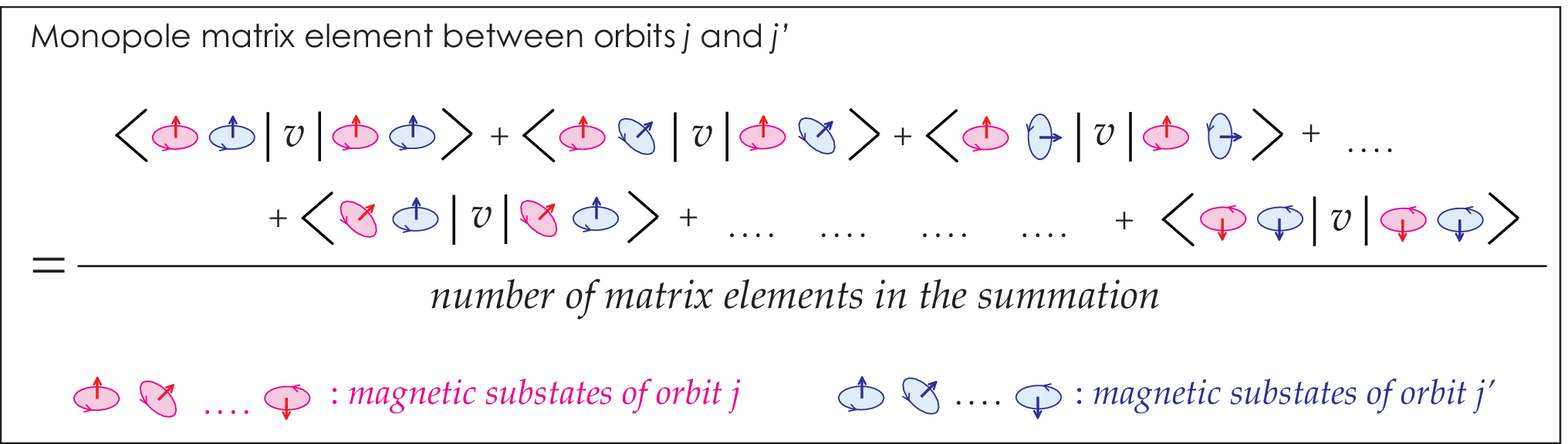}  
\caption{A schematic visualization of monopole matrix elements for the two-body interaction $v$.  }
\label{fig:mono}
\end{figure*}

The product state with the first and second neutron in the states $j,m$ and $j',m'$, respectively, 
is written as
\begin{equation} 
\mid j,m \otimes j',m' \,). 
\label{eq:direct}
\end{equation} 
Their antisymmetrized state is indicated by 
\begin{flalign}
\label{eq:nn-anti}
  &\,\,\,\,\,\,\,\,\,\,\,\,\, |j,m \,;\, j',m' >  & \nonumber \\
  & \,\,\,\,\,\,\,\,\,\,\,\,\, =  \, \Bigl\{ |j,m \otimes j',m' \,) - |j',m' \otimes j,m \,) \Bigr\} / \sqrt{2}. & 
\end{flalign}   

A two-body interaction between two neutrons can be written as 
\begin{eqnarray}
\label{eq:vnn}
\hat{v}_{nn} & = & \Sigma_{(j_1, m_1\,;\, j_1', m_1'),(j_2, m_2 \,;\, j_2', m_2') } \, \, \nonumber \\
 & &  \, \, \langle j_1, m_1\,;\, j_1', m_1' | \, \hat{v}_{nn} \, | j_2, m_2 \,;\, j_2', m_2' \rangle  \nonumber \\
 & & \, \, a^{\dagger}_{j_1, m_1} a^{\dagger}_{j_1', m_1'}  a_{j_2', m_2'} a_{j_2, m_2},
\end{eqnarray}
where $(j, m \,;\, j', m')$ in the summation is an ordered pair of two states $j, m$ and $j', m'$, 
$\langle ... | \, \hat{v} \, | ... \rangle$ denotes an antisymmetrized two-body matrix element, and 
$a^{\dagger}_{j, m}$ ($a_{j, m}$) implies the creation (annihilation) operator of the state $j, m$.
Regarding the ordered pair $(j_1, m_1\,;\, j_1', m_1')$, we can assume without generality that 
$m_1 < m_1'$ if $j_1 = j_1'$ or $j_1 \, < \, j_1'$ in their prefixed ordering as mentioned above.

The monopole interaction is defined as a component extracted from a given interaction, $\hat{v}_{nn}$, so that 
it represents the effect averaged over all possible orientations of two neutrons in the orbits $j$ and $j'$.
Here, orientations refer to various combinations of $m$ and $m'$ within the orbits $j$ and $j'$.
Figure~\ref{fig:mono} provides a general visualization of the monopole matrix element, exhibiting
different orientations by differently tilted orbiting planes.
In order to formulate this, 
the monopole matrix element for the orbits $j$ and $j'$ is defined as
\begin{eqnarray}
\label{eq:m_nn}
V_{nn}^m (j,j') \, = \, \frac{\sum_{(m, m')}  \langle j,m\,;\, j',m' | \hat{v}_{nn} | j,m \,;\, j',m' \rangle }{\sum_{(m,m')} 1}, \nonumber \\ 
\end{eqnarray}
where the summation over $m, m'$ is taken for all ordered pairs allowed by the Pauli principle. 

As the denominator counts the number of allowed states,   
this is exactly the average mentioned above.  
The monopole interaction as an operator is then expressed as
\begin{equation}
\hat{v}_{nn,mono} \,  =  \, \sum_{j \le j'} \, \hat{v}_{nn}^m (j,j') \,  
\label{eq:Vm_nnop}
\end{equation}
with 
\begin{equation}
\label{eq:Vm_nnop2}
\hat{v}_{nn}^m (j,j') \, = \, V_{nn}^m (j,j') \,\Sigma_{m,m'} \,  a^{\dagger}_{j, m} a^{\dagger}_{j', m'}  
a_{j', m'} a_{j, m} .
\end{equation}
After simple algebra, this turns out to be
\begin{eqnarray}
\label{eq:Vm_nnop3}
 \hat{v}_{nn}^m (j,j') \, & =  & \left\{ \begin{array} {ll} 
 \, V_{nn}^m (j,j) \, \frac{1}{2} \, \hat{n}_j \, (\hat{n}_j -1) & {\rm for}\,\, j=j'  \\
 \\
 \, V_{nn}^m (j,j') \,  \hat{n}_j \, \hat{n}_{j'} & {\rm for}\,\, j \neq j'  \\
\end{array} \right. 
\end{eqnarray}
where $\hat{n}_j$ stands for the number operator for the orbit $j$.  
The form in eq.~(\ref{eq:Vm_nnop3}) appears to be in accordance with what can be expected 
intuitively, from the concept of average, for identical fermions.  
The two neutrons in the orbits $j$ and $j'$ can be coupled to the total angular momentum, $J$, 
where $\vec{J} = \vec{j} + \vec{j'}$, and 
the wave function with a good $J$ value is given by a particular superposition of the states in 
eq.~(\ref{eq:Vm_nnop}) over all possible values of $m$ and $m'$.  
It is obvious that the effects of the monopole interaction in eq.~(\ref{eq:Vm_nnop3}) is independent of 
the total angular momentum, $J$.
We emphasize again that  the monopole interaction is 
simply an average of a given general interaction over all possible orientations, and its effect can be
expressed by the orbital number operator as in eq.~(\ref{eq:Vm_nnop3}) for the neutron-neutron
interaction.  

We next discuss systems composed of protons and neutrons.
The total Hamiltonian is then written as
\begin{equation}
\hat{H} \,  =  \, \hat{H_{n}} \,+ \, \hat{H_{p}} \,+ \,  \hat{v}_{pn} \,  ,
\label{eq:H}
\end{equation}
where $\hat{H_{p}}$ stands for the proton Hamiltonian defined similarly to 
eq.~(\ref{eq:Hn}) and $\hat{v}_{pn}$ means the proton-neutron effective interaction. 

The proton and neutron number operators in the orbit $j$ are denoted, respectively, as $\hat{n}^p_j$ and $\hat{n}^n_j$.
We introduce the isospin operators in the orbit $j$:   
$\hat{\tau}^+_{j}$, $\hat{\tau}^-_j$ and $\hat{\tau}^z_j$.  
We adopt the convention that protons are in the state of isospin z-component $\tau_z =+1/2$, 
whereas neutrons are in $\tau_z =-1/2$.   
Here, $\hat{\tau}^+_{j}$ ($\hat{\tau}^-_j$) denotes the operator changing a neutron (proton) to a proton (neutron) in the same ($j,m$) state,  
and $\hat{\tau}^z_j$ equals $(\hat{n}^p_j - \hat{n}^n_j)/2$.  
In other words, $\hat{\tau}^+_{j}$ and $\hat{\tau}^-_j$ are nothing but  
the isospin raising and lowering operators restricted to the orbit $j$, while $\hat{\tau}^0_j$ is its z component.

The magnitude of the usual isospin, {\it i.e.}, not specific to an orbit, is denoted by $T$, including that of two nucleons interacting through the $NN$ interaction.

\begin{figure}[tb]     
\includegraphics[width=8.0cm]{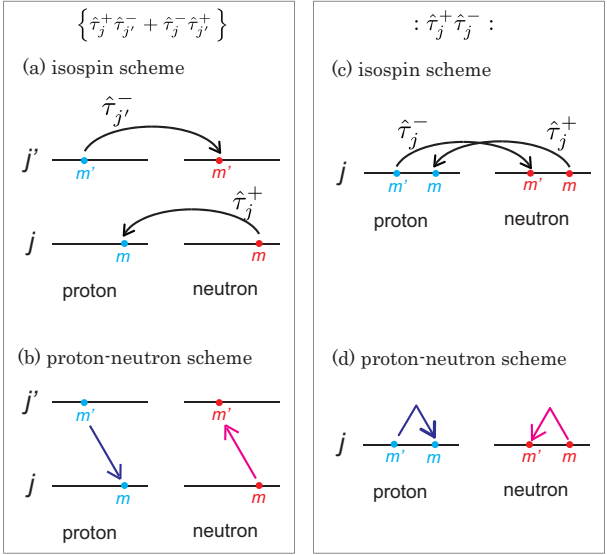}  
\caption{Implication of $\hat{\tau}^+_{j} \hat{\tau}^-_{j'}$ terms. 
Panels (a) and (c) are for the $\{ \hat{\tau}^+_{j} \hat{\tau}^-_{j'} +  \hat{\tau}^-_{j} \hat{\tau}^+_{j'} \, \} $ and 
$: \hat{\tau}^+_{j} \hat{\tau}^-_{j} :$ cases in the isospin scheme, respectively.
Panels (b) and (d) are similar to (a) and (c), respectively, in the proton-neutron scheme.
The magnetic substates are indicated by $m$ and $m'$. }
\label{fig:tautau}
\end{figure}

We now discuss the proton-neutron monopole interaction.  Although the basic idea remains the same as the previous case for two neutrons, certain differences arise.  To be precise, a proton and a neutron are  
coupled in a symmetric way for the $T$=0 case, whereas in an antisymmetric way for the $T$=1 case.  Details of the discussions are presented in 
Appendix~\ref{app:pn mono}, 
and we show here only major points, referring to the corresponding parts there.   
 
The monopole interaction due to the $T$=0 part of $\hat{v}_{pn}$ is expressed as
(see discussions in  
Appendix~\ref{app:pn mono}   
up to eq.~(\ref{eq:Vm_jj'T0})),  
\begin{flalign}
 \hat{v}_{pn,mono,T=0} \, &=  \, \sum_{\, j,\, j'} \, V_{T=0}^m (j, \, j') \, \frac{1}{2} \, \hat{n}^p_j \, \hat{n}^n_{j'} \, \nonumber \\
 & -  \sum_{\, j <  j'} \, V_{T=0}^m (j, \, j') \, \frac{1}{2} \, \Bigl\{ \hat{\tau}^+_{j} \hat{\tau}^-_{j'} + 
 \hat{\tau}^-_{j} \hat{\tau}^+_{j'} \, \Bigr\}  \nonumber & \\
 &  - \sum_{j} \, V_{T=0}^m (j, \, j) \, \frac{1}{2} : \hat{\tau}^+_{j} \hat{\tau}^-_{j} : \, , & 
\label{eq:Vm_jj'T0_main}
\end{flalign}
where the symbol $: ... :$ denotes a normal product and 
the $T$=0 monopole matrix element is defined as an average over states 
with all possible orientations with the symmetric coupling of proton and neutron, as indicated by ${\cal S}$
(see discussions in  
Appendix~\ref{app:pn mono}   
linked to 
eqs.~(\ref{eq:m_pn}, \ref{eq:mono_T=0})):   
\begin{eqnarray}
\label{eq:m_pn_main}
V_{T=0}^m (j,j') \, & = &\, \frac{\sum_{(m, m')}  ( m; m' :{\cal S}| \hat{v}_{pn} | m; m' :{\cal S} ) }{\sum_{m,m'} 1}.
\end{eqnarray}

If the interaction $\hat{v}$ is isospin invariant as usual, 
the monopole matrix element in eq.~(\ref{eq:m_nn}) is nothing but the $T=1$ monopole matrix element, 
\begin{equation}
V_{T=1}^m (j,j') \,=\, V_{nn}^m (j,j') \, .
\label{eq:mono_T=1_main}
\end{equation}

Coming back to antisymmetric couplings of a proton and a neutron, we apply the procedures similar to those for $T=0$ states, and obtain (see discussions in 
Appendix~\ref{app:pn mono}  
linked to eq.~(\ref{eq:Vm_jj'T1}))  
\begin{flalign}
 \hat{v}_{pn,mono,T=1} &= \sum_{\, j ,\, j'} \, V_{T=1}^m (j, \, j') \, \frac{1}{2} \, \hat{n}^p_j \, \hat{n}^n_{j'} \, \nonumber &\\
& + \sum_{\, j <  j'} V_{T=1}^m (j, \, j') \, \frac{1}{2} \, \Bigl\{\, \hat{\tau}^+_{j} \hat{\tau}^-_{j'} \, + 
 \hat{\tau}^-_{j} \hat{\tau}^+_{j'} \, \Bigr\}   \nonumber & \nonumber \\
& + \sum_{j} \, V_{T=1}^m (j, \, j) \, \frac{1}{2} : \hat{\tau}^+_{j} \hat{\tau}^-_{j} : \, .& 
\label{eq:Vm_jj'T1_main}
\end{flalign}
By combining eqs.~(\ref{eq:Vm_jj'T0_main}) and (\ref{eq:Vm_jj'T1_main}), the whole expression of the proton-neutron monopole interaction becomes
\begin{eqnarray}
 \hat{v}_{pn,mono} & = & \sum_{\, j , \,  j'}   \frac{1}{2} \, \Bigl\{V_{T=0}^m (j, \, j')\, + \, V_{T=1}^m (j, \, j') \Bigr\}
                                            \,  \hat{n}^p_j \, \hat{n}^n_{j'}   \nonumber \\
                           & - & \sum_{\, j < \,  j'}  \frac{1}{2} \, \Bigl\{V_{T=0}^m (j, \, j')\, - \,V_{T=1}^m (j, \, j') \Bigr\} \nonumber  \\  
   &  & \,\,\,\,\,\,\,\,\,\,\,\,\, \Bigl\{ \hat{\tau}^+_{j} \hat{\tau}^-_{j'} \,+\, \hat{\tau}^-_{j} \hat{\tau}^+_{j'}  \Bigr\} \nonumber  \\
                                  & - & \sum_{\, j }  \frac{1}{2} \, \Bigl\{V_{T=0}^m (j, \, j)\, - \,V_{T=1}^m (j, \, j) \Bigr\}   
                                     \,  :\hat{\tau}^+_{j} \hat{\tau}^-_{j}:  . \nonumber \\                                    
\label{eq:m_pntot_main}
\end{eqnarray}
Although the meaning of the first term  on the right-hand side of eq.~(\ref{eq:m_pntot_main}) is straightforward, 
it needs some explanations to understand the other two terms in depth.  Figure~\ref{fig:tautau} may help, 
by showing how they work.  
In the case of $j \ne j'$, panels (a) and (b) indicate the same process
in the isospin and proton-neutron schemes, respectively.   Panel (a) indicates that 
the $\{ \hat{\tau}^+_{j} \hat{\tau}^-_{j'} +  \hat{\tau}^-_{j} \hat{\tau}^+_{j'} \, \} $ term produces a monopole 
interaction with a charge exchange process, whereas the same process may look differently in panel (b).  
Panels (c) and (d) are for the case of one orbit $j$ with similar implications.  One thus see, from Fig.~\ref{fig:tautau}, how the charge exchange processes can be incorporated into the monopole interaction.  We will come back to this figure.  
We note that the $T=0$ and $T=1$ monopole matrix elements contribute with opposite sign relations 
as compared to the first term.   

The neutron-neutron and proton-proton monopole interactions can be re-written in similar ways, as 
\begin{eqnarray}
\label{eq:m_nn-2}
 \hat{v}_{nn,mono}  \,  & = & \sum_{j} V_{T=1}^m (j,j) \, \frac{1}{2} \, \hat{n}^n_j \, (\hat{n}^n_j -1)    \nonumber \\
               & + & \sum_{j < j'} V_{T=1}^m (j,j') \,  \hat{n}^n_j \, \hat{n}^n_{j'} \, ,  
\end{eqnarray}
and 
\begin{eqnarray}
\label{eq:m_pp}
 \hat{v}_{pp,mono}  \, & = & \sum_{j} V_{T=1}^m (j,j) \, \frac{1}{2} \, \hat{n}^p_j \, (\hat{n}^p_j -1)  \nonumber \\
               & + & \sum_{j < j'} V_{T=1}^m (j,j') \,  \hat{n}^p_j \, \hat{n}^p_{j'} \,.   
 \end{eqnarray}
We thus gain the complete expression for the total monopole interaction,
\begin{equation}
\label{eq:m_tot}
\hat{v}_{mono}  \, = \, \hat{v}_{pp,mono}  \,+ \, \hat{v}_{nn,mono}  \,+\, \hat{v}_{pn,mono} \,\, .
\end{equation}

\subsection{Multipole interaction} \label{subsect:multipole}

We have discussed the monopole interaction which is a part of the $NN$ interaction.
The remaining part of the $NN$ interaction is called the {\it multipole interaction}.
The multipole interaction is often expressed as
$\hat{v}_M$, and it includes, in particular the quadrupole interaction.  In this article, we 
denote the multipole interaction as $\hat{v}_{multi}$, being defined by
\begin{equation}
\label{eq:full}
\hat{v}_{multi} \,=\, \hat{v} \, - \, \hat{v}_{mono}  \,\,,
\end{equation}
where $\hat{v}$ stands for the full interaction, and $\hat{v}_{mono}$ is defined in eq.~(\ref{eq:m_tot}).  
The multipole interaction may have subscript $pp, nn$, or $pn$, if necessary. 

We note that although the notion of the multipole interaction has appeared, for instance, in \citep{Brown1967}, including the importance of the quadrupole and hexadecupole forces, the multipole interaction in the present sense was introduced, as ``non monopole'', in \citep{Poves1981}.  A model of the multipole interaction was introduced and developed in a global description of collective states in \citep{Dufour1996}.   

\subsection{Monopole matrix element in the $j-j$ coupling scheme} \label{subsect:mono_jj}

The monopole matrix element is defined, in some cases, by an alternative but equivalent expression.
\begin{eqnarray}
\label{eq:mono_J-2}
V_{T}^m (j, \, j') &=& \frac{\sum_{J}  (2J+1) \langle j, j' ; J,T | \hat{v} | j, j' ; J,T \rangle }{\sum_{J} (2J+1)} 
   \nonumber \\
   & & \,\,\, {\rm for} \,\,\, T=0 \,\, {\rm and} \,\, 1, 
\end{eqnarray}
where $J$ takes only even (odd) integers for $j = j'$ with $T=1$ $(T=0)$.
Appendix~\ref{app:alternative}   
shows that this expression is indeed equivalent to the one presented here.

The closed-shell properties are derived from the expressions shown so far.
The actual derivations and results are given in  
Appendix~\ref{app:closed}. 


\subsection{Effective single particle energy \label{subsec:espe}}

We discuss, in this subsection, effective single particle energy and its derivation from the monopole interaction.

As one moves on the Segr\`e chart,  
the proton number, $Z$, and the neutron number, $N$, change, and the single-particle energy $\epsilon^0_j$ mentioned in Sec.~\ref{subsec:mono} will change also.  
This change has the following two aspects. 
One is due to the kinetic energy:  as $A$ increases, the radius of the nucleus becomes larger, and 
consequently the radial wave function of each orbit 
becomes wider.  This lowers the kinetic energy.  
The other aspect is the variation in the effects from nucleons in the inert core.   
As $A$ increases, the radial wave functions
of the orbits in the inert core also become stretched out radially.  This can reduce the 
magnitude of their effects. 
While these two changes can be of relevance, for instance, over a long chain of isotopes, 
they are considered to be rather minor within each region of current interest on the Segr\`e chart \cite{BM1},   
and we do not take them into account in this article.  

The single-particle energy has another origin: the contribution from other nucleons outside 
the inert core, {\it i.e.}, valence nucleons.  This valence contribution to the orbit $j$ is referred to as 
$\hat{\epsilon}_j$ hereafter.  
The total single-particle energy, called {\it effective single particle energy (ESPE)} usually, 
is denoted as, 
\begin{eqnarray}
\label{eq:espe}
\epsilon_j  \,= \, \epsilon^0_j \, + \,  \hat{\epsilon}_j  \,\, .
\end{eqnarray}
We shall discuss, in this subsection, the valence contribution, $\hat{\epsilon}_j$, in some detail.  
Note that $\epsilon^0_j$ is a constant as stated, whereas $\hat{\epsilon}_j$ 
is an operator by nature because of its dependence on the states of other valence nucleons. 

The magnetic substates of the orbits $j$ and $j'$ are denoted, respectively,  by $m$ ($m = j, j-1, ..., -j+1, -j$) 
and $m'$ ($m' = j', j'-1, ..., -j'+1, -j'$). 
The matrix element $\langle m, m' | \, \hat{v} \, |  m, m' \rangle$ varies   
for different combinations of $m$ and $m'$.  
On the other hand, as $\hat{\epsilon}_j$ is a part of the single-particle energy of the orbit $j$, 
it should be independent of $m$.  
We therefore extract the $m$-independent component from these matrix elements,     
in order to evaluate their contribution to $\hat{\epsilon}_j$. 
Because of the $m$ and $m'$ dependences, this can be done by taking  
the average over all possible combinations of $m$ and $m'$,    
which is nothing but the monopole 
interaction discussed in Sec.~\ref{subsec:mono}. 

In the case of two neutrons in the same orbit $j$,  
the monopole interaction is included in eq.~(\ref{eq:m_nn-2}).   
The difference due to the addition of one neutron, $\hat{n}^n_j  \rightarrow \hat{n}^n_j + 1$, gives the contribution 
to $\hat{\epsilon}_j$ as
\begin{eqnarray}
\label{eq:Delta_e}
\Delta^{(j,nn)} \epsilon_j  \,&=& \, V_{T=1}^m (j,j) \frac{1}{2} \bigl\{ (\hat{n}^n_j + 1)\hat{n}^n_j - \hat{n}^n_j(\hat{n}^n_j - 1) 
    \bigl\}    \, \nonumber \\
&=&\, V_{T=1}^m (j,j) \,  \hat{n}^n_j  \,\, .
\end{eqnarray}
The difference due to the increase, $\hat{n}^n_{j}  \rightarrow \hat{n}^n_{j}+ 1$, for $j \ne j'$ is written as
\begin{eqnarray}
\label{eq:Delta_e-2}
\Delta^{(j',nn)} \epsilon_j  \,&=& \, V_{T=1}^m (j,j') \bigl\{\hat{n}^n_{j'}(\hat{n}^n_j + 1)\,-\, \hat{n}^n_{j'} \, \hat{n}^n_j  
    \bigl\}    \, \nonumber \\
&=&\, V_{T=1}^m (j,j') \,  \hat{n}^n_{j'}  \,\, . 
\end{eqnarray}
Thus, the contribution from neutron-neutron interaction results in  
\begin{eqnarray}
\label{eq:enn}
\hat{\epsilon}^{n  \rightarrow n}_j  \,= \, \sum_{j'} \, V_{T=1}^m (j,j') \, \hat{n}^n_{j'}  \,\, . 
\end{eqnarray}
The contribution from the proton-proton interaction can be shown similarly,   
\begin{eqnarray}
\label{eq:epp}
\hat{\epsilon}^{p  \rightarrow p}_j  \,= \, \sum_{j'} \, V_{T=1}^m (j,j') \, \hat{n}^p_{j'}  \,\, . 
\end{eqnarray}

In the case of the proton-neutron interaction, the monopole interaction is shown in eq.~(\ref{eq:m_pntot_main}).
We first discuss the effect from the first term on the right-hand-side. 
The difference due to the increase, $\hat{n}^n_{j}  \rightarrow \hat{n}^n_{j} + 1$, gives the contribution 
to $\hat{\epsilon}_j$ (of neutrons) as
\begin{eqnarray}
\label{eq:epn0}
\hat{\epsilon}^{p  \rightarrow n;0}_j  &=& \sum_{j'} \, \frac{1}{2}  \Bigl\{V_{T=0}^m (j', \, j)\, + \, V_{T=1}^m (j', \, j) \Bigr\}  \nonumber \\
       &  & \,\,\,\,\,\,\,\,\,\,\,\, \times \Bigl\{\hat{n}^p_{j'}(\hat{n}^n_{j} + 1)\,-\, \hat{n}^p_{j'} \, \hat{n}^n_{j} \Bigl\}  \nonumber \\       
       &=& \sum_{j'}  \frac{1}{2}  \Bigl\{V_{T=0}^m (j', \, j)\, + \, V_{T=1}^m (j', \, j) \Bigr\} \,  \hat{n}^p_{j'}  \, .  
\end{eqnarray}
Likewise, 
the difference due to the increase, $\hat{n}^p_{j}  \rightarrow \hat{n}^p_{j} + 1$, gives the contribution 
to $\hat{\epsilon}_j$ (of protons) as
\begin{eqnarray}
\label{eq:enp0}
\hat{\epsilon}^{n \rightarrow p;0}_j  
   &=& \sum_{j'}  \frac{1}{2} \Bigl\{V_{T=0}^m (j, \, j')\, + \, V_{T=1}^m (j, \, j') \Bigr\} \,  \hat{n}^n_{j'}  \, . 
\end{eqnarray}

We next discuss the effect from the second and third terms on the right-hand-side of eq.~(\ref{eq:m_pntot_main}).
Because the operator 
$\hat{\tau}^+_{j} \hat{\tau}^-_{j'} \,+\, \hat{\tau}^-_{j} \hat{\tau}^+_{j'}$ working between $j \ne j'$ shifts 
a proton $j' \rightarrow j$  and a neutron $j \rightarrow j'$ and vice versa (see Fig.~\ref{fig:tautau} (a,b)),  
the second term does not contribute to the ESPE.   
Note that effects of this term are fully included when the Hamiltonian is diagonalized.

The situation is different for the last term on the right-hand-side of eq.~(\ref{eq:m_pntot_main}), 
 $:\hat{\tau}^+_{j} \hat{\tau}^-_{j}:$.  Note that the protons and neutrons occupy the same orbit $j$ now.
Since the term, $-:\hat{\tau}^+_{j} \hat{\tau}^-_{j}:$, exchanges a proton and a neutron, 
a subset of its effect is relevant now, if this term  
annihilates a proton and a neutron both in the {\it same} magnetic substate $m$, and creates them in exactly the same substate.   
Formally speaking, 
this process cannot be written like the first term on the right-hand side of eq.~(\ref{eq:m_pntot_main}).
We, however, can introduce a practical approximation. 
If there are $n^n_j$ neutrons in the orbit $j$, they can be assumed, in first approximation, to be equally distributed over all possible $m$-states.   In this equal distribution approximation, a proton in the 
magnetic substate $m$ can feel an interaction with a neutron in the substate $m$ with a probability 
$\hat{n}^n_{j} / (2j+1)$.
This approximation can be expressed as 
\begin{eqnarray}
\label{eq:eq_approx}
- :\hat{\tau}^+_{j} \hat{\tau}^-_{j}: &\,\sim\,  &  \frac{\hat{n}^p_{j}\hat{n}^n_{j}}{2j+1} \,\, .  
\end{eqnarray}
This approximation can be understood also by considering the case of $m=m'$ in Fig.~\ref{fig:tautau} (d).  
By combining eq.~(\ref{eq:eq_approx}) with the first term on the right-hand side of eq.~(\ref{eq:m_pntot_main}), we define the {\it effective} proton-neutron monopole interaction as
\begin{flalign}
\,& \hat{v}_{pn,mono-eff} &\nonumber \\ 
&\,\, = \sum_{\, j \ne \,  j'}   
      \frac{1}{2} \Bigl\{V_{T=0}^m (j, \, j')\, + \, V_{T=1}^m (j, \, j') \Bigr\} \,  \hat{n}^p_j \, \hat{n}^n_{j'}   \nonumber \\
&\,\, + \sum_{j}  \frac{1}{2} \Bigl\{ \, V_{T=0}^m (j, \, j) \frac{2j+2}{2j+1} 
             + V_{T=1}^m (j, \, j) \frac{2j}{2j+1} \Bigr\} \, \hat{n}^p_{j}\hat{n}^n_{j} \,  .                                                                      
\label{eq:m_pntot-1}
\end{flalign}
The ESPE is evaluated with this effective monopole interaction hereafter.

The proton-neutron interaction thus contributes to the ESPE of the neutron orbit $j$ as
\begin{eqnarray}
\label{eq:epn0-2}
\hat{\epsilon}^{p \rightarrow n}_j  \,&=& \sum_{j'}  \frac{1}{2} \, \Bigl\{\tilde{V}_{T=0}^m (j', \, j)\, + \, \tilde{V}_{T=1}^m (j', \, j) \Bigr\} \,  \hat{n}^p_{j'}  \,\, ,  \,\,
\end{eqnarray}
while to the ESPE of the proton orbit $j$ as
\begin{eqnarray}
\label{eq:enp0a}
\hat{\epsilon}^{n \rightarrow p}_j  \, &=&
    \sum_{j'}  \frac{1}{2} \, \Bigl\{\tilde{V}_{T=0}^m (j, \, j')\, + \, \tilde{V}_{T=1}^m (j, \, j') \Bigr\} \,  \hat{n}^n_{j'}  \,\,  ,  \,\,
\end{eqnarray}
where $\tilde{V}$'s are modified monopole matrix elements defined by
\begin{eqnarray}
\label{eq:hatV-1}
\tilde{V}_{T=0,1}^m (j, \, j') \, &=& V_{T=0,1}^m (j, \, j') \,\,\,\,\,\, \text{for}  \,\,\,\, j \ne\ j'  \,\, , 
\end{eqnarray}
\begin{eqnarray}
\label{eq:hatV-2}
\tilde{V}_{T=0}^m (j, \, j) \, &=& V_{T=0}^m (j, \, j)\, \frac{2j+2}{2j+1} \,\, , 
\end{eqnarray}
and
\begin{eqnarray}
\label{eq:hatV-3}
\tilde{V}_{T=1}^m (j, \, j) \, &=& V_{T=1}^m (j, \, j)\, \frac{2j}{2j+1} \,\, .
\end{eqnarray}
We note that this substitution of $V_{T=1}^m (j, \, j)$ by $\tilde{V}_{T=1}^m (j, \, j)$ is only for 
the proton-neutron interaction, keeping eqs.~(\ref{eq:enn},\ref{eq:epp}) unchanged.   
It is worth mentioning that the effective monopole interaction in eq.~(\ref{eq:m_pntot-1}) produces the energy 
exactly for a closed shell, $\langle  \hat{n}^p_{j} \rangle = 2j+1$ or  $\langle  \hat{n}^n_{j} \rangle = 2j+1$,
because the equal distribution approximation turns out to be exact.   

We express the valence contribution to the ESPE from  
eqs.~(\ref{eq:enn},\ref{eq:epp},\ref{eq:enp0},\ref{eq:epn0-2}), by introducing
\begin{eqnarray}
\label{eq:ESPEvpn}
\tilde{V}_{pn}^m (j,j') \, = \frac{1}{2} \Bigl\{\tilde{V}_{T=0}^m (j, \, j')\, + \, \tilde{V}_{T=1}^m (j, \, j') \Bigr\} \,\,   .
\end{eqnarray}
It is then for the proton orbit $j$, 
\begin{eqnarray}
\label{eq:epj_int}
\hat{\epsilon}^{p}_j  &\,= \,& \sum_{j'} \, V_{T=1}^m (j,j') \, \hat{n}^p_{j'}  \,
   + \, \sum_{j'}  \, \tilde{V}_{pn}^m (j, \, j')\, \,  \hat{n}^n_{j'}  \,\,  ,  \,\,
\end{eqnarray}
and for the neutron orbit $j$, 
\begin{eqnarray}
\label{eq:enj_int}
\hat{\epsilon}^{n}_j  &\,= \, &\sum_{j'} \, V_{T=1}^m (j,j') \, \hat{n}^n_{j'}  \, 
  + \, \sum_{j'}  \tilde{V}_{pn}^m (j', \, j)\,  \hat{n}^p_{j'}  \,\, .  \,\,
\end{eqnarray}
Note that one can use $V_{x}^m (j,j') = V_{x}^m (j',j)$ for any subscript $x$, if more convenient.

We point out that for the closed-shell-plus-one-nucleon systems, 
the results shown in eqs.~(\ref{eq:epj_int},\ref{eq:enj_int}) produce the exact 
energy for a single proton state $j$, 
\begin{eqnarray}
\label{eq:epj-c}
\epsilon^{p}_j  &\, =\, & \sum_{occ. \, j'_p} V_{T=1}^m (j,j'_p) \, (2j'_p+1)  \, \nonumber \\
   & \, + & \, \sum_{occ. \, j'_n}  \, \tilde{V}_{pn}^m (j, \, j'_n)\,  (2j'_n+1)  \,  ,  
\end{eqnarray}
where the summation of $j'_p$ or $j'_n$ is taken for all fully occupied orbits in the valence 
space and the ESPEs are treated as c-numbers.  
This is because the approximation in eq.~(\ref{eq:eq_approx})
becomes an equality relation due to the apparent equal distribution in the closed shell.  
Single-hole states can be treated in the same way.  
A similar expression is obtained for neutrons.

   
\subsection{Short summary and relations to earlier works \label{subsec:short_summary}}

We first summarize some properties relevant to subsequent discussions.  
The variation of ESPE is more relevant than the ESPE itself, in many applications. 
The difference can be taken between different nuclei, or between different states of the same nucleus.     
It can be expressed conveniently, based on eqs.~(\ref{eq:epj_int},\ref{eq:enj_int}), as,
\begin{equation}
\label{eq:epj}
\Delta \hat{\epsilon}^{p}_j  \,=  \sum_{j'} \, V_{T=1}^m (j,j') \, \Delta \hat{n}^p_{j'}  \,
   + \sum_{j'}  \, \tilde{V}_{pn}^m (j, \, j')\, \,  \Delta \hat{n}^n_{j'}  \, , 
\end{equation}
and 
\begin{equation}
\label{eq:enj}
\Delta \hat{\epsilon}^{n}_j  \,= \sum_{j'} \, V_{T=1}^m (j,j') \,\Delta  \hat{n}^n_{j'}  \, 
  + \sum_{j'}  \tilde{V}_{pn}^m (j', \, j)\, \Delta  \hat{n}^p_{j'}  \, .  
\end{equation} 
Here $\Delta$ refers to the difference like $\langle \Psi \, | \hat{\epsilon}^{p}_j | \,  \Psi \rangle - \langle \Psi' \, | \hat{\epsilon}^{p}_j | \,  \Psi' \rangle$ between two states $\Psi$ and $\Psi'$.   

While the occupation numbers, $\hat{n}^p_{j}$ and  $\hat{n}^n_{j}$, in eq.~(\ref{eq:epj_int},\ref{eq:enj_int}) are operators, relevant are their expectation values in many cases.    
Thus, although the ESPE (of an orbit) is an operator, its expectation value (with respect to some state, {\it e.g.} the ground state) is sometimes called ESPE also.  
The same is true for their differences in eq.~(\ref{eq:epj},\ref{eq:enj}).
Likewise, in the filling scheme where  
nucleons are put into the possible lowest orbit one by one, these operators
are c-numbers for a given nucleus, and the ESPEs become c-numbers also.   We omit 
the symbol  $\,\hat{\,\,}\,$ in those cases.

The coefficients in these equations are given by the monopole matrix elements and their slight modifications $\tilde{V}_{pn}^m (j,j')$ (see eqs.~(\ref{eq:hatV-1},\ref{eq:hatV-2},\ref{eq:hatV-3},\ref{eq:ESPEvpn})).   In practical studies,  the expression in eq.~(\ref{eq:mono_J-2}) is more convenient than the definition with m-scheme states, because the values can be taken directly from shell-model interactions.

We next comment on relations of the present approach to earlier ones.  Based on some initial shell-model works, for instance \cite{deShalit1963,French1966,French1969}, 
\citet{Bansal} introduced ``the average two-body interaction energy (taken with a (2$J$+1)(2$T$+1) 
weighting)" and also ``Another average, taken without  the (2$T$+1) weighting".
Thus, Bansal and French regarded these approaches as two different schemes.  The former is
basically suitable for a closed shell where both proton and neutron shells are completely
occupied.  The averaging of all two-body matrix elements is carried out for all 
neutron-neutron, proton-proton and proton-neutron pairs, and the weighting factor (2$J$+1)(2$T$+1)
arises.  As the $T$=0 and 1 two-body matrix elements are very different in size, another 
parameter was introduced to account for it \citep{Bansal,Zamick1965}.  
The formulation of the present work is based on an averaging also.  
But this is the averaging over all possible orientations of a given two-nucleon configuration
$j \otimes j'$, and the idea is visualized in Fig.~\ref{fig:mono} with the definition in eq.~(\ref{eq:m_nn}) and in 
other related equations.   
The derived monopole interaction is shown in Sec.~\ref{subsec:mono} with 
eqs.~(\ref{eq:m_pntot_main},\ref{eq:m_nn-2},\ref{eq:m_pp},\ref{eq:m_tot}). 
These equations include the terms proportional to $\hat{n}^p_j \, \hat{n}^n_{j'}$, 
which may be related to Bansal-French's second scheme mentioned above. 
This second scheme is described further in \citep{Bansal} as 
``This is the average which one encounters in an n-p formalism 
(one in which neutrons and protons are separately numbered) in those cases where the neutron 
is necessarily in one orbit, the proton in the other".  
Equations~(\ref{eq:m_pntot_main},\ref{eq:m_nn-2},\ref{eq:m_pp},\ref{eq:m_tot}) include terms 
dependent on isospin operators as illustrated in Fig.~\ref{fig:tautau}, which enables us to remove 
such a restriction of the orbits and allow protons and neutrons to be in the same orbit. 

 \citet{Poves1981} developed the scheme of Bansal and French, 
 stating  ``$H_{\rm m}$ and $H_{\rm mT}$ can be thought of as generalization of the French-Bansal formulae".  
The weighting factors (2$J$+1)(2$T$+1) are included in $H_{\rm mT}$ (see also \cite{Caurier05}), while the isospin is not considered in $H_{\rm m}$ (see also \cite{Zuker1994}).   
 The monopole interaction $H_{\rm mT}$ presented in \citep{Poves1981} 
 produces the same energy for closed-shell states as the present approach.
 So, the result of \citep{Poves1981} and the 
 relevant result of the present approach are obtained, most likely,  
 by different procedures with consistent outcome.   
 This consistency may be supported by the fact that the monopole interaction can be composed of the number 
 and isospin operators of individual orbits, and closed shells can give sufficient constraints on the values of their parameters.   
The use of the monopole Hamiltonian of \citep{Poves1981} has been developed and applied to properties of closed-shell nuclei and their neighbors with $\pm$ one particle, producing precisely the global systematics of nuclear masses \cite{Zuker1994,Duflo1995,Duflo1999,Caurier05,Zuker2005}.  A review of them is given in \citep{Caurier05}.  
 
One thus sees that the two approaches mentioned by Bansal and French are 
 basically two facets of one common monopole interaction derived from the orientation averaging in the present scheme, keeping isospin properties.   
 In this way, we can settle a long-standing question 
 on the definition and uniqueness of 
 the monopole interaction, finding that basically all those arguments are along the same line.  
 The additional $\tau\tau$ term of eq.~(\ref{eq:m_pntot_main}) is of interest.   
 
Another interest can be in the variational approach with monopole interaction in open-shell nuclei as discussed in \cite{Yazaki1977}.
 

\subsection{Equivalence to ESPE as defined by Baranger \label{subsec:Baranger}}

We discuss the definition of the ESPE by \citet{Baranger1970}.
The ESPE of the orbit $j$ on top of the eigenstate $|0\rangle$ is considered by referring to the $n$-th ($N$-th) eigenstate, $| n \rangle$ ($| N \rangle$), of the nucleus with one more (less) particle of interest.  
The ESPE is then expressed as,
\begin{equation}
\epsilon_j  \,=  \sum_{n} \, (E_n - E_0) S^+_n + \sum_{N} \, (E_0 - E_N) S^-_N \, ,
\label{eq:baranger}
\end{equation}
where $E_0$ is the energy of the state $| 0 \rangle$, and $E_n$ ($E_N$) denotes the energy of the state $| n \rangle$
($| N \rangle$).  
Here, $S^+_n$ ($S^-_N$) stands for spectroscopic factors $| \langle n |a^{\dagger}_{q} | 0 \rangle|^2$
($| \langle N | a_{q} | 0 \rangle|^2$) with $q$ being a magnetic substate of the orbit $j$.  
Equation~(\ref{eq:baranger}) implies that 
the ESPE is comprised not only of energy gains in going from $| 0 \rangle$ to $| n \rangle$ weighted by the spectroscopic factors, but also of minus times energy losses from $| 0 \rangle$ to $| N \rangle$ weighted similarly.  Note that the latter contributes if the orbit $j$ is occupied in $| 0 \rangle$.   
We discuss now the relation between this definition and the one discussed so far.  Note that the state $| 0 \rangle$ is assumed to be the ground state of a double-closed-shell nucleus in \citep{Baranger1970}, but we can generalize it 
to a 0$^+$ state.  However, if its spin/parity is not 0$^+$, eq.~(\ref{eq:baranger}) does not represent the ESPE.

Equation~(\ref{eq:baranger}) can be rewritten
\begin{equation}
\label{eq:bar1}
\epsilon_j  \,= \langle 0 | a_{q} (H - E_0)  a^{\dagger}_{q} \, | 0  \rangle + 
                       \langle 0 | a^{\dagger}_{q} (H - E_0)  a_{q} \, | 0  \rangle \, ,
\end{equation}
where $H$ is the Hamiltonian.  This is identical to eq.~(6) of \citep{Baranger1970}, even though they look different.  After algebraic processing, we come to
\begin{equation}
\label{eq:baran2}
\epsilon_j  \,=  \, \epsilon^0_j 
                   + \sum_{\beta, \delta} \, v_{q \beta q \delta} \langle 0  | a^{\dagger}_{\beta} a_{ \delta} \, | 0  \rangle \, ,
\end{equation}
where $ \epsilon^0_j$ is defined in eq.~(\ref{eq:Hn}) and $v_{q \beta q \delta}$ denotes an antisymmetrized matrix element of the two-body interaction. 

Because $| 0 \rangle$ being a 0$^+$ state, the following relations hold in eq.~(\ref{eq:baran2}) for the magnetic quantum number, $m_{\beta}$=$m_{\delta}$, and for the angular momentum, $j_{\beta}$=$j_{\delta}$ denoted by $j'$.  We assume that the states $\beta$ and $\delta$ are the same for the sake of simplicity, while a more general treatment is possible. We note that this assumption is valid with two HO major shells or in other similar cases.  The matrix element $\langle 0  | a^{\dagger}_{\beta} a_{ \delta} \, | 0  \rangle$ can be replaced with $\langle 0  | \hat{n}_{j'} \, | 0  \rangle / (2j'+1)$, which is independent of $m_\beta$.  Here, $\hat{n}_{j'}$ is the number operator of the orbit $j'$.  Although $\epsilon_j$ in eq.~(\ref{eq:baranger}) is independent of $q$, 
we sum $v_{q \beta q \beta}$ over $q$, and the sum can be expressed as the monopole matrix element $V^m (j,j')$ multiplied by the number of relevant antisymmetrized states of $j$ and $j'$.  Note that the difference in this number between $j$=$j'$ and $j \neq j'$ cases is incorporated.  
We finally obtain the following unified expression,
\begin{equation}
\label{eq:baran3}
\epsilon_j  \,=  \, \epsilon^0_j + \sum_{j'} \, V^m_{j, j'} \, \langle 0  | \hat{n}_{j'} | 0  \rangle \, .
\end{equation}
This is nothing but the ESPE discussed so far with the substitution of $\hat{n}^p_{j'}$ and $\hat{n}^n_{j'}$ in eq.~(\ref{eq:epj_int}) and eq.~(\ref{eq:enj_int}) by their expectation values with respect to the eigenstate $| 0 \rangle $.
Namely, the ESPE formulation by \citet{Baranger1970} is included in the present monopole formulation as a specific  case with $| 0 \rangle $ being a 0$^+$ state, while the present one is applicable to the other states as well.  We mention that the present approach has a modification due to the isospin
(see eqs.~(\ref{eq:hatV-2},\ref{eq:hatV-3})), and it is of interest how to include this in the above discussion. 

We can thus present a unified formulation on the monopole properties, starting from the natural construction of the monopole interaction.  In \cite{Duguet2015},  it is stated in footnote 1 ``In the traditional shell model, ESPE usually refers to single-particle energies obtained by averaging over the monopole part of
the Hamiltonian on the basis of a naive filling ...  The latter denotes an approximate version of the
full Baranger-French definition ...''.  This footnote remark may be applicable to some earlier works using the filling scheme for defining the ESPE, but is not relevant to the present formulation.  

The ESPE, as the expectation value of the 0$^+$ ground state (see the texts below eq.~(\ref{eq:enj})), can be extracted from experiment if all relevant spectroscopic factors in eq.~(\ref{eq:baranger}) are obtained to a great precision.  Because one-nucleon addition {\it and} removal experiments are required, this is a very difficult task in general, except for cases where the spectroscopic factors are negligible in either direction, {\it e.g.} at a closed shell.   Despite this experimental challenge, the ESPE is useful for understanding and explaining phenomena and mechanisms.


\subsection{Illustration by an example \label{subsec:TalmiCNO}}

\begin{figure}[tb]     
\begin{center}
\hspace*{-0.5cm}
\includegraphics[width=6cm]{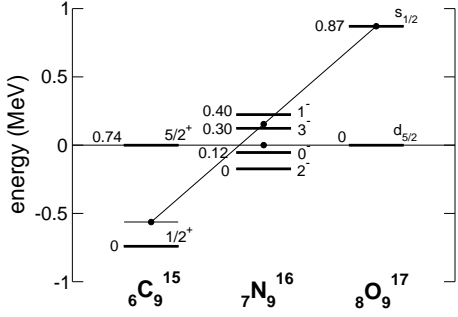}  
\caption{
Experimental energy levels of $N$=9 isotones, some of which are regarded 
as neutron 2$s_{1/2}$ and 1$d_{5/2}$ single-particle energies. See the text.
Based on \citet{Talmi1960}. 
}
\label{CNO}
\end{center}
\end{figure}

We present an example of the change of ESPE's of the $N=9$ isotones as shown in Fig.~\ref{CNO}.  
This figure is taken from Fig. 2 of Ref. \citep{Talmi1960}, as one of the earliest related papers.  
We discuss here how the changes shown in Fig.~\ref{CNO} can be described 
within the framework presented in the previous subsection.
The discussions are somewhat detailed because this is the first actual example.
   
\begin{figure}[tb]     
\begin{center}
\includegraphics[scale=0.42]{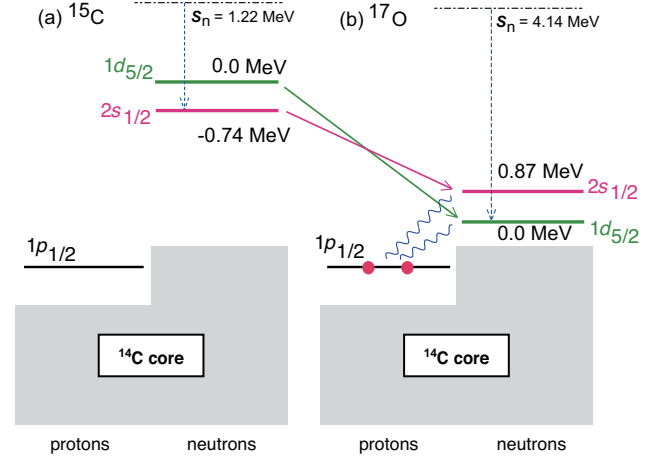}  
\end{center}
\caption{(a) Schematic picture of shell structure (a) on top of the $^{14}$C core. 
 (b) Two more protons (red solid circles) are added into the 1$p_{1/2}$ orbit.  
Experimental levels are identified as single-particle states: green level for 1$d_{5/2}$, and 
pink level for 2$s_{1/2}$.  Numbers near the levels are energies relative to 1$d_{5/2}$.
The wavy lines imply proton-neutron interactions.  Solid arrows indicate the changes of 
SPE's.  The dashed-dotted line denotes neutron threshold, and downward dashed arrows 
mean one neutron separation energy ($S_{\rm n}$). }
\label{c14core}
\end{figure}

We assume the $^{14}$C core with $Z=6$ and $N=8$. 
Figure~\ref{c14core} illustrates the shell structure on top of this $^{14}$C core.
The levels shown in Fig.~\ref{c14core} are taken from experimental data \citep{ensdf},
assuming that the observed lowest levels are of single-particle nature, and are almost the same as
the corresponding ones in Fig.~\ref{CNO}.    
Figure~\ref{c14core} (a)  
indicates somewhat schematically neutron 2$s_{1/2}$ and 1$d_{5/2}$ orbits 
on top of the $^{14}$C core. 
Note that in Fig.~\ref{c14core} (a), the 2$s_{1/2}$ orbit is 0.74 MeV below 1$d_{5/2}$.

We then add protons into the 1$p_{1/2}$ orbit, as shown in Fig.~\ref{c14core} (b).
The proton 1$p_{1/2}$ orbit is fully occupied, or closed now. 
The ESPE's of neutron 2$s_{1/2}$ and 1$d_{5/2}$ orbits are both lowered, but more
interestingly, their order is reversed due to protons in the 1$p_{1/2}$ orbit, 
following eq.~(\ref{eq:enj_int}) with $j' = 1p_{1/2}$, and $j=2s_{1/2}$ or 1$d_{5/2}$.
The ESPEs are treated here as c-numbers because the wave function of the other nucleons 
is fixed, as mentioned in Sec.~\ref{subsec:espe}.
The difference of ESPEs can be written as
\begin{flalign}
\label{eq:c15o17diff}
 & \epsilon^{n}_{2s_{1/2}}(^{17}{\rm O}) - \epsilon^{n}_{1d_{5/2}}(^{17} {\rm O}) \nonumber \\
& \,\,\,\,\,=  \, \epsilon^{n}_{2s_{1/2}}(^{15}{\rm C}) - \epsilon^{n}_{1d_{5/2}}(^{15}{\rm C}) \nonumber \\
&  \,\,\,\,\,+ \bigl \{ V_{pn}^m (1p_{1/2}, \, 2s_{1/2})\,-\, V_{pn}^m (1p_{1/2}, \, 1d_{5/2}) \bigr\}  \times 2\,\, .  
\end{flalign}

The proton sector of the Hamiltonian 
produces the common effect between the $J^{\pi}$=$1/2^+$ and $5/2^+$ states.
Thus, the above difference of ESPEs corresponds to the difference of experimental levels
in the assumption that these states are of single-particle nature (which will be re-examined 
in Sec.~\ref{subsec:2-body LS} with Fig.~\ref{fig:COsd}), 
and the monopole matrix elements satisfying 
\begin{flalign}
\label{eq:c15o17Vm}
\,\,V_{pn}^m (1p_{1/2}, \, 2s_{1/2})\,-\, V_{pn}^m (1p_{1/2}, \, 1d_{5/2})  \nonumber \\
\,\,\,\,\,\,  = (0.87 + 0.74) / 2 = 0.805 \, {\rm (MeV)} 
\end{flalign}
explain the change in Fig.~\ref{c14core}.
This result indicates that 
$V_{pn}^m (1p_{1/2}, \, 1d_{5/2})$ is more attractive by $\sim$0.8 MeV than $V_{pn}^m (1p_{1/2}, \, 2s_{1/2})$.
Thus, what actually occurs is more rapid lowering of the neutron 1$d_{5/2}$ orbit than that of the 
neutron 2$s_{1/2}$ orbit, as protons fill the 1$p_{1/2}$ orbit.
This can be explained as a consequence of very important and general features of the monopole 
interactions of nuclear forces as discussed in Sec.~\ref{sec:SE} extensively. 

If the energy is measured relative to the neutron 1$d_{5/2}$ orbit, the monopole-matrix-element difference in eq.~(\ref{eq:c15o17Vm}) pushes up the neutron 2$s_{1/2}$ orbit from  \nuc{15}{C} to \nuc{17}{O}.   
We shall describe the approach by \citet{Talmi1960} with this convention such that the energy is measured from the neutron 1$d_{5/2}$ ESPE with the filling scheme.     
The energy levels in Fig.~\ref{c14core} can be viewed in this convention, including 
\nuc{16}{N} ($Z$=7 and $N$=9) with one proton in the 1$p_{1/2}$ orbit.
This proton is coupled with a neutron either in 2$s_{1/2}$ or 1$d_{5/2}$.
The former coupling yields $J^{\pi}$=$0^-$ and $1^-$ states, while the latter $J^{\pi}$=$2^-$ and $3^-$ states.   
In this simple configuration, the proton-neutron interaction, $\hat{v}_{pn}$, shifts the energies of these states by 
\begin{equation}
\label{eq:N16}
\frac{1}{2}  \; \Sigma_{T=0, 1}  \, \langle j, j' ; J,T | \hat{v}_{pn} | j, j' ; J,T \rangle  .
\end{equation}
The $(2J+1)$-weighted average of the quantities in eq.~(\ref{eq:N16}) is nothing but the corresponding monopole matrix element, 
because of 
eqs.~(\ref{eq:mono_J-2}, \ref{eq:hatV-1}, S2).  
Thus, those averages can be discussed as the ESPEs driven by the monopole interaction, and  the aforementioned convention can be adopted.
As the change from \nuc{15}{C} to \nuc{17}{O} is then twice the monopole-matrix-element difference due to two additional protons, the middle point of the line connecting the states of the same spin/parity of \nuc{15}{C} and \nuc{17}{O} represents the corresponding monopole quantity. 
Thus, if the present scheme works ideally, the 1/2$^+$ levels and the relevant average quantity of \nuc{16}{N} should be on a straight line. 
Talmi and Unna did, in \citep{Talmi1960}, this analysis in a slightly different way: they took
the observed energy levels of \nuc{17}{O} and the weighted averages for the observed levels of \nuc{16}{N}, and extrapolated to \nuc{15}{C}.  The extrapolated value appeared rather close to the observed one, implying the validity of this picture, which will be re-visited in Sec.~\ref{subsec:2-body LS}.


Talmi and Unna discussed another case with $^{11}$Be - $^{12}$B - $^{13}$C
($N$=7 isotones with $Z$=4, 5 and 6)  \citep{Talmi1960}.
Although the $1/2^+$ levels change almost linearly as a function of $N$, the mechanism 
is different from the $N$=9 isotone case discussed above.   Since protons occupy the 
1$p_{3/2}$ orbit now, one has to take into account the coupling of two protons.
It was taken to be $J=0$ (see eq. (1) in \citep{Talmi1960}), which enables us to connect 
the change of the structure to the monopole interaction, because multipole interactions are 
completely suppressed.  
Note that the terminology of monopole interaction was not used then, but the same quantity was used.
This restriction to the $J=0$ coupling, however, may not be appropriate, because the deformation 
of the shape is crucial and simultaneously configuration mixings occur even between the  
1$p_{3/2}$ and 1$p_{1/2}$ proton orbits.  The single-particle nature is broken also on the neutron side due to 
configuration mixing between the 2$s_{1/2}$ and 1$d_{5/2}$ orbits.  Thus, the
$N$=7 isotones may not be a good example of the change of single-particle energies.
In fact, the magnitude of the change is twice larger than the $N$=9 isotone case, 
which may be indicative of dominant additional effects. 

An example of the ESPE change due to the $T$=1 interaction is given in 
Appendix~\ref{app:N=28 gap}.  

\section{Shell evolution, Monopole Interaction and Nuclear Forces 
\label{sec:SE}}

The effective single-particle energy (ESPE) is shown to be varied according to the relations in 
eqs.~(\ref{eq:epj}, \ref{eq:enj}).
Since it depends linearly on the proton or neutron number operators of a particular orbit $j$, denoted respectively as 
$\hat{n}^p_j$ and $\hat{n}^n_j$, the ESPE can be changed to a large extent
if the occupation number of a given orbit becomes large.   This further can result in a substantial change 
of the shell structure, called {\it shell evolution}.   Thus, the shell evolution can occur, for instance, as a function of $N$ 
along an isotopic chain.  We shall discuss, in this section, some basic points of the shell evolution 
in close relations to nuclear forces.

We note here that the multipole interaction defined in Sec.~\ref{subsec:mono} produces a variety 
of correlations, for instance, quadrupole deformation, and that the final structure is 
determined jointly by the monopole and multipole interactions, 
as is done automatically when the Hamiltonian is diagonalized.  
Although there is no {\it a priori} separation of effects of the monopole interaction from those of 
the multipole interaction, 
the monopole effects, particularly the shell evolution, can be made visible in many cases.
We shall focus, in this section, on such effects of the monopole interactions due to various 
constituents of the $NN$ interaction, such as central, tensor and two-body spin-orbit.

\subsection{Contributions from the central force \label{subsec:central}}

The central-force component of the nuclear force is the main driving force of the formation of
the nuclear structure.  

Let us start with an extreme case, if the effective nucleon-nucleon ($NN$) interaction, $\hat{v}$, 
is a central force with infinite range and no dependence on spin, 
the values of monopole matrix elements $V^m_{T=1} (j,j')$ and $V^m_{pn} (j,j')$
become independent of $j$ and $j'$, being constants.   If this $\hat{v}$ is attractive,  $V^m_{T=1} (j,j')$ and  $V^m_{pn} (j,j')$ take separate constant negative values.
This implies, for instance, that if more neutrons occupy the orbit $j'$, all proton orbits $j$ become more bound
to the same extent.   In other words, the proton shell structure is conserved but becomes more deeply bound.

On the other hand, if $\hat{v}$ is given by a $\delta$-function with a certain strength parameter, 
the values of $V^m_{T=1} (j,j')$ and  $V^m_{pn} (j,j')$  
become sensitive to the overlap between the wave functions of the orbit $j$ and that of the orbit $j'$. 
This implies, for instance, that if more neutrons occupy the orbit $j'$, proton ESPE for the orbit $j$ in 
eq.~(\ref{eq:epj}) become more bound, but the amount of the change is not uniform.   
In other words, the pattern of the proton single-particle orbits may change 
while they all become more bound as a whole.   

The actual situation is certainly somewhere in between. 
We here show how monopole matrix elements look like for a central Gaussian
interaction given by 
\begin{eqnarray}
\label{eq:v_gauss}
v_c = \sum_{S,T} \, f_{S,T} \, P_{S,T} \, \exp( - (r / \mu)^2)\, ,
\end{eqnarray}
where $S(T)$ means spin (isospin), $P$ denotes the projection operator onto the channels $(S,T)$
with strength $f$, and $r$ and $\mu$ are the internucleon distance and Gaussian parameter, 
respectively.  We fix here all $f$ parameters to a common value of $+$166 MeV just in order to see 
the effects of the four terms on the right-hand side. 
The $\mu=1$ fm is used as we shall discuss in Sec.~\ref{subsec:VMU} too.  
Note that we shall use this $v_c$ extensively hereafter with realistic values of parameters such as $f_{1,0} =f_{0,0} = - 166$ MeV ({\it i.e.}, the same magnitude with the opposite sign from the 
above value),
and $f_{0,1} =0.6f_{1,0}$ and $f_{1,1} =-0.8f_{1,0}$.  
Such $v_c$ gives basic features of effective $NN$ interaction of the shell-model calculation, 
as called $V_{MU}$ \citep{Otsuka2010b}.

\begin{figure}[tb]     
\begin{center}
\includegraphics[width=6.5cm]{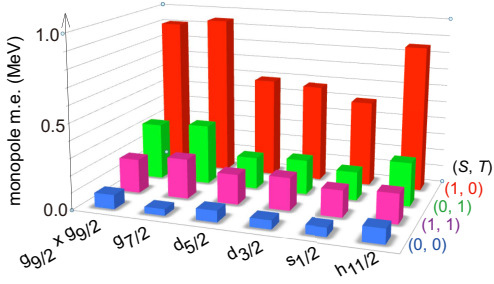}   
\caption{Monopole matrix elements of central Gaussian interactions of eq.~(\ref{eq:v_gauss})
for all $(S,T)$ channels with an equal strength parameter ({\it i.e.}, +166 MeV, see the text).  
One of the orbits is 1$g_{9/2}$, and the other is shown.  }
\label{3dim_A100}
\end{center}
\end{figure}

Figure~\ref{3dim_A100} shows monopole matrix elements thus obtained. 
The harmonic oscillator wave functions are used as single-particle wave functions hereafter.
We take $A=100$ in Fig.~\ref{3dim_A100}.  

Figure~\ref{3dim_A100} indicates that the $(S=1, \, T=0)$ channel produces 
major contributions, apart from the actual $f_{S,T}$ values.   
Furthermore, as mentioned just above, 
the actual value of $f_{S=1,T=0}$ appears to be the largest among the four $(S, T)$ 
channels ({\it e.g.}, in the $V_{MU}$ \citep{Otsuka2010b}), and this 
dominance becomes enhanced after considering the actual $f_{S,T}$ values.

Within the $(S=1, \, T=0)$ channel, Fig.~\ref{3dim_A100} demonstrates that the coupling between 
orbits with $n=1$ ({\it i.e.}, no node in the radial wave function) like 1$g_{9/2}$-1$g_{9/2,7/2}$ or 1$g_{9/2}$-1$h_{11/2}$
are stronger than the others.   
This can be understood in terms of the larger overlap between 
their radial wave functions than those in the other categories in Fig.~\ref{3dim_A100}.

\begin{figure}[tb]     
\begin{center}
\includegraphics[width=6.0cm]{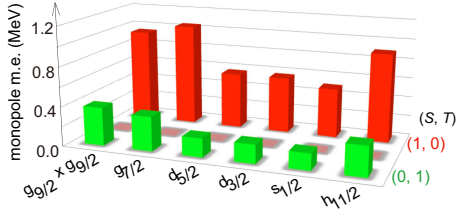}
\end{center}
\caption{Monopole matrix elements of delta interactions
for (S,T) channels.  See the caption of Fig.~\ref{3dim_A100}.  }
\label{3dim_A100_delta}
\end{figure}

We show similar histograms for the $\delta$-function interaction in Fig~\ref{3dim_A100_delta}.  
One finds rather good overall similarity to Fig.~\ref{3dim_A100}.  On the other hand, 
the monopole matrix elements vanish for $(S=1, \, T=1)$ or $(S=0, \, T=0)$ channel, 
as not shown in Fig~\ref{3dim_A100_delta}.   This is a consequence of the Pauli principle 
which forbids two nucleons at the same place for $S=0, \, T=0$ or  $S=1, \, T=1$. 

\begin{figure}[tb]     
\begin{center}
\includegraphics[width=7.5cm]{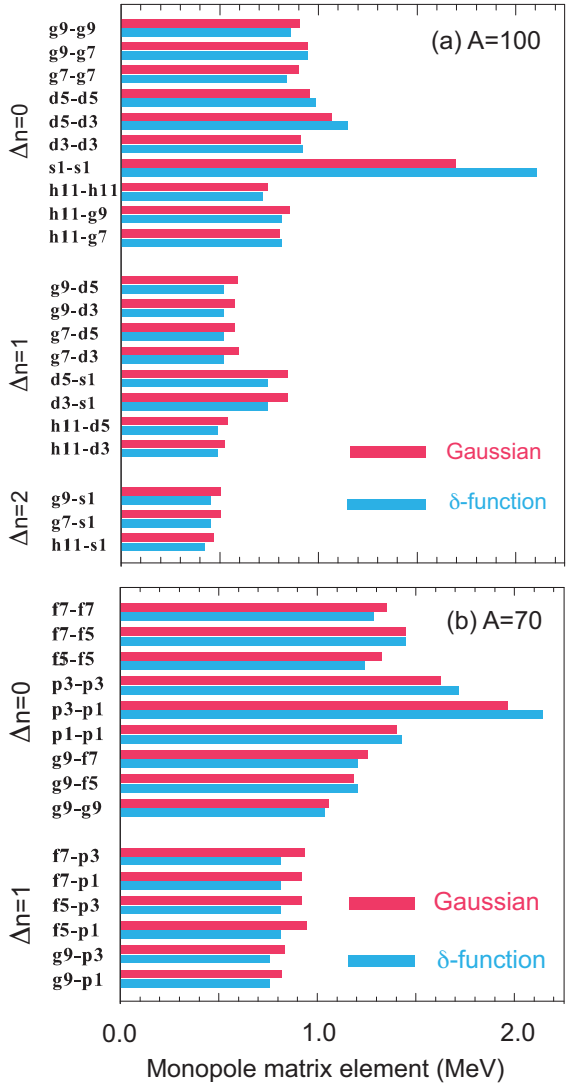}
\caption{Monopole matrix elements of central Gaussian and $\delta$ interactions
for the $(S=1, \, T=0)$ channel.  The opposite sign should be taken for 
actual values.  The orbit labeling is abbreviated like g9 for 1$g_{9/2}$, {\it etc}.
 The orbits are from the valence shell for (a) $A=100$ and (b) $A=70$.}  
\label{fig:monoGD}
\end{center}
\end{figure}

It is now of interest to survey the overall dependence of the monopole matrix element on the 
nodal structure of the radial wave function.  
Figure~\ref{fig:monoGD} shows monopole matrix elements in the $(S=1, \, T=0)$ channel for 
the Gaussian and $\delta$-function interactions.
In Fig.~\ref{fig:monoGD}, various pairs of orbits are taken for the valence shells around 
(a) $A=100$ and also (b) $A=70$, with their labels abbreviated like g7 for 1$g_{7/2}$.
The strength of the $\delta$-function interaction is adjusted so that the monopole matrix 
element becomes equal to that given by the Gaussian interaction for 
(a) the g9-g7 pair and (b) the f7-f5 pair.

The orbital pairs are classified into categories according to the 
difference of the number of the nodes in their radial wave functions, as denoted by  
$\Delta n$ in Fig,~\ref{fig:monoGD}. 
It is noticed that the monopole matrix elements are generally large when the radial wave functions  
have the same number of the nodes ({\it i.e.}, $\Delta n=0$).   
The monopole matrix elements become smaller as $\Delta n$ increases, while  
the difference between the two categories $\Delta n=1$ and $\Delta n=2$ is much smaller.
On the other hand, the monopole matrix element varies within the $\Delta n=0$ category.
The large value of the s1-s1 in Fig,~\ref{fig:monoGD} (a) is exceptional.  Among the others
with $\Delta n=0$ in Fig,~\ref{fig:monoGD} (a), stronger coupling can be found 
between the following orbits:
\begin{equation}
 j_> \, = \, \ell \, + \, 1/2 \,\,\,\, {\rm and} \,\,\,\,  j_< \, = \, \ell \, - \, 1/2 \, ,
\label{j>j<} 
\end{equation}
where $\ell$ stands for the orbital angular momentum and $1/2$ represents the spin.
In other words, 
$j_>$ and $j_<$ are spin-orbit partners having the same radial wave functions
in the Harmonic Oscillator scheme, and therefore the central force, both the Gaussian and 
the $\delta$-function interactions, produces stronger monopole interactions between them.
This feature is seen in the cases of (i) g9 and g7, (ii) d5 and d3, (iii) f7 and f5, 
and (iv) p3 and p1 in Fig,~\ref{fig:monoGD}.  
We note that this type of enhanced coupling becomes weaker with the 
Gaussian interaction than with the $\delta$-function interaction.

We emphasize that the monopole matrix elements of the central force, as modeled by the 
Gaussian interaction in eq.~(\ref{eq:v_gauss}), vary considerably, and can produce
sizable shell evolution depending on the occupation pattern over relevant single-particle orbits.
Concrete examples are shown in Secs.~\ref{subsec:VMU} and \ref{subsect:tensorSE}.

Regarding the dependence on the mass number, $A$, 
the monopole matrix elements of $A=100$ in Fig,~\ref{fig:monoGD} (a) is, as a whole, about 2/3 of those of 
$A=70$ in Fig,~\ref{fig:monoGD} (b).  
This feature can be expressed by a $1/A$ dependence in a rough approximation if wished.
This approximate scaling law appears to be reasonable because the probability to find the partner 
of a pair of interacting nucleons inside the interaction range is inversely proportional to the nuclear volume, 
as far as the density saturation holds.  Note that this overall trend is seen in experimentally 
extracted data, while other $A$-dependences can be found locally in certain groups (see Fig. 8 of \citep{Sorlin2008}). 

Stronger couplings between particular orbits are a natural idea, and were, in fact, 
discussed in earlier works, for instance, 
by Federman, Pittel, {\it et al.} \citep{Federman1977,Federman1979a,Federman1979b,Federman1984,Pittel1993}. 
It has been argued by \citet{Federman1977} that the
proton-neutron central force 
in the $^3S_1$ channel, where $S$ stands for the $s$ wave ($L$=0) 
with the spin triplet 
and the relative orbital (total) angular momentum $L$=0 ($J$=1), 
gives rise to a strong attraction between two orbitals $(n_P, l_P, j_P)$ and $(n_N, l_N, j_N)$ 
when the relations $n_P = n_N$ and $l_P \approx l_N$ are satisfied because of a large spatial 
overlap \citep{deShalit1953}.  It certainly contributes to the present $(S=1, \, T=0)$ channel.  
Figure~\ref{fig:monoGD} (a) suggests that the monopole matrix element $V^m_{T=0} ({\rm 1}g_{7/2}, {\rm 1}g_{9/2})$
is about 0.3 MeV more attractive than $V^m_{T=0} ({\rm 2}d_{5/2}, {\rm 1}g_{9/2})$ with the realistic 
sign of the parameter mentioned above.  
Equation~(\ref{eq:enj}) combined with eqs.~(\ref{eq:hatV-1},\ref{eq:ESPEvpn}) indicates that the present Gaussian central force 
lowers the ESPE of the neutron 1$g_{7/2}$ orbit relative to the 2$d_{5/2}$ orbit   
by $\sim \frac{1}{2} \times 0.3 \times$ 10 = 1.5 MeV,  in going from $Z=40$ to $Z=50$. 
Here we assumed that the $Z=40$ and $N=50$ closed shells are kept, the neutron 
1$g_{7/2}$ and 2$d_{5/2}$ orbits are on top of this closed shell, and additional 10 protons occupy the 1$g_{9/2}$ orbit.
This change is quite sizable, but will be shown to be about a half of what has been known experimentally,
which hints that the central force is responsible only for a part of the story.
\citet{Smirnova2004} compared $\delta$-function and G-matrix interactions.  The former is a central force, but the latter contains other components.  The reported difference is therefore consistent with the present observation on the deficiency of the central force.
Relevant further studies were reported in \citep{Umeya2004,Umeya2006}.  

We here come back to the limit of long-range interaction, but include  
dependences on the spin and isospin \citep{magic}.
If there is no spin dependence, an infinite-range interaction gives a constant shift as discussed above.  
Let us now take a spin-isospin interaction such as
\begin{eqnarray}
\label{eq:v_sstt}
v_{\tau \tau \sigma \sigma} = \vec{\tau} \cdot  \vec{\tau} \,\,\, \vec{\sigma} \cdot  \vec{\sigma} \,  f(r) \, ,
\end{eqnarray}
where $f(r)$ represents the dependence on the relative distance $r$, ``$\cdot$'' implies a scalar product, 
and $\vec{\sigma}$ ($\vec{\tau}$) refers to spin (isospin) operators.   

\begin{figure}[tb]     
\includegraphics[width=7.0cm]{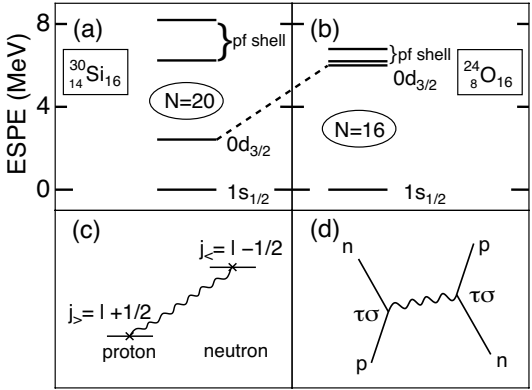}
\caption{Neutron ESPEs for (a) $^{30}$Si and (b) $^{24}$O, relative to
the 2$s_{1/2}$ (shown as 1$s_{1/2}$) orbit. The dotted line connecting (a) and (b) is drawn to indicate
the change of the 1$d_{3/2}$ (shown as 0$d_{3/2}$) level.  (c) The major interaction
producing the basic change between (a) and (b). (d) The process
relevant to the interaction in (c).
From \citet{magic}.
}
\label{magic}  
\end{figure}

The matrix element of the term $\vec{\tau} \cdot  \vec{\tau}$ is trivial, being $-\frac{3}{4}$ and $\frac{1}{4}$
for $T$=0 and 1, respectively.  
The monopole matrix elements of this interaction with $f(r) \equiv$1 show an interesting 
analytic property, and we shall
discuss it now.  We consider antisymmetric states in 
eq.~(\ref{eq:nn-anti}) or eq.~(\ref{eq:pn-anti})  
and 
symmetric states in  eq.~(\ref{eq:pn-sym1})  
or  eq.~(\ref{eq:pn-sym2}).     
The monopole matrix element consists of direct and exchange contributions.
The direct contribution from the $\vec{\sigma} \cdot  \vec{\sigma}$ term is 
\begin{flalign}
\label{eq:ss-dir}
\Sigma_{m,m'} \, (\, j,m | \sigma_z |  j,m \,) (\, j',m'  | \sigma_z |  j',m' \,) =0, 
\end{flalign}   
where $\sigma_z$ stands for the $z-$component of $\vec{\sigma}$ and  
$\sum_{m} \, (\, j,m | \sigma_z |  j,m \,)\,=\,0$ is used.
On the other hand, the exchange contribution is expressed as   
\begin{flalign}
\label{eq:ss-ex}
\mp \Sigma_{m,m'} \, \bigl\{ & (1/2) \, \{ \, (\, j,m | \sigma_{+} |  j',m' \,) (\, j',m' | \sigma_{-} |  j,m \,) \nonumber \\
                                         & \,\,\,+  (\, j,m | \sigma_{-} |  j',m' \,) (\, j',m' | \sigma_{+} |  j,m \,) \}   \nonumber \\
  &  + (\, j,m | \sigma_{z} |  j',m' \,) (\, j',m' | \sigma_{z} |  j,m \,) \bigr\}    \, ,
\end{flalign}   
where $\sigma_{+}$ and $\sigma_{-}$ stand for the raising or lowering operator of $\vec{\sigma}$, and the 
overall sign $\mp$ corresponds to the antisymmetric and symmetric states, respectively.
Thus, direct terms do not contribute, and only exchange contributions remain.  
We point out that for interactions without the $\vec{\sigma} \cdot  \vec{\sigma}$ term, the situation is
very different as the direct term is the major source of the monopole interaction. 
In order to have finite values in eq.~(\ref{eq:ss-ex}), $j$ and $j'$ must have the same $\ell$, implying that 
$j$ and $j'$ are either $j_>$ or $j_<$ for the same $\ell$. 
After some algebra of angular momentum, the final results are tabulated in Table~\ref{table:ttss}.
The $j_> - j_<$ coupling appears to be about twice stronger than  the $j_> - j_>$ 
or  the $j_< - j_<$ couplings.  This is precisely due to larger matrix elements of spin-flip transitions,
like $( \, j_> \, | \vec{\sigma} |\, j_< \, ) $ or $( \, j_< \, | \vec{\sigma} |\, j_> \, ) $, than 
spin-nonflip transitions like $( \, j_> \, | \vec{\sigma} | \, j_>\, )$ or 
$( \, j_< \, | \vec{\sigma} | \, j_< \, )$ \citep{magic}.   The same mathematical feature 
applies to the isospin matrix elements, enhancing charge exchange processes like the one shown in 
Fig.~\ref{magic} (d).  
The most important outcome of these features is the strong proton-neutron coupling between $j_>$ or $j_<$ with the same $\ell$, or between $\ell + 1/2$ and $\ell - 1/2$ (see Fig.~\ref{magic} (c)).  

\begin{table}[htb]
\caption{%
Monopole matrix elements of the $\tau\tau\sigma\sigma$ interaction with $f(r)\equiv$1. Based on Table 1 of \cite{Otsuka2002}.
}
\begin{ruledtabular}
\renewcommand{\arraystretch}{2.0}
\begin{tabular}{|c|c|c|c|}
\label{table:ttss}
\textrm{$j_1$}&
\textrm{$j_2$}&
\textrm{$T$=0}&
\textrm{$T$=1}\\
\colrule
$\ell +\frac{1}{2}$ & $\ell +\frac{1}{2}$ & -3/16(2$\ell$+1) & -(2$\ell$+3)/16(2$\ell$+1)$^2$\\
\hline
$\ell +\frac{1}{2}$ & $\ell-\frac{1}{2}$ & -3/8(2$\ell$+1) & -1/8(2$\ell$+1)\\
\hline
$\ell-\frac{1}{2}$ & $\ell-\frac{1}{2}$ & -3(2$\ell-$1)/16(2$\ell$+1)$^2$ & -1/16(2$\ell$+1)\\
\end{tabular}
\end{ruledtabular}
\end{table}

A concrete example is shown in Fig.~\ref{magic} (a,b).
We assume here the simple filling configuration that the last six protons in $^{30}$Si are in 
the 1$d_{5/2}$ (shown as 0$d_{5/2}$) orbit in Fig.~\ref{magic} (a). 
On the other hand, $^{24}$O has no proton in the 1$d_{5/2}$ orbit, and 
shows a large gap between neutron 1$d_{3/2}$ and 2$s_{1/2}$ orbits, consistent with
experiment \citep{Hofmann2008,Kanungo2009}.    

The monopole matrix element of the $\tau\tau\sigma\sigma$ interaction with $f(r) \equiv$1 
vanishes for 
any pair involving an $s_{1/2}$ orbit.  Thus, these last six protons in $^{30}$Si lower the ESPE of 
the neutron 1$d_{3/2}$ orbit relative to the 2$s_{1/2}$ orbit through the monopole matrix element,  
\begin{equation}
 V_{pn}^m ({\rm 1}d_{5/2}, \, {\rm 1}d_{3/2}) = - 1\,/ \, \{ 4 \, (2 \times 2+1) \}  \, ,
\end{equation} 
obtained from Table~\ref{table:ttss}, following eq.~(\ref{eq:enj}).   
Although not shown in Fig.~\ref{magic}, 
the ESPE of the neutron 1$d_{5/2}$ orbit is lowered by about half of the change of the 
neutron 1$d_{3/2}$ ESPE, as can be seen in Table~\ref{table:ttss} with $\ell \gg 1$.
Thus, while the $\tau\tau\sigma\sigma$ interaction can change the spin-orbit splitting, 
both spin-orbit partners are shifted in the same direction.

The above argument on a more attractive monopole matrix element between $\ell + 1/2$ 
and $\ell - 1/2$ orbits can be extended, with certain modifications, to finite-range and zero-range central 
interactions as we have seen numerically in Fig.~\ref{fig:monoGD}.   
We note that in the case of the zero-range $\delta$-function central interaction, the total spin of two 
interacting nucleons is restricted to $S$=0 for $T$=1 and $S$=1 for $T$=0, and this induces some 
spin-spin effects even for a simple $\delta$-function interaction without explicit spin dependence.   
We point out also that within the central forces, the coupling between orbits with $\ell$ and 
$\ell'$ with $\ell \ne \ell'$ is not enhanced as can be understood from Fig.~\ref{magic} (d) and as can be
confirmed numerically from Fig.~\ref{fig:monoGD}. 
We will come back to these features after discussing the tensor-force effect.


\subsection{Shell Evolution due to the Tensor force}
\label{subsec:tensor}

\subsubsection{Tensor force}

We now study the shell evolution due to another major component of the nuclear force, the tensor force.  
Yukawa proposed the meson exchange process as the origin of the nuclear forces \citep{Yukawa1935}. 
Although this was on the exchange of a scalar meson and is not directly related to the tensor force, 
the meson exchange theory was developed further, and Bethe demonstrated that the tensor force is formulated
with the coupling due to another kind of meson (i.e., referred to as 
$\pi$-meson (or pion) presently), 
with explicit reference to the tensor force and its effect on the deuteron property  \cite{Bethe1940a,Bethe1940b}.  
We can thus identify the tensor force with its unique features as one of the most important and visible manifestations 
of the meson exchange process initiated by Yukawa.

We start our discussion with 
the one-$\pi$ exchange potential between the i-th and j-th nucleons,  
\begin{equation}
 V_{\pi} \,= f \,  (\vec{\tau}_i \, \cdot \, \vec{\tau}_j  )
 (\vec{\sigma}_i \cdot \nabla) (\vec{\sigma}_j \cdot \nabla)
 \frac{e^{-m_{\pi}r}}{r} ,
\label{eq:yukawa_1}
\end{equation}
where $\vec{\tau}_i $ and $\vec{\sigma}_i$ indicate, respectively, the isospin and spin operators of the i-th nucleon, 
$\vec{r}$ denotes the relative displacement between these two nucleons with $r= |\vec{r}|$, and 
$\nabla$ stands for the derivative by $\vec{r}$.  
Here, $f$ and $m_{\pi}$ are the coupling constant and  the $\pi$-meson mass, respectively.   
Equation~(\ref{eq:yukawa_1}) is rewritten as 
\begin{flalign}  
 \,\, V_{\pi} &= \, \frac{f\,m_{\pi}^2}{3} \, (\vec{\tau}_i \, \cdot \, \vec{\tau}_j  )  \nonumber \\
            & \times \bigl\{ (\vec{\sigma}_i \cdot \vec{\sigma}_j)
                      + S_{ij} \,   \{ 1 + \frac{3}{m_\pi r} + \frac{3}{(m_\pi r)^2} \} \bigr\} \,  \frac{e^{-m_{\pi} r}}{r},
\label{eq:OPEP}
\end{flalign}
with   
\begin{equation}
  S_{ij}=3(\vec{\sigma}_{i}\cdot \vec{r}) (\vec{\sigma}_{j}\cdot \vec{r})/r^2 \,
            - \, (\vec{\sigma}_{i}\cdot \vec{\sigma}_{j}) \, .
\label{eq:tensor S}
\end{equation}
Here an additional $\delta$ function term is omitted in eq.~(\ref{eq:OPEP}) as usual 
(because there are other processes at short distances). 
The first term within $ \bigl\{  \,\, \bigr\}$ on the right-hand side of eq.~(\ref{eq:OPEP}) produces a central force, 
and is not considered hereafter.
The second term within this $ \bigl\{  \,\, \bigr\}$ generates the tensor force from the one-$\pi$ exchange process.   

\begin{figure}[tb]    
\begin{center}
\includegraphics[width=5cm]{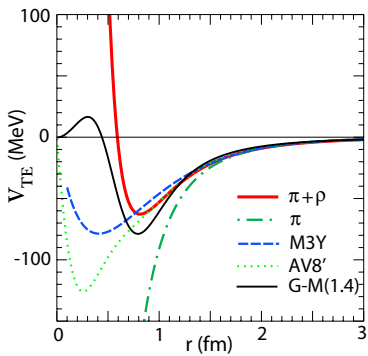}
\end{center}
\caption{Triplet-even potential due to the tensor force for various interaction models.
Adapted from \citep{Otsuka2005}.
}
\label{fig:tensor_potentials}
\end{figure}

As an example of the radial dependence of actual tensor potentials, 
Fig.~\ref{fig:tensor_potentials} shows the triplet-even (TE) potential due to the tensor potentials in some approaches
(see \citep{Otsuka2005} for details).  
Except for the $\pi$-meson exchange case (no $\rho$ meson), the TE potentials exhibit rather similar
behaviors outside $\sim$ 0.6 fm.   While differences arise inside, the relative-motion wave functions of 
two interacting nucleons are suppressed there because of forbidden coupling between $S$-wave bra and
ket states. 

The tensor force has been known for a long time in connection to the one-$\pi$ exchange potential 
as stated here, and its effects were studied extensively from many angles.
Early studies in connection to the nuclear structure include an extraction of the tensor-force component 
in the empirical $NN$ interaction by Schiffer and True \cite{schiffer}, 
a derivation of microscopic effective $NN$ interaction ({\it i.e.}, so-called ``G-matrix interaction'') 
including second-order 
effects of the tensor force by Kuo and Brown \cite{KuoBrown}, a calculation of magnetic moments 
also including second-order tensor-force contributions by Arima and his collaborators \cite{KShimizu} and by Towner \cite{Towner}, a review on its effects in light nuclei \cite{Fayache1997}, {\it etc}.  
 
Besides such effects, the tensor force produces another effect on the shell structure
in its lowest order, or, by the one-$\pi$ exchange process.
This effect must have been contained in numerical results, 
but its simple, robust and general features had not been mentioned 
or discussed until the work done in \cite{Otsuka2005}, where
the change of the shell structure, {\it i.e.}, the shell evolution, 
due to the tensor force was presented for the first time. 

We now present the monopole interaction of the tensor force first, in order to clarify such 
tensor-force driven shell evolution.    
Because the $S$ operator in eq.~(\ref{eq:tensor S}) between nucleons ``1'' and ``2'' can be rewritten as
\begin{equation}
  S_{12} \, = \,  \sqrt{24 \, \pi}  \, [ \, [\vec{\sigma_1} \times \vec{\sigma_2}]^{(2)} \times Y^{(2)} (\theta, \phi) ]^{(0)} \, ,
\label{eq:tensor S-2}
\end{equation}
where $[\, \times \,]^{(K)}$ means the coupling of two operators in the brackets
to an angular momentum (or rank) $K$, and $Y$ denotes the spherical harmonics of the given rank 
for the Euler angles, $\theta$ and $\phi$, of the relative coordinate.
The tensor force can then be rewritten in general as
\begin{equation}
   V^{ten} = (\vec{\tau}_1 \cdot \vec{\tau}_2) \, 
         ( \, [\vec{\sigma_1} \times \vec{\sigma_2}]^{(2)} \cdot Y^{(2)} (\theta, \phi) )  f^{ten} (r),
\label{eq:tensor}
\end{equation}
where $f^{ten} (r)$ is an appropriate function of the relative distance, $r$.  
Note that the scalar product is taken instead of $[\, \times \,]^{(K)}$. 
Eq.~(\ref{eq:tensor}) is equivalent to the usual expression containing
the $S_{12}$ function.  
Because the spins $\vec{\sigma_1}$ and $\vec{\sigma_2}$ are dipole
operators and are coupled to rank 2, the
total spin $S$ (magnitude of $\vec{S}=\vec{s}_1 + \vec{s}_2$) 
of two interacting nucleons must be $S$=1. 
If both of the bra and ket states of $V^{ten}$ have $L$=0, with $L$
being the relative orbital angular momentum, their matrix element vanishes  
because of the $Y^{(2)}$ coupling.  
The crucial roles of these properties will be shown in the rest of this subsection.

Besides the $\pi$-meson exchange, the $\rho$ meson
contributes to the tensor force.  In the following, we use the  
$\pi$+$\rho$ meson exchange potential with the coupling constants taken from \cite{Osterfeld}.
The function $f^{ten} (r)$ therefore corresponds to the sum of these exchange processes.  
The magnitude of the tensor-force effects to be discussed becomes about three quarters 
as compared to the results by the one-$\pi$ exchange only.
The basic physics will not be changed.  We will compare the $\pi$+$\rho$ meson results with 
those by modern theories of nuclear forces.  

\subsubsection{Tensor force and two-nucleon system}

Having these setups, we first recall the basic properties of the tensor force, by taking a two-nucleon system.
From the previous subsubsection, we know $S=1$ for two nucleons interacting through the tensor force.  
We therefore assign $s_z=1/2$ for each nucleon, taking the z-axis in the direction of the spin. 

\begin{figure}[tbh]     
\includegraphics[width=0.34\textwidth,clip=]{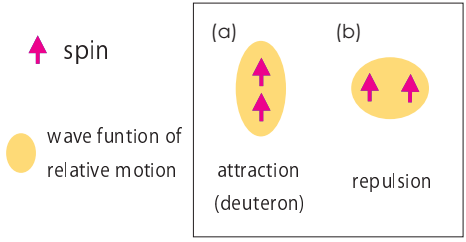}
 \caption{Intuitive picture of the tensor force acting on two nucleons.
    }
  \label{tensor_NN}
\end{figure}

Figure~\ref{tensor_NN} displays schematically this system in two different situations.
The spins are shown by arrows pointing upwards, and are placed where two nucleons are placed at rest.
In other words, two nucleons are displaced (a) in the direction of the spin or (b) in the perpendicular direction.   
This is certainly a modeling of the actual situation of which the wave function of the relative motion is shown schematically by yellowish shaded areas in Fig.~\ref{tensor_NN}.   

We now consider the effect of the tensor force 
in these two cases, by denoting the value of the operator in eq.~(\ref{eq:tensor S}) by $\mathcal{S}_{ij}$.
In the case of Fig.~\ref{tensor_NN} (a), we obtain
\begin{equation}
(\vec{\sigma}_{i}\cdot \vec{r}) /r \,\times\, (\vec{\sigma}_{j}\cdot \vec{r}) /r \,=\, \frac{1}{2} \times \frac{1}{2},
\end{equation}
while this quantity vanishes for Fig.~\ref{tensor_NN} (b) because of the orthogonality 
between $\vec{\sigma}$ and $\vec{r}$.
Because of $S=1$,
\begin{equation}
(\vec{\sigma}_{i}\cdot \vec{\sigma}_{j}) \,=\,\frac{1}{4}
\end{equation}
holds.  Combining these, we obtain
\begin{eqnarray}
\mathcal{S}_{ij} \,&=\, \left\{ \begin{array} {llll} 
 \frac{3}{4} &-  \frac{1}{4} &=\,\,\,\,\, \frac{1}{2} \,&\,\, \mathrm{ for\, (a)}  \\
 &  &  \\
0  & -  \frac{1}{4} &= - \frac{1}{4}\, & \,\, \mathrm{ for\, (b)}  \\
\end{array}  \right.
\label{eq:tensor NN splitting}
\end{eqnarray}
The tensor force works for the two cases in Fig.~\ref{tensor_NN}  (a) and (b) with opposite signs.
The actual sign of $f$ in eq.~(\ref{eq:OPEP}) is positive, while the $(\vec{\tau}_i \, \cdot \, \vec{\tau}_j  )$
term becomes -3/4 for $T=0$ where $T$ stands for the coupled isospin of the nucleons.
The case in Fig.~\ref{tensor_NN}  (a) gains the binding energy from the tensor force, and indeed 
corresponds to the deuteron.  The other case is actually unbound.
In the case of $T=1$, the attractive effect from the tensor force is three times weaker than in 
the $T=0$ case, in a na\"ive approximation.

\subsubsection{Tensor-force effect and orbital motion: intuitive picture  \label{subsubsec:tensor-intuitive}}

\begin{figure}[tb]     
\includegraphics[width=7.0cm]{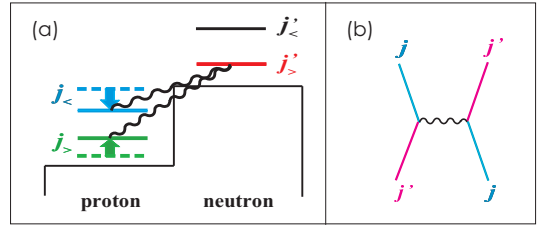}
\caption{(a) Schematic picture of the monopole interaction
    produced by the tensor force between a proton in    
$j_{>,<}$\,=\,$l \pm 1/2$ and a 
neutron in $j'_{>,<}$\,=\,$l' \pm 1/2$.
(b) Exchange processes contributing to the monopole interaction of
the tensor force. 
From \citet{Otsuka2005}. 
}
\label{fig:tensor-1}
\end{figure}

We next consider tensor-force effects on the ESPEs in nuclei: the reduction of the spin-orbit splitting.
As will be shown in this and subsequent sections, the monopole interaction of the tensor force is always attractive between $j_>$ and $j'_<$ 
orbits, whereas it is always repulsive between $j_>$ and $j'_>$ as well as between $j_<$ and $j'_<$.  
Figure~\ref{fig:tensor-1} shows a typical case that the occupation of the neutron $j'_>$ orbit changes 
the splitting between the proton $j_>$ and $j_<$ orbits, as expected by applying these monopole matrix 
elements to eq.~(\ref{eq:epj}).  Such changes lead us to the significant
variation of the shell structure, {\it i.e.}, shell evolution, in association with sizable occupations of a particular orbit.
This basic feature has been presented in \citep{Otsuka2005} followed by further developments.    
We shall discuss here the mechanism and consequence of such tensor-force driven shell evolution in some detail
including those developments.

\begin{figure}[tbh]     
\includegraphics[width=0.36\textwidth,clip=]{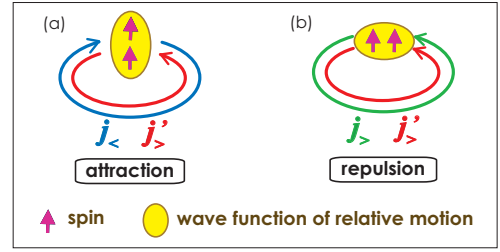}
 \caption{Intuitive picture of the tensor force acting on two nucleons
   on orbits $j$ and $j'$.
From \citet{Otsuka2005}.
    }
\label{intuition}
\end{figure}

Figure~\ref{intuition} shows, in an intuitive way, the phenomena we are looking into.  
Spins are shown by arrows, and they are set to be both up, because of $S=1$ for the tensor force.
We compare two cases:  (a) the tensor-force coupling between $j_>$ and $j'_<$ 
orbits, (b) the one between $j_>$ and $j'_>$ (and also $j_<$ and $j'_<$).

Before evaluating quantitatively these couplings, we present a simplified picture.
This is based on the argument first shown briefly in \citep{Otsuka2005}, followed by an elaborated
description in \citep{otsuka_nobel} and by a further extended version with a figure in \citep{OtsukaGENSHIKAKUKENKYU}.
As the last one is most extensive but in Japanese, we provide a slightly revised text and figure here.

We begin with the case shown in Fig.~\ref{intuition}(a) where a nucleon in $j_<$ is 
interacting with another in $j'_>$ through the tensor force. 
Since the spin of each nucleon is fixed to be up,   
two nucleons must rotate on their orbits in opposite ways.  
We shall look into the relative motion of the two interacting nucleons, 
as the interaction between them is relevant only to their relative 
motion but not to their center-of-mass motion.
We model the relative motion by a linear motion on the $x$ axis.
When two nucleons are close to each other within the interaction range, which is shorter than 
the scale of the orbital motion, the motion of two nucleons can be approximated by a linear motion, and 
the interaction works only within this region.  It is also assumed that the two nucleons continue to move on the 
$x$ axis, which is fulfilled in the present case.  As the tensor-force potential becomes quite damped at  
the distance $\gtrsim$ 2 fm, this is a reasonable modeling for nuclei with larger radii.

\begin{figure}[bt]     
\begin{center}
\includegraphics[width=8.0cm,clip=]{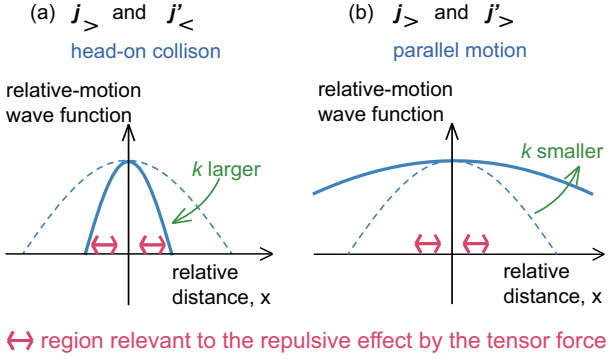} 
\caption{Intuitive picture of the tensor force acting between two nucleons
   in a one-dimensional model.  The relative-motion wave function is shown for (a) head-on collision and 
   (b) parallel motion cases.  In (a), the change is shown as the relative momentum $k$ becomes larger. 
   See the text for explanation.
Adapted from \citet{OtsukaGENSHIKAKUKENKYU}. 
    }
\label{collision_1dim}
\end{center}
\end{figure}

In this linear motion model, 
the wave functions of the two nucleons are approximated by plane waves.  
The case (a), shown in Fig.~\ref{intuition} (a), correspond to the ``head-on collision'' in the linear motion model.  
The case (b), shown in Fig.~\ref{intuition}, corresponds on the other hand to the parallel linear motion of  
the two nucleons. 
We assign indices 1 and 2 to the two nucleons.  Their wave numbers on the $x$ axis  
are denoted by $k_1$ and $k_2$, while their coordinates are denoted by $x_1$ and $x_2$.  
The wave function, $\Psi$, consists of products of two plane waves.  
We now take a system of 
a proton and a neutron in the total isospin $T$=0, which is antisymmetric with respect to the exchange
of the nucleons 1 and 2.
The spin part is $S$=1, which is symmetric.  
As the total wave function must be antisymmetric, 
the coordinate wave function has to be symmetric, taking the form such as,
\begin{eqnarray}
 \Psi \, &\propto & \, 
  e^{ik_1 x_1}e^{ik_2 x_2} \, + \, e^{ik_2 x_1}e^{ik_1 x_2} \,
   = \, e^{iKX} \, \{e^{ikx} \, + e^{-ikx} \} \nonumber \\
   &= & \, 2 \, e^{iKX} \, \cos(kx),  
\label{Psi}
\end{eqnarray}
where center-of mass and relative momenta are defined, respectively, as, 
\begin{equation}
 K, \, = \, k_1 + k_2, \,\,\,\,\, k \, = \, k_1 - k_2, 
\label{Kk}
\end{equation}
and center-of mass and relative coordinates are likewise as,
\begin{equation}
 X, \, = \, (x_1 + x_2)/2, \,\,\,\,\,  x \, = \, (x_1 - x_2)/2.
\label{Xx}
\end{equation}
With these definitions, we see that the relative motion is 
expressed by the wave function 
\begin{equation}
\phi (x) \propto \cos (kx)
\label{rel}
\end{equation}
and the center-of-mass motion has a wave number $K$ for $e^{iKX}$.

In the present case ({\it i.e.}, Fig.~\ref{intuition}(a)),  
$k_1 \sim -k_2$ can be assumed.
The relative motion then has a large momentum, 
$k \sim 2 \, k_1$. 
Its wave function $\phi (x)$ is shown in Fig.~\ref{collision_1dim} (a), with 
the trend with increasing $k$.
We note $K \sim 0$ with $k_1 \sim -k_2$, implying   
the center of mass being almost at rest, or a nearly
uniform wave function of the center-of-mass motion.

Based on Fermi momentum in nuclei, $k$ is considered to be 
of the order of magnitude 1 fm$^{-1}$, but not to 
exceed $\sim$ 1.5 fm$^{-1}$.  From the range of the force,
the area inside $x \sim$ 1 fm is relevant, as the tensor-force potential becomes very weak 
beyond 2 fm.
Thus, the relevant range of $kx$ in Eq.~(\ref{rel})  
is $|kx| \lesssim \pi /2$.   Because of this, Fig.~\ref{collision_1dim} (a) displays 
up to the first zeros in both directions.  
The wave function $\phi (x)$ in Eq.~(\ref{rel})  
is damped more quickly for $k$ larger within this range (see Fig.~\ref{collision_1dim} (a)).    
Figure~\ref{tensor_NN} shows that two nucleons attract each other if they are displaced 
in the direction of spin, but repel each other if they are displaced 
in the direction perpendicular to the spin, {\it i.e.}, the $x$ axis now.
We point out that if two nucleons are at a very short distance without high momenta, 
the tensor force does not work because its angular dependence comes from the spherical
harmonics $Y^{(2)}$ prohibiting a finite probability at zero distance.
The two nucleons should have a certain distance in order to experience some effects, attractive or
repulsive, from the tensor force.  If the distance is too large, the effect is diminished also.
Thus, although schematically, the region shown by bi-directional arrows in Fig.~\ref{collision_1dim} 
is relevant to the tensor force, which is repulsive presently. 
With larger relative momentum $k$, 
Fig.~\ref{collision_1dim} (a) suggests that the wave function is damped faster or 
the region of sizable probability amplitude is more compressed, along the $x$ axis.  
This occurs in the region where the tensor force works repulsively.  Thus the reduction of the repulsion 
takes place more strongly with larger $k$.   This means that as $k$ becomes larger, the repulsion becomes weaker, 
but the attraction remains basically unchanged.  This is nothing but the net effect becoming more attractive.
 
We now come back from one-dimensional modeling to the three-dimensional orbital motion. 
The relative-motion wave function is discussed in a similar manner. 
The yellow shaded area in Fig.~\ref{intuition}(a) indicates, schematically, the region with a sizable probability 
amplitude of the relative-motion wave function, as discussed above.  Its vertically stretched shape implies
the attractive net effect, being consistent with the deuteron case.

We now move on to the case of Fig.~\ref{intuition}(b).
The corresponding case in the linear motion model is shown in Fig.~\ref{collision_1dim} (b).
The parallel motion of the two nucleons occurs, and $k_1 \sim k_2$ can be assumed.
The relative motion then has a small momentum, 
$k \sim 0$, implying a stretched wave function of the relative motion along the $x$ axis,
as shown in Fig.~\ref{collision_1dim} (b).  The probability amplitude then turns out to be large in the region 
of the repulsive effect of the tensor force, yielding the net repulsive effect, assuming that
the net effect has vanished before this repulsive enhancement.     
This case corresponds to Fig.~\ref{intuition}(b), where the two nucleons are 
apart from each other in the direction perpendicular to the total spin.   
The region of larger probability amplitude of the relative wave function (shown by the yellow area) 
is stretched horizontally, which is consistent with the case different from the bound deuteron
shown in Fig.~\ref{tensor_NN} (b).  

In the above linear-motion model, the wave functions in the $y$ and $z$ directions are not
discussed.  The probability amplitude in the $z$ direction contributes to the attraction, whereas those in the $y$ direction to the repulsion.
Those amplitudes are not constant, unlike the ideal plane-wave modeling.  
But they are not affected by the mechanism based on the relative momentum discussed so far, and hence do not differ between the
two cases represented by Figs.~\ref{tensor_NN} (a) and (b).
In short, by having $k$ high enough for the case (a), the linear-motion wave function is pushed 
into the region with no sensitivity to the tensor force, and only the attractive effect remains. 
On the contrary, $k$ becomes $\sim$0 for the case (b), and the full repulsion works out.  
  
Thus, we obtain a robust picture that $j_<$ and $j'_>$ (or vice versa)
orbits attract each other, whereas $j_>$ and $j'_>$ (or $j_<$ and $j'_<$) 
repel each other.    As the monopole interaction represents average effects, 
it is natural that they follow the same trend.  
We will discuss below analytically and numerically how the monopole matrix elements behave.
Note that the essence of the above one-dimensional explanation 
can also be considered as Heisenberg's uncertainty principle.

We make some remarks on the findings made so far.
The coordinate wave function is symmetric in the above cases,  
corresponding the coupling between S and D waves of the relative
motion.  If the total isospin is $T$=1, the antisymmetric 
coordinate wave function is taken, corresponding to P waves.
In this case, the wave function in Eq.~(\ref{rel}) is replaced 
by $\sin (kx)$.  This wave function produces horizontally 
stretched wave function, reversing the above argument for
the case in Fig.~\ref{intuition}(a).  However, because of the 
isospin dependence (see $(\vec{\tau}_1 \cdot \vec{\tau}_2) $ in eq.~(\ref{eq:tensor})), 
there is another sign change, producing an 
attractive effect in total.  Thus, $j_>$-$j'_<$ and $j_<$-$j'_>$ 
couplings give us always attractive effect, 
whereas $j_>$-$j'_>$ and $j_<$-$j'_<$ couplings repulsive.

The radial wave functions of the two orbits must be similar in order to have 
sizable monopole matrix elements.  In addition, a narrow distribution in the radial direction 
is favored in order to have a ``deuteron-like'' shape for the relative-motion wave function.
This is fulfilled if the two orbits are both near the Fermi energy, because 
their radial wave functions have rather sharp peaks around the surface.
If the radial distributions of the two orbits differ, 
not only their overlap becomes smaller but also the relative spatial wave 
function is stretched in the radial direction, which weakens the 
deuteron-like shape, making the effect less pronounced.
Note that for the same radial condition, larger $\ell$ and $\ell'$ 
enhance the tensor monopole effect in general, as their relative momentum 
increases (See Fig.~\ref{intuition}).

\subsubsection{Tensor-force effect and orbital motion: analytic relations}
\label{subsubsec:tensor analytic}

We now move on to the analytic expression on the monopole matrix element. 
An identity on the monopole matrix element  
of the tensor force has been derived in \citep{Otsuka2005}, showing
the properties consistent with the discussions in the previous subsubsection.   
For the orbits $j$ and $j'$, the following identity has been derived
for the tensor force in \citep{Otsuka2005},
\begin{eqnarray} 
(2j_>+1) \, V^{ten;m}_{T} (j_>,\,j') \, +\, (2j_<+1) \, V^{ten;m}_{T} (j_<,\,j') = 0 \,, \nonumber \\
\label{eq:iden}
\end{eqnarray}
where $j'$ is either $j'_>$ or $j'_<$.  
The identity in eq.~(\ref{eq:iden}) can be proved, for instance, with angular momentum algebra 
by summing all spin and orbital magnetic substates for the given $\ell$, where $j_{>,<} = \ell \pm 1/2$. 
The quickest but somewhat more mathematical proof is described here: The left hand side of eq.~(\ref{eq:iden}) 
is equivalent to the total effect of the $T=0$ or $1$ tensor force 
from the fully occupied $j_>$ and $j_<$ orbits coupled with a nucleon in the orbit $j'$.
In the state comprised of fully occupied $j_>$ and $j_<$ orbits, all magnetic substates 
of $\ell$ and those of spin 1/2 are fully occupied, respectively.  
This means that the total spin should be zero. The sole nucleon in the  $j'$ orbit has a spin 1/2, which then constitutes the total spin $0 + 1/2 = 1/2$.  The spin sector of the tensor force in eq.~(\ref{eq:tensor}) is 
$[\vec{\sigma_1} \vec{\sigma_2}]^{(2)}$, which has a rank 2 (angular momentum carried by the operator).
If this operator is sandwiched by the states of spin 1/2, the angular momentum can not be matched, 
and the outcome is zero.  Thus, one can prove the identity.    
The proof can also be made through the re-coupling of angular momenta in the monopole matrix
elements and the explicit form of  the tensor force.  In all these proofs, 
it is assumed that the radial wave function is the 
same for $j_>$ and $j_<$ orbits, which is exactly fulfilled in the
harmonic oscillator 
and practically so in other models if the orbits are well bound.

We make some remarks on this identity.

\begin{itemize}

\item
By moving the second term to the right-hand side of eq.~(\ref{eq:iden}), one sees that 
the $j_>$-$j'$ and $j_<$-$j'$ couplings have the opposite signs always, being perfectly   
consistent with the intuitive explanation in Sec.~\ref{subsubsec:tensor-intuitive}.
There is no exception.   On the other hand, the identity in eq.~(\ref{eq:iden}) does not 
suggest which sign is positive and vice versa.  The intuition explained in Sec.~\ref{subsubsec:tensor-intuitive}
plays a crucial role for the general argument.

\item
Although this identity is not applicable to the cases with $j_>$ or $j_<$ = $j'$ in eq.~(\ref{eq:iden})
with a good isospin ($T$=0 or 1), quite similar behavior is found numerically.   
We note that despite this feature, this identity holds exactly for the proton-neutron interaction in the proton-neutron
formalism.    
Thus, the opposite sign is a really universal feature of the monopole matrix elements 
of the tensor force, and can be used in all cases.

\item
One can prove that $V^{ten;m}_{T} (j_>,\,j')\,=\,0$ for $j$ or $j'=s_{1/2}$.  This is reasonable as
one cannot define $j_>$ or $j_<$ for an $s$ orbit.

\item
As already mentioned, eq.~(\ref{eq:iden}) suggests that if both $j_>$ and $j_<$ orbits are 
fully occupied, there is no monopole effect from the tensor force on any orbit. 
Consequently, $LS$ closed shells produce no monopole effect from the tensor force.

\item
The above derivation indicates also that 
only exchange processes shown in Fig.~\ref{fig:tensor-1}(b) contribute to the monopole matrix 
elements of the tensor force, while the contribution of direct processes vanishes.  
The same property holds for a spin-isospin central interaction discussed in Sec.~\ref{subsec:central}.
This can be understood from the point of view that the vertex 
$(\vec{\sigma} \cdot \nabla)$ in eq.(\ref{eq:yukawa_1})
does not allow a monopole direct process.
If only exchange terms remain, the spin-coordinate contributions of 
$T$=0 and 1 are just opposite. 
Combining this property with 
$(\vec{\tau}_1 \cdot \vec{\tau}_2)$ in eq.~(\ref{eq:tensor}), one obtains
\begin{equation} 
\label{iden_isospin}
\,\,\,\, \,\,\,\,\,\,\, V^{ten;m}_{T=0} (j,\,j') \,=\,3 \times V^{ten;m}_{T=1} (j,\,j') \,\, \textrm{for} \,\, j \neq j' \, . \\
\end{equation}
Thus, the proton-neutron tensor monopole interaction is twice as 
strong as the $T$=1 monopole interaction.  
This implies also that the monopole effect from the tensor force has the same sign 
between $T$=0 and 1, provided that the $(\vec{\tau}_1 \cdot \vec{\tau}_2)$ is included
in the potential.

\end{itemize}

\begin{figure}[tb]     
\includegraphics[width=7.0cm]{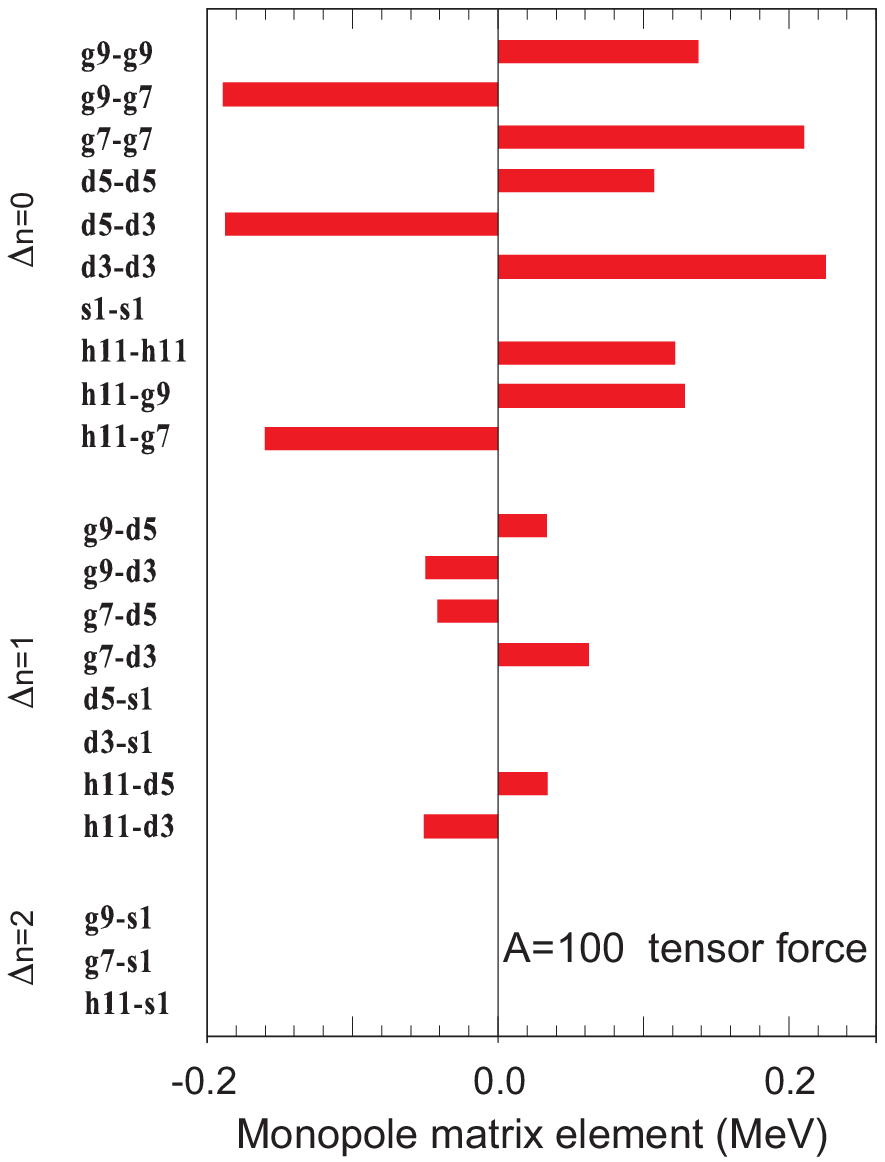}
\caption{Monopole matrix elements of the tensor force in the $T$=0 channel.  
The orbit labeling is abbreviated like g9 for 1$g_{9/2}$, {\it etc}.  
The orbits are from valence shell for $A=100$.
  }
\label{fig:ten_mono_A100}
\end{figure}

\begin{figure}[tb]    
\includegraphics[width=7.0cm]{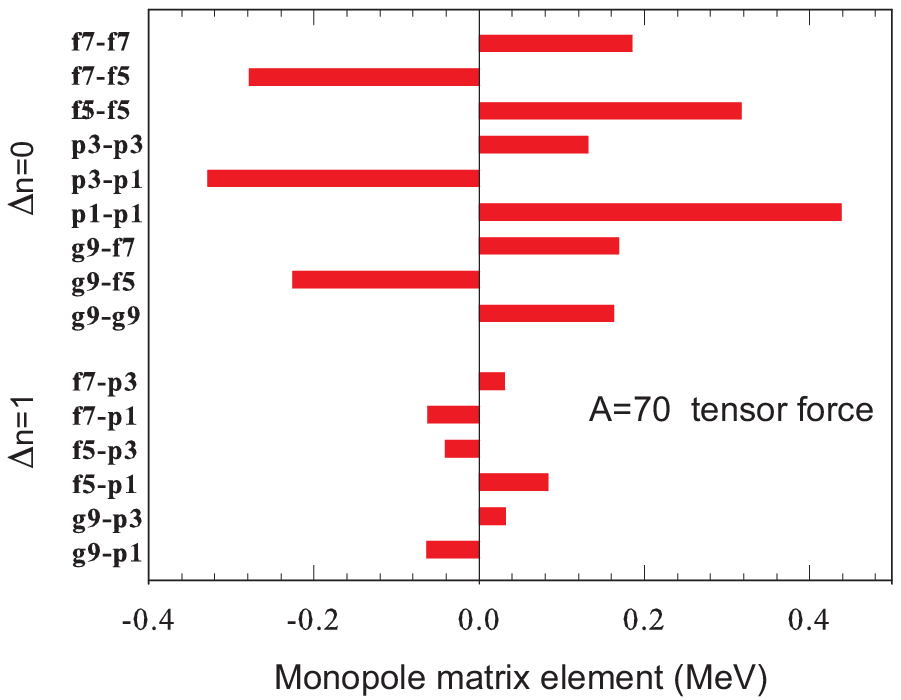} 
\caption{Monopole matrix elements of the tensor force in the $T$=0 channel.  
The orbit labeling is abbreviated like f7 for 1$f_{7/2}$, {\it etc}.  
The orbits are from the valence shell for $A=70$. 
  }
\label{fig:ten_mono_A70}
\end{figure}

Figures~\ref{fig:ten_mono_A100} and \ref{fig:ten_mono_A70} display 
some examples of the monopole matrix elements 
of the $\pi$-meson + $\rho$-meson exchange tensor force with the parameters of \citep{Osterfeld}. 
The same set of single-particle orbits are taken as in Fig.~\ref{fig:monoGD}.  
The identity in eq.~(\ref{eq:iden}) is exactly fulfilled.  
The magnitude of the monopole matrix elements is generally larger for the central force, while the 
variations, for instance within the spin-orbit partners, are of the same order of magnitude 
between the central and the tensor forces.
Their competition produces intriguing phenomena in many cases.

\subsection{Combination of the central and tensor forces}
\label{subsec:VMU}

The previous two subsections presented the monopole interactions from the central and tensor forces. 
We combine them in this subsection.

  
\begin{figure}[tbh]    
\begin{center}
\includegraphics[width=8.6cm]{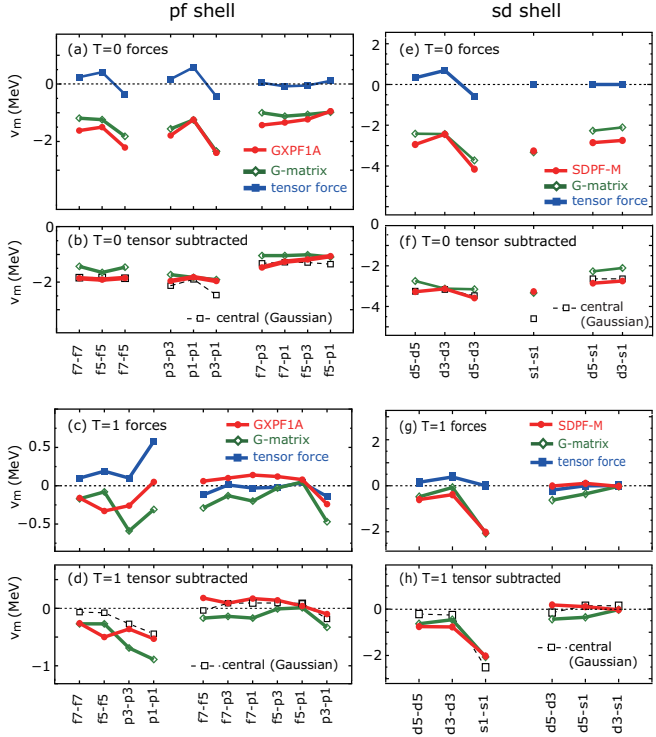}
\end{center}
\caption{Monopole matrix elements of various forces for (a)-(d) pf and (e)-(h) sd shells. In (b),(d),(f ),(h), the tensor force effect is subtracted from the others, and results from a Gaussian central force are shown.
Adapted from \citet{Otsuka2010b}. 
}
\label{fig:VMU-1}
\end{figure}

The central and tensor forces are major components in the effective $NN$ interaction used for 
nuclear structure studies.  
As typical examples of such effective $NN$ interactions, we take the interactions,  
SDPF-M, GXPF1A and G-matrix, described, respectively, 
in \citep{Utsuno1999}, \citep{Honma2005}, and \citep{HJensen1995}.
The former two have been obtained by fitting some two-body matrix elements to experimental energy levels 
using microscopically derived interactions as initial input.  
In addition, the sd-shell part of SDPF-M was obtained by modifying the USD interaction \citep{usd}.   
The G-matrix interaction refers to G-matrix + in-medium corrections by the Q-box formalism \citep{HJensen1995}, 
and will be called this way hereafter, for the sake of brevity.  

The monopole matrix elements of these interactions are shown in Fig.~\ref{fig:VMU-1}, 
where the panels (a)-(d) are for the $pf$ shell, while (e)-(h) for the $sd$ shell.
The $T$=0 matrix elements are shown in the panels (a), (b), (e), and (f), while the $T$=1 are in 
(c), (d) (g) and (h).

In the panel (a), the monopole matrix elements from GXPF1A and G-matrix interactions are 
shown as well as those obtained from the $\pi$-meson + $\rho$-meson exchange tensor
force with the parameters in \citep{Osterfeld}.  

\begin{figure}[tb]    
\begin{center}
\includegraphics[width=6.5cm]{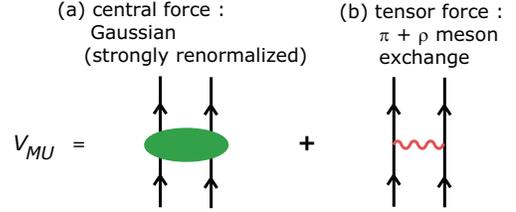} 
\end{center}
\caption{(Color online) Diagrams for the $V_{\rm MU}$ interaction.
From \citet{Otsuka2010b}. 
}
\label{fig:VMU-2}
\end{figure}

It was pointed out in \citep{Otsuka2010b} that the kink pattern is quite similar among 
the GXPF1A, the G-matrix, and the tensor-force monopole matrix elements.   This similarity is indeed
remarkable, and is indicative of the tensor-force origin of the kinks of the other two.  
We can subtract this tensor-force contribution from the GXPF1A or G-matrix
results, as shown in the panel (b).   It was also noted in \citep{Otsuka2010b} that the remaining 
monopole matrix elements are surprisingly flat.  In order to reproduce such monopole matrix 
elements, a Gaussian central force was introduced in \citep{Otsuka2010b}.  
This interaction is called the monopole based universal interaction, or $V_{\mathrm MU}$, and it was already mentioned in eq.~(\ref{eq:v_gauss}).
The parameters selected in  \citep{Otsuka2010b} are $f_{1,0} =f_{0,0} = - 166$ MeV, 
$f_{0,1} =0.6f_{1,0}$ and $f_{1,1} =-0.8f_{1,0}$, and  $\mu=1$ fm.
The $V_{\mathrm MU}$ interaction was described in \citep{Otsuka2010b} as 
``we can describe the monopole component by two simple terms: the tensor force generates ``local'' variations,
while the Gaussian central force produces a flat ``global'' contribution.'', as illustrated graphically 
in Fig.~\ref{fig:VMU-2}.  Here, ``local'' refers to the strong dependences on the single-particle orbits up to sign changes, whereas ``global''  to the weak dependences with large
magnitudes.  

The $T$=1 monopole matrix elements are shown in panel (c).   One notices that they are much weaker
than the $T$=0  monopole matrix elements, by a factor of about 1/10.   This large difference is a general trend.
Within such small monopole matrix elements, the pattern is not so simple.  Panel (d) shows that 
the repulsive $T$=1 monopole interaction in the central Gaussian potential is important.

Moving from the $pf$-shell to the $sd$-shell, 
quite similar properties can be found in panels (e)-(h).   Note that the parameters of
the $V_{\mathrm MU}$ potential are independent of the orbits or the shells.     
The good description is remarkable in this respect.   The reason for this with respect to the tensor force
will be presented in Sec.~\ref{sec:RP}.


\subsection{Shell evolution driven by the central and tensor forces in actual nuclei}
\label{subsect:tensorSE}

This subsection demonstrates how the shell evolution occurs due to the central and tensor forces.  The tensor force is taken from the $\pi$-meson + $\rho$-meson exchange potential in all cases 
(as in the $V_{\mathrm MU}$ interaction), in order to clarify the underlying mechanism.  Likewise, the central force is modeled by the $V_{\mathrm MU}$ interaction 
for all cases, exhibiting different roles of these interactions in a consistent manner.

\subsubsection{Inversion of proton 1$f_{5/2}$ and 2$p_{3/2}$ in Cu isotopes}
\label{subsubsect:f5p3}

One of the most visible examples of the shell evolution driven by the tensor force 
is the change of the proton 1$f_{7/2}$ - 1$f_{5/2}$ splitting due to neutron occupations of the 1$g_{9/2}$ orbit, from the theoretical viewpoint.   
We shall describe this case in some detail, as it is one of the early examples regarding the tensor force.    
The underlying mechanism can be understood by Fig.~\ref{intuition} in a straightforward way. 
Namely, in this case, the neutron $j'_>$ orbit is 1$g_{9/2}$, and it is occupied by more neutrons as
we move on the Segr\`e chart from $^{69}$Cu to heavier Cu isotopes.    
The changes of the ESPEs of the proton $j_> = 1f_{7/2}$ and $j_< = 1f_{5/2}$ orbits  
are given, by following eq.~(\ref{eq:epj}), as
\begin{equation}
\label{eq:epj-f5}
\Delta  \epsilon^{p}_{f_{5/2}} \, = \,  V_{pn}^m (f_{5/2}, \, g_{9/2})\, \Delta n^n_{g_{9/2}} \, ,  
\end{equation}
and
\begin{equation}
\label{eq:epj-f7}
\Delta  \epsilon^{p}_{f_{7/2}} \, = \,  V_{pn}^m (f_{7/2}, \, g_{9/2})\, \Delta n^n_{g_{9/2}}  \,  ,  
\end{equation}
where the symbol $\hat{\,}\,$ is omitted because the occupation number of 
the neutron 1$g_{9/2}$ orbit  is treated as a c-number here.  Note that such a simpler treatment was 
mentioned as a possible option in Sec.~\ref{subsec:espe}.
From the $V_{\mathrm MU}$ interaction,  
the monopole matrix elements have two sources: one from the central force and the 
other from the tensor force.  Table \ref{table:vmu-fpg9} shows their corresponding values.  One sees that 
between the two couplings 1$f_{5/2}$ - 1$g_{9/2}$ and 1$f_{7/2}$ - 1$g_{9/2}$, the central force 
gives a somewhat stronger attraction to the latter, as can be expected from Fig.~\ref{fig:monoGD} (b).  
On the other hand, the tensor force pushes the  
1$f_{5/2}$ orbit down with more neutrons in the 1$g_{9/2}$ orbit, whereas it pulls the 
1$f_{7/2}$ orbit up at the same time.   

\begin{table}[bht]
\caption{
Monopole matrix elements from the central and tensor forces. The unit is MeV.  
The mass number $A$=70 is taken for the Harmonic Oscillator Wave Function of
the single-particle orbit. 
}
\begin{ruledtabular}
\begin{tabular}{|c|c|c|cl}
\renewcommand{\arraystretch}{2.0}
proton orbit  &  neutron orbit  &central& tensor \\
\colrule
1$f_{5/2}$ & 1$g_{9/2}$ & -0.63 & -0.15 \\
\hline
1$f_{7/2}$ & 1$g_{9/2}$ & -0.70 & +0.11 \\
\hline
\multicolumn{2}{|c|}{ difference between 1$f_{5/2}$ and 1$f_{7/2}$ }& +0.07 & -0.26 \\ 
\hline
2$p_{3/2}$ & 1$g_{9/2}$ & -0.46 & +0.02 \\
\hline
\multicolumn{2}{|c|}{ difference between 1$f_{5/2}$ and 2$p_{3/2}$ }& -0.17 & -0.17 \\ 
\end{tabular}
\end{ruledtabular}
\label{table:vmu-fpg9}
\end{table}

\begin{figure}[tb]    
\begin{center}
\includegraphics[width=8.5cm]{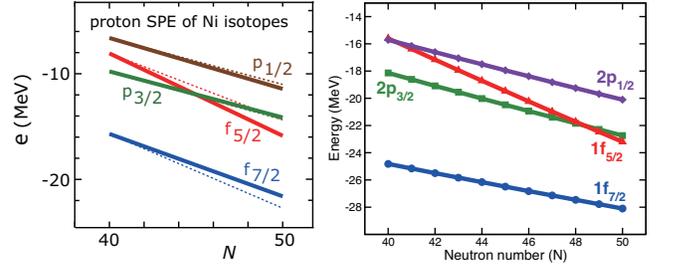}
\caption{(left) Proton ESPEs of Cu isotopes as predicted by the $V_{\mathrm MU}$ interaction (solid lines).  Dashed lines are obtained only with the central-force part.  
The neutron number in the 1$g_{9/2}$ orbit is equal to $N-40$, as the filling scheme is assumed. 
(right) Same quantities by the A3DA-m Hamiltonian used in \citep{Sahin2017}. 
From \citet{Otsuka2010b} (left) and \citet{Sahin2017} (right).  
\label{fig:CuESPE} 
}
\end{center}
\end{figure}

The ESPEs provided by eqs.~(\ref{eq:epj-f5}) and (\ref{eq:epj-f7}) are shown in the left panel of Fig.~\ref{fig:CuESPE},  
where the number of neutrons in the 1$g_{9/2}$ orbit is given by $N - 40$ as the filling scheme 
is taken.  
The ESPEs at $N$=40 are obtained from empirical values \citep{Grawe2005,Otsuka2010b}.  
The full $V_{\mathrm MU}$ interaction is taken for the ESPEs displayed by the solid lines
in the left panel of Fig.~\ref{fig:CuESPE}, while the dashed lines depict results only with the central force.  
One confirms the same trends
as discussed above: the central force (with neutrons in the 1$g_{9/2}$ orbit) slightly repels the 1$f_{5/2}$ and 1$f_{7/2}$ orbits from each  
other, while the tensor force brings them distinctly closer.  The 1$f_{5/2}$ - 1$f_{7/2}$ splitting is $\sim$8 MeV at $N$=40, but is decreased to $\sim$6 MeV at $N$=50.  The $Z$=28 gap is between 
the 2$p_{3/2}$ and 1$f_{7/2}$ orbits at $N$=40 with a gap of $\sim$6 MeV, whereas it is   
between the 1$f_{5/2}$ and 1$f_{7/2}$ orbits at $N$=50 with a gap of $\sim$6 MeV.

The lowering of the proton 1$f_{5/2}$ orbit produces another significant consequence.
The left panel of Fig.~\ref{fig:CuESPE}  shows that the proton 2$p_{3/2}$ orbit comes down, as a function of $N$ 
more slowly than the 1$f_{5/2}$ orbit, and their order is inverted around $N$=45.
In fact, Table~\ref{table:vmu-fpg9} suggests that the central-force contribution to the 
lowering of the proton 2$p_{3/2}$ orbit is $\sim$2/3 of the one 
for 1$f_{7/2}$ or 1$f_{5/2}$ and the tensor-force contribution almost negligible.  
These properties are quite natural due to differences in the radial wave functions.   
Table~\ref{table:vmu-fpg9} shows also how the 1$f_{5/2}$-2$p_{3/2}$ gap is changed by the central and
tensor forces.  Both contribute to the inversion equally, and the total effect is large enough.  

The change of the ESPE of the proton 1$f_{5/2}$ orbit has been investigated experimentally.  
The earlier ones 
\citep{Franchoo1998,Franchoo2001} were made for $^{69,71,73}$Cu 
prior to the theoretical studies presented above.  The experimental findings were compared 
to shell-model calculations \citep{Ji1989}, \citep{S3V1992}.
The main message may be found in the quoted statements as 
``unexpected and sharp lowering of the $\pi$f$_{5/2}$ orbital'' and 
``the energy shift originates from the residual proton-neutron interaction, while its magnitude 
is proportional to the overlap of the proton and neutron wave function'' \citep{Franchoo2001}.
There was no mention of the tensor force, and the lowering of the proton 1$f_{5/2}$ orbit 
seems to have been attributed to the central force.  We can see from Table~\ref{table:vmu-fpg9} 
that the central force accounts for one half of the effect.   Note that spectroscopic factors have been deduced for $^{69,71}$Cu \citep{Morfouace2015}.  
We point out that the interplay of collective and single-particle behavior is discussed for $^{67-73}$Cu in \citep{Stefanescu2008}.

\begin{figure}[tb]    
\begin{center}
\includegraphics[width=7.5cm]{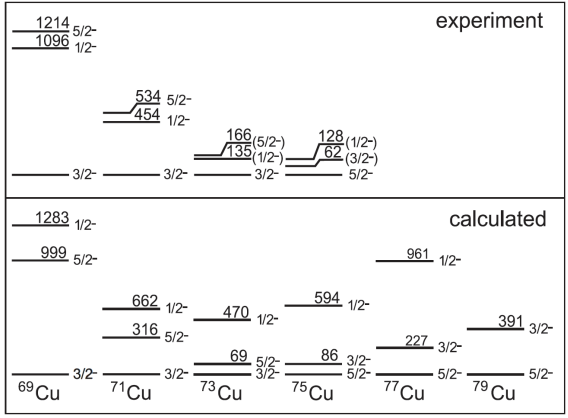} 
\end{center}
\caption{Energy of the lowest levels from experiment and shell-model calculations.
Reprinted with permission from \citet{Flanagan2009}. 
}
\label{fig:Flanagan-Cu}
\end{figure}

The experimental studies were further extended in \citep{Flanagan2009} up to $^{75}$Cu, 
as shown in Fig.~\ref{fig:Flanagan-Cu}.   The inversion between the lowest $5/2^-$ and 
$3/2^-$ levels was observed for the first time in the Cu isotopic chain.  The role of the 
tensor force was known then, and the work was recognized as 
``a crucial step in the study of the shell evolution'' \citep{Flanagan2009}.
The observed levels were compared to shell-model calculations by \citep{Brown_priv}
with a reasonable agreement.  It is very likely that a proper amount of the tensor force was 
included in the shell-model interaction as a result of the fit well-done \citep{Lisetskiy2004,Lisetskiy2005}.  The single-particle nature of the lowest 
$5/2^-$ being the 1$f_{5/2}$ single-particle state in $^{75}$Cu was confirmed
by the measured magnetic moment and by the shell-model calculation, as well as
the lowest $3/2^-$ being the 2$p_{3/2}$ single-particle state in $^{69}$Cu.
On the other hand, the ground state ({\it i.e.}, $3/2^-$ state) of $^{71,73}$Cu was shown 
to have mixed nature.  Besides such intermediate situations, an inversion between 
the 1$f_{5/2}$ and the 2$p_{3/2}$ states has thus been suggested \citep{Flanagan2009}, and the trend  
was extended to heavier Cu isotopes \citep{Daugas2010,Koster2011}.  
This series of experiments showing a clear signal of the lowering of the 1$f_{5/2}$ orbit   
can be considered as a major milestone in establishing the shell evolution.

Effects of various correlations including collective ones were investigated both theoretically and experimentally, but the main conclusion will remain apart from minor changes.  For instance, the precise point of the inversion is
sensitive theoretically to the adopted values of ESPEs at $N$=40 which are not known so accurately constrained to date (see Fig.~\ref{fig:CuESPE}).

The structures of neutron-rich $^{77}$Cu and $^{79}$Cu isotopes have recently been studied experimentally \citep{Sahin2017} and \citep{Olivier2017}, respectively, 
and experimental data were compared well to the results of the shell-model calculation with the A3DA-m Hamiltonian.    
The right panel of Fig.~\ref{fig:CuESPE} indicates the ESPEs obtained from this Hamiltonian including the 1$f_{5/2}$-2$p_{3/2}$ crossing, and shows also that the $Z$=28 gap becomes smaller but remains still greater than 4 MeV up to $N$=50.  As the shell-model calculations contain correlations in the configuration space, the 5/2$^-$ and $3/2^-$ levels are inverted  at $N$=46 in agreement with experiment.  Because the A3DA-m interaction contains empirical corrections (for the given model space) as compared to the VMU interaction, they produce somewhat different reduction of the 1$f_{7/2}$-1$f_{5/2}$ splitting, but the substantial reduction is common, being consistent with the tensor-force driven shell evolution. 
 
\subsubsection{Shell Evolution from $^{90}$Zr to $^{100}$Sn} 
\label{subsubsect:ZrSn}

\begin{figure}[tb]    
\begin{center}
\includegraphics[width=8cm]{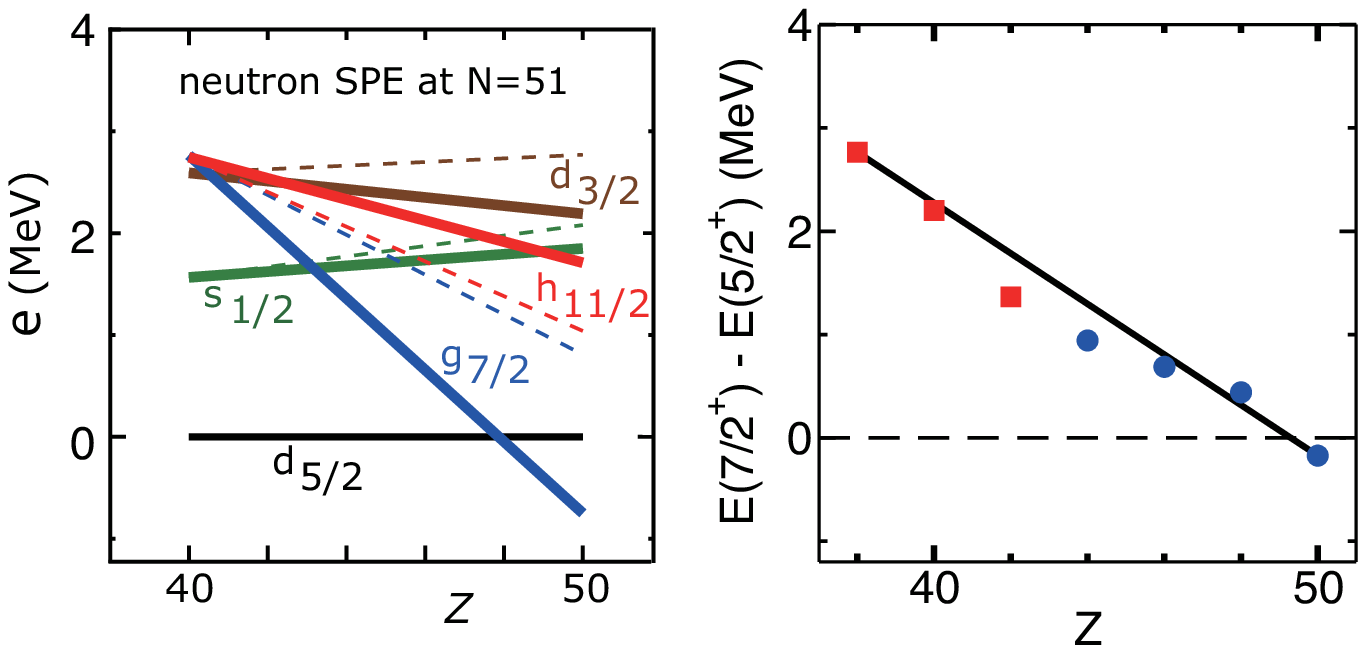}
\end{center}
\caption{
(left) Neutron ESPEs relative to the 2$d_{5/2}$ orbit as a function of $Z$. 
Dashed and full lines have the same meaning 
as in Fig.~\ref{fig:CuESPE}.  Adapted from \citet{Otsuka2010b}.  
(right) Measured energies of the $7/2^+$ level relative to the $5/2^+_1$ states for $N=51$ isotones.  The squares show states with large $1g_{7/2}$ single-neutron strength, as quoted in \citep{Federman1977}.  The circles stand for the lowest observed $7/2^+$ level \cite{ensdf}.  
The present assignment for $^{101}$Sn is 
by \citet{Darby2010}. The straight line connects the points at $Z$=38 and 50.}
 
\label{fig:SEZ4050}
\end{figure}

Another typical case of the tensor-force-driven shell evolution 
is discussed with Fig.~\ref{fig:SEZ4050}, in the filling scheme on top of the $Z$=40 and $N$=50 closed shell.  In its left panel, the ESPEs of neutrons are displayed relative to the one for the 2$d_{5/2}$ orbit, 
where the number of protons in the $1g_{9/2}$ orbit is increased from 0 to 10.  This represents the change from $^{90}$Zr to $^{100}$Sn.   
The neutron ESPEs for $Z$=40 were adjusted to experimental data including the fragmentation of single-particle strengths \citep{ensdf}, and their evolution for larger $Z$'s follows eq.~(\ref{eq:enj})  with the $V_{\mathrm MU}$ interaction.  
The corresponding experimental data, including those mentioned in \citep{Federman1977}, are shown in the right panel. 

One finds, in the left panel, two
sets of calculated results: one (solid lines) is obtained with the full $V_{\mathrm MU}$ interaction, while the other (dashed lines) is only with the central-force part of $V_{\mathrm MU}$.  
A sharp drop of the $1g_{7/2}$ ESPE with the full $V_{\mathrm MU}$ interaction is remarkable, 
ending up with an ESPE below 2$d_{5/2}$.  A similar behavior is seen in 
experimental data (see the right panel). 
This drop is largely due to a strong proton-neutron 
monopole interaction on a proton in the $1g_{7/2}$ orbit generated by neutrons in the 1$g_{9/2}$ orbit.  
The actual values of the relevant central-force monopole matrix elements are $-0.51$ MeV for the proton-neutron $1g_{9/2}$-$1g_{7/2}$ coupling and $-0.32$ MeV for the $1g_{9/2}$-2$d_{5/2}$ coupling (see Fig.~\ref{fig:monoGD} for $T$=0 contribution), while the tensor force contribution $-0.13$ MeV for the former and $+0.02$ MeV for the latter (see Fig.~\ref{fig:ten_mono_A100} for $T$=0 contribution). Thus, the difference between two couplings is $-0.19$ MeV from central, while $-0.15$ MeV from tensor.
It may be worth mentioning that 
the notable central-force contribution was suggested by \citet{Federman1977} as stated in Sec.~\ref{subsec:central}.


The left panel of Fig.~\ref{fig:SEZ4050} shows also that if the central-force part only is taken,  
the 1$g_{7/2}$ and 1$h_{11/2}$ ESPEs come down together (dashed lines).  These two ESPEs, 
however, repel each other towards $Z$=50, if the tensor force is included (solid lines).  
This is because a repulsive monopole interaction works on the $h_{11/2}$ orbit due to the
$j_>$ - $j'_>$ coupling, whereas the tensor interaction is attractive on the 
1$g_{7/2}$ as discussed above.  This attraction produces 
an additional lowering of $1g_{7/2}$, letting it reach below 2$d_{5/2}$ at $Z$=50. 
We note that a similar trend in the 1$g_{7/2}$ - 1$h_{11/2}$ splitting was shown with a monopole-corrected G-matrix interaction in \cite{Sieja2009}.
The energy levels of $^{101}$Sn have been investigated experimentally 
\cite{100Sn1,Darby2010}, which show different ground-state spins but are consistent with the lowering of the $1g_{7/2}$ orbit.   

As the second interesting point, we mention that the bunching of three orbits, 1$h_{11/2}$, 2$d_{3/2}$ and 3$s_{1/2}$, seems to be consistent with the shell structure of the Sn isotopes. 
The left panel of Fig.~\ref{fig:SEZ4050} demonstrates that the tensor force plays a crucial role for obtaining it.
The tensor-force contribution  is thus essential for the shell structure of $^{100}$Sn, which 
has further relevance to various issues of exotic nuclei.



\begin{figure}[tb]    
\begin{center}
\includegraphics[width=8.2cm]{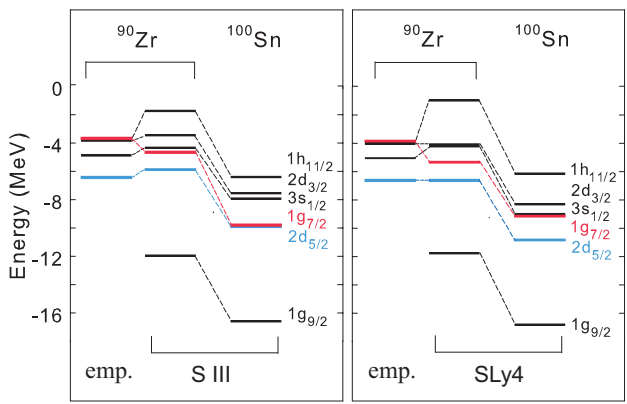}
\end{center}
\caption{Single-particle energies of $^{90}$Zr and $^{100}$Sn calculated by
(left) SIII and (right) SLy4 Skyrme interactions.
}
\label{fig:HFZrSn}
\end{figure}

Another feature of interest is the relation to the Skyrme Hartree-Fock calculation as shown in Fig.~\ref{fig:HFZrSn}.
With the SIII interaction, the 1$g_{7/2}$ - 2$d_{5/2}$ gap is decreased by about 1.2 MeV.
This change is
comparable to the corresponding shift by the central force of the $V_{\mathrm MU}$ interaction.
Note that the ESPE of  the 1$g_{7/2}$ (2$d_{5/2}$) orbit is predicted to be lower (higher) than the empirical value, and thereby the change of their splitting is smaller than that shown in Fig.~\ref{fig:SEZ4050}.  Thus, the tensor force is needed to account for a larger relative 
change between the 2$d_{5/2}$ and 1$g_{7/2}$ orbits, eventually leading to their crossing in
Sn.
Furthermore, Fig.~\ref{fig:HFZrSn} (b) indicates that the gap between the 1$g_{7/2}$ and 2$d_{5/2}$ orbits is even increased with the SLy4 interaction.\\

\subsubsection{Appearance of $N$=16 magic number and disappearance of $N$=20\label{subsubsect:N1620}}

The change of the neutron 1$d_{3/2}$ ESPE was discussed in Sec.~\ref{subsec:central}.
Figure~\ref{magic} shows that the neutron 1$d_{3/2}$ ESPE is about 6 MeV above the 2$s_{1/2}$ ESPE 
in $^{24}$O, but comes down by about 4 MeV in $^{30}$Si.   

This change was discussed, in Sec.~\ref{subsec:central}, as a consequence of the strong attractive monopole matrix element between $\ell + 1/2$ and $\ell - 1/2$ orbits.   
This strong coupling is included in shell-model effective interactions, {\it e.g.}, SDPF-M 
\citep{Utsuno1999} and in the G-matrix \citep{KuoBrown}, whereas it was weakened in some others, {\it e.g.}, USD  \citep{usd}.
It was indicated in Sec.~\ref{subsec:central} that the $\tau\tau\sigma\sigma$ interaction in eq.~(\ref{eq:v_gauss}) can lower, in principle,  
the neutron 1$d_{3/2}$ orbit as protons occupy the 1$d_{5/2}$ orbit.   Although this coupling is strongest 
in Table~\ref{table:ttss}, if the $\tau\tau\sigma\sigma$ interaction is taken, the neutron 1$d_{5/2}$ orbit is lowered
by about half the amount, implying some difficulty.
On the other hand, the strong attraction between the $\ell + 1/2$ and $\ell - 1/2$ orbit was suggested, leading us to a sizable spin-isospin coupling \citep{magic}. 

Four years later \citep{Otsuka2005}, another origin in nuclear forces was proposed for this spin-isospin coupling, the tensor force.   
In fact, the tensor force provides the relevant monopole matrix elements 
being $V_{pn}^{ten;m} ({\rm 1}d_{5/2}, \, {\rm 1}d_{3/2})$ = -0.37 MeV and
$V_{pn}^{ten;m} ({\rm 1}d_{5/2}, \, {\rm 2}s_{1/2})$ = 0 MeV.
By having six protons in the 1$d_{5/2}$ orbit, the 1$d_{3/2}$ orbit is then lowered by 2.2 MeV 
relative to the 2$s_{1/2}$ orbit.
This implies that one half of the lowering of the 1$d_{3/2}$ orbit is due to the tensor force. 
We stress also that the neutron 1$d_{5/2}$ orbit is pushed up by the tensor force with 
six protons in 1$d_{5/2}$, in contrast to the $\tau\tau\sigma\sigma$ interaction. 

The neutron 1$d_{3/2}$ orbit needs to be shifted down by another 2 MeV relative to the 2$s_{1/2}$ orbit
by the central force, because the tensor-force effect is robust (not tunable much) as we shall discuss in Sec.~\ref{sec:RP}.  
In fact, the central force should 
produce a weaker monopole matrix element for the 1$d_{5/2}$ - 2$s_{1/2}$ coupling, 
and this is the case.  

We mention here that some features of the $\tau\tau\sigma\sigma$ interaction are shared by the
tensor force, for instance, the spin-isospin operator $\tau\sigma$ acts on the vertex 
in favor of spin-isospin-flip process, and 
only the exchange process contributes to the monopole matrix element (see Fig.~\ref{magic} and 
Fig.~\ref{fig:tensor-1}).  These properties produce stronger couplings between $j_>$ and $j_<$ 
orbits with a common $\ell$ with both the central and tensor forces, but only the tensor force does it also for 
$j_>$ and $j'_<$, {\it i.e.}, $\ell \,\ne\, \ell'$.  
In this sense, the special importance of the spin-isospin interaction in the shell evolution in exotic nuclei
was pointed out as an initial study in \citep{magic}, being a precursor to more comprehensive studies including the tensor force.   

The relative raising of the neutron 1$d_{3/2}$ orbit from Fig.~\ref{magic} (a) to (b) occurs as
$Z$ is reduced from 14 to 8, while $N$ is kept at 16.  This isotonic change from a stable to an exotic nucleus creates an $N$=16 gap, and 
diminishes the $N$=20 conventional gap.   Thus, the present shell evolution 
can change the magic numbers.  
We mention here that the large $N$=16 gap was recognized in an earlier shell model study 
within the systematics of the oxygen isotopes \citep{Brown1993}.  
The gap was pointed out based on experimental data on the masses and radii \citep{Ozawa2000}. 

We will come back to the $N$=20 magic number in Secs.~\ref{subsec:ab initio} and ~\ref{subsec:key}. 

\subsubsection{Appearance of $N$=34 magic number in the isotonic chain\label{subsubsect:34}}    

\begin{figure}[tb]    
\begin{center}
\includegraphics[width=7cm]{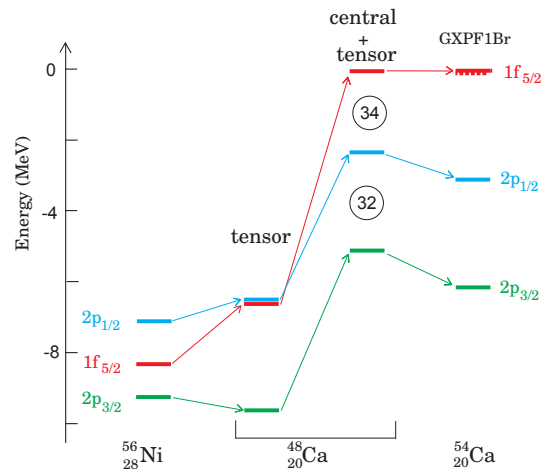}
\end{center}
\caption{Change of ESPEs from $^{56}$Ni to $^{48}$Ca, and to $^{54}$Ca.  
The arrows indicate the change of the ESPE of each orbit.
The arising magic numbers, $N$=32 and 34, are shown in black circles.
The dashed line just below the 1$f_{5/2}$ ESPE (red bar) in the right column means the 1$f_{5/2}$ ESPE calculated with one neutron hole in the 2$p_{1/2}$ orbit.
}
\label{fig:CaNi}
\end{figure}

Another case of new magic numbers has been found in the Ca isotopes, with $N$=32 and 34.  
A remarkable proton-neutron $j_>$ - $j_<$ coupling within a major shell is seen 
in the shell evolution between Ca and Ni.
Figure~\ref{fig:CaNi} displays this shell evolution concretely in terms of the $V_{\mathrm MU}$ interaction.  
We take the filling scheme, where no proton occupies the 1$f_{7/2}$ orbit  in $^{48}$Ca.
In $^{56}$Ni, on the other side, eight protons occupy the 1$f_{7/2}$ orbit, changing  
the ESPEs of the neutron orbits 1$f_{5/2}$, 2$p_{3/2}$ and 2$p_{1/2}$.    
Figure~\ref{fig:CaNi} then indicates how much each orbit is moved 
with the decomposition into the tensor- and central-force monopole contributions.
It is found that  in going from $^{48}$Ca to $^{56}$Ni, both forces contribute additively to the sharp rise of the 1$f_{5/2}$ orbit relative to the  2$p_{3/2}$ and 2$p_{1/2}$ orbits.   The splitting between the  2$p_{3/2}$ and 2$p_{1/2}$ orbits is slightly increased, and becomes a (sub-) magic gap as the 1$f_{5/2}$ orbit is not in between any longer.

Figure~\ref{fig:CaNi} exhibits, the changes of the ESPEs with the decomposition 
into the tensor- and central-force monopole contributions.  
We point out that the monopole components of the tensor and central forces contribute to 
the evolution of the 1$f_{5/2}$ ESPE, showing its sharp rise.  
The splitting between the  2$p_{3/2}$ and 2$p_{1/2}$ orbits remains almost unchanged, 
and becomes a (sub-)magic gap after the 1$f_{5/2}$ orbit is shifted above 2$p_{1/2}$.
Thus, the $N$=32 gap corresponds to the 2$p_{3/2}$ - 2$p_{1/2}$ spin-orbit gap, but its effect is hidden if the 1$f_{5/2}$ is lying between the 2$p_{3/2}$ and 2$p_{1/2}$ orbits.
So, the evolution of the 1$f_{5/2}$ orbit crucially affects the appearance of the $N$=32 magic number.
It is noted that the tensor force enlarges the gap between the 2$p_{3/2}$ and 2$p_{1/2}$.  
The magic numbers 32 and 34 thus appear in going from Ni to Ca as indicated also in Fig.~\ref{fig:CaNi}, 
as the eight protons in $^{56}$Ni are taken away.  
We emphasize that the $N$=34 gap basically vanishes when the tensor-force effect is taken away, 
and that is, since the present shell-evolution effect is linearly dependent on the number of proton holes 
in the 1$f_{7/2}$ orbit, as $Z$ decreases, the $N$=34 (sub-)magic structure fades away first and 
the $N$=32 also disappears eventually.  

The far-right part of Fig.~\ref{fig:CaNi} shows the shell evolution from $^{48}$Ca to $^{54}$Ca 
due to the neutron-neutron interaction, adding six more neutrons still in the filling scheme.
The GXPF1Br shell-model interaction \citep{54Ca} is used, as more fine details are relevant now. 
The neutron-neutron effective interaction produces the shell evolution with patterns very different from those of the proton-neutron interaction.
Four and two neutrons occupy the 2$p_{3/2}$ and 2$p_{1/2}$ orbits, respectively, in $^{54}$Ca. 
The ESPE is shown for the 1$f_{5/2}$ orbit on top of the $^{54}$Ca core, with a very small change from 
$^{48}$Ca.   
As the 2$p_{3/2}$ and 2$p_{1/2}$ orbits are occupied in the $^{54}$Ca core, we show the ESPE for the 
last neutron to occupy these orbits.  
The 2$p_{3/2}$ ESPE is calculated for $^{54}$Ca by assuming a fully occupied 2$p_{1/2}$ orbit.
In order to assess the energy needed for particle-hole excitation, in the far-right part of Fig.~\ref{fig:CaNi},  
the dashed line below the 1$f_{5/2}$ level shows the ESPE 
calculated with one neutron hole in the 2$p_{1/2}$ orbit, which is very close to the solid line.
Thus, the effects of the neutron-neutron monopole interaction is minor and can be repulsive.
The 2$p_{3/2}$ and 2$p_{1/2}$ ESPEs are somewhat lowered due to the pairing component between the same orbit, when they are occupied.   

\begin{figure}[tb]    
\begin{center}
\includegraphics[width=7.5cm]{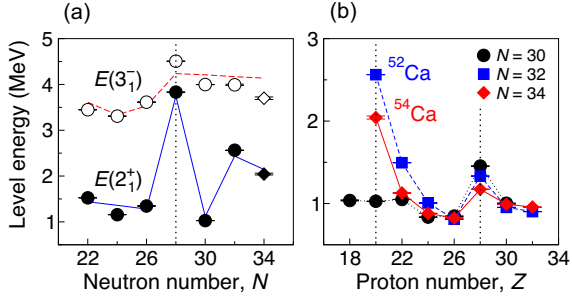}
\end{center}
\caption{Observed first 2$^+$ levels as a function of (a) $N$ and (b) $Z$.
Observed first 3$^-$ levels are shown in panel (a).
Adapted from \citep{54Ca}.
}
\label{fig:54Ca_levels}
\end{figure}

We can thus see the basic mechanism of the appearance of the $N$=32 and 34 gaps.  
This was the prediction in Ref.\citep{magic}, being a consequence of the 
strong attractive coupling between the $\ell + 1/2$ and $\ell - 1/2$ orbit with $\ell = 3$, 
analogous to a similar coupling with $\ell = 2$ leading to the $N=16$ new magic number.   
The corresponding text in \citep{magic} is quoted as 
``we can predict other magic numbers, for instance, $N$=34 associated 
with the $0f_{7/2}-0f_{5/2}$ interaction'', where $0f_{7/2,5/2}$ means $1f_{7/2,5/2}$ in the present notation.  The experimental investigation of the $N$=34 magic number 
in the Ca isotopes had not been feasible for more than a decade, casting doubt over this magic
number \cite{Janssens_Nature}.  In 2013, finally, the 2$^+$ excitation energy was measured
at the RIBF \citep{54Ca} to be significantly higher than in heavier isotones consistent with an $N$=34 gap, as shown in Fig.~\ref{fig:54Ca_levels}.   
A sharp rise of the 2$^+$ excitation energy as a function of $Z$ was thus confirmed experimentally 
for $^{54}$Ca, as shown in panel (b) of Fig.~\ref{fig:54Ca_levels}, 
in accordance with the rise of the 1$f_{5/2}$ orbit from $^{56}$Ni to $^{48}$Ca (see  Fig.~\ref{fig:CaNi}).  
The intermediate situation between the Ca and Ni isotopes is discussed in \citep{54Ca}. 
The $N$=32 gap 
in the Ca isotopes was investigated experimentally at ISOLDE in 1985 in terms of the 2$^+$ excitation energy \cite{Huck}.  The magic structures of Ca isotopes attracted much attention in recent years
\cite{Prisciandaro,Janssens,Liddick,Burger,Dinca2005,Gade2006b,Rodriguez,Rejmund,Honma,Coraggio,
Crawford,kaneko2011,Holt,Utsuno,Hagen2012a,Wienholtz,Steppenbeck2015,GarciaRuiz2016,Perrot2006}.

Figure~\ref{fig:54Ca_levels} (b) exhibits that raising pattern towards $Z$=20 of the 2$^+_1$ level 
differs between $N$=32 and 34 isotonic chains.   
Significant experimental efforts were made, particularly for Ti ($Z$=22),  
for instance \cite{Janssens2002, Fornal2004, Liddick2004, Fornal2005,Dinca2005}, 
partly because Ti is only $\Delta Z$=2 away from Ca.
As three quarters of the shift from Ni to Ca occurs in Ti in Fig.~\ref{fig:CaNi}, the 1$f_{5/2}$ level is located near the 2$p_{1/2}$ level, making the $N$=32 gap rather visible but not the $N$=34 gap, consistent with these experiments.  Thus, studies on Ti and Sc isotopes \cite{steppenbeck2017}, rather support the appearance mechanism of $N$=32 and 34 magic numbers in Ca isotopes, .

The levels of single-particle-like 
states on top of the $N(Z)=28$ and 50 closure were discussed systematically by \citep{Grawe2004}, with 
sharp decreases of (a) neutron $1f_{5/2}$ with proton $1f_{7/2}$ filled, 
(b) proton $1f_{5/2}$ with neutron $1g_{9/2}$ filled, 
(c) proton $1g_{7/2}$ with neutron $1h_{11/2}$ filled, and 
(d) neutron $1g_{7/2}$ with proton $1g_{9/2}$ filled. 
The case (a) is nothing but the change from $^{48}$Ca to $^{56}$Ni depicted in Fig.~\ref{fig:CaNi}.
All of them are of the $j_>$-$j'_<$ coupling with large $j$ and $j'$, and hence the sharp decreases can be understood in terms of the coherent effects of the central and tensor forces discussed so far.  
Related systematic trends of the monopole matrix elements are obtained empirically in \cite{sorlin2014}, indicating that the proton-neutron $1d_{5/2}$- $1d_{3/2}$ monopole matrix element is more attractive than the $1d_{5/2}$- $1f_{7/2}$ one, which is more attractive than the 
$1d_{5/2}$- $2p_{3/2}$ one.  This is consistent with the monopole properties discussed and supports them.


\subsubsection{Repulsion between proton 1$h_{11/2}$ and 1$g_{7/2}$ orbits in the Sb isotopes
\label{subsubsect:Sb}}

\begin{figure}[tb]    
\begin{center}
\includegraphics[width=5.5cm]{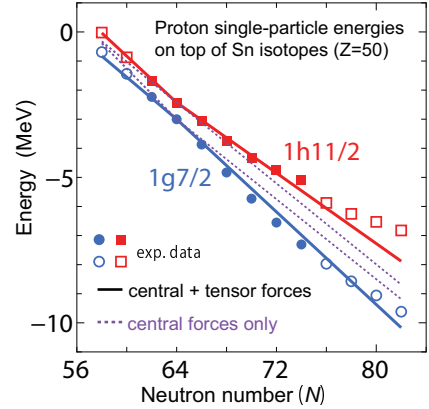}
\caption{ESPEs of proton 1$h_{11/2}$ and 1$g_{7/2}$ orbits in Sb isotopes as functions of $N$. Assuming Sn-isotope core, 
the solid lines are calculated with the tensor-force effect, whereas the dotted lines
are without it.  Symbols are based on experimental values from \citep{Schiffer2004}: fragmentation of
single-particle strength is considered for filled circles, while bare
energies are used for open symbols.  See the text for relevant arguments on those
values.
From \citep{otsuka_nobel}.  
}
\label{fig:Sb}
\end{center}
\end{figure}

Figure~\ref{fig:Sb} shows the ESPEs of the proton 1$h_{11/2}$ and 
1$g_{7/2}$ orbits in Sb isotopes as a function of $N$.
There are 51 protons in the Sb isotopes: one proton on top of the $Z$=50 magic core
in the filling scheme.   This last proton can be either in the 1$h_{11/2}$ orbit or 
the 1$g_{7/2}$ orbit.
The experimental values are taken from \citep{Schiffer2004}, which report that the centroid of 
fragmented single-particle strengths are evaluated as much as possible.   Some questions on the validity of this analysis have been raised, for instance in \citep{Sorlin2008}, in connection to the couplings to
various collective modes including the octupole one.   While this remains an open problem both 
experimentally and theoretically, we discuss it here from the viewpoint of the monopole effect, to explore 
what can be presented with such a simple argument.  We
expect more developments for further clarifications.  

Around the middle ($N \sim$ 66) of the major shell between $N$=50 and 82, the 1$h_{11/2}$ and 
1$g_{7/2}$ orbits 
(or two corresponding experimental states) are
close to each other with a gap of less than 1 MeV.   The gap increases with $N$, as 
seen in Fig.~\ref{fig:Sb}.  It was quite difficult to reproduce this enlargement of the gap 
within mean field models when the experimental values were published \citep{Schiffer2004}.   
In those Sb isotopes, the neutron  
1$h_{11/2}$ orbit is filled more and more as $N$ increases.   The monopole interaction from 
the tensor force is repulsive between the proton 1$h_{11/2}$ orbit and the neutron 
1$h_{11/2}$ orbit (see Fig.~ \ref{intuition}).   Its effect is, on the other hand, attractive between 
the proton 1$g_{7/2}$ orbit and the neutron 1$h_{11/2}$ orbit (see also Fig.~ \ref{intuition}).
In fact, the theoretical ESPEs in Fig.~\ref{fig:Sb} are calculated from the monopole matrix elements 
(see eq.~(\ref{eq:enj})) of the $V_{\rm MU}$ interaction, which consists of the $\pi$+$\rho$ 
meson-exchange tensor force 
and the Gaussian central force, as discussed in Sec.~\ref{subsec:VMU}.

The ESPEs in Fig.~\ref{fig:Sb} are calculated with the monotonic
increase of the uniform occupation probabilities of the neutron 1$h_{11/2}$, 2$d_{3/2}$ and 3$s_{1/2}$ orbits 
starting from $N$=64, for the sake of simplicity.
Figure~\ref{fig:Sb} shows the ESPEs calculated without the tensor force (dashed lines), 
indicating that the two ESPEs come down together with the gap even slightly narrowing.
Once the tensor force is included (solid lines in Fig.~\ref{fig:Sb}), however, it moves the two orbit more apart 
from each other as $N$ increases.   A similar but simpler figure was shown in Fig. 4 (d) of \citep{Otsuka2005}, 
which was published immediately after \citep{Schiffer2004}, demonstrating the explanation of the anomalous 
gap-widening in terms of the tensor force for the first time.
It was remarkable that the gap increase can be explained almost perfectly once the tensor force
is incorporated without adjustment of the tensor-force strength.  
This point has to be clarified with more precise calculations including other correlations.
We also point out the upbending curvature towards $N$=82 shown in 
Fig.~\ref{fig:Sb} may suggest some effects beyond the monopole effect.

It is thus important and essential to examine to what extent other effects, for instance, couplings to collective excitations, 
affect the observed energy levels, while the tensor-force effects seem to remain as a major mechanism.\\

\subsection{Mean-field approaches to the tensor-force driven shell evolution
\label{subsubsect:mean-feild} } 

The effects of the tensor-force has been included in various studies of nuclear models by now, 
those based on the mean-field models \cite{Otsuka2006,Brown2006,Brink2007,Lesinski2007,
Bender2009,Colo,Lalazissis2009,Long2006}.  
Regarding the inclusion of the tensor force into Skyrme-based mean field approaches, 
rather few studies have been done before these works, probably in considerations of some issues 
pointed out, for instance, in \cite{BenderRMP}.

The importance of the tensor force was anticipated by Skyrme, 
when the original form of the Skyrme interaction was proposed \citep{SKYRME1958615}.
The tensor force was, however, not much studied within the Skyrme-model calculations for a while, 
with probably only the exception of the work of Stancu, Brink and Flocard \citep{Stancu},  
who adopted the zero-range approximate form for the tensor force with terms mixing   
$S$ and $D$ waves of the relative motion as well as $P$ waves  
\citep{SKYRME1958615,VAUTHERIN1970149}. 
This form can be written as 
\begin{eqnarray}
 v_{T} &=& \frac{1}{2}T \{[(\vec{\sigma_1}\cdot \vec{k'})(\vec{\sigma_2}\cdot \vec{k'}) -\frac{1}{3}(\vec{\sigma_1}\cdot \vec{\sigma_2})k'^{2}] \delta(\vec{r_1}-\vec{r_2})\nonumber\\
 &+&\delta(\vec{r_1}-\vec{r_2})[(\vec{\sigma_1}\cdot \vec{k})(\vec{\sigma_2}\cdot \vec{k}) -\frac{1}{3}(\vec{\sigma_1}\cdot \vec{\sigma_2})k^{2}]\}\nonumber\\
 &+& U \{(\vec{\sigma_1}\cdot \vec{k'}) \delta(\vec{r_1}-\vec{r_2})(\vec{\sigma_1}\cdot \vec{k})\nonumber\\
&-&\frac{1}{3}(\vec{\sigma_1}\cdot \vec{\sigma_2})[\vec{k'}\delta(\vec{r_1}-\vec{r_2})\vec{k}]\}
\end{eqnarray}
where $\vec{k}$ =$(\vec{\nabla}_1 -\vec{\nabla}_2)/2i$ acts on the right and  $\vec{k'} = -(\vec{\nabla}_1 -\vec{\nabla}_2)/2i$ on the left.  

The tensor term gives rise to additional spin-orbit strengths   
written as
\begin{equation}
\begin{array}{rcl}
 \Delta W_n &=& \alpha_T J_n + \beta_T J_p \\
 \Delta W_p &=& \alpha_T J_p + \beta_T J_n, 
\end{array} 
\end{equation}
where $J_{\rho}$ ($\rho=p,n$) are the spin densities 
given by 
\begin{equation}
\begin{array}{rl}
 J_{\rho}(r)= & \displaystyle \frac{1}{4\pi r^3}\sum_{a} (2j_a+1) \\
& \times \left[ j_a(j_a+1)-\ell_a(\ell_a+1)-\frac{3}{4} \right]R_a^2(r) 
\end{array}
\end{equation}
with occupied orbitals $\{a\}$. Since the spin-orbit potential for 
$\rho=p,n$ is $W_{\rho}\vec{\ell}\cdot\vec{\sigma}/r$, a large negative 
$W_{\rho}$ gives a strong spin-orbit splitting. 
From the sign of 
$j_a(j_a+1)-\ell_a(\ell_a+1)-\frac{3}{4}$, one can see that 
$J_{\rho}$ increases and decreases with the occupation of 
$\alpha=j_>$ and $j_<$ orbitals, respectively, and that 
changing $J_{\rho}$ causes the evolution of the spin-orbit 
coupling as discussed already. 
The $\alpha_T$ and $\beta_T$ correspond to like-particle 
and proton-neutron tensor forces. 
The equality $\beta_T$ =2$\alpha_T$ holds if the tensor force has
the same isospin structure, $\tau_1 \tau_2$, as the $\pi$+$\rho$ meson-exchange potential.  
Monopole terms of the tensor forces given by the two parameters are compared with those of the tensor forces by  
$\pi+\rho$ meson exchanges to study the validity of the use of the approximate zero-range form. 
 
\begin{figure}[tbh]    
\begin{center}
\includegraphics[width=6.5cm,clip]{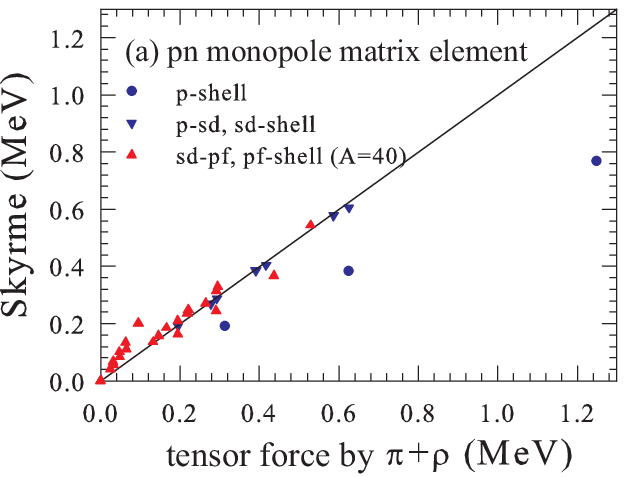}
\\
\vspace{0.5cm}
\includegraphics[width=7.0cm,clip]{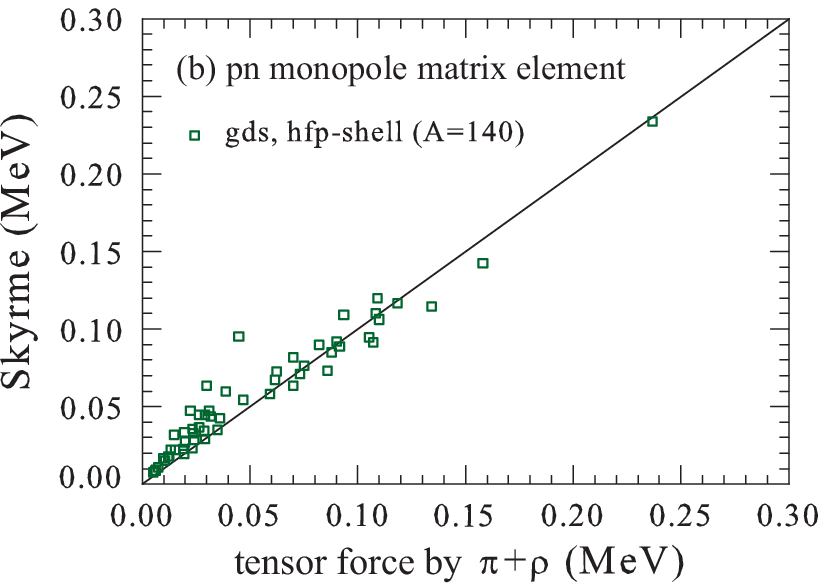}
\caption{Comparison of monopole matrix elements of the zero-range tensor force of \citep{Stancu} to those of 
the $\pi+\rho$ meson-exchange tensor force.  The $p$, $sd$ and $pf$ shells are covered in the upper panel, 
while the valence shell relevant to $A \sim$140 is considered in the lower panel.  
The parameter $\beta_T$=128.75 MeV$\cdot$fm$^5$ is used.
}
 \label{fig:chimono}
 \end{center}
\end{figure}

Figure~\ref{fig:chimono} depicts a comparison between monopole matrix elements of the zero-range tensor force of \citep{Stancu} to those of the $\pi+\rho$ meson-exchange tensor force.  
For the former, the parameters obtained by a chi-square fitting to the monopole matrix elements of the $\pi+\rho$ tensor forces for $A\approx$40 mass region are used with actual values 
$(\alpha_T, \beta_T)=(64.38, 128.75)$ MeV$\cdot$fm$^5$. 
As shown in Fig.~\ref{fig:chimono}, the zero-range form for the tensor 
force can simulate the monopole interaction of the $\pi +\rho$ tensor force to a certain extent, 
but there are rather large fluctuations and deviations, especially in case of light nuclei.
We note that the present $(\alpha_T, \beta_T)$ values are close to the G-matrix ones
$(\alpha_T, \beta_T)=(60, 110)$ MeV$\cdot$fm$^5$  \citep{Brown2006}.
For lighter nuclei, larger parameters become necessary to reproduce the monopole matrix elements 
of the $\pi +\rho$ tensor force, whereas the deviation is opposite in heavy nuclei.   
This variation of the parameters is not much in accordance of the Skyrme phenomenology where
constant parameters for all nuclei are a major advantage. 

\begin{figure}[tb]    
 \begin{center}
 \includegraphics[width=8.5cm,clip]{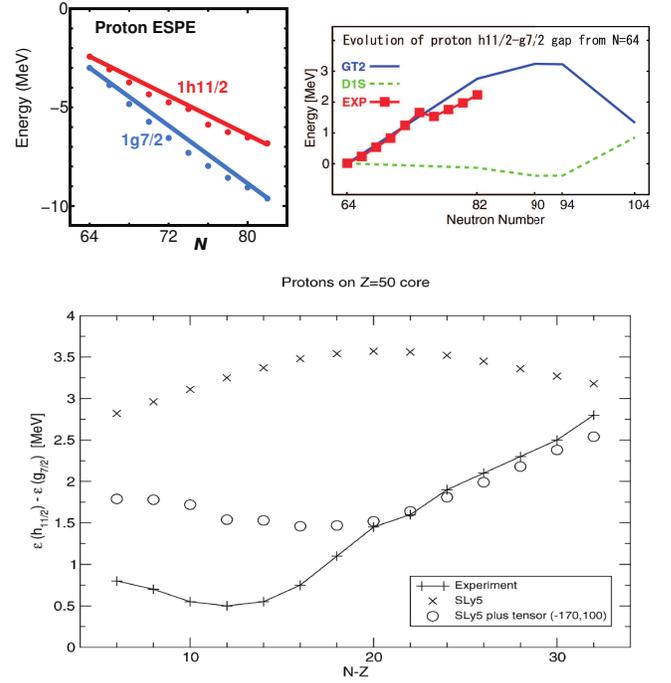}
 \caption{ Evolution of the proton $h_{11/2}$-$g_{7/2}$ gap 
in the Sb isotopes with and without the tensor term.
(upper left) $\pi +\rho$ meson-exchange tensor force on top of 
the usual Woods-Saxon potential.  
(upper right) A Gogny-type calculation with the tensor force (GT2) and without it (D1S).
(lower) A zero-range tensor force calculation added to the SLy5 force.  From \citet{Otsuka2005} (upper left) and \citet{Otsuka2006} (upper right), and reprinted with permission from \citet{Colo} (lower). 
}
 \label{fig:Sb5}
 \end{center}
\end{figure}

\begin{figure}[tbh]    
 \begin{center}
 \includegraphics[width=8cm,clip]{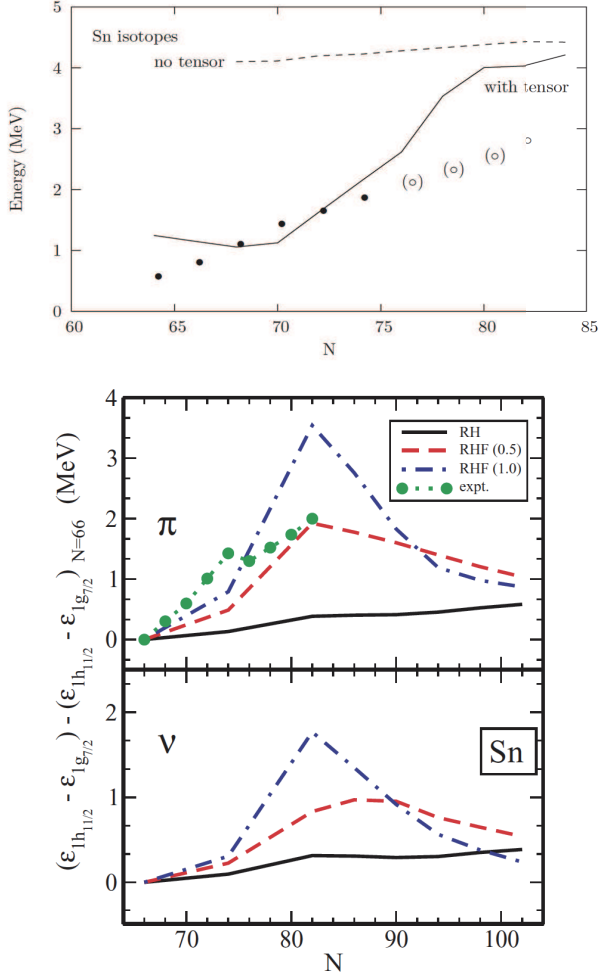}
 \caption{ Same as Fig.=\ref{fig:Sb5}.  
(upper) A zero-range tensor force calculation.
(lower) A relativistic mean field calculation.
Reprinted with permission from \citet{Brink2007} and from \citet{Lalazissis2009}. 
}
 \label{fig:Sb5b}
 \end{center}
\end{figure}

Although the effect of the tensor force on the spin-orbit potential 
was thus recognized in the
1970s, the tensor term has been dropped in 
most of the Skyrme parameterizations until recently. 
One of the probable reasons for this is that the inclusion of 
the tensor term does not lead to significant improvement in 
the single-particle spectra for doubly magic nuclei \citep{Stancu}. 
In addition, as pointed out by \citet{Sagawa2014}, 
not much attention was paid to the {\it evolution} of 
shells with successive mass numbers, likely due to the missing expectation of the shell evolution. 
We point out also that the meaning of the zero-range approximation of the tensor force remains  
to be investigated.

Following the work of \citep{Otsuka2005}, the tensor term 
in the Skyrme forces has been revisited in terms of the shell evolution. 
For instance, \citet{Brown2006} reported the first investigation of the effects of the inclusion of tensor forces into the shell evolution 
based on the Skyrme density functionals, 
employing empirical values $(\alpha_T, \beta_T)=(-118, 110)$ MeV$\cdot$fm$^5$.  
Brink and Stancu \citep{Brink2007} re-investigated, after their work in \citep{Stancu}, the ESPE gaps
between the proton $1h_{11/2}$ and $1g_{9/2}$ single-particle levels in Sb ($Z$=51)  
isotopes as well as those between the neutron $1i_{13/2}$ and $1h_{9/2}$ single-particle levels
in $N$=83 isotones.   
Figures~\ref{fig:Sb5} and \ref{fig:Sb5b} depict results of various calculations 
on the proton $1h_{11/2}$ - $1g_{9/2}$ gap in Sb isotopes.  
While this gap was discussed in case (5) in Sec.~\ref{subsect:tensorSE} with Fig.~\ref{fig:Sb}
from the viewpoint of the $V_{\rm MU}$ interaction, we shall survey other approaches, 
in some of which other correlation effects were investigated.  
Upper panel of Fig.~\ref{fig:Sb5} shows, as a reference, the monopole effect by the $\pi +\rho$ meson-exchange 
tensor force on top of 
the usual mean potential effect like a Woods-Saxon potential \citep{Otsuka2005}.  
\citet{Colo} has examined this shell evolution as shown in the right panel of Fig.~\ref{fig:Sb5}, 
confirming that the inclusion of the tensor term clearly improves the agreement with experimental data
with the adopted values $(\alpha_T, \beta_T)=(-170, 100)$ MeV$\cdot$fm$^5$.
Upper panel of Fig.~\ref{fig:Sb5b} displays a similar calculation by \citep{Brink2007} 
with $(\alpha_T, \beta_T)=(-118.75, 120)$ MeV$\cdot$fm$^5$. 
To the present shell evolution, the proton-neutron monopole interaction matters, which is 
controlled by the $\beta_T$ parameter.  We notice that three works \citep{Brown2006}, 
 \citep{Colo}, and \citep{Brink2007} use, respectively, rather close values, 
$\beta_T$ = 110, 100, and 120 MeV$\cdot$fm$^5$.

Besides the extension of Skyrme phenomenology, there was another early attempt based on the Gogny force plus 
Gaussian-type finite-range tensor force \citep{Otsuka2006} like the  AV8' interaction \citep{Pudliner1997}.
The result for the shell evolution discussed above is shown in middle panel of Fig.~\ref{fig:Sb5}, 
exhibiting a good reproduction of observed systematics. 
A relevant systematic study with the M3Y-type interactions was reported \citep{Nakada2008}.
Combining these works with Skyrme-based calculations, the tensor-force driven shell evolution has been confirmed quite well \citep{Dobaczewski2006,Zou2008,Zalewski2008,Bartel2008,Zalewski2009a,Zalewski2009b,Tarpanov2008,Moreno-Torres2010,Dong2011,Anguiano2011,Anguiano2012,Wang2013,Shi2017}.

We comment on $\alpha_T$ values empirically determined.  They causes the opposite direction of the 
evolution to the ones of the $\pi+\rho$ meson exchange potential and 
the $G$-matrix results. 
The justification of using such negative $\alpha_T$ values is not clear 
in connection to the nucleon-nucleon forces. Regarding open problems with Skyrme-based
approaches, we quote a comment 
``the currently used central and spin-orbit parts of the Skyrme energy
density functional are not flexible enough to allow for the presence of 
large tensor terms" from \citep{Lesinski2007}, and another remark  
``Studies of tensor terms are extended to the case with deformations for
future construction of improved density functionals" from \citep{Bender2009}.

In relativistic mean-field models, $\pi$-meson degrees of freedom were taken into account in relativistic Hartree-Fock (RHF) method by its exchange contributions \citep{PhysRevC.36.380,Long2006,Lalazissis2009}.  
Lower panel of Fig.~\ref{fig:Sb5b} depicts an example of such calculations for the proton $1h_{11/2}$ - $1g_{9/2}$ gap in Sb isotopes, presenting the tensor-force effect within the relativistic framework and the more explicit
treatment of $\pi$ meson in contrast to Skyrme zero-range tensor force. 
Contributions from $\rho$ meson were found to cure the pseudo-shell closures 
at $N$ or $Z$=58 and 92 leading to realistic subshell closure at 64 \citep{Long2007}. 
In recent RHF models, density dependent meson-nucleon couplings \citep{Long2006,Long2007} or softened parametrized couplings \citep{Lalazissis2009} are adopted, which result in smaller effects of tensor forces from $\pi$ or $\pi +\rho$ meson exchanges compared to non-relativistic models \citep{PhysRevLett.101.122502,Lalazissis2009}.
Inclusion of many-body correlations beyond RHF+RPA is still in progress \citep{Litvinova2016138, PhysRevC.73.044328} and left to future investigations. 


\subsection{Contributions from the 2-body LS force}
\label{subsec:2-body LS}

The 2-body LS (2b-$LS$) force is another substantial source of the monopole interaction.
Although it has been proposed in \cite{Elliott1954} based on an earlier work \cite{Blanchard1950}, 
its monopole component has never appeared explicitly in literatures. 
We sketch its major monopole features here with more detailed discussions presented in 
Appendix~\ref{app:LS}.  

The monopole matrix elements of the 2b-$LS$ force contribute, 
in many cases, to the spin-orbit splitting in the usual sense.    
Schematic explanations on their basic properties are shown in  
Appendix~\ref{app:LS}.  
and the obtained characteristic features are listed below. 

(1) The monopole interaction from the 2b-$LS$ force turns out to be consistent with the usual 
one-body spin-orbit splitting (see, {\it e.g.}, \citep{BM1}) in many cases, 
as discussed below.

(2) A schematic semi-classical picture can be drawn for the intuitive understanding of the general and basic properties of the 2b-$LS$ monopole interaction 
(see Fig.~\ref{fig:2bLS}).  
The usual one-body spin-orbit interaction (see \citep{BM1}) 
includes the radial derivative of the density, $\partial \rho / \partial r$, with $\rho$ and $r$ being, respectively, the nucleon density and the distance from the center of the nucleus.   
The present picture leads us to an explanation of this dependence in terms of the difference between the monopole contributions from nucleons inside $r$ and those from nucleons outside $r$. 

(3) Based on this feature, a standard value for each 2b-$LS$ monopole matrix element can be introduced (see the text for eq.~(\ref{eq:LS-normalized})).   
The actual values of the 2b-$LS$ monopole matrix elements are not far from the corresponding standard values in many cases.  This property may be related to the empirical systematics suggested in \citep{Mairle1993}.   

(4) Sizable deviations are found in some cases, however.  
Among them, 
the coupling between an $s$ and a $p$ orbits can be quite strong with a large magnitude of monopole 
matrix element, (see Fig.~\ref{fig:LSmono_spf}  
for example).
This anomaly can be explained in a simple quantum mechanical manner 
based on the range 
of the 2b-$LS$ force and the relative motion of two interacting nucleons, being a robust effect.

Although this effect has been presented orally since 2004, the first publication of one of its outcome was as late as in \cite{Suzuki2008}, where a notable enlargement of
the proton 1$p_{3/2}$-1$p_{1/2}$ splitting due to neutrons in the 2$s_{1/2}$ orbit was shown
as a consequence of the 2b-$LS$ force 
(see Fig. 1 of \cite{Suzuki2008} and relevant texts).  
Another example was presented in \citep{Burgunder2014} for the effect of the proton 2$s_{1/2}$ occupation on the neutron 2$p_{3/2}$-2$p_{1/2}$ splitting in comparison to experiment as
will be discussed in Secs.~\ref{subsubsec:direct},~\ref{subsubsec:spin-orbit}.   
A trend consistent with the present effect can be seen in the spin-tensor decomposition, for instance, in Sect.~\ref{subsect:spin-tensor}.

(5) In some other cases, the sign of the monopole interaction can be opposite from the standard one mentioned
above, due to the radial wave functions (see  Fig.~\ref{fig:LSmono_pd}).

(6) If two nucleons are in the same orbit, the semi-classical picture is inapplicable, and another type of large deviation occurs, for instance,  
between two nucleons in the same 2$p_{1/2}$ orbit 
(see  Fig.~\ref{fig:LSmono_pf}).  
This case is very interesting and important.   In fact, 
the monopole matrix element represents the whole interaction for two neutrons in the 2$p_{1/2}$ orbit, and the tensor and 2b-$LS$ forces produce strong repulsion (see Figs.~\ref{fig:VMU-1} (c),  \ref{fig:VMU-4} (b) and \ref{fig:LSmono_pf}).   
This feature lowers the 2$^+$ level of $^{54}$Ca discussed in Sec.~\ref{subsubsect:34}, by reducing the pairing gap and thereby shifting the ground-state energy upward.  Thus, the actual $N=34$ shell gap is likely larger than what is expected from the actual 2$^+$ level.  
The present repulsive effect also gives a natural explanation to the unusually weak or even repulsive value of the 2$p_{1/2}^2$ pairing matrix element mentioned in \citep{Brown2013}.  While the tensor-force effect was suggested earlier \citep{Otsuka2010b}, another argument was given in \citep{Brown2013}. \\


\begin{figure}[tb]    
\begin{center}
\includegraphics[width=8.0cm]{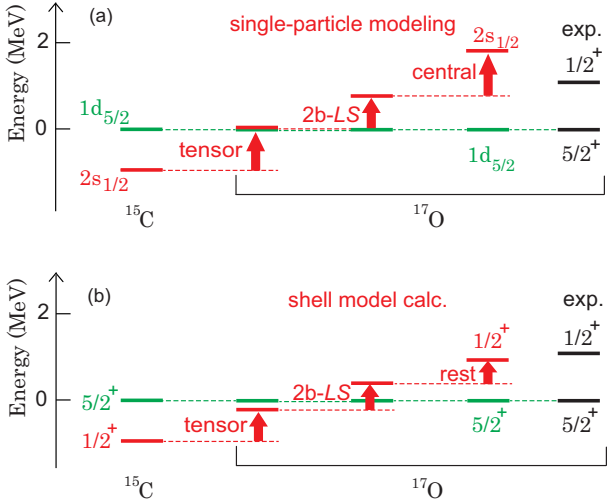}
\end{center}
\caption{(a) Energy of the neutron 2$s_{1/2}$ orbit relative to the 1$d_{5/2}$ orbit
in $^{15}$C and $^{17}$O,  
calculated within the single-particle scheme using the $V_{\rm MU}$ interaction plus  
the M3Y 2b-$LS$ force (see the text).  Contributions from the tensor, 2b-$LS$ and central
forces are decomposed.  The changes are added to the experimental $1/2^+_1$ level 
placed relative to the experimental $5/2^+_1$ level.
The experimentally observed level of $^{17}$O is shown far right. 
 (b) Analysis similar to panel (a) in terms of the shell-model calculation with the 
YSOX \citep{Yuan2012} interaction.
 Some expectation values obtained by the YSOX calculation are shown correspondingly to (a)
 with respect to the shell-model eigenstates. 
 See the caption of (a).}
\label{fig:COsd}
\end{figure}

We now look into the shell structure of $^{15}$C and $^{17}$O, as an example of notable 
contributions of the 2b-$LS$ monopole interaction.    
Although this case has been discussed in ~\ref{subsec:TalmiCNO}, we revisit it.  Figure~\ref{CNO} 
depicts the inversion between neutron 2$s_{1/2}$ and 1$d_{5/2}$ orbits between 
$^{15}$C and $^{17}$O, and 
Fig.~\ref{fig:COsd} (a) shows how the monopole matrix elements work for this inversion.
We now illustrate the origins of those monopole matrix elements in Fig.~\ref{fig:COsd}
in terms of the tensor, 2b-$LS$ and central forces between protons and neutrons.
Here it is assumed that from $^{15}$C to $^{17}$O, the proton 1$p_{1/2}$ orbit is fully occupied, and 
the last neutron is either in the 2$s_{1/2}$ or 1$d_{5/2}$ orbit.  
Figure~\ref{fig:COsd} (a) displays how the neutron 2$s_{1/2}$ orbit is shifted relative to
the 1$d_{5/2}$ orbit in going from $^{15}$C to $^{17}$O in this genuine single-particle limit.

Figure~\ref{fig:COsd} (b) shows a related analysis.  This is similar to Fig.~\ref{fig:COsd} (a), but 
the contributions of the tensor, 2b-$LS$ and the rest of Hamiltonian are shown with respect to 
the shell-model eigenstates obtained by the diagonalization of the Hamiltonian. 
These energy levels can be calculated by shell-model Hamiltonians recently developed, 
SFO-tls \citep{Suzuki2008} and YSOX \citep{Yuan2012}.  The latter is taken in Fig.~\ref{fig:COsd} (b), 
while the former gives a similar result.    
The contributions here mean the expectation values.  The tensor and 2b-$LS$ values are
about 80\% of the corresponding values in Fig.~\ref{fig:COsd} (a), which appear to be 
very similar to the probability of the lowest configuration in the shell-model full
wave functions of $^{17}$O.   Thus, the discussions in terms of the monopole matrix elements
and ESPEs are further proven to be sensible.  On the other hand, the contributions from the central force 
in Fig.~\ref{fig:COsd} (a) is reduced much in the rest part of  Fig.~\ref{fig:COsd} (b).  Here the rest
includes not only effects of the central force but also effects of the (bare) single-particle energies 
due to excitations from lower to higher orbits.  It is clear that various correlations due to the
rest part decrease the raising of the $1/2^+_1$ level.  The general aspect of this feature is of certain interest. 
The importance of non-central forces is thus confirmed in the case shown in Figure~\ref{fig:COsd}, consistent with earlier remark \citep{Millener1975}.
 

\section{Related features of nuclear forces \label{sec:force}}

In this section, we discuss some features of nuclear forces related to the shell evolution.

\subsection{Renormalization persistency of the tensor force\label{sec:RP}}

The effects of the tensor force have been discussed in previous subsections in terms of 
the $\pi$+$\rho$-meson exchange potential.  This potential is derived in the free space, and 
one has to investigate the changes due to various renormalization procedures
for the short-range repulsion and the in-medium corrections.  This study has been done in Refs. 
\cite{Otsuka2010b,Tsunoda2011}, which suggest that the changes are quite small for the tensor force, 
referred to as {\it renormalization persistency}.  

\begin{figure}[tb]    
\begin{center}
\includegraphics[width=7cm]{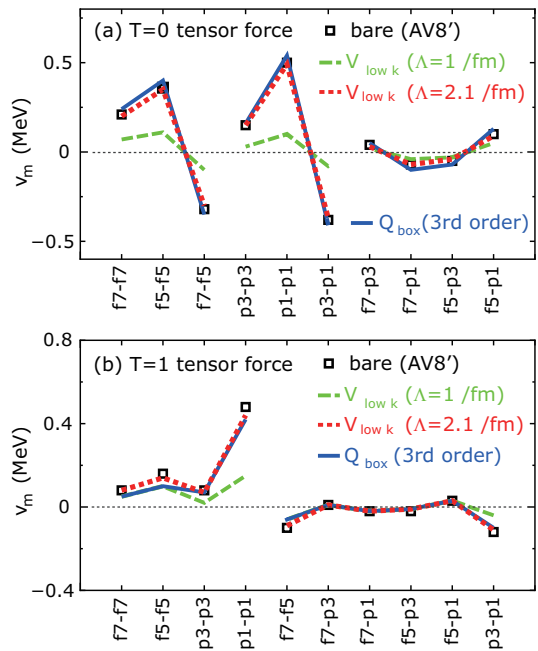}
\end{center}
\caption{Monopole matrix elements from tensor forces in the AV8' interaction \citep{Pudliner1997}, 
in low momentum interactions obtained from the AV8', 
and in the third order Q$_{box}$ interaction for (a) $T$=0 and (b) $T$=1.  
From \citet{Otsuka2010b}. 
}
\label{fig:VMU-4}
\end{figure}

An example is shown in Fig.~\ref{fig:VMU-4} \citep{Otsuka2010b}, 
where the AV8' interaction \citep{Pudliner1997} was used as the starting nuclear force in the free space.   
A low-momentum interaction V$_{low\,k}$ \citep{Bogner2003} was derived in order to treat short-range correlations, 
and the third order Q-box calculation with folded diagram corrections \citep{HJensen1995} 
was performed in order to include medium effects like core polarization.

The spin-tensor decomposition has been carried out over decades  \citep{Elliott,Kirson,Klingenbeck,Yoro,Brown1988,Osnes}, 
in order to extract the tensor-force component.
Here, the spin-tensor decomposition serves as a very useful classification technique of a given two-body interaction 
into several pieces such as the scalar- (central force), axial-vector- (two-body LS (spin-orbit) force), and 
tensor-coupled spin components. 

We shall outline this now.  A given two-body interaction can be rewritten in general as 
\begin{equation}
V = \sum_{k=0,1,2} V_k = \sum_{k=0,1,2} U^k\cdot X^k, 
\label{eq:k-decomp-1}
\end{equation}
where $U^k$ and $X^k$ are tensor operators of rank $k$ in the coordinate and spin spaces, respectively.  One can thus uniquely extract the $LS$-coupled matrix elements of each $k$ component: 
\begin{equation}
 \begin{array}{cl}
  & \displaystyle \langle n_a \ell_a n_b \ell_b LSJT | V_k |  n_c \ell_c n_d \ell_d L'S'JT \rangle \\
= & \displaystyle (-1)^J(2k+1)
\left\{ 
\begin{array}{ccc}
 L & S & J \\
 S' & L' & k
\end{array}
\right\} \\ 
& \displaystyle \times \sum_{J'} (-1)^{J'}(2J'+1) 
\left\{ 
\begin{array}{ccc}
 L & S & J' \\
 S' & L' & k
\end{array}
\right\} \\
& \times \displaystyle \langle n_a \ell_a n_b \ell_b LSJ'T | V |  
n_c \ell_c n_d \ell_d L'S'J'T \rangle. 
\end{array}
\label{eq:k-decomp-2} 
\end{equation}
The $k=0,1,2$ matrix elements correspond, as mentioned above, to the central force, 
spin-orbit (plus antisymmetric spin-orbit) force(s), and 
tensor force, respectively. 

These are all possible components for interactions with the dependence on relative coordinates.  
If dependence on the center-of-mass coordinate is allowed for some reason, other terms 
like antisymmetric LS appear.
Since the shells being considered are full harmonic oscillator shells containing all 
spin-orbit partners, this spin-tensor decomposition is possible.  
We note that the tensor component here is obtained from a given interaction, and can differ from 
the one form the $\pi$+$\rho$-meson exchange potential.  We will see that this turns out to be 
a minor difference in the present discussion with realistic interactions.

Figure~\ref{fig:VMU-4} displays monopole matrix elements thus calculated for $T=0$ and 	1 in the $pf$ shell, 
starting with the AV8' interaction \citep{Pudliner1997} and 
varying the cutoff parameter in the V$_{low \,k}$ process.   For the usual value 2.1 fm$^{-1}$,  
the result is very close to those obtained directly from the bare AV8' tensor-force.

This feature that a nuclear-force component remains unchanged to a good extent by the renormalization
processes has been referred to as {\it renormalization persistency} \citep{Tsunoda2011}.  
The renormalization persistency was studied particularly well for the monopole interaction of 
the tensor force with a variety of the shell, the original interaction and the renormalization methods.
Such studies, not only the earliest one \citep{Tsunoda2011} but also more recent ones with $\chi$EFT forces \citep{syoshida2017}, indicate that 
the tensor force fulfills the renormalization persistency 
at least at the level of the monopole interaction.  
The renormalization persistency  therefore provides us with a good rationale to discuss general features of the monopole effects of the tensor force
in terms of the  $\pi$+$\rho$-meson exchange potential, as was done so far.

\subsection{Spin-tensor decomposition of shell-model interaction}
\label{subsect:spin-tensor}

\begin{figure}[t]    
 \begin{center}
 \includegraphics[width=8.5cm,clip]{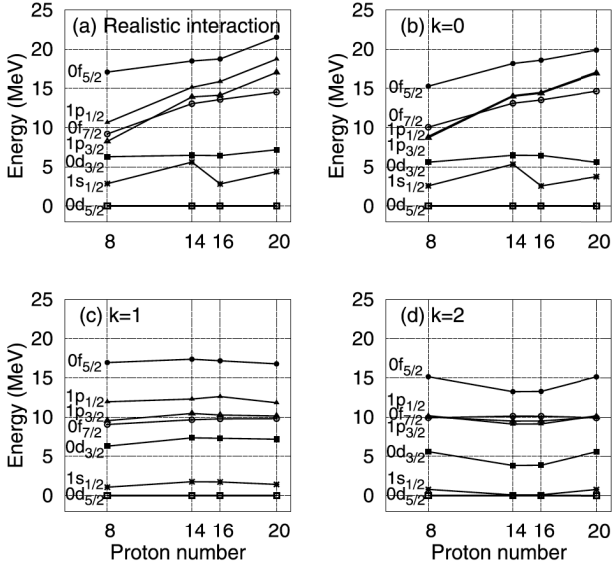}
 \caption{(a) Neutron effective single-particle energies 
of the SDPF-U interaction \citep{Nowacki2009} and their 
(b) $k=0$, (c)  $k=1$, and (d) $k=2$ contributions 
with increasing proton number. 
Reprinted with permission from \citet{Smirnova2010}. 
}
 \label{fig:smirnova}
 \end{center}
\end{figure} 

The spin-tensor decomposition, discussed in the previous subsection, is a useful tool to analyze the amount of the tensor and other components contained in the shell-model interaction. 
\citet{Smirnova2010,Smirnova2012} applied the spin-tensor decomposition technique 
\citep{Elliott,Kirson,Klingenbeck,Yoro,Brown1988,Osnes} to 
a realistic interaction for the $sd$-$pf$ shell \citep{Nowacki2009}
and examined $k=0,1,2$ contributions in eq.~(\ref{eq:k-decomp-1}) to the ESPEs.   
Figure~\ref{fig:smirnova} shows the evolution of the neutron 
effective single-particle energies with protons occupying 
$d_{5/2}$ ($Z=8-14$), $s_{1/2}$ ($Z=14-16$) and $d_{3/2}$ 
($Z=16-20$).  
It is demonstrated that the spin-orbit splittings, especially 
those of $f_{7/2}$-$f_{5/2}$ and $d_{5/2}$-$d_{3/2}$,  
are changed notably by the tensor component 
and that their increase from $Z=16$ to 20 
and decrease from $Z=8$ to 14 follow the way 
we have presented already, which can be regarded as a confirmation of 
the appropriateness of the empirically fitted shell-model interaction used in \citep{Smirnova2010,Smirnova2012}. 
The tensor component also accounts for nearly 
half of the reduction of the $N=20$ shell gap 
(i.e. $d_{3/2}$-$f_{7/2}$ gap) when going from $Z=14$ to 8. 
These behaviors are in accordance with what the $V_{\rm MU}$ 
interaction gives \citep{Otsuka2010b} (see the left panel of Fig.~\ref{fig:N20_spe}).

\subsection{Fujita-Miyazawa three-body force and the shell evolution}
\label{subsec:FM3NF}  
 
\begin{figure}[t]    
\begin{center}
\includegraphics[width=6.0cm,clip=]{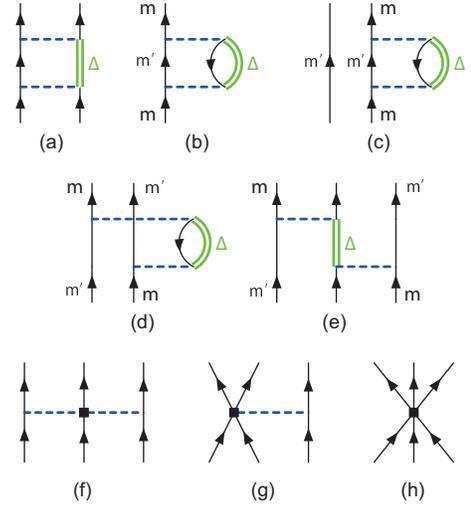}
\end{center}
\caption{Processes involved in the discussion of 3N forces and their
contributions to the monopole components of the effective
interactions between two valence neutrons. The solid lines denote
nucleons, the dashed lines denote $\pi$-mesons, and the thick lines denote
$\Delta$ excitations. Nucleon-hole lines are indicated by downward
arrows. The leading $\chi$EFT 3N forces include the long-range
two-$\pi$-exchange parts, diagram~(f), which take into account the
excitation to a $\Delta$ and other resonances, plus shorter-range
one-$\pi$ exchange, diagram~(g), and 3N contact interactions,
diagram~(h).  
From \cite{Otsuka2010a}. 
\label{3NF_diagrams}}  
\end{figure}

We now turn to 
three-nucleon forces (3NF), shedding light on their contributions to the shell evolution.
Three-nucleon forces were introduced in the pioneering work of
~\citet{FM}  (FM).  One of the main sources of 3NF is the fact that nucleons are
composite particles. In fact, the FM 3N mechanism is due to one nucleon virtually exciting
a second nucleon to the $\Delta$(1232 MeV) resonance, which is
de-excited by scattering off a third nucleon, see Fig.~\ref{3NF_diagrams}(e).

The quantitative role of FM 3N interactions has been pointed out in ab initio calculations for $A \leq 12$ by  
the Green Function Monte Carlo (GFMC) method  \cite{GFMC1,GFMC2,Pudliner1997} and by the No-Core Shell Model (NCSM) method \cite{Navratil2000a,Navratil2000b,Navratil2007}.  These works have been reviewed in \cite{Carlson2014} and in \cite{Barrett2013}, respectively.
Three-nucleon interactions arise naturally also in the chiral effective field
theory ($\chi$EFT) (see a review in \cite{Hammer2013}) as will be discussed in the next subsection.

We here focus on the monopole effect from the FM 3NF with the actual example of the oxygen anomaly
\cite{Otsuka2010a}.
We sketch first the mechanism for the monopole effect presented in \cite{Otsuka2010a}.
Figure~\ref{3NF_diagrams}~(a) depicts the leading contribution
to $NN$ forces due to $\Delta$-resonance excitation, induced by the
exchange of $\pi$-mesons between nucleons. Because this is a second-order
perturbation approach, its contribution to the monopole interaction is attractive. 
The same process changes the SPE of the state $j, m$, as illustrated 
in Fig.~\ref{3NF_diagrams}~(b), by the
$\Delta$--nucleon-hole loop where the initial nucleon in the state $j, m$ is virtually
excited to another state $j', m'$. This lowers the
energy of the state $j, m$. However, if another nucleon of the same kind occupies 
the state $j', m'$ as shown in Fig.~\ref{3NF_diagrams}~(c), 
this process is forbidden by the Pauli exclusion principle. 
The corresponding contribution must be subtracted from the SPE change. 
This is taken into account by the inclusion of the exchange diagram shown in 
Fig.~\ref{3NF_diagrams}~(d), where the nucleons in the intermediate state
are exchanged and this leads to the exchange of the final (or
initial) labels $j, m$ and $j', m'$. Because this process
reflects a cancellation of the lowering of the SPE, the contribution
from Fig.~\ref{3NF_diagrams}~(d) has to be repulsive.
Finally, we can rewrite Fig.~\ref{3NF_diagrams}~(d) as the FM 3N force of
Fig.~\ref{3NF_diagrams}~(e), where the middle nucleon is summed over all
nucleons in the core.
We thus obtain robustly repulsive monopole interactions between the valence nucleons
originating in the FM 3NF.   It is clear that only the monopole component is produced 
by this particular process, without touching on multipole components.

\begin{figure}[t]    
\begin{center}
\includegraphics[width=8cm,clip=]{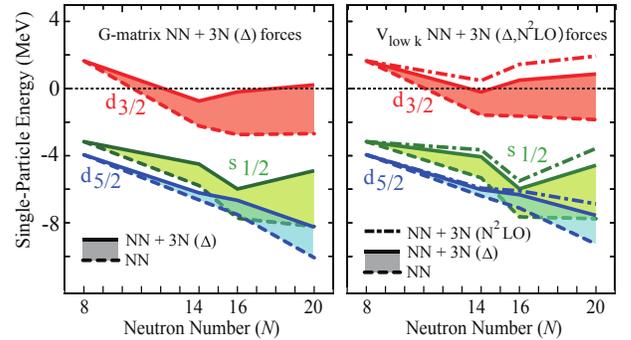} 
\end{center}
\caption{ESPE of neutron 1$d_{5/2}$, 2$s_{1/2}$
and 1$d_{3/2}$ orbitals measured from the energy of $^{16}$O as a
function of $N$. 
The ESPEs calculated (left) from a $G$ matrix and (right) from
low-momentum interactions V$_{{\rm low}\,k}$ are shown. 
The changes due to 3N forces based on $\Delta$ excitations are highlighted by the shaded areas.  
Based on \cite{Otsuka2010a}.
\label{3NF_Fig2}}
\end{figure}

\begin{figure*}[t]    
\begin{center}
\includegraphics[width=17.8cm,clip=]{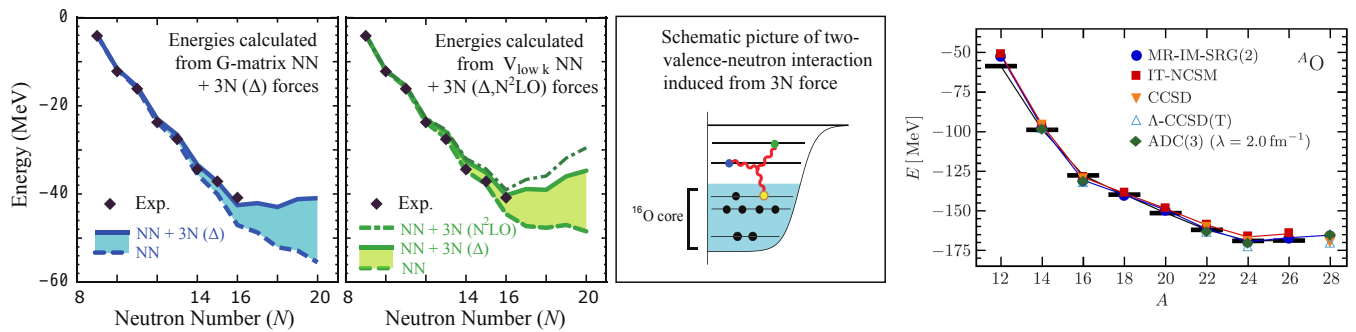}
\end{center}
\caption{(Left, 2nd left)  Ground state energies of oxygen isotopes including processes shown in (2nd right).  Based on \cite{Otsuka2010a}. (right) The ground state energies calculated in several $\chi$EFT approaches\cite{Hergert2016}.  
Reprinted with permission from \cite{Hergert2016}.  
\label{3NF_Fig4}}
\end{figure*}

Figure~\ref{3NF_Fig2} shows, as an example, neutron ESPEs of the oxygen isotopes  
starting from the stable $^{16}$O to heavier ones with more neutrons. 
The ESPEs calculated \cite{Otsuka2010a} with $NN$ interactions in the $G$ matrix
formalism~\cite{HJensen1995}.  A similar result with $\chi$EFT forces will be discussed in the next subsection.
The $d_{3/2}$ ESPE decreases rapidly as
neutrons occupy the $d_{5/2}$ orbit, and remains well-bound from
$N=14$ on. This leads to bound oxygen isotopes out to $N=20$ and puts
the neutron drip-line incorrectly beyond $^{28}$O.  

The changes in the ESPE evolution due to the addition of FM 3NF are included 
in the left panel of Fig.~\ref{3NF_Fig2}.  
The repulsive FM 3N contributions become significant with increasing $N$.
Figure~\ref{fig:VMU-1}(g,h) indicates that monopole 
components are modified to be more repulsive from G-matrix to
SDPF-M in the $sd$ shell, except for the case with $j$=$j'$=$d_{3/2}$.
Since SDPF-M reproduces the experimental data rather well, this general trend 
seems to suggest that a good fraction of the effects of the FM 3NF, and perhaps other 3NFs in general, 
are included empirically in shell-model interactions.
It was argued ~\cite{Zuker2003,Zuker2005} that effective $NN$ interaction was nearly perfect  
and any deviation suggested by experiment should be due to some three-body force.  

The ground-state energies of oxygen isotopes are shown in Fig.~\ref{3NF_Fig4}, where 
the 3NF changes them to be very close to experimental values and places the dripline 
correctly.  
Figure~\ref{3NF_Fig2} shows the key role of the FM 3NF for new magic numbers
$N=14$ between the 1$d_{5/2}$ and 2$s_{1/2}$ orbits~\cite{Stanoiu2004}, 
and $N=16$ between the 2$s_{1/2}$ 
and 1$d_{3/2}$ orbits~\cite{Ozawa2000,Hofmann2008,Kanungo2009}.

\subsection{Ab-initio approaches to nuclear structure}
\label{subsec:ab initio}   


We discuss ab-initio approaches to the nuclear structure in this subsection.  As there have been many activities on this topic recently, a devoted review is needed, and we mainly discuss certain recent outcomes related to the shell and structure evolutions in exotic nuclei.   
Quite naturally, few-body systems have been studied in ab-initio ways as reviewed in \cite{Leidemann2012}.
The GFMC \cite{Carlson2014,GFMC1,GFMC2,Pudliner1997} and the NCSM \cite{Barrett2013,Navratil2000a,Navratil2000b,Navratil2007} calculations were started around the year 2000, showing that the structure of light nuclei (up to $A \sim 10$) can be described well from the nucleon-nucleon forces (2NF) determined by the nucleon-nucleon scattering combined with the 3NF appropriately determined.  
In the mean time, the $\chi$EFT ~\cite{Kolck1994,Epelbaum2002} was developed to construct  nuclear forces in a systematic expansion from leading to successively higher orders
~\cite{Entem2003,chiral1,Epelbaum2009}, which are visualized by diagrams showing 
nucleons interacting via $\pi$ exchanges and shorter-range contact terms
(see a review \cite{Machleidt2011}).
The interactions from the $\chi$EFT are modified to be applicable to low-momentum phenomena by using the low-momentum interactions V$_{{\rm low}\,k}$ \cite{Bogner2003} or by the similarity renormalization method (SRG) \cite{Bogner2007b}.    


The right panel of Fig.~\ref{3NF_Fig2} displays the ESPE calculated 
from chiral low-momentum interactions V$_{{\rm low}\,k}$ 
including the changes
due to the leading (N$^2$LO) 3N forces in $\chi$EFT~\cite{Kolck1994,Epelbaum2002},
(see Fig.~\ref{3NF_diagrams}~(f)--(h)), as well as due to $\Delta$
excitations~\cite{Bogner2009}.   The second left panel of Fig.~\ref{3NF_Fig4} shows the 
ground-state energy of oxygen isotopes calculated with these interactions, depicting a
good agreement with experiment \cite{Otsuka2010a}. 

  
A similar shell evolution is seen in exotic Ca isotopes, where the inclusion of 3NF effects
raises ESPE's of the $pf$-shell neutron orbits  \cite{Holt2012,Otsuka2013}.

\begin{figure}[t]    
\begin{center}
\includegraphics[width=6.5cm,clip=]{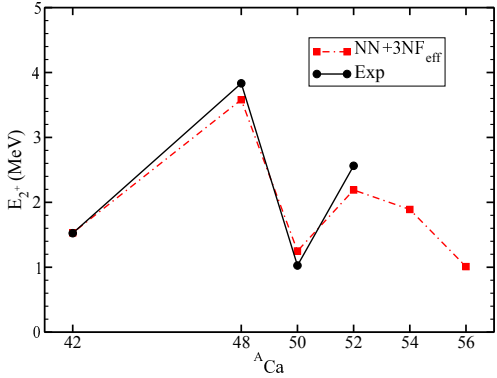}
\end{center}
\caption{ 2$^+_1$ level of Ca isotopes calculated by the CC method.
Reprinted with permission from \cite{Hagen2012b}.}
\label{HagenCa_fig}
\end{figure}

The Coupled-Cluster (CC) calculations \cite{Hagen2008,Hagen2009,Hagen2010} started with the 2NF obtained as the N$^3$LO $\chi$EFT interaction.  The N$^2$LO 3NF was included in the CC calculation \cite{Hagen2012a,Hagen2012b} for O and Ca isotopes, with results consistent with those mentioned just above.  Figure~\ref{HagenCa_fig} shows the 2$^+_1$ level of Ca isotopes calculated by the CC method \cite{Hagen2012b}, showing results consistent with the shell evolution in Ca isotopes discussed in \ref{subsubsect:34}, including the $^{54}$Ca 2$^+$ level (see Fig.~\ref{fig:54Ca_levels}).
  
The 3NF is converted into an effective 2NF by the normal ordering combined with a reference state, which is Fermi gas or a Hartree-Fock state.   The In-Medium SRG (IM-SRG) was introduced and developed in \cite{Tsukiyama2011,Tsukiyama2012,Hergert2013a,Hergert2013b,Hergert2014} (see a review in \cite{Hergert2016}) so as to renormalize in-medium effects into effective interactions.  

A frequently used interaction (called A for brevity) has been introduced \cite{Hebeler2011} by  the SRG transformation of the N$^3$LO 2NF of \cite{Entem2003} with the cut-off parameter 500 MeV/c combined with the N$^2$LO 3NF where the parameters c$_D$ and c$_E$ (shown in Fig.~\ref{3NF_diagrams} (g,h), respectively) are fitted to the triton binding energy and the $^4$He charge radius.  This set A interaction was shown to produce larger radii of proton distribution by 
the CC calculations \cite{Hagen2016a}.  Since then, this interaction has been used in many works; for magic nuclei \cite{Hagen2016b}, for $sd$-shell nuclei \cite{Simonis2016}, and for density saturation in finite nuclei \cite{Simonis2017}.   
The CC calculations show larger charge radii of heavy Ca isotopes, being consistent with recent measurement made up to $^{52}$Ca \cite{GarciaRuiz2016}. 

\begin{figure}[t]    
\begin{center}
\includegraphics[width=7cm,clip=]{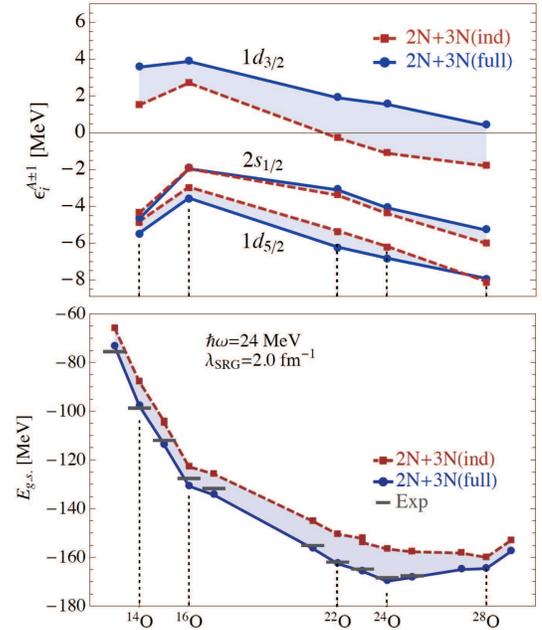}
\end{center}
\caption{(Upper panel) ESPEs of neutrons calculated \cite{Cipollone2013} at sub-shell closures of oxygen isotopes.  
(Lower panel) Similarly calculated ground-state energies compared to experiment (bars).  
Reprinted with permission from \cite{Cipollone2013}.
}
\label{Cipollone_fig}
\end{figure}

There is another frequently used interaction (called B for brevity) introduced by \cite{Roth2012}, where 
the 3NF is different from the set A in a local form with the cut-off parameter 400 MeV/c.  
This set B interaction has been used in \cite{Binder2013,Binder2014,Tichal2014,Hergert2014} for ground-state properties of Ca, Ni, Sn {\it etc}.  The Self-Consistent Green's Function theory also provided ground-state energies \cite{Soma2011,Soma2014}, and furthermore, the ESPEs \cite{Cipollone2013,Cipollone2015} as shown in Fig.~\ref{Cipollone_fig}, which indicates results consistent with those shown in the previous subsection.  We point out that the ESPE in \cite{Cipollone2013,Cipollone2015}, based on the formulation by \citet{Baranger1970}, is consistent with the ESPE discussed in this article, as illustrated in \ref{subsec:Baranger}.   

\begin{figure}[t]    
\begin{center}
\includegraphics[width=7.5cm,clip=]{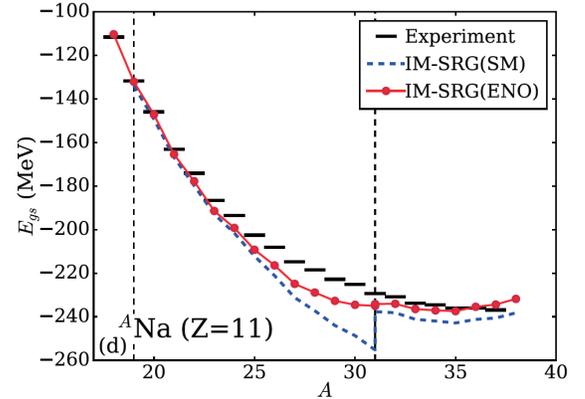}
\end{center}
\caption{Ground-state energies of Na isotopes calculated with IM-SRG (SM).  The IM-SRG (SM) curves use a core reference, while the curves labeled IM-SRG (ENO) use an ensemble reference.
Reprinted with permission from \cite{Stroberg2017}.
}
\label{Stroberg_fig}
\end{figure}

The procedures with the sets A and B can be summarized:\\
(1) The Hamiltonian consisting of N$^3$LO 2NF and N$^2$LO 3NF is obtained from the $\chi$EFT.
For set B, the values of the parameters $c_D$ and $c_E$ are fitted to the triton and $^4$He properties by performing a few-body calculation. \\
(2) Short-range correlations are processed by the SRG, being truncated up to three-nucleon terms.
These are 2NF and 3NF for set A, with $c_D$ and $c_E$ fitted in the same way at this stage.\\
(3) HF calculation is carried out with 2NF and 3NF thus derived/fitted as the reference state(s).\\
(4) The Hamiltonian is truncated up to two-nucleon terms by the normal-ordering with the reference state(s).\\
(5) With such two-nucleon interactions, the CC, IM-SRG, MBPT, {\it etc.} are carried out.\\

\begin{figure}[t]    
\begin{center}
\includegraphics[width=7.5cm,clip=]{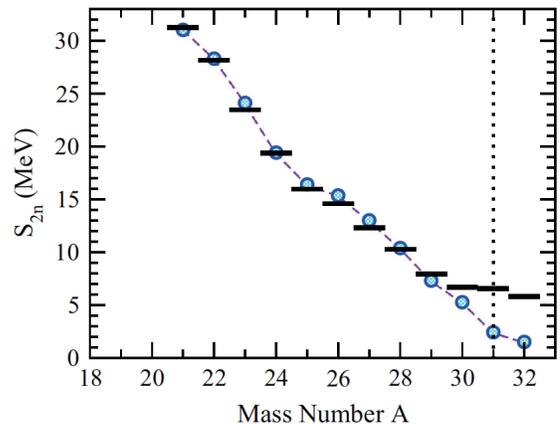}
\end{center}
\caption{ Results of \cite{Simonis2017}.  The two-neutron separation energies of Na isotopes calculated with IM-SRG.
Reprinted with permission from \cite{Simonis2017}.
}
\label{Simonis_fig}
\end{figure}

The right panel of Fig.~\ref{3NF_Fig4} shows that ab-initio calculations based on the $\chi$EFT reproduce the ground state energies of oxygen isotopes well \cite{Hergert2016}, being consistent with other works in the left panels.  In going to proton-neutron open-shell nuclei, 
further developments are made to obtain the shell-model interactions so that their eigenvalues are calculated.
A shell-model interaction has been calculated in \cite{Lisetskiy2008} based on the NCSM.   With the IM-SRG
\cite{Stroberg2016,Stroberg2017,Simonis2017}, the reference state was improved so that two reference states are considered with the ensemble normal ordering (ENO) in going through an open shell taking a weighted average.  Figure~\ref{Stroberg_fig} and ~\ref{Simonis_fig} display, respectively, the ground-state energies \cite{Stroberg2017} and the two-neutron separation energies  \cite{Simonis2017} of Na isotopes.
The agreement with experiment was improved, with certain differences between the two calculations.  
The difference is mainly due to the different interactions sets A and B.  In the latter, the experimental values are reproduced up to $N \sim$ 16, but some deviations arise over the neutron magic number 20 probably because of substantial mixings of intruder configurations.  

It is worth mentioning that the radius is predicted often too small in ab-initio calculations, but this problem was
avoided by the so-called N$^2$LO$_{sat}$ interaction where the parameters are taken only up to the N$^2$LO
being fitted to properties of heavier nuclei such as $^{14}$C and $^{16,23,24,25}$O \cite{Ekstrom2015}. 

Despite these significant improvements in ab-initio approaches in general, the discrepancy with experiment remains at present.  For instance, the extra binding due to intruder configurations may not be 
reproduced well, as discussed for Figs.~\ref{Stroberg_fig} and \ref{Simonis_fig}.  On the other hand, this is one of the most crucial features of exotic nuclei, as emphasized also in the next section.
As a possible breakthrough, the Extended-Kuo-Krenciglowa (EKK) method has been proposed and developed \cite{Takayanagi2011a,Takayanagi2011b,Tsunoda2014}.  The EKK method is one of the Many-Body Perturbation Theories (MBPT).  The other MBPT calculations have a possibility of divergence when applied to two or more major shells, but the EKK method is free from this difficulty.  As two major shells merge or the shell gap becomes smaller in exotic nuclei rather often, it is crucial to include two or more shells properly.  
 
\begin{figure}[t]    
\begin{center}
\includegraphics[width=8.5cm,clip=]{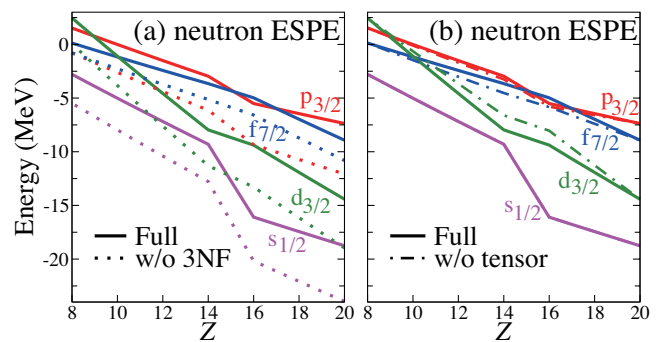}
\end{center}
\caption{
 ESPEs of $N$=20 isotones for neutrons obtained in the normal filling scheme.
 Solid (dotted) lines in (a) show the case with (without) three-nucleon forces, while  the  
 solid (dot-dashed) lines in (b) represent the case with (without) the tensor component.
From \cite{Tsunoda2017}.
}
\label{EKK_Fig5}
\end{figure}

The EEdf1 interaction was obtained for the $sd-pf$ shell from a $\chi$EFT $NN$ interaction at N$^3$LO with the EKK treatment of in-medium effects and from the FM 3NF (Sec. \ref{subsec:FM3NF}) \cite{Tsunoda2014}.
Figure~\ref{EKK_Fig5} shows ESPE calculated from the EEdf1 interaction, for $N$=20 isotones as a function of $Z$.
Figure~\ref{EKK_Fig5} (a) shows the ESPEs obtained by the full calculation and those obtained after 
removing the FM 3NF.   One finds that this 3NF shifts the SPEs upwards, and that the shifts become larger as $Z$ increases.    
Figure~\ref{EKK_Fig5} (b) depicts the ESPEs obtained by the full calculation and those obtained after removing the tensor component from the EEdf1 interaction.
Although the magnitude of the tensor-force effects 
is smaller than that of the 3NF as a whole, the tensor force effects are not monotonic and produce more 
rapid changes in the shell structure in contrast to the 3NF effects.
We note that the tensor component is quite minor in the effective $NN$ interaction originating in the FM 3NF.
In those calculations, although the one-body SPEs are fitted at certain nuclei, 
the evolution of the ESPEs is given by the interaction thus derived, and the resulting changes  as a function of $Z$ or $N$ have nothing to do with the fit.  In this sense,  Figure~\ref{EKK_Fig5} (a,b) confirms the shell evolution at $N$=20, appearing consistent with earlier results to be 
discussed in the next section.    
Some results of the EEdf1 interaction will be presented in the next section.

\section{Examples of structural change manifested in experimental observables}
\label{sec:actual}

We discuss in this section how theoretical results are confronted with a variety of experimental measurements.   
In such cases, both shell-evolution effects and other many-body correlations arise and can mutually affect each other.  For the examples, we explain the mechanisms of shell evolution at play.

\subsection{Measuring the key indicators of shell evolution in the island of inversion  \label{subsec:key}} 

Since short-lived ``exotic'' nuclei cannot be made into targets, measurements of
their 
properties have to start from an ion beam which is subjected to an in-beam
measurement in inverse kinematics, implanted into an active or passive stopper
to observe its decay, or manipulated for ion trapping or laser spectroscopic
approaches, for example.  

The first challenge of any experiment with short-lived, ``exotic'' nuclei is
their production. Today, a broad range of rare isotopes is available for
experiments in the form of ion beams. Two main production and separation mechanisms
have 
emerged as the workhorse techniques in rare-isotope beam production and are
employed in nuclear physics laboratories around the world: 
\begin{itemize}
\item Beams of short-lived nuclei are produced and separated {\it in-flight} and are directly used for experiments (in-flight
  separation).   
\item Exotic nuclei are produced and thermalized in a thick target, extracted,
  ionized, transported or reaccelerated (isotope
  separation on-line -- ISOL).
\end{itemize}

The production strategies for rare-isotope beams and the different types of
rare-isotope facilities around the 
world were recently reviewed~\citep{Blumenfeld2013}. 

\begin{figure} [tb]    
\includegraphics[width=8.5cm] {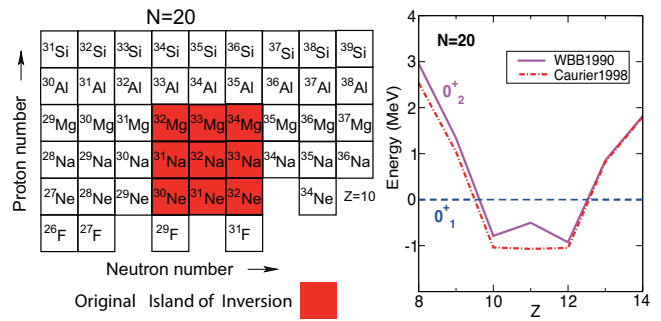} 
\caption {(Left) Part of the Segr\`e chart around the IoI.  
Red area indicates its original picture as of \citep{Warburton1990}. 
By now, the IoI has extended much further (see the text).  
(Right) Relative energy of the lowest normal and lowest neutron 2p-2h intruder 0$^+$  states, resulting from diagonalizing in the separate subspace based on \cite{Warburton1990} and \cite{Caurier1998}.
}
\label{fig:island}
\end{figure}

In this subsection, we use the example of the ``Island of Inversion'' (IoI) centered around \nuc{32}{Na} (see Fig.~\ref{fig:island}), in order to describe 
how typical observables 
are measured and interpreted as indicators of structure changes. 

\subsubsection{Sketch of the Island of Inversion \label{subsubsec:island}}

We first sketch the IoI mainly from the shell-evolution viewpoint,   
and will be brief because dedicated reviews exist \citep{Caurier05}. 
The IoI was named by \citep{Warburton1990}, after earlier experimental studies had reported 
various anomalous features, for example, \cite{Thibault1975} followed by \cite{Huber1978, Detraz1979, Guillemaud-Mueller1984}.  
It is characterized by deformation-related neutron particle-hole excitations from the $sd$ shell into the $pf$ shell across the $N=20$ shell gap.
Such particle-hole excitations across a shell gap are often referred to as ``intruder configurations'', which can be energetically favored over the normal configurations and dominate the ground states of the nuclei in the IoI, as shown in the right panel of Fig.~\ref{fig:island}.  States comprised mainly of intruder configurations are called ``intruder states'' or ``intruders''.  
Most of the binding-energy gains are due to the deformation from a sphere to an ellipsoid.   Thus, an intruder at zero or low excitation energy implies shape coexistence with states based on spherical normal-order configurations, which is seen in many exotic nuclei.

\begin{figure}[tb]    
\begin{center}
\includegraphics[width=8.5cm,clip=]{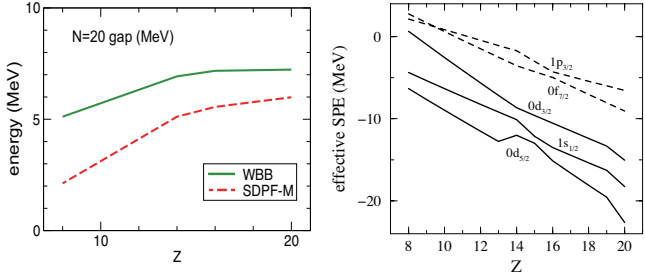}
\caption{(Left) (Green solid line) $N$=20 shell gap from \citep{Warburton1990} and (red dashed line) from the sdpf-m interaction \cite{Utsuno1999}.  
 (Right) ESPEs of N=20 isotones for neutrons obtained in the normal filling scheme
 from SDPF-M interaction.
 From \citet{Utsuno1999}.
}v
\label{fig:N20_intruder}
\end{center}
\end{figure}

\begin{figure}[tb]    
\begin{center}
\includegraphics[width=8.5cm,clip=]{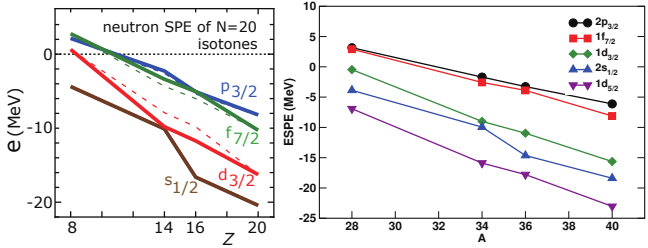}
\caption{
 ESPEs of N=20 isotones for neutrons obtained in the normal filling scheme
 (left) from $V_{\rm MU}$ interaction and (right) from SDPF-NR interaction.
From \citet{Otsuka2010b} (left) and reprinted with permission from \citet{Caurier05} (right). 
}
\label{fig:N20_spe}
\end{center}
\end{figure}

Early theoretical studies 
also indicated that the ground states can be deformed for nuclei in the IoI, such as 
a deformed Hartree-Fock solution for $^{31}$Na in \cite{Campi1975}, 
and intruder shell-model ground states despite the rather constant $N$=20 gap in \cite{Poves1987}.  
Regarding the shell evolution, so-called ``modified single-particle energy'' was introduced in \cite{Storm1983}, 
corresponding to the present ESPE in the special case of a (sub)shell closure $\pm$ one particle (see the texts around eq.~(\ref{eq:epj-c})). 
Although the monopole interaction was not mentioned, this may be the first appearance of the ESPE.
The changes of the neutron shell structure was then discussed in \cite{Storm1983} with some differences from the current  picture.  

The left panel of Fig.~\ref{fig:N20_intruder} displays the $N$=20 shell gap of $N$=20 isotones obtained
from \cite{Warburton1990}, with values $>$ 5 MeV.   
The right panel of Fig.~\ref{fig:N20_intruder} presents ESPEs calculated from the SDPF-M interaction
\cite{Utsuno1999}, and the resulting $N$=20 gap is included in the left panel 
of Fig.~\ref{fig:N20_intruder}.     
The gaps now vary more, and become as low as 2 MeV for $Z$=8.  A quite similar evolution of the ESPEs of the $d_{3/2}$ and $f_{7/2}$ orbits were obtained in \cite{Fukunishi1992}, where large-scale shell-model calculations made successful predictions.   The $d_{3/2}$ ESPE changes more steeply with the SDPF-M interaction, however.  This is because \cite{Fukunishi1992} uses the USD interaction where some change was made from G-matrix \cite{Kuo1967}.  This change appeared to be rather inappropriate \cite{magic}, and was removed in the SDPF-M interaction, 
resulting in a better description.  This is an example of the importance of nuclear forces to the shell evolution.           
This $N$=20 gap reduction was schematically shown earlier in \cite{Heyde1991} in terms of the proton-neutron monopole interaction of a $\delta$-function interaction, while the obtained pattern is too monotonic partly due to missing tensor force.  
The intruder states stay higher towards $Z$=8 in \cite{Caurier1998}, as exhibited in the right panel of Fig.~\ref{fig:island}.  
Thus, although the breakdown of the $N$=20 magicity in the IoI was commonly accepted, in the 1990's, the vanishing of the $N$=20 gap towards $Z$=8 was suggested (in a quantitative way) rather uniquely in \cite{Fukunishi1992,Utsuno1999}.  The situation is changed now, and other calculations also suggest a similar reduction,  (see Figs.~\ref{fig:N20_spe} and~\ref{EKK_Fig5}),   
as a trend with more realistic interactions, particularly with the tensor force.
Note that such reduction of the gap facilitates more particle-hole excitations, which can enhance quadrupole deformation and pairing correlations.
Thus, anomalous features around $N$=20 have been intensely studied, providing a strong motivation to clarify, both experimentally and theoretically, how the gap evolution occurs and what consequences arise.   
We here refer to other related works from mean-field or clustering viewpoints \cite{Campi1975,Peru2000,Rodriguez2000,Kimura2007,Peru2014,Ren1996,Terasaki1997,Reinhard1999,
Stoitsov2000,Stevenson2002,Yao2011,Hinohara2011}, some of which have been or will be discussed concretely.    
Those anomalous features are still very much contemporary subjects, as we shall see also below.   

\subsubsection{Masses and separation energies}

The mass of a nucleus is among the most basic properties directly
accessible to measurements. Masses and derived quantities, {\it e.g.} one- and two-nucleon
separation energies, frequently provide the first hints for the evolution of
shell structure and signal the onset of deformation. 

Experimental methods for the determination of atomic masses basically fall into
two broad categories. Approaches that measure the $Q$ values 
in decays or reactions make use of Einstein's mass-energy equivalence; mass
measurements that are based on the deflection of ions in electromagnetic fields
determine the mass-to-charge ratio. The most precise mass spectrometry
is accomplished through frequency measurements~\citep{Myers2013}. The cyclotron
or revolution  
frequencies of ions in a magnetic field are measured to determine the
mass-to-charge ratio in a Penning trap~\citep{Blaum2006} or in a storage
ring~\citep{Franzke2008}.  
A recent example for a mass measurement at the northern boundary of
the IoI is found out to $A=34$~\citep{Kwiatkowski2015} from Penning-trap mass
spectrometry at TITAN facility~\citep{Dilling2006}. 

\begin{figure} [tbh]    
\includegraphics[width=7cm] {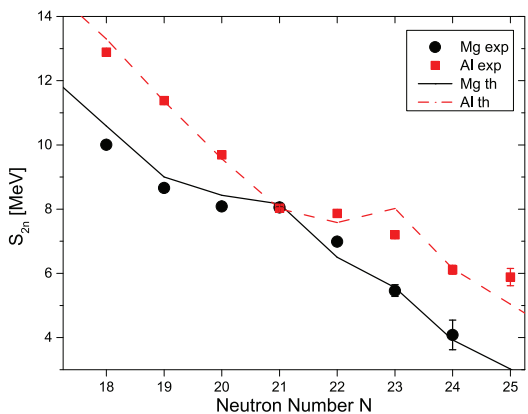}
\caption {Symbols indicate two-neutron separation energies for the Mg and Al isotopic chains from
  the 2015 TITAN experiment and the mass compilation by
  \citet{Audi2012}.   
Shell-model calculations in the $sd$-$pf$ shell ~\citep{Caurier2014} are shown also 
by the solid and dashed lines.
Reprinted with permission from \citet{Kwiatkowski2015}. 
}
\label{fig:island_mass}
\end{figure}

The two-neutron separation energies ($S_{2n}$ values) for the Al and Mg isotopic
chains are shown in 
Fig.~\ref{fig:island_mass} with overlaid shell-model calculations in the
$sd$-$pf$ model space using the SDPF-U-MIX interaction ~\citep{Caurier2014}. 
Typically, along an isotopic chain, the two-neutron separation energy, 
$S_{2n}$, decreases steadily towards the neutron dripline. 
We remind the reader that, in the presence of a large spherical shell gap at N=20, the $S_{2n}$ values would drop at $N$=21 when the shells above the gap start to be filled.
The flattening in the trend for $N=19-21$ in the Mg chain is contradictory to this, and indicates the increased correlation energy of these
deformed nuclei relative to their neighbors with two neutrons less. In Al, a
hint of this effect only appears beyond $N=21$, putting \nuc{31-34}{Al} outside
and \nuc{35-37}{Al} at the very
boundary if not inside the IoI. Of interest is the unique
crossing of $S_{2n}$ in 
the Mg and Al isotopic chains at \nuc{34}{Al}, which -- in comparison to shell
model -- is attributed to Mg significantly gaining
correlation energy upon entrance into the IoI
between $N=20$ and 21, while the $S_{2n}$ in the Al chain is still on its almost
linear downward trend up to $N=22$.         

\subsubsection{Magnetic Dipole and Electric Quadrupole moments}



The deviation from sphericity of nucleus that has non-zero spin can be
quantified through its electric quadrupole moment.
The electric quadrupole moment was measured for the ground state of Al isotopes
at the LISE spectrometer at GANIL~\citep{DeRydt2009}, 
with spectroscopic quadrupole moment $|Q_s|$ extracted for
\nuc{31,33}{Al} relative to \nuc{27}{Al}~\citep{Heylen2016}. 


   
\begin{figure} [tbh]    
\includegraphics[width=7cm] {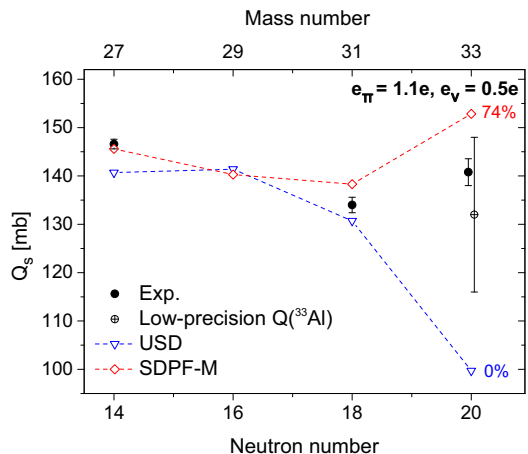}
\caption{Measured spectroscopic quadrupole moments of Al isotopes compared to
  shell-model calculations limited to the neutron $sd$ shell (USD) and allowing
  for particle-hole excitations across $N=20$ (SDPF-M). The given percentage
  signifies the amount of ground-state intruder configurations in the respective
  shell-model approach. 
Reprinted with permission from \citet{Heylen2016}. 
}
\label{fig:island_Q}
\end{figure}

The implications for the structure of \nuc{33}{Al} are shown in
Fig.~\ref{fig:island_Q}. The improved uncertainty of
$|Q_s(\nuc{33}{Al})|$ 
compared to that of the previous measurement \citep{Shimada2012}
led to argue the presence of neutron 
intruder configurations in comparison to shell-model calculations that are
restricted to the $sd$ shell only (USD) and that allow for neutron intruder configurations 
across the $N=20$ shell gap (SDPF-M) \citep{Utsuno1999,Utsuno2004}. 
It is noted that these conclusions
contradict the ones from the mass measurements reviewed above, where
\nuc{33}{Al} was placed outside of the IoI and they are at odds with
shell-model calculations using the SDPF-U-MIX effective interaction~\citep{Caurier2014} that also
allows for neutrons in the $pf$ shell. This may highlight the
different levels of detail probed, with the moment measurement more sensitive to
the very details of the configurations, or point to a puzzle in our
understanding of \nuc{33}{Al} at the northern border of the IoI.  
Spectroscopic data on $^{33}$Al, obtained for example using direct reactions, 
may identify the energies of intruder states, assessing in a complementary way 
the degree of intruder admixtures to the low-lying level structure of this nucleus.


Measuring hyperfine structure using laser spectroscopy is a powerful method to 
unambiguously determine the spin and magnetic moment of the ground state. 
A good example applied to the IoI is \nuc{31}{Mg}, whose ground state was assigned to be 
$1/2^+$ \cite{Neyens2005}.
This measurement clearly shows that the $N=19$ nucleus $^{31}$Mg belongs to the IoI, 
because the normal state, dominated by neutron $1d_{3/2}^{-1}$, must have $J^{\pi}$=$3/2^+$. 
The spectroscopy of $^{31,33}$Mg and their particle-hole structure were reviewed \citep{Neyens2011}, indicating a variety of intruders coexist at low-excitation energies.
The properties of low-lying states of odd-$A$ nuclei, including their spin/parities, can thus be related to the shell evolution, sometimes, up to the gap between two major shells. 

\begin{figure}[tb]    
\begin{center}
\includegraphics[width=8.5cm,clip=]{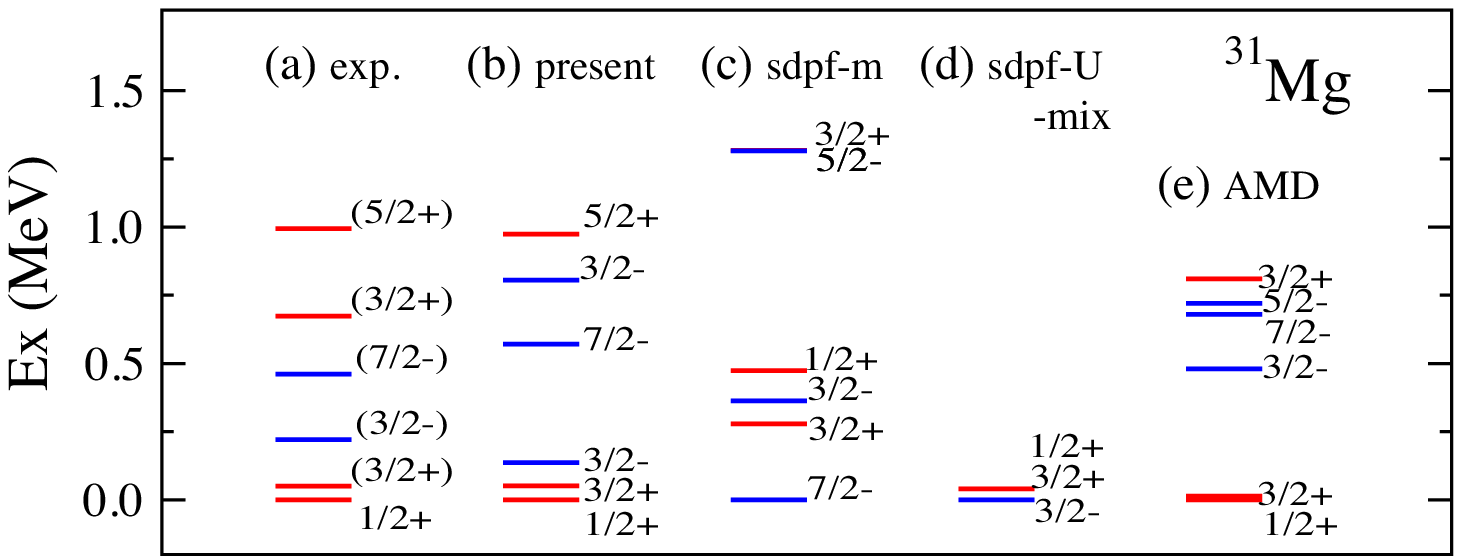}
\end{center}
\caption{
 Energy levels of \nuc{31}{Mg}. (a) experimental values, (b) EEdf1 \cite{Tsunoda2017}, (c) 
 SDPF-M~\cite{Utsuno1999},
 (d) SDPF-U-MIX~\cite{Caurier2014} and (e) AMD+GCM calculation~\cite{Kimura2007}, respectively.
From \citet{Tsunoda2017}.
\label{EKK_31Mg_Fig4}}
\end{figure}

\subsubsection{Excitation energy}

Energies of excited nuclear states are often among the first quantities
accessible in experiments \cite{Gade2015}. They can be measured directly and without any
model dependence and are thus some of the key observables that can be tracked to
unravel changes in the nuclear structure. 
For instance, the systematics of the lowest 2$^+$ energies was discussed in Sec.~\ref{sec:introduction} as one of the indicators of the magic structure (Fig.~\ref{fig:ex2+}). 
For excited states
below the nucleon separation energies, prompt or delayed $\gamma$-ray
spectroscopy is frequently used to extract excitation energies of rare isotopes
with great precision, measured from the spectroscopy of the $\gamma$-ray
transitions that connect different states. Electric monopole transitions between
$0^+$ states~\citep{Wood1999}, of $E0$ character, proceed to a large
extent through conversion electron emission and electron spectroscopy or other
charged-particle 
spectroscopy techniques, e.g. in transfer reactions, are required \cite{GadeLiddick2016}. 
Excited states can be populated in nuclear 
reactions~\citep{Gade2008a} or $\beta$ decay~\citep{Rubio2009}, exploiting the
selectivities inherent to the different population mechanisms. 
For instance, the coexistence of normal and intruder states in 
$^{29}$Na was found through $\beta$-delayed $\gamma$-ray spectroscopy 
\citep{Tripathi2005}.
The energies of very long-lived isomeric states can be accessed, for example, with
Penning-trap~\citep{Block2008} or storage-ring~\citep{Reed2010} mass
spectrometry. For states that are unbound with respect to neutron or proton
emission, excited-state energies can be deduced from invariant mass or missing
mass spectroscopy. The spectroscopy of bound~\citep{Gade2015} and
unbound~\citep{Baumann2012} excited states was reviewed recently.

\begin{figure} [tb]    
\includegraphics[width=8.5cm] {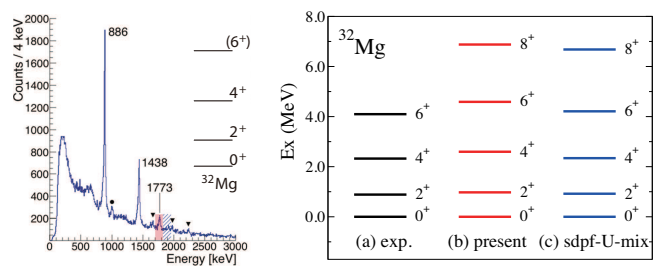}  
\caption {(Left) Prompt $\gamma$-ray spectrum detected with GRETINA in coincidence with
  the \nuc{32}{Mg} projectile-like fragmentation residues identified in the S800
spectrograph. Reprinted with permission from \citet{Crawford2016}.
(Right) Comparison to theoretical calculations with EEdf1 and SDPF-U-MIX interactions. From \citep{Tsunoda2017}.}
\label{fig:Mg32_spectrum}
\end{figure}

The most recent spectroscopy inside the $N=20$ IoI addressed one
of the hallmark nuclei in this region of shell evolution, \nuc{32}{Mg}, that has
been subject to experimental study since its low-lying $2^+_1$ energy
contradicted the presence of the $N=20$ magic number in this isotopic
chain~\citep{Detraz1979}. Using the advanced $\gamma$-ray tracking array
GRETINA~\citep{Paschalis2013},  
excited states in \nuc{32}{Mg} ~\citep{Crawford2016} were populated in the
secondary fragmentation of an \nuc{46}{Ar} rare-isotope beam at NSCL.
The $\gamma$ rays spectrum is displayed in
Fig.~\ref{fig:Mg32_spectrum}. Aside from the previously known $\gamma$-ray
transitions at 885 keV and 1438 keV that are attributed to the $2^+_1
\rightarrow 0^+_1$ and $4^+_1 \rightarrow 2^+_1$ transitions, respectively, a
new transition at 1773~keV was observed that is proposed to connect the $6^+_1$
and $4^+_1$ states~\citep{Crawford2016}. With \nuc{32}{Mg} suspected to be
well-deformed, this would establish the lowest part of the yrast rotational
band.  
Figure~\ref{fig:Mg32_spectrum} shows good agreement with shell model calculation with the SDPF-U-MIX interaction 
as well as that with the EEdf1 interaction which is of {\it ab initio} type.  

\subsubsection{Electromagnetic transition strength}

Nuclear structure can be probed experimentally in quantitative ways by a variety
of nuclear reactions that are selective to specific degrees of freedom. Inelastic
scattering, in particular Coulomb excitation, of nuclei has long been used to
investigate collective degrees of freedom that involve the coherent motion of
many nucleons. $B(\sigma\lambda)$ reduced electromagnetic transition matrix
elements are extracted from measured cross sections to quantify the degree of
collectivity~\citep{Alder1956,Cline1986,Glasmacher1998}. Reduced electromagnetic
transition strength can alternatively be 
deduced from excited-state lifetime measurement, extracted from Doppler
energy shifts or lineshapes in $\gamma$-ray spectroscopy~\citep{Dewald2012}.    

\begin{figure}[tb]    
\begin{center}
\includegraphics[width=8.6cm,clip=]{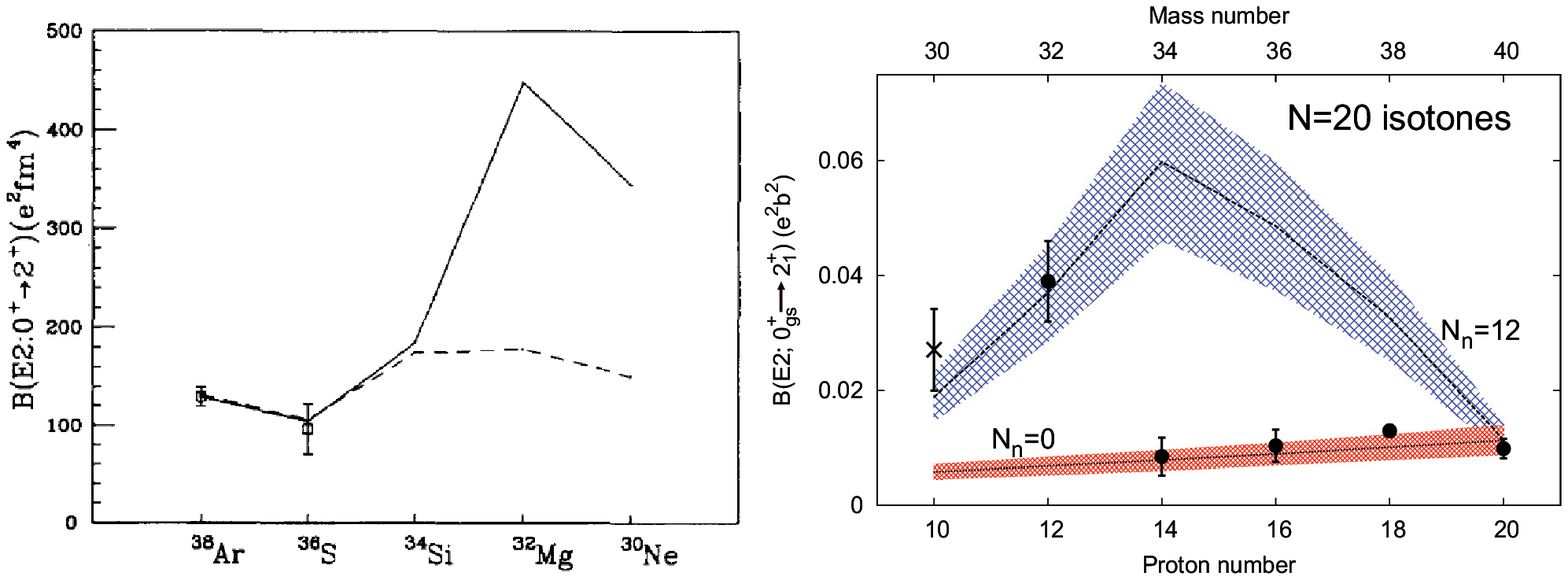}
\caption {$B(E2)$ value of the $N=20$
isotones plotted as function of $Z$.  
(Left) Shell-model calculation by \cite{Fukunishi1992} where the solid line includes neutron
excitations across the $N=20$ gap but the dashed line does not.
(Right) 
The measured values
are confronted with the $N_pN_n$
scheme calculation~\citep{Casten1993} for 
$N_n=0$ ($N=20$ closed) and $N_n=12$ ($sd$ and lower $pf$ shells
 combined). See the text for details.  
From \citet{Fukunishi1992} (left) and adapted with permission from \citep{Doornenbal2016} (right). 
}
\label{fig:island_transition}
\end{center}
\end{figure}

At collision energies below the Coulomb barrier, the excitation
probabilities and interaction times are large enough to allow for
multistep excitations and the determination of quadrupole moments and their
signs, giving a glimpse at the degree and the character of
deformation~\citep{Cline1986}. In the  
regime of intermediate-energy or relativistic projectile 
energies, multistep processes are suppressed by several orders of
magnitude. This greatly simplifies the analysis of the 
resulting excitation spectra, and the $B(E2;0^+_1 \rightarrow 2^+_1)$ value 
has been measured for \nuc{32}{Mg} \cite{Motobayashi1995}, establishing the strong deformation 
of this nucleus, for instance, a prediction by \cite{Fukunishi1992} 
shown in the left panel of Fig.~\ref{fig:island_transition}.
The higher-lying states of collective bands, on the other hand, 
remain out of reach with this technique in typical experiments lasting
a few days with beam rates of a few per second~\citep{Glasmacher1998,Gade2008a}. Excited-state lifetime
measurements on the other hand do not require nuclear models to extract
transition strengths but can suffer from observed and unobserved feeding from
higher-lying states depending on the population mechanism of the excited
states. 

In a recent inelastic scattering experiment at RIBF in RIKEN 
\citep{Nakamura2017}, the quadrupole
collectivity or deformation of \nuc{36}{Mg} and \nuc{30}{Ne} was determined from measured
$0^+_1 \rightarrow 2^+_1$ excitation cross sections.   
The beams of \nuc{30}{Ne} and \nuc{36}{Mg}
impinged upon Pb and C targets.   
Inelastic scattering off C and relativistic
Coulomb excitation on a Pb target revealed a $B(E2)$ value and deformation
length, respectively, that indicates a quadrupole deformation parameter of
$\beta_2 \approx 0.5$ for both, showing that the quadrupole deformation in the
Mg chain persists towards the neutron dripline and that neutron excitations 
across $N=20$ are critical for reproducing the collectivity of $N=20$
\nuc{30}{Ne}~\citep{Doornenbal2016}. The telltale nature of the reduced $B(E2;0^+_1
\rightarrow 2^+_1)$ value as nuclear structure observable is illustrated 
in the right panel of
Fig.~\ref{fig:island_transition}, where the $B(E2)$ strength of the $N=20$
isotones is plotted as function of $Z$. 
 The measured values show good agreement with the earlier shell-model prediction \cite{Fukunishi1992}.  Note that the order is inverted between the left and right
panels.  In addition, the measured values
are confronted with the phenomenology of the $N_pN_n$ scheme~\citep{Casten1993}
for $N_n=0$ ($N=20$ shell closure intact and no valence neutrons) and $N_n=12$
($sd$ shell+$f_{7/2}$+$p_{3/2}$ combined as the neutron shell)). The sharp onset of
collectivity for $Z \leq 12$ is consistent with the picture of dominant neutron
particle-hole excitations across the $N=20$ shell gap for the Mg and Ne $N=20$
isotones, a hallmark of the IoI, at least for its northern boundary.    
When moving from $Z$=12 to 14, the proton quadrupole collectivity is likely 
reduced due to the closure of the 1$d_{5/2}$ orbit, and the $N$=20 shell gap becomes wider
(Figs.\ref{fig:N20_intruder}, \ref{fig:N20_spe}, \ref{EKK_Fig5}).   

\subsubsection{Shape coexistence in the Island of Inversion and at its boundaries: additional evidences from $\beta$-decay and $E0$ transition}

Outside the IoI, excited intruder states   
can coexist with the still spherical ground
states~\citep{GadeLiddick2016}. So far, shape-coexisting $0^+$ states have been
identified in 
\nuc{34}{Si}~\citep{Rotaru2012} and \nuc{30}{Mg}~\citep{Sch2009}. 
Along the line of the $N=20$ isotones, \nuc{34}{Si} is situated at the northern
boundary of the IoI.  In a pioneering measurement at
GANIL, the $\beta$ decay of the $1^+$ isomer of \nuc{34}{Al} was used to
selectively feed $0^+$ states in \nuc{34}{Si}, including the previously
unobserved excited $0^+_2$
state at 2719(3)~keV~\citep{Rotaru2012}. This state is located below the $2^+_1$,
presenting an experimental challenge. Since $\gamma$-ray decays between $0^+$ states are
angular-momentum forbidden, this low-lying $0^+$ state can only de-excite via
electron conversion or internal pair  
formation, where an electron-positron $e^+e^-$ pair is released with a total 
energy of $E_{e^-} + E_{e^+} = E(0^+_2) - 2 \times 511$ keV. 
From the difference timing between the
$\beta$-decay events and the $e^+e^-$ pair signals, a half-life of 19.4(7)~ns
was determined for the 
$0^+_2$ state~\citep{Rotaru2012}. The resulting low $E0$ transition strength indicates
only weak mixing between the $0^+_1$  ground state and the $0^+_2$ excited
state. Combining all spectroscopic 
information, including $B(E2;2^+_1 \rightarrow 0^+_2)=61(40)e^2$fm$^4$, as
extracted from a small $\gamma$-ray branch and the $2^+_1$  lifetime, results in a
quadrupole deformation parameter for the $0^+_2$ state of $\beta=0.29(4)$, in agreement
with SDPF-U-MIX shell-model calculations~\citep{Rotaru2012}. 
All these properties are consistent with the argument presented at the end of
the previous subsection.  Once the ground state becomes closed-shell like, the shape coexistence often arises \cite{heyde2011}. 

For \nuc{32}{Mg}, at the heart of the IoI, a $(t,p)$
neutron-pair transfer reaction was used in reverse kinematics 
to -- for the first time -- identify the $0^+_2$ state in
\nuc{32}{Mg} at 1058(2)~keV at the REX-ISOLDE facility (CERN) ~\citep{Wimmer2010,Bildstein2012}. 
The proton angular distributions of both states were shown to display the shape of an
angular-momentum transfer of  $\Delta L=0$ onto the ground state of
\nuc{30}{Mg}. 
It was thus concluded that both states in \nuc{32}{Mg} populated in the $(t,p)$ transfer have spin 0. 

Coincident $\gamma$-ray transitions detected, a new transition with an energy of 172~keV and the well-known $2^+_1
\rightarrow 0^+_1$ transition at 886~keV, allowed to put the newly-discovered excited
$0^+$ state at the more precise energy of 1058(2)~keV. Based on DWBA
calculations, it was concluded that the ground state is comprised of
$(f_{7/2})^2$ and $(p_{3/2})^2$ intruder configurations and the excited $0^+$
state could be largely described with the assumption of $sd$-shell normal-order
configurations, such as $(d_{5/2})^2$, however, with a
small $(p_{3/2})^2$ intruder contribution necessary~\citep{Wimmer2010}. These findings support the 
picture of a deformed $fp$-shell intruder ground state and an $sd$-shell
dominated (spherical) first excited 
$0^+$ state. The approximately equal cross sections for the formation of the two
$0^+$ states in $(t,p)$ were used to infer significant mixing between the two
states. A measurement of the electric monopole strength connecting the two state
remains a challenge for future experiments. The $0^+_2$ excitation energy of about 1~MeV was found significantly
below available model predictions at the time~\citep{Wimmer2010}. 
These properties of the $0^+_2$ state of $^{32}$Mg pose a formidable challenge 
for theory including beyond-mean-field models \citep{Rodriguez2000,Peru2014}. 

Recently, shell-model calculations that allow for the mixing of
configurations that have 2, 4 and 6 neutrons promoted across the $N=20$ shell
gap (SDPF-U-MIX) reproduce the reported, low
$0^+_2$ energy and suggest a rather unique character of this $0^+$
state~\citep{Caurier2014}. A ground-state neutron configuration of 9\% 0p-0h, 54\% 2p-2h, 35\%
4p-4h, and 1\% 6p-6h emerges and suggests a mixture of deformed and
superdeformed configurations.  
The excited $0^+$ 
state is calculated to be comprised of 33\% 0p-0h, 12\% 2p-2h, 54\% 4p-4h, and 1\% 6p-6h
neutron particle-hole configurations, painting a rather complex picture of
\nuc{32}{Mg} where the second $0^+$ state carries significant 
spherical as well as superdeformed configurations, 
rendering the simple concept of a deformed ground state and a spherical excited $0^+$ as
too simplistic. The confirmation and further  
characterization of the very interesting $0^+_2$ state of \nuc{32}{Mg} appears
warranted to clarify the important phenomenon of shape coexistence inside the $N=20$
IoI.  

 Seemingly contradictory conclusions to what was inferred by \citet{Wimmer2010}, termed the
\nuc{32}{Mg} puzzle, were drawn from a simple two-level mixing
model~\citep{For11,For12} and resolved recently using a three-level mixing
approach~\citep{Macchiavelli2016}, in line with a more complicated structure
that has been suggested by the shell-model calculations mentioned above.    

\subsubsection{Spectroscopy of the nuclear wave function through direct reactions}
\label{subsubsec:direct}

Direct nuclear reactions have proven to be a vital tool for the spectroscopy of
the single-particle components in the nuclear wave function,
showing direct relevance to the probing of the shell evolution. In a glancing
collision of a projectile and a target nucleus, one or a few nucleons are
transferred directly without formation of an intermediate compound system.

The classic low-energy transfer reactions, that for stable target nuclei use a
variety of light projectiles to probe occupied single-particle levels and
valence states~\citep{Macfarlane1960}, e.g., the $(d,p)$ neutron-adding and the $(d,\nuc{3}{He})$ proton-removing transfers, are now employed at low-energy rare-isotope facilities in inverse kinematics when low-emittance, high-intensity rare-isotope
beams are available (see
\citep{Gaudefroy2006,Wimmer2010,Catford2010,Kanungo2010,Fernandez2011,Burgunder2014} for
examples from different facilities). At intermediate beam energies ($\sim$100~MeV/u), thick-target
$\gamma$-ray tagged one- and two-nucleon knockout reactions on \nuc{9}{Be} or
\nuc{12}{C} 
targets have  been developed into spectroscopic tools to study single-nucleon
hole states and correlations of two like nucleons in exotic
nuclei~\citep{Hansen2003,Gade2008b,Bazin2003,Tostevin2004,Yoneda2006,Simpson2009a,Simpson2009b}. 

By comparing cross sections with C and Pb targets, 
it is also possible to extract Coulomb reaction cross sections, which are used to look into neutron shell structure through the halo formation in \nuc{31}{Ne} and \nuc{37}{Mg} 
\citep{Nakamura2009, Nakamura2014, Kobayashi2014}.

At the high beam energies, typically exceeding 70~MeV/nucleon, a theoretical 
description \citep{Tostevin1999} in the framework of eikonal trajectories  and sudden
approximation is applicable. Therefore, the model 
dependence is limited as compared to the classical low-energy transfer
reactions, whose description involves the Distorted Wave Born Approximation
(DWBA) or higher-order formalisms, that depend strongly on entrance- and
exit-channel optical model potentials~\citep{Kramer1988}, which have not been
established yet for nuclei with extreme neutron-to-proton ratios. It was shown
recently that low-energy transfer reactions and nucleon removal reactions can be
analyzed to give consistent results~\citep{Mutschler2016a}. Both knockout and transfer reactions have been used to
track the descent of intruder states along isotopic chains approaching the
IoI. Two complementary examples are reviewed in the
following. 

\begin{figure} [tb]    
\includegraphics[width=6.5cm] {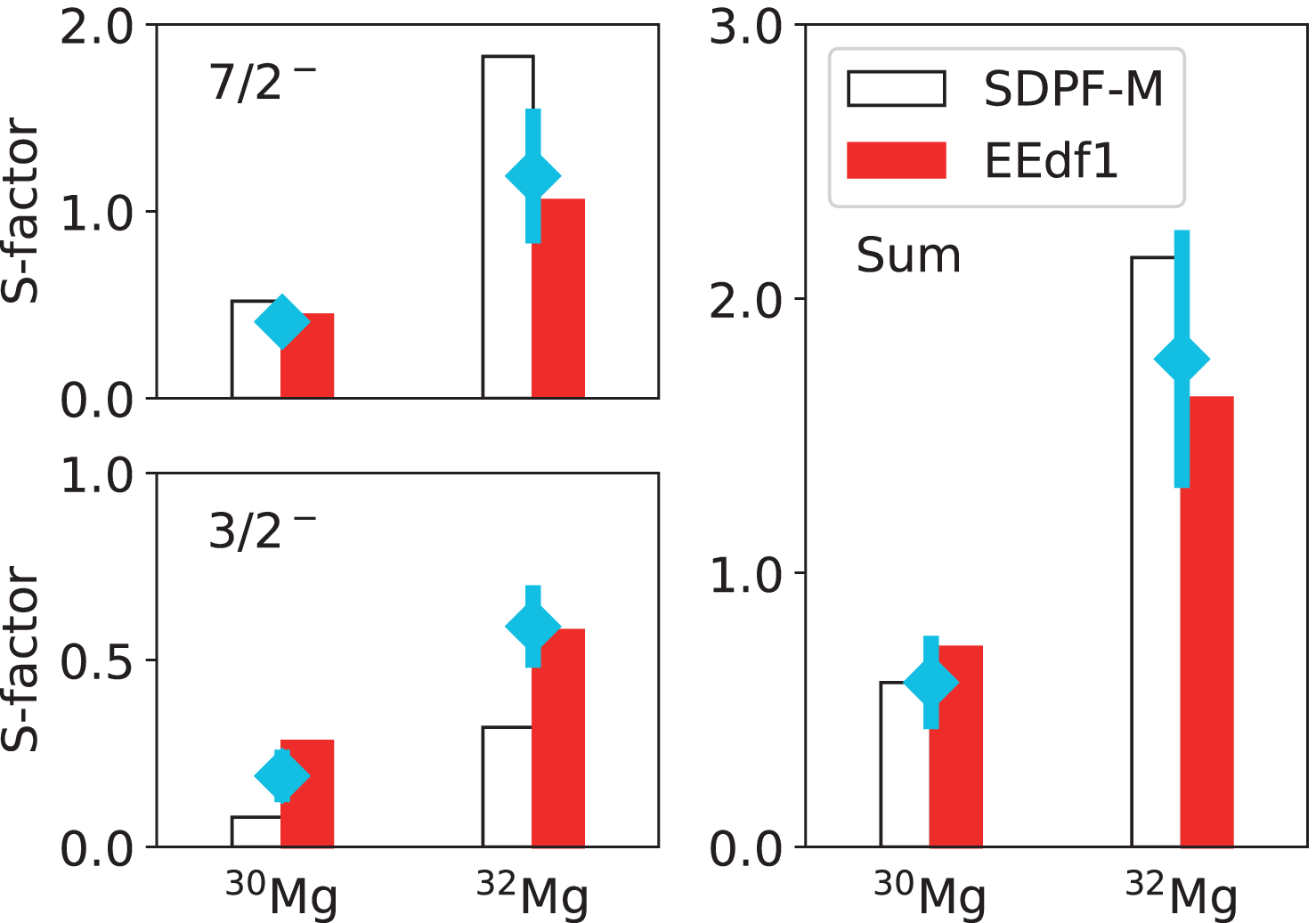}
\caption {Spectroscopic factors from \nuc{9}{Be}(\nuc{32,30}{Mg},\nuc{31,29}{Mg})X knockout reactions to 
two negative-parity states of \nuc{32,30}{Mg}.
Deduced spectroscopic factors are indicated by blue point with error bar~\citep{Terry2008}.  
Single-particle occupancies obtained from SDPF-M shell model \cite{Utsuno1999} are represented by 
blanc histograms.
Spectroscopic factors calculated with the EEdf1 interaction are shown by red histograms \cite{Tsunoda2018}. 
}
\label{fig:Mg_intruder}
\end{figure}

The onset of $pf$ shell intruder configurations was quantified along
the Mg chain with $\gamma$-ray tagged one-neutron removal measurements,
\nuc{9}{Be}(\nuc{30}{Mg},\nuc{29}{Mg}+$\gamma$)X and
\nuc{9}{Be}(\nuc{32}{Mg},\nuc{31}{Mg}+$\gamma$)X, performed at NSCL
 ~\citep{Terry2008}. From the shapes of the \nuc{29,31}{Mg} parallel momentum
distributions
gated on the individual $\gamma$-ray transitions, 
the 1.095 and 1.431~MeV  states in \nuc{29}{Mg}
and the 0.221  and 0.461~MeV levels in \nuc{31}{Mg} were shown to be of $\ell=1$
and $\ell=3$ orbital angular momentum, respectively, identifying $p_{3/2}$  and
$f_{7/2}$ single-neutron configurations in the ground states of both
\nuc{30}{Mg} and \nuc{32}{Mg}.
From the partial cross sections for the
population of the negative-parity states in the knockout residues, $f_{7/2}$ and $p_{3/2}$ spectroscopic
factors were deduced.  
The resulting quantification of the onset of $f$ and $p$
intruder configurations in the ground states of \nuc{30}{Mg} and \nuc{32}{Mg} is
seen in Fig.~\ref{fig:Mg_intruder}: the neutron $pf$-shell strengths
increase significantly at $N=20$, signaling a dramatic shift in the nuclear
structure of \nuc{32}{Mg} as compared to \nuc{30}{Mg}.  
The spectroscopic factors calculated with EEdf1 interaction \cite{Tsunoda2018,Tsunoda2017} show a good agreement with experiment, when this calculation includes a reduction factor of 0.75 that is inherent to knockout reactions \cite{Tostevin2014}.  The occupation numbers obtained with the SDPF-M interaction~\cite{Utsuno1999} 
were used in the analysis of \citep{Terry2008} as shown in Fig.~\ref{fig:Mg_intruder}, depicting a similar trend.    
Compared to the SDPF-M interaction, the EEdf1 interaction gives a better  
description for the energy levels for \nuc{32,31}{Mg} in Figs.~\ref{fig:Mg32_spectrum} and \ref{EKK_31Mg_Fig4}, respectively, as well as for the spectroscopic factors in Fig.~\ref{fig:Mg_intruder}.  The latter illustrates the amount of the excitations across the $N$=20 magic gap.  We stress that the shell evolution through this  EEdf1 interaction (see Fig.~\ref{EKK_Fig5}) exhibits similarities to earlier results shown in Figs.~\ref{fig:N20_intruder} and \ref{fig:N20_spe}.  

\begin{figure} [tb]    
\includegraphics[width=7cm] {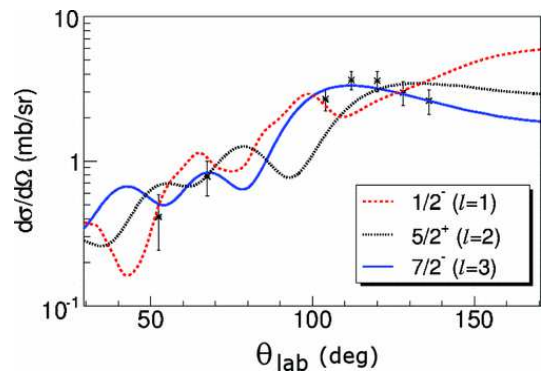} 
\caption {Proton angular distribution for the new neutron-unbound state
  discovered at 1.74~MeV in \nuc{27}{Ne}. In comparison to reaction theory,
  $\ell=3$ orbital angular momentum was assigned~\citep{Brown2012}. 
Reprinted with permission from \citet{Brown2012}.
}
\label{fig:Ne_intruder}
\end{figure}

In the Ne isotopic chain, a $\gamma$-ray tagged neutron-adding transfer
reaction, $d$(\nuc{26}{Ne},\nuc{27}{Ne}+$\gamma$)$p$, 
performed at GANIL,
identified for the
first time the (neutron unbound) $7/2^-_1$ state at 1.74(9)~MeV in \nuc{27}{Ne}~\citep{Brown2012}. 
The $\ell=3$ orbital angular momentum of the state was concluded from the proton
angular distribution in comparison to ADWA transfer reaction calculations
(see Figure~\ref{fig:Ne_intruder}).
        
The $3/2^-$ state could be identified at 0.765~MeV, confirming earlier
work that could only restrict the orbital angular momentum of this state to
$\ell=0,1$~\citep{Terry2006}. The fact that the $7/2^-$ state is higher in
energy than the $3/2^-$ level presents a remarkable inversion from the ordering
closer to stability and disagrees with the sequence predicted by the SDPF-M
Hamiltonian~\citep{Brown2012}. This result will serve as an important benchmark
for new effective shell-model Hamiltonians in the region in their quest to
describe the shell evolution in and around the IoI.

\subsubsection{More on direct reactions: Tracking single-particle strengths to learn about the spin-orbit force}
\label{subsubsec:spin-orbit}

The spin-orbit splitting is a corner stone of the nuclear shell model. 
Recent work using inverse-kinematics transfer reactions~\citep{Burgunder2014} and one-proton knockout
reactions~\citep{Mutschler2016b} on the key nucleus \nuc{34}{Si}, located 
at the boundary of the island of
inversion, explored the signatures and evolution of the spin-orbit splitting in
neutron-rich nuclei. 

\begin{figure}[tb]    
\includegraphics[width=8.8cm]{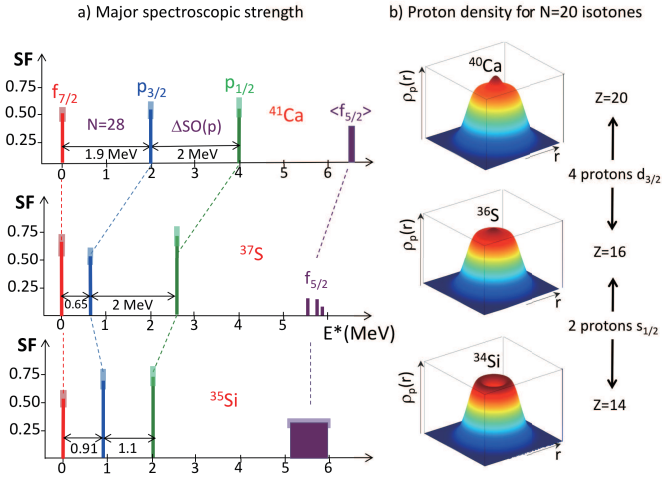}
\caption{(a) Evidence for a reduction of the
  2$p_{3/2}$-2$p_{1/2}$ spin-orbit splitting in the $N=21$ isotonic chain at
  \nuc{35}{Si}. For comparison, the spin-orbit splitting remains unchanged
  between \nuc{41}{Ca} and \nuc{37}{S}. 
Reprinted with permission from \citet{Burgunder2014}.
(b) Change of the proton density along the $N=20$
  isotone line from density functional theory (relativistic mean-field with the
  DDME2 interaction). The vanishing proton occupation of the $s_{1/2}$ orbital
  leads to a central depletion in the density that has been likened to a ``bubble'' (Figure by O. Sorlin, J. P. Ebran).
}
\label{fig:so_reduction} 
\end{figure}

At GANIL,  
the single-particle nature of states in \nuc{37}{S} and \nuc{35}{Si} and the
associated spectroscopic strengths were obtained for the first time by inverse-kinematics $(d,p)$ 
reactions~\citep{Burgunder2014}. 
In comparison to reaction theory, 
the proton angular distributions were measured
(i) to assign $\ell$ values for the transferred neutrons from their shape and
(ii) to extract spectroscopic factors from their absolute scale. By tracking the
location of the dominant 2$p_{1/2}$ and 2$p_{3/2}$ fragments, it was reported  
that the spin-orbit splitting between the 2$p_{3/2}$ and 2$p_{1/2}$ 
neutron orbits decreases by 25\% in \nuc{35}{Si} relative to the less exotic isotone
\nuc{37}{S}, while almost no change was found for the 
neutron 1$f_{7/2}$ - 1$f_{5/2}$ spin-orbit
splitting (Figure~\ref{fig:so_reduction} (a))~\citep{Burgunder2014}.  
We can understand this feature as explained below.   
The major difference from \nuc{35}{Si} to \nuc{37}{S} is the occupancy of the proton 
2$s_{1/2}$ orbit, which has a large effect on the 2$p_{3/2}$-2$p_{1/2}$ splitting due to the 2b-$LS$ 
force (see Sec.~\ref{subsec:2-body LS}).  
On the other hand, the 2$s_{1/2}$ occupancy has a weak (vanishing) effect on the 1$f_{7/2}$-1$f_{5/2}$ splitting
due to the 2b-LS (tensor) force (see the third item of the remarks in Sec.~\ref{subsubsec:tensor analytic}). 

Further studies on the neutron 2$p_{3/2}$-2$p_{1/2}$ splitting of the same nuclei has been made recently \citep{Kay2017}, where the change of this splitting was interpreted in terms of loose binding effects. 
It, however, can be described in terms of the monopole effect of the 2b-LS force, as described 
in 
Appendix~\ref{app:p-split}.  
Further studies are of great interest.

Electron scattering off stable nuclei demonstrated that their central
densities are saturated, as for a liquid drop, for example. In rare isotopes at
the extreme of isospin, the possibility of a depleted central density, or a
``bubble'' structure, has been discussed for more than 40 years. If observed, 
it will be of much interest. 
In general, central depletions will arise from the reduced
occupation of low-$\ell$ single-particle orbits, as exemplified in Fig.~\ref{fig:so_reduction} (b) 
for the $N=20$ isotones \nuc{40}{Ca}, \nuc{36}{S}, and
\nuc{34}{Si} with calculated proton density distributions from
a relativistic mean-field functional (DDME2). The central depletion in the
proton density for \nuc{34}{Si} is attributed to a vanishing occupancy of the
proton 2$s_{1/2}$ orbital. A one-proton knockout
measurement from a \nuc{34}{Si} projectile beam at NSCL, 
combined with in-beam $\gamma$-ray spectroscopy using GRETINA, 
revealed indeed that the proton
2$s_{1/2}$ orbital in this nucleus is depleted, possibly leading to a depleted
central proton density or ``bubble'' inside of neutron-rich \nuc{34}{Si}, making
this the best candidate for this phenomenon to date~\citep{Mutschler2016b}. 
In knockout
reactions, the shape of the parallel momentum distributions of the knockout
residues is sensitive to the $\ell$-value of the removed
nucleon and the partial cross sections for the population of individual final
states can be used to extract spectroscopic factors in comparison to reaction
theory~\citep{Hansen2003}. 
With this approach, the cross section for the removal of an $\ell=0$ proton from \nuc{34}{Si} was found to be only 10\% of that for the proton
removal from \nuc{36}{S}~\citep{Mutschler2016a,Mutschler2016b}. Since the cross
section for the removal of protons 
from an orbit is proportional to the orbit's proton occupancy, 
this difference in cross section was interpreted as evidence for a depleted
2$s_{1/2}$ proton orbital in \nuc{34}{Si}, in striking contrast to the same
orbital being fully occupied in the \nuc{36}{S}
isotone~\citep{Khan1985,Mutschler2016a}. 

\subsubsection{At the southern border: continuum and shell evolution \\ 
\;\;\;\; - cases with multi-nucleon transfer reaction -}
\label{subsubsec:southern border}

On the nuclear chart, two protons south of the island-of-inversion nucleus
\nuc{30}{Ne} lies \nuc{28}{O}. The $N=20$ nucleus \nuc{28}{O} has been suspected
to be unbound with respect to neutron decay based on cross section or
yield systematics established in its attempted production in the
fragmentation of intermediate-energy \nuc{36}{S} and \nuc{40}{Ar} 
beams at GANIL and RIKEN, respectively~\citep{Tarasov1997,Sakurai1999}. 
The neutron-rich oxygen isotopes at the southern border of the island of
inversion have been a formidable testing ground for nuclear theory,  
where the particularly visible feature is that \nuc{24}{O} is the last bound oxygen isotope,
while the fluorine isotopes with just one more proton exist out to
at least mass number $A=31$, as sometimes called the ``oxygen anomaly'' in~\citep{Otsuka2010a}.
Shell-model approaches~\citep{Volya2005,Otsuka2010a,Tsukiyama2015},
mean-field theory~\citep{Erler2012,Co2012} and ab-initio type
calculations~\citep{Hagen2009,Hagen2010,Duguet2012,Cipollone2013,Bogner2014,Simonis2016}
have been made in the quest for new physics in nuclei near driplines. 
The incorporation of the continuum is an ongoing effort in the development of many-body approaches.

The nucleus \nuc{26}{O} is a unique three-body system since it was found to be
barely unbound, only able to decay by two-neutron emission with an energy of
less than 20 keV~\citep{Kondo2016}. Two early measurements at NSCL and GSI provided the
first evidence for the ground-state resonance of \nuc{26}{O} at
$150^{+50}_{-150}$~keV~\citep{Lunderberg2012} and
25$\pm$25~keV~\citep{Caesar2013}, respectively. In
all measurements, the experimental scheme was very similar. 
In kinematically complete
measurements, the energy of decaying resonances was reconstructed in invariant
mass spectroscopy from the momentum vectors of the two emitted neutrons and the
residue in \nuc{24}{O}+$n$+$n$. The highest-statistics
measurement yet was performed at RIBF/RIKEN with the SAMURAI spectrometer 
\citep{Kobayashi2013,Kondo2016,Nakamura2017}. 
From reconstruction of the invariant
mass, the ground state of \nuc{26}{O} was found at only 
18$\pm$3(stat)$\pm$4(syst)~keV above the two-neutron decay
threshold~\citep{Kondo2016}. In addition, a candidate for the excited $2^+_1$
state at 1.28$^{+0.11}_{-0.08}$~MeV was identified for the first time.

\begin{figure}[t]    
\includegraphics[width=8cm]{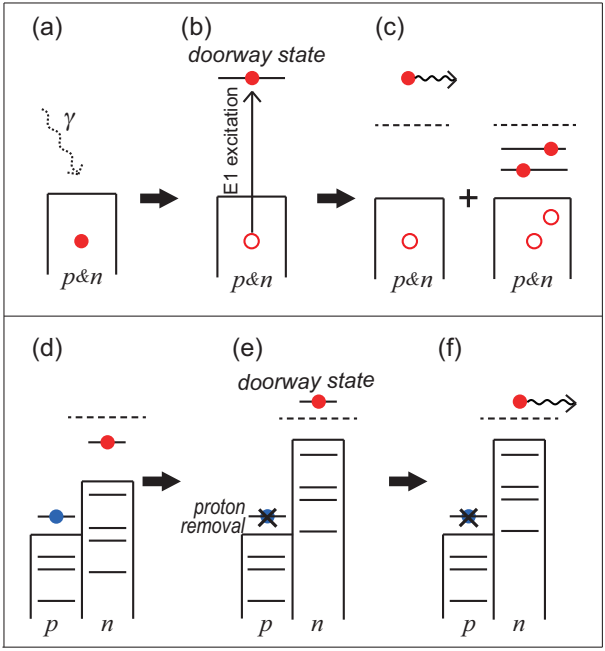}
\caption{Schematic pictures of the doorway state for (a)-(c) E1
 excitation and (d)-(f) a reaction induced by the removal of a
 proton. Dashed lines indicate the neutron threshold. Red filled circles
 indicate the neutron being discussed, while red open circles are
 neutron holes. Blue circles are protons, and crossed blue circles are
 absent after the initial impact of the reaction.  
From~\citep{Tsukiyama2015}. 
}
\label{fig:doorway}
\end{figure}

Regarding the shell evolution, one finds, in \citep{Kondo2016}, ``The structure of $^{26}$O may be influenced by shell evolution, $nn$ correlations, and continuum effects.''   It is, however, not trivial how and what ``resonance'' states can be created in various transfer reactions including those involving heavy ions.  Using Fig.~\ref{fig:doorway}, we shall explain schematically the relation between the shell evolution and the neutron emission after such reactions: panels (a)-(c) exhibit the doorway state in a ($\gamma, n$) process, while panels (d)-(f) depict a similar doorway state due to a sudden removal of a proton by a transfer reaction.  The removal of the proton lifts up neutron ESPEs by the amount of its monopole effect (see panel (d)).  If this single-particle state is in the continuum, it becomes a doorway state as shown in panel (e).  
Its wave function is the same as the corresponding state before the reaction.   
The neutron in the doorway state goes away through one of the continuum state, with the probability given basically by
the squared overlap between the doorway state and such continuum states.  
The shape of the energy spectrum is determined by this probability, 
with the peak shifted by continuum couplings.  Thus, the neutron spectrum indicates the combined effect of the shell evolution and the continuum (see details in \cite{Tsukiyama2015}).   Although actual situations may contain different details, the basic picture is expected to remain.                   
   

A possible long lifetime of the ground-state resonance that would allow for the
term of ``two-neutron radioactivity'' is controversially
discussed~\citep{Kohley2013,Caesar2013,Grigorenko2013,Kondo2016} and remains an
interesting possibility for a new phenomenon beyond the neutron dripline.


\subsection{Neutron halo observed in exotic C isotopes and $N$=16 magic number}
\label{sec:pshell}

Halo nuclei have been identified through their greatly enhanced interaction cross section measured in the bombardment with a variety of targets. With the example of the C isotopes, we discuss in the following the relationship between halo formation and shell evolution.
The SFO-tls \citep{Suzuki2008} Hamiltonian is used, 
while the CK (\citet{Cohen1965}) and MK (\citet{Millener1975}) Hamiltonians were employed earlier. 
The SFO-tls Hamiltonian is designed for $p$-$sd$ shell nuclei with 
the cross-shell tensor and 2b-$LS$ parts taken, respectively, from the $V_{\rm MU}$ interaction (see Sec.\ref{subsec:VMU}) and the M3Y 2b-$LS$ interaction (see  
Secs.~\ref{subsec:2-body LS} and 
S5, 
so as to include shell-evolution effects in a manner quantitatively similar to the results presented so far. 
The $sd$-shell part is improved also 
by taking into account the effects of three-body forces (see Secs.~\ref{subsec:FM3NF} and \ref{subsec:ab initio}).  
Calculations with this Hamiltonian reproduce well the shell evolution in the $^{15}$C-$^{16}$N-$^{17}$O isotones 
including the 5/2$^+$-1/2$^+$ inversion (see Sec.~\ref{subsec:2-body LS}).   

\begin{figure}[tb]    
\includegraphics[width=8.5cm]{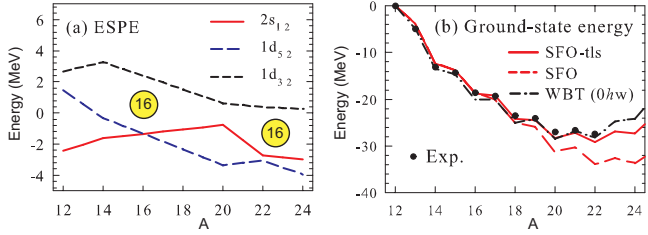} 
\caption{(a) ESPEs for neutron orbits in C isotopes obtained with SFO-tls interaction, 
(b) ground-state energies of C isotopes obtained with SFO-tls, SFO
and WBP as well as experimental data.
In (a), the filling scheme is taken in the order of 1$p_{1/2}$, 1$d_{5/2}$, 2$s_{1/2}$ and 
1$d_{3/2}$, as this order represents rather well the configurations of actual eigenstates.  The ESPEs at their closures are connected.
The $N$=16 (sub-)magic gap is highlighted by the yellowish circle. 
\label{fig:Carb}}
\end{figure}

Figure~\ref{fig:Carb} (a) depicts neutron ESPEs of C isotopes 
obtained from the SFO-tls Hamiltonian in the filling scheme.   
While the 2$s_{1/2}$ orbit is below the 1$d_{5/2}$ orbit in $^{12}$C, 
the 2$s_{1/2}$ ESPE is raised through $A$=20, crossing the 1$d_{5/2}$ orbit. 
This is because the neutron-neutron 1$p_{1/2}$-2$s_{1/2}$ and 1$d_{5/2}$-2$s_{1/2}$ monopole interactions are both repulsive, and push up the 2$s_{1/2}$ orbit     
as neutrons occupy the 1$p_{1/2}$ and 1$d_{5/2}$ orbits.
This disappearance of the gap at $N$=14 in C isotopes around $A$=16 was reported in \cite{Stanoiu2008}.
This shell evolution produces the 1/2$^{+}$ ground state in $^{15}$C, and the 3/2$^{+}$ ground state in $^{17}$C which is natural with dominant neutron 1$d_{5/2}^3$ configuration.  The present irregular variation of the ground-state spin can thus be understood.
Figure~\ref{fig:Carb} (a) indicates that the $N$=16 magic gap appears around $A$=16.  It then disappears around $A$=20 because of the raise of the 2$s_{1/2}$ orbit.  Interestingly the 2$s_{1/2}$ orbit becomes loosely bound.  Because this is an $s$ orbit, a neutron halo occurs in the 1/2$^+$ ground state of $^{19}$C with the $s_{1/2}^1$ $d_{5/2}^4$ neutron configuration, consistently with experiments \citep{Nakamura1999,Kanungo2016}. This shows how the shell evolution is related to the neutron halo formation.  We note that the 2$s_{1/2}$ orbit is raised by a repulsive effect simulating the three-body-force effect as mentioned above.   As the $^{20}$C ground state consists, to a large extent, of the sub-shell closure of the $d_{5/2}$ orbit in the shell-model calculation \cite{Suzuki2016}, no neutron halo is expected there.  

\begin{figure}[tb]    
\includegraphics[width=8cm]{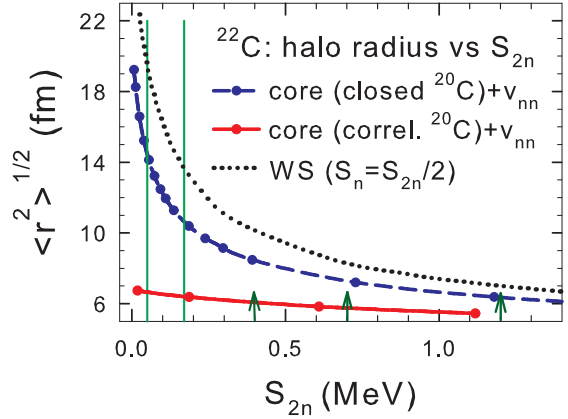} 
\caption{RMS radius of the halo neutron as a function of two-neutron
 separation energy, S$_{2n}$. Blue dashed line and filled circle
 indicate the result obtained with the core of the closed-shell
 $^{20}$C, while red solid line and filled circle the result with the
 core of the correlated $^{20}$C. The result obtained from WS potential
 (S$_n$=S$_{2n}$/2) without v$_{nn}$ is shown by the black dotted
 line. The range of S$_{2n}$ obtained from \cite{ensdf} is shown by
 green thin vertical lines. Green arrows denote values discussed in
 \cite{Kobayashi2012}.  
From \citet{Suzuki2016}.
\label{fig:C22_halo}}
\end{figure}

Figure~\ref{fig:Carb} (a) indicates that the $N$=16 magic number appears again around 
$A$=22, which brings about another interplay between the shell evolution and the 
neutron halo. 
Figure~\ref{fig:Carb} (b) shows the ground-state energies of C isotopes relative to that of $^{12}$C 
for SFO-tls, SFO and WBT \citep{Warburton1992} Hamiltonians in comparison to experiment.  
A repulsive neutron-neutron monopole interaction contained in the SFO-tls interaction pushes up the energy in the neutron-rich region, reproducing the experimental data, similarly to O isotopes discussed in Secs.~\ref{subsec:FM3NF} and \ref{subsec:ab initio}.  Figure~\ref{fig:Carb} (a) shows that the 2$s_{1/2}$ orbit is rather well bound with an ESPE below -2 MeV at $A$=22 in the filling scheme, indicative of a situation opposing a two-neutron halo.  On the other hand, Fig.~\ref{fig:Carb} (b) displays that $^{22}$C is barely bound with respect to $^{20}$C as far as the total binding energy is concerned.  The many-body correlations in $^{22}$C bring about the formation of two-neutron halo, which is unlikely from the viewpoint of the mean potential. The neutron halo of $^{22}$C was reported experimentally in \cite{Tanaka2010, Kobayashi2012, Togano2016}, 
while theoretical studies were performed with three-body models \cite{Horiuchi2006,Yamashita2011,Kucuk2014}.  We report here a rather different approach: The extended shell-model calculation is performed not only by including usual shell-model correlations but also by taking into account the interaction between the halo neutrons taken from the low-energy limit of neutron-neutron scattering \cite{Suzuki2016}.  
Figure~\ref{fig:C22_halo} depicts the radius of the two-neutron halo ($\sim$6-7 fm) consistently with experiment \cite{Togano2016}; the halo radius deduced from the matter radius \cite{Togano2016} appears to be $6.74^{+0.71}_{-0.48}$ fm, which is well below the value obtained for such a small separation energy by the usual simple relation (halo radius$>$10 fm for S$_{2n}<$0.3 MeV)  \cite{Suzuki2016}.  Thus, 
the combination of shell evolution and dynamical correlations can 
give a proper description of this unusual formation of a two-neutron halo. It is of interest that the ground-state neutron halo seems to occur in $^{19}$C as a single-particle phenomena and in $^{22}$C as a result of correlations. 

As $Z$ becomes smaller, below $Z$=6, the neutron 1$p_{1/2}$ orbit is raised due to weakened attraction with the proton 1$p_{3/2}$ orbit, and approaches the 2$s_{1/2}$ orbit.   
This shell evolution leads to the vanishing of the shell closure at $N$=8 and 
the SO magic number $N$=6 becomes reinforced (see Fig.~\ref{fig:HO-LS}).   
The decrease of the gap between the 1$p_{1/2}$ and 2$s_{1/2}$ orbits enhances large admixture of $sd$-shell components in the ground states of nuclei 
such as \nuc{12}{Be} as well as in the dripline nucleus \nuc{11}{Li}. 


\subsection{Shell evolution examined by (e,e'p) experiment.} 

\begin{figure}[t]    
 \begin{center}
  \includegraphics[width=8.5cm,clip]{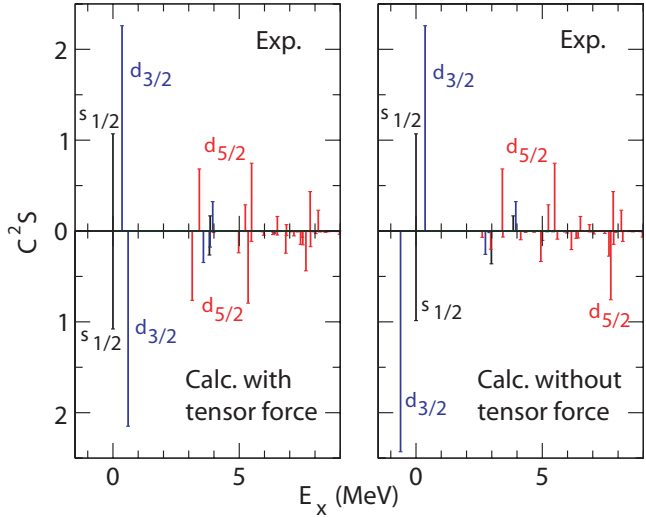}
 \caption{
Distribution of proton hole strengths in $^{48}$Ca compared 
between the $(e,e'p)$ data \citep{Kramer2001} and 
shell-model calculations with the SDPF-MU interaction. 
The left and right panels show the calculations with and 
without the cross-shell tensor force, respectively. 
The calculated overall spectroscopic factors are quenched by 0.7. 
From \citet{Utsuno2012}. 
}
 \label{fig:ca48}
 \end{center}
\end{figure}

The electron scattering enables us to carry out a model-independent analysis of obtained  data, and therefore provides us with an excellent and unique tool to see the nuclear structure, apart from the limitation due to low cross sections and the limited applicability only to stable nuclei at present.   
Among various types of experiments, the (e,e'p) experiment is a superb method to investigate proton single-particle properties including the shell structure.  Figure~\ref{fig:ca48} shows, in the upper panels, the distribution of the proton-hole strengths with respect to the $^{48}$Ca nucleus measured 
with the $^{48}$Ca$(e,e'p)^{47}$K reaction \citep{Kramer2001}. 
In the same figure, the measured distribution is compared to shell-model calculations using the SDPF-MU interaction, with (left lower panel) and without (right lower panel) the tensor force. The $sd$-$pf$ cross-shell part of the SDPF-MU interaction is the $V_{\mathrm MU}$ interaction with which many theoretical analyses have been carried as mentioned earlier in this article, and contains the same tensor force as is used throughout this article.   
Figure~\ref{fig:ca48} exhibits that calculations by the SDPF-MU interaction with the tensor force reproduce quite well the measurement for both energies and strengths.   The proton 1$d_{5/2}$-1$d_{3/2}$ splitting is calculated to be 5.1 MeV with the SDPF-MU interaction. 
Once the tensor part of the interaction is switched off, the 1$d_{5/2}$ strengths are shifted to higher energies by the absence of the mechanism shown in Fig.~\ref{fig:tensor-1}.
Note that the hole energy goes up when the corresponding orbit is lowered. 

The spin-orbit splitting in $^{40}$Ca is estimated to be 
$\sim 6.7$~MeV on the basis of the centroid energy using the 
$(d, ^3$He$)$ reaction data \citep{Doll1976}, where the $1d_{5/2}$ 
strengths are highly fragmented in the $E_x > 5$ MeV region. 
A precise measurement for $^{40}$Ca 
similarly to the one for $^{48}$Ca is of much interest. 
See \citet{Sorlin2008} for details of deducing proton-hole energies 
in the K isotopes from the $(d, ^3$He$)$ data. 


\subsection{Other cases in heavy nuclei}
Some of other relevant studies on heavier nuclei are worked out in 
Appendix~\ref{app:others}   
\cite{Federman1977,Federman1979a,Federman1984,Pittel1993,Goodman,Zeldes,Kay2008,Kay2011,Ogawa1978,Santamaria2015,Schiffer2013}.  

\section{Summary}    
\label{sec:summary}

This article presents a review of the structure of exotic nuclei mainly 
from the viewpoint of the shell evolution driven by nuclear forces.
While shell evolution implies changes of the shell/magic structure, such changes, 
in particular, substantial and/or systematic ones, were not expected several decades ago.  In fact,
the shell/magic structure proposed by Mayer and Jensen was shown to be extremely successful in the description of the structure of nuclei.  A few exceptional cases of notable changes were known, with their example mentioned in Sec.~\ref{sec:primer} and  Sec.~\ref{subsec:TalmiCNO}.
Certain changes of the shell structure have gradually been noticed and some empirical analyses were made, as reviewed, for instance, in  \cite{Sorlin2008,Sorlin_nobel,Grawe2004}.
However, over the past two decades, many cases of substantial and systematic changes of the shell/magic structure have been clarified with underlying theoretical mechanisms and experimental data thanks mostly to Rare-Isotope (RI) beam experiments.  Among the various outcomes and phenomena, particularly visible ones are the identification of new magic numbers (16, 32, 34, ...) and the recognition of dimished traditional magic numbers (8, 20, 28, ...), occurring in certain regions of the Segr\'e (nuclear) chart.
Thus, the shell evolution turned out to be a distinctive phenomenon, visible particularly in exotic nuclei.

The shell evolution is driven by the monopole interaction, which is a component of the nuclear force in nuclei.   
The monopole interaction has been discussed in various ways since 1964 (Sec.~\ref{subsec:short_summary}), and we review, throughout this article, 
its underlying mechanism and its appearance in a variety of physics phenomena.

After a brief survey of earlier works in Secs.~\ref{sec:introduction} and \ref{sec:primer}, 
we start with a possible definition of the monopole interaction in Sec.~\ref{sec:monopole}, which is applicable for closed-shell and open-shell nuclei. 
In the case of atomic nuclei, rotational invariance is imposed as a symmetry constraint, and this symmetry produces degeneracy with respect to the magnetic substates of each single-particle orbit.  The ``monopole'' interaction then arises for a given two-body interaction from this degeneracy: the motion of two interacting particles in given single-particle orbits $j$ and $j'$ can take various two-body quantum states.
The monopole matrix element is an average with respect to them (Sec.~\ref{sec:monopole}).  
The effective single-particle energy (ESPE) is obtained by combining this monopole interaction and a given configuration (an occupation pattern over all single-particle orbits)
(Sec.~\ref{subsec:espe}).  The ESPEs are operators, but can be c-numbers if the configuration is fixed.   The ESPEs calculated for a typical configuration provide us with a clear and simple perspective of nuclear structure, for example, as neutrons are added to a specific orbit in an isotopic chain.  While the definition or meaning of the ESPE might look different among different formulations, they are shown to be consist
(Sec.~\ref{subsec:short_summary},  Sec.~\ref{subsec:Baranger}).  

The monopole interactions of the central, tensor, 2-body spin-orbit and three-nucleon forces produce different characteristic features in the variations of the ESPEs ({\it i.e.}, shell evolution) as illustrated in Secs.~\ref{sec:SE} and \ref{sec:force}.  
The tensor and 2-body spin-orbit forces provide unique and notable effects because of their spin dependences (Sec.~\ref{sec:SE}).  Many of the underlying properties of these many-body effects were clarified rather recently both theoretically and experimentally, although  these forces have been known for several decades.  
Because of the renormalization persistency, the monopole effect of the tensor force can be evaluated in a simple way (Sec.~\ref{sec:RP}).

The central force basically senses similarities of radial single-particle wave functions (Sec.~\ref{subsec:central}), and produces important contributions; in many of the cases of shell evolution, the central and tensor forces work coherently with similar magnitudes. 
For instance, this coherence is directly related to the appearance of the $N$=34 magic number, (Sec.~\ref{subsubsect:34}) as well as the shell structure on top of the $^{100}$Sn closed shell (Sec.~\ref{subsubsect:ZrSn}), for which extensive experimental studies are ongoing.   
Some aspects of central-force effects have been discussed since its early days (Secs.~\ref{subsec:central} and \ref{subsect:tensorSE}).   
A wide variety of mean-field approaches, non-relativistic and relativistic, have been proposed for the description of the shell structure including various functionals for the tensor-force effects (Sec.~\ref{subsubsect:mean-feild}). 

Modern {\it ab initio} approaches are expected to derive effective $NN$ interaction from the QCD level (Sec.~\ref{sec:force}), including three-nucleon-force effects.  
The monopole effect from the three-nucleon forces has been shown to be crucial for nuclear binding, including the dripline of the oxygen isotopes (Sec.~\ref{sec:force}).  

The shell evolution was evaluated in many analyses presented in this review in terms of the V$_{\mathrm MU}$ interaction and the 2-body spin-orbit force in the M3Y interaction.    These are given in simple analytic forms, and provide us with a consistent assessment in a unified way.   Although these interactions can be improved for fine details, we focused on overall trends.  

Further studies on the effective $NN$ interactions, including those of the origin in the three-nucleon forces, are on-going with various approaches, but definitely more studies are needed, to develop and deepen the physics of exotic nuclei up to driplines.   The shell evolution is expected to play a major role as it reflects an average property.  

The effects of the shell evolution in actual nuclei have been examined and explored experimentally as discussed in Sec.~\ref{sec:actual} and other parts.
The usage of a variety of experimental probes, from the $\gamma$-ray spectroscopy to transfer reactions to electron scattering, are demonstrated in Sec~\ref{sec:actual} with a  focus on the Island of Inversion.  


The lowering of intruder states containing particle-hole excitations across a magic gap is a dominant phenomenon in the island of inversion or in the shape coexistence in general (also Secs.~\ref{sec:introduction} and \ref{sec:primer}), and has naturally strong connections to the shell evolution.  Various experimental probes clarify different aspects of it.  

The interplay of the shell evolution with the continuum physics and weakly bound states, {\it etc}, is mentioned in Sec~\ref{subsubsec:southern border}.  This subject is being developed, 
with great interest both theoretically and experimentally, and will be a major trend in the forth-coming studies.   In those states, substantial changes may appear in the effective interaction, single-particle wave function, {\it etc.}, and the field continues to devise  innovative experimental approaches to investigate them.  After all, it is of much interest how the shell evolution changes/persists at the dripline as well as for loosely bound states. 


As the shell evolution will keep unveiling static and dynamic features of exotic nuclei not expected within the conventional view, there will be intriguing, diverse and glorious frontiers emerging in many ways in nuclear structure physics.   Such frontiers do include heavy nuclei eventually up to the nuclei of superheavy elements, where improvements to predictive power will also contribute.
Furthermore, such changes in the understanding and properties of exotic nuclei may impact also other disciplines of science, for instance, astrophysics and astronomy, and nuclear engineering,
as neutron-rich exotic nuclei are intermediate products in explosive stellar processes and  nuclear reactors.

\vspace{1cm} 


\section{Acknowledgments}
Useful discussions with Drs. J.P. Schiffer, B. Kay, A. Poves, F. Nowacki, H. Grawe, M. Gorska and P. Ring are acknowledged.    T.O. thanks Drs. Y. Tsunoda and J. Menendez for valuable discussions on the monopole interaction, and Dr. T. Miyagi for his great contributions to the overview of ab-initio approaches.  T.O. is grateful also to Drs. M. Honma, R. Fujimoto, T. Matsuo, D. Abe and K. Tsukiyama, and  Profs. Y. Akaishi and A. Schwenk for many relevant productive collaborations.  He acknowledges Dr. N. Tsunoda for private communications and a related figure besides fruitful collaborations.    
This work was supported in part by the HPCI Strategic Program (The origin of matter and the universe)
and ``Priority Issue on Post-K computer'' (Elucidation of the Fundamental Laws and Evolution of the Universe)
from MEXT and JICFuS (hp140210, hp150224, hp160211,hp170230), and is a part of the RIKEN-CNS joint research project on large-scale nuclear-structure calculations.
T.O. acknowledges support in part by the Grant-in-Aids for Scientific Research (A) 20244022 of the JSPS. 
A.G. acknowledges support from the US National Science Foundation under Grant
No. PHY-1102511 and PHY-1565546 (NSCL). 
T.S. acknowledges support in part by the Grant-in-Aids for Scientific Research under Grant No. JP15K05090 of the JSPS. 
Y.U. acknowledges support in part by the Grant-in-Aids for Scientific Research under Grant
No. JP15K05094 of the JSPS.
\\
\vspace{1cm}


\begin{appendix}

\noindent
Note: The following Appendices will appear as Supplemental Materials in the published version.

\section{Proton-neutron monopole interaction}
\label{app:pn mono}

The basic idea of the proton-neutron monopole interaction remains the same as in the case for two neutrons.
A proton in the state $j, m$ and a neutron in $j', m'$ are considered.    
We first treat the proton and the neutron as different kinds of fermions with no mutual relation such as the isospin.
A two-body state is expressed as 
\begin{equation}
\label{eq:pn-state}
   |\,\pi(j, m) \otimes \,\nu(j', m') \, ),
\end{equation}
similarly to eq.~(\ref{eq:direct}).
We use, when appropriate, indices $\pi$ and $\nu$ for quantities 
related to protons and neutrons, respectively.
Note that the proton state comes first and the neutron state second.
The state $(j, m)$ is abbreviated as $m$ hereafter for brevity, except for the cases that explicit 
expressions are needed or would help.  
The proton-neutron interaction can be written as
\begin{eqnarray}
\label{eq:vpn}
 \hat{v}_{pn} & = & \Sigma_{\pi m_1, \nu m_2, \pi m'_1, \nu m'_2 } \, \,
  ( \,\pi m_1 \otimes  \,\nu m_2 | \, \hat{v}_{pn} \, | \,\pi m'_1 \otimes \,\nu m'_2 )  \nonumber \\
 & & \, \, a^{\dagger}_{\pi m_1} a^{\dagger}_{\nu m_2}  a_{\nu m'_2} a_{\pi m'_1} .
\end{eqnarray}
Here, 
$( \,\pi m_1 \otimes  \,\nu m_2 | \, \hat{v}_{pn} \, | \,\pi m'_1 \otimes \,\nu m'_2 )$ denotes 
two-body matrix element.  

We shall show that the monopole matrix element for the proton-neutron system has a different feature from that for the neutron-neutron system, because the states comprised of a proton and 
a neutron can be decomposed into two groups according to the symmetry with respect to the exchange between proton and neutron.   
The symmetric state is defined by
\begin{flalign}
\label{eq:pn-sym1}
\,\,\, &|\, \pi m; \, \nu m' : {\cal S})  =     
 \bigl\{ |\, \pi m \otimes \nu m' ) + |\, \pi m' \otimes \nu m ) \bigr\}/\sqrt{2}, & \nonumber \\
 & \,\,\,\,\,{\rm for} \,\,\pi j \ne \nu j' , \,\,\,\,
  {\rm or \,\, for}  \,\, \pi j=\nu j' \,\, {\rm but} \,\, \pi m \ne \nu m', &  
\end{flalign}
and for $\pi j = \nu j'$ and $\pi m=\nu m'$ by,
\begin{flalign}
\label{eq:pn-sym2}
 |\, \pi m; \, \nu m' : {\cal S})  = |\, \pi m \otimes \nu m' ).  
\end{flalign}

The monopole matrix element is defined, in this case, as an average over all symmetric states 
with all possible orientations.  It is therefore given by 
\begin{eqnarray}
\label{eq:m_pn}
V_{pn,s}^m (j,j') \, & = &\, \frac{\sum_{(m, m')}  ( m; m' :{\cal S}| \hat{v}_{pn} | m; m' :{\cal S} ) }{\sum_{m,m'} 1}.
\end{eqnarray}
We shall denote the creation operators for a proton and a neutron in the state $j, m$  
as $c^{\dagger}_{j, m}$ and $a^{\dagger}_{j, m}$, respectively.  Likewise, the annihilation 
operators are denoted as $c_{j, m}$ and $a_{j, m}$.  
Here, the proton operators, creation and annihilation, commute with the neutron operators, 
creation and annihilation. 
The state in eq.~(\ref{eq:pn-sym1}) is then created by the operator 
\begin{equation}
\label{eq:sym1-op}
\frac{1}{\sqrt{2}}\Bigl\{ c^{\dagger}_{j, m} a^{\dagger}_{j', m'}  + c^{\dagger}_{j', m'} a^{\dagger}_{j, m}  \Bigr\},
\end{equation}
while the state in eq.~(\ref{eq:pn-sym2}) is created by the operator
\begin{equation}
\label{eq:sym2-op}
c^{\dagger}_{j, m} a^{\dagger}_{j', m'}   .
\end{equation}
Pair annihilation operators can be introduced similarly as hermitian conjugates of 
eqs.~(\ref{eq:sym1-op}) and (\ref{eq:sym2-op}).

By combining eqs.~(\ref{eq:m_pn}), (\ref{eq:sym1-op}), and (\ref{eq:sym2-op}), 
the monopole-interaction operator is written as
\begin{equation}
\hat{v}_{pn,mono,s} \,  =  \, \sum_{j \le  j'} \, \hat{v}_{pn,s}^m (\, j, \, j') \,  .
\label{eq:Vm_pns}
\end{equation} 
Here, $\hat{v}_{pn,s}^{m} (j,j')$ is given for $ j \ne  j'$ as 
\begin{flalign}
\label{eq:Vm_pns1}
& \hat{v}_{pn,s}^{m} (j,j') \,= \, V_{pn,s}^m (j,j') \,\,\, \frac{1}{2}\Sigma_{m,m'} & \nonumber \\
&\,\,\,\,\,\,\,\,\,\,\,\,\,\, \bigl( c^{\dagger}_{j, m} a^{\dagger}_{j', m'}  + c^{\dagger}_{j', m'} a^{\dagger}_{j, m}  \bigr) 
    \bigl( a_{j', m'} c_{j, m}  + a_{j, m} c_{j', m'} \bigr) .&
\end{flalign}
Based on eqs.~(\ref{eq:sym1-op},\ref{eq:sym2-op}), it is written for $ j =  j'$ as 
\begin{flalign}
\label{eq:Vm_pns2}
& \hat{v}_{pn,s}^{m} (j,j) \,= \, V_{pn,s}^m (j,j) \,\,& \nonumber \\
& \,\, \Bigl\{ \frac{1}{2}\Sigma_{m < m'}  \bigl( c^{\dagger}_{j, m} a^{\dagger}_{j, m'}  
+ c^{\dagger}_{j, m'} a^{\dagger}_{j, m}  \bigr)  \bigl( a_{j, m'} c_{j, m}  + a_{j, m} c_{j, m'} \bigr) & \nonumber \\
& \,\,\,\, + \Sigma_{m} \, c^{\dagger}_{j,m} \, a^{\dagger}_{j,m} a_{j,m} c_{j,m} \,\,\Bigr\} .&
\end{flalign}

We introduce the proton and neutron number operators in the orbit $j$ as,
\begin{equation}
\hat{n}^p_j \, = \, \sum_{m} \, c^{\dagger}_{j,m} c_{j,m} \,\,\, {\rm and}\,\, \hat{n}^n_j \, = \, \sum_{m} \, a^{\dagger}_{j,m} a_{j,m}.
\label{eq:nj}
\end{equation}
In addition, we introduce the following operators,
\begin{equation}
\hat{\tau}^+_{j} \, = \, \sum_{m} \, c^{\dagger}_{j,m} a_{j,m},  \,\,\, {\rm and}\,\, 
\hat{\tau}^-_j \, = \, \sum_{m} \, a^{\dagger}_{j,m} c_{j,m}.
\label{eq:tau}
\end{equation}
We note here that the operators 
in eq.~(\ref{eq:tau}) are nothing but  
the isospin raising and lowering operators restricted to the orbit $j$.
Here, we take the convention that protons are in the state of isospin $z$-component $\tau_z =$+1/2, 
whereas neutrons are in $\tau_z =$-1/2. 

With these operators, the monopole interaction can be rewritten as
\begin{flalign}
 \hat{v}_{pn,mono,s} \, =&  \, \sum_{\, j <  j'} \, V_{pn,s}^m (j, \, j')   \frac{1}{2} \nonumber & \\
 & \,\, \Bigl\{ \hat{n}^p_j \, \hat{n}^n_{j'}  + \hat{n}^p_{j'} \, \hat{n}^n_j  - \hat{\tau}^+_{j} \hat{\tau}^-_{j'} -
 \hat{\tau}^+_{j'} \hat{\tau}^-_{j} \, \Bigr\}  \nonumber & \\
 &  + \sum_{j} \, V_{pn,s}^m (j, \, j)   \frac{1}{2} \Bigl\{ \hat{n}^p_j \, \hat{n}^n_{j} \, 
            -  : \hat{\tau}^+_{j} \hat{\tau}^-_{j} : \, \Bigr\} ,&
\label{eq:Vm_jj'}
\end{flalign}
where the symbol $: ... :$ denotes a normal product.
This equation can be rewritten as
\begin{flalign}
 \hat{v}_{pn,mono,s} \, =& \, \sum_{\, j \, j'} \, V_{pn,s}^m (j, \, j')   \frac{1}{2} \,\,  \hat{n}^p_j \, \hat{n}^n_{j'} \, \nonumber  &\\
& - \sum_{\, j <  j'} \, V_{pn,s}^m (j, \, j')   \frac{1}{2} \Bigl\{ \hat{\tau}^+_{j} \hat{\tau}^-_{j'} + 
 \hat{\tau}^-_{j} \hat{\tau}^+_{j'} \, \Bigr\}  \nonumber & \\
 &  - \sum_{j} \, V_{pn,s}^m (j, \, j)   \frac{1}{2}  \, : \hat{\tau}^+_{j} \hat{\tau}^-_{j} : \, .&
\label{eq:Vm_jj'-2}
\end{flalign}
Some visual explanation of the last two terms on the right-hand side are displayed in Fig.~\ref{fig:tautau}. 

We shall now move to the isospin scheme.   
The total wave function must be antisymmetric with respect to the exchange of 
any pair of nucleons in the isospin scheme, and the total wave function is a product of the
coordinate-spin wave function and the isospin wave function.   
The present case, where the coordinate-spin wave function is symmetric, corresponds to the isospin 
wave function being antisymmetric, which means that
the proton and neutron couple to isospin $T=0$.
Thus, the states belonging to eqs.~(\ref{eq:pn-sym1},\ref{eq:pn-sym2}) are $T=0$ states.
Equation~(\ref{eq:Vm_jj'-2}) represents the $T=0$ part of the proton-neutron monopole interaction 
with  
the $T=0$ monopole matrix elements given by,
\begin{equation}
V_{T=0}^m (j,j') \,=\, V_{pn,s}^m (j,j') ,
\label{eq:mono_T=0}
\end{equation}
where $V_{pn,s}^m (j,j')$ is defined in eq.~(\ref{eq:m_pn}).

Equation~(\ref{eq:Vm_jj'-2}) can then be expressed as,
\begin{flalign}
 \hat{v}_{pn,mono,T=0} \, &=  \, \sum_{\, j,\, j'} \, V_{T=0}^m (j, \, j') \, \frac{1}{2} \, \hat{n}^p_j \, \hat{n}^n_{j'} \, \nonumber \\
 & -  \sum_{\, j <  j'} \, V_{T=0}^m (j, \, j') \, \frac{1}{2} \, \Bigl\{ \hat{\tau}^+_{j} \hat{\tau}^-_{j'} + 
 \hat{\tau}^-_{j} \hat{\tau}^+_{j'} \, \Bigr\}  \nonumber & \\
 &  - \sum_{j} \, V_{T=0}^m (j, \, j) \, \frac{1}{2} : \hat{\tau}^+_{j} \hat{\tau}^-_{j} : \, . & 
\label{eq:Vm_jj'T0}   
\end{flalign}
This is the $T=0$ monopole interaction within the present scheme.

We next discuss antisymmetric proton-neutron states as
\begin{eqnarray}
\label{eq:pn-anti}
   |m; \, m' : {\cal A}) & = & \{ |m \otimes m' ) - |m' \otimes m )\}/\sqrt{2} \,\,\,\,  . 
\end{eqnarray}
The state in eq.~(\ref{eq:pn-anti}) is then created by the operator 
\begin{equation}
\label{eq:anti-op}
\frac{1}{\sqrt{2}}\Bigl\{ c^{\dagger}_{j, m} a^{\dagger}_{j', m'}  - c^{\dagger}_{j', m'} a^{\dagger}_{j, m}  \Bigr\} ,
\end{equation}
acting on the appropriate vacuum ({\it i.e.,} a closed shell).

The state in this category is antisymmetric with respect to the exchange between proton and 
neutron, which means that the isospin part should be symmetric.
If the proton and neutron form a state with total isospin $T=1$, its wave function does not change
the sign upon exchange of proton and neutron.  Thus, the states in 
eq.~(\ref{eq:pn-anti}) are $T=1$ states.   

The state in eq.~(\ref{eq:nn-anti}) can be generated by changing a proton in the state
of eq.~(\ref{eq:anti-op}) into a neutron.  Indeed, by using the isospin-lowering operator mentioned above,
we obtain 
\begin{flalign}
\label{eq:pn-anti-2}
 & \,\,\,\,\,\,\,\, |j,m \,;\, j',m' >  & \nonumber \\
 & \,\,\,\,\,\,\,\,\,\,\,\,=\, a^{\dagger}_{j, m} a^{\dagger}_{j', m'} |0>  & \nonumber \\
 & \,\,\,\,\,\,\,\,\,\,\,\,\propto \, \bigl\{ \tau^-_j + \tau^-_{j'}  \bigr\} 
 \Bigl\{ c^{\dagger}_{j, m} a^{\dagger}_{j', m'}  - c^{\dagger}_{j', m'} a^{\dagger}_{j, m}  \Bigr\} 
 |0> , & 
\end{flalign}   
where $|0>$ indicates the relevant vacuum as usual.
We can include states of two protons, 
\begin{equation}
\label{eq:pp} 
c^{\dagger}_{j, m} c^{\dagger}_{j', m'} |0> ,
\end{equation}   
in the same way.
We note that the three states in eqs.~(\ref{eq:nn-anti},\ref{eq:anti-op},\ref{eq:pp})
form the isospin multiplet with $T=1$.  
If the interaction $\hat{v}$ is isospin invariant as usual, 
the isospin operators $\tau^{+}$ and $\tau^{-}$ commute with $\hat{v}$, and 
the monopole matrix element in eq.~(\ref{eq:m_nn}) is the $T=1$ monopole matrix element
to be used commonly for these isospin multiplet states:
\begin{equation}
V_{T=1}^m (j,j') \,=\, V_{nn}^m (j,j') \, .
\label{eq:mono_T=1}
\end{equation}

Coming back to the proton-neutron state, we apply procedures similar to those for the $T=0$ states, and obtain
\begin{flalign}
 \hat{v}_{pn,mono,T=1} &= \sum_{\, j ,\, j'} \, V_{T=1}^m (j, \, j') \, \frac{1}{2} \, \hat{n}^p_j \, \hat{n}^n_{j'} \, \nonumber &\\
& + \sum_{\, j <  j'} V_{T=1}^m (j, \, j') \, \frac{1}{2} \, \Bigl\{\, \hat{\tau}^+_{j} \hat{\tau}^-_{j'} \, + 
 \hat{\tau}^-_{j} \hat{\tau}^+_{j'} \, \Bigr\}   \nonumber & \nonumber \\
& + \sum_{j} \, V_{T=1}^m (j, \, j) \, \frac{1}{2} : \hat{\tau}^+_{j} \hat{\tau}^-_{j} : \, .& 
\label{eq:Vm_jj'T1}
\end{flalign}

\section{Alternative definition of the monopole interaction}
\label{app:alternative}

An alternative but equivalent expression of the monopole matrix element is indicated here.  
In the case of $j < j'$, eq.~(\ref{eq:nn-anti}) can be rewritten as
\begin{flalign}
\label{eq:nn-anti-2}
\,\,\,\,\,\, & |j,m \,;\, j',m' >  & \nonumber \\
  &  =  \, \sum_{J_1} (j m j' m' | J_1 M=m+m') & \nonumber \\
  &  \,\,\,\,\,\,\,\,\,\,\,\, \sum_{m_1 m'_1} (j m_1 j' m'_1 | J_1 M=m+m')  | m_1 m'_1 >, & 
\end{flalign}   
where $J_1$ in the summation runs through all possible values ({\it i.e.}, from $J_1 = |j-j'|$ up to $j+j'$), and 
the symbols like $(j mj m' | J_1 M_1)$ are Clebsch-Gordan coefficients. 
We apply this expansion to the ket state of the numerator of eq.~(\ref{eq:m_nn}) as well as a similar expansion 
to its bra state, and come up with the following equation by using the orthonormal relation of the 
Clebsch-Gordan coefficient: 
\begin{flalign}
\label{eq:m_nn-3}
\,\,\,\, &\sum_{m, m'}  \langle j,m\,;\, j',m' | \hat{v}_{nn} | j,m \,;\, j',m' \rangle  &   \nonumber \\
&= \sum_{J_1}\sum_{J_2}\sum_{m, m'}  \, (j m j' m' | J_2 M=m+m') &\nonumber \\
& \,\,\,\,\,\,\,\,\,\,   (j m j' m' | J_1 M=m+m') \, \langle j, j' ; J_2 | \hat{v}_{nn} | j, j' ; J_1 \rangle &\nonumber \\
&= \sum_{J_1} (2J_1 + 1) \langle j, j' ; J_1 | \hat{v}_{nn} | j, j' ; J_1 \rangle. &
\end{flalign}
Likewise, the denominator of eq.~(\ref{eq:m_nn}) is rewritten as
\begin{equation}
\label{eq:m_nn-4}
\sum_{m, m'}  1 \,= \, \sum_{J_1} (2J_1 + 1), 
\end{equation}
where the factor $(2J+1)$ is the degeneracy of the two-particle states
having the same value of $J$.  
We finally obtain
\begin{eqnarray}
\label{eq:mono_J}
V_{nn}^m (j, \, j') = \frac{\sum_{J}  (2J+1) \langle j, j' ; J | \hat{v}_{nn} | j, j' ; J \rangle }{\sum_{J} (2J+1)}.
\end{eqnarray}

In the case of $j$ = $j'$, $J_1$ in the summation takes only even integers due to the antisymmetrization.
The neutron-neutron interaction corresponds to the $T=1$ interaction.
The $T=0$ interaction can be treated similarly as symmetric combinations between proton and neutron.
Combining all these relations, we can obtain 
\begin{eqnarray}
\label{eq:mono_J-2_app}
V_{T}^m (j, \, j') &=& \frac{\sum_{J}  (2J+1) \langle j, j' ; J,T | \hat{v} | j, j' ; J,T \rangle }{\sum_{J} (2J+1)} 
   \nonumber \\
   & & \,\,\, {\rm for} \,\,\, T=0 \,\, {\rm and} \,\, 1, 
\end{eqnarray}
where $J$ takes only even (odd) integers for $j = j'$ with $T=1$ ($T=0$).
This is nothing but eq.~(\ref{eq:mono_J-2}) in \ref{subsect:mono_jj}.

\section{Closed-shell properties}
\label{app:closed}

Here we discuss properties of closed-shell states.
We assume that the orbits $j_1, j_2, ..., j_k$ form the shell for both protons and neutrons. 

We start with the neutron closed shell where all these orbits are completely filled by neutrons.
The expectation value of $\hat{v}_{nn}$ is then given by the straightforward calculation as
\begin{eqnarray}
& E_{nn} &= \, \sum_{j} \sum_{m_1 < m_2}  \langle j,m_1\,;\, j,m_2 | \hat{v}_{nn} | j,m_1 \,;\, j,m_2 \rangle  \nonumber \\
&  & + \, \sum_{j < j'} \sum_{m_1, m_2}  \langle j,m_1\,;\, j',m_2 | \hat{v}_{nn} | j,m_1 \,;\, j',m_2 \rangle .  
\nonumber \\
\label{eq:E_nn}
\end{eqnarray}
Considering $\sum_{m<m'} 1 \,=\, (2j+1) j$ as well as  
$\sum_{m,m'} 1 \,=\, (2j \,+\,1) (2j'\,+\, 1)$ for $j \, \ne \, j'$, 
we can rewrite $E_{nn}$ in eq.~(\ref{eq:E_nn}) as
\begin{eqnarray}
& E_{nn} &\,= \, \sum_{j} \, (2j \,+\,1) \, j   \,\, V_{T=1}^m (j,j) \nonumber \\
&  & \,+ \, \sum_{j < j'}  \, (2j \,+\,1) (2j'\,+\, 1) \,\,V_{T=1}^m (j, \,j' \,)  \, ,  
\label{eq:E_nn-1}
\end{eqnarray}
where $V_{T=1}^m (j, \,j' \,)$ is defined in eq.~(\ref{eq:mono_T=1}).  
We compare this result with the expectation value of $\hat{v}_{nn,mono}$ in eq.~(\ref{eq:m_nn-2}) 
with respect to the present closed-shell state:
\begin{flalign}
\label{eq:E_nn-2}
& \,\, \langle \hat{v}_{nn,mono} \rangle \, & \nonumber \\
 & \,\,\, = \sum_{j} V_{T=1}^m (j,j) \,  (2j \,+\,1) \, j   & \nonumber \\
  &    \,\,\,  \,\,\,   + \sum_{j < j'} V_{T=1}^m (j,j') \,  (2j \,+\,1) \, (2j' \,+\,1) \, ,&  
\end{flalign}
where the following substitution is made: 
\begin{equation}
\langle \hat{n}^n_j \rangle \,=\, (2j \,+\,1) \,
\label{eq:n_2j+1}
\end{equation}
The quantity in eq.~(\ref{eq:E_nn-2}) is exactly the same as the one in eq.~(\ref{eq:E_nn-1}).   
We point out that the closed-shell properties are not used in the derivation of the present monopole interaction.

We next consider the case with both protons and neutrons.   To start, we assume that 
protons and neutrons occupy the same orbit $j$.   More general cases will then be discussed.
The contributions from the proton-proton and neutron-neutron interactions have been indicated 
just above.  The remaining point is the contribution from the proton-neutron interaction.
The proton-neutron interaction depends on the symmetry with respect to the exchange of  
proton and neutron, as discussed in the previous subsection.  For a closed shell with 
$2j+1$ protons and $2j+1$ neutrons, each state with a pair of a proton and neutron is specified by
$m, m'$, where $m$ ($m'$) refers to a proton (neutron) state.  As $m$ runs from $m=-j$ to $m=+j$, 
the total number of those states is $(2j+1)^2$.
The direct product $| m \otimes m' \rangle$ spans those states.  In order to incorporate the symmetry 
dependence of the nuclear forces, such direct-product states are 
transformed by an appropriate unitary transformation into symmetric and antisymmetric combinations.
The symmetric states are given by eq.~(\ref{eq:pn-sym1}) for $m \ne m'$ (with $j =j'$) and by 
eq.~(\ref{eq:pn-sym2}) for $m = m'$.   
The total contribution from those symmetric states is nothing but 
\begin{equation}
\label{eq:E_pn_s}
E_{pn,s} = \sum_{(m, m')}  ( m; m' :{\cal S}| \hat{v}_{pn} | m; m' :{\cal S} ) \, .
\end{equation}
As the quantity on the right hand side is used in eq.~(\ref{eq:m_pn}) and 
the number of symmetric pairs turns out to be 
\begin{equation}
\sum_{m,m'} 1 \,=\, \frac{1}{2} \, (2j+1+1)(2j+1) \,=\, (j+1)(2j+1),
\end{equation}
eq.~(\ref{eq:m_pn}) leads us to
\begin{equation}
E_{pn,s} = V_{pn,s}^m (j,j) \, (j+1)(2j+1) \, .
\label{eq:E_pn_s2}
\end{equation}

The antisymmetric combinations are treated similarly.   The antisymmetric state is
shown in eq.~(\ref{eq:pn-anti}).   
Due to the isospin invariance of the interaction and the isospin multiplet property, 
their total contribution can be taken from eq.~(\ref{eq:E_nn-1}), and is explicitly given as 
\begin{equation}
\label{eq:E_pn_a}
E_{pn,a} \,= \,  (2j \,+\,1) \, j   \,\, V_{T=1}^m (j,j)  \, .
\end{equation}
Combining eqs.~(\ref{eq:E_pn_s2},\ref{eq:E_pn_a}), we obtain
\begin{equation}
\label{eq:E_pn}
E_{pn} \,= \,  V_{T=0}^m (j,j) \, (j+1)(2j+1) \,+\,  V_{T=1}^m (j,j)  \, (2j \,+\,1) \, j   \,\, .
\end{equation}

We compare this result with the expectation value of $\hat{v}_{pn,mono}$ in eq.~(\ref{eq:m_nn-2}) 
with respect to the present closed-shell state :
\begin{eqnarray}
 \langle \hat{v}_{pn,mono} \rangle & = & \frac{1}{2} \, \Bigl\{V_{T=0}^m (j, \, j)\, + \, V_{T=1}^m (j, \, j) \Bigr\}
                                            \,  (2j \,+\,1)^2 \,   \nonumber \\
                                  & - &\frac{1}{2} \, \Bigl\{V_{T=0}^m (j, \, j)\, - \,V_{T=1}^m (j, \, j) \Bigr\}   
                                     \,  \langle :\hat{\tau}^+_{j} \hat{\tau}^-_{j}: \rangle . \nonumber \\                                    
\label{eq:E_pn-2}
\end{eqnarray}
where eq.~(\ref{eq:n_2j+1}) is used for protons as well as for neutrons. 
The expectation value of the second term on the right hand side, $- :\hat{\tau}^+_{j} \hat{\tau}^-_{j}:$, 
counts the number of the pairs of 
a proton and a neutron both in the state $m$, and its value is obviously $(2j \,+\,1)$ for the closed shell.  
By substituting this, we finally obtain
\begin{eqnarray}
\langle \hat{v}_{pn,mono} \rangle & = & \, V_{T=0}^m (j, \, j)\, \bigl( 2j^2 \,+\, 3j \,+\,1 \bigr) \,   \nonumber \\
& + & \, V_{T=1}^m (j, \, j)\, \bigl( 2j^2 \,+\, j \, \bigr) \,    .                                    
\label{eq:E_pn-3}
\end{eqnarray}
This is exactly the same as the result of eq.~(\ref{eq:E_pn}).  

More general cases can be formulated in a straightforward way based on the arguments here.
The second term ($\hat{\tau}^+_{j} \hat{\tau}^-_{j'} \, +  \hat{\tau}^-_{j} \hat{\tau}^+_{j'}$) 
on the right hand side of eq.~(\ref{eq:Vm_jj'T1}) does not contribute to the energy of the 
closed shell, because it transforms a proton into a neutron or vice versa.
Thus, a generalization of eq.~(\ref{eq:E_pn}) (or equivalently eq.~(\ref{eq:E_pn-3}))  becomes 
\begin{eqnarray}
E_{pn} &=&   \sum_{\, j , \,  j'}   \frac{1}{2}  \bigl\{V_{T=0}^m (j, j')+ V_{T=1}^m (j,  j') \bigr\}
                                            \,  (2j +1) \, (2j' +1)   \nonumber \\
            & + & \sum_{\, j }  \frac{1}{2}  \bigl\{V_{T=0}^m (j, \, j)\, - \,V_{T=1}^m (j, \, j) \bigr\}   
                                     \,  (2j \,+\,1)  . \nonumber \\                                    
\label{eq:E_pn-4}
\end{eqnarray}

The total contribution is obtained by summation as
\begin{equation}
\label{eq:E_tot}
E_{tot} \,= \,  E_{pp} \,+\, E_{nn} \,+\, E_{pn}. 
\end{equation}
This becomes for the case of the orbit $j$ only 
\begin{eqnarray}
\label{eq:E_tot_j}
E_{tot} &\,= &\,  V_{T=0}^m (j,j) \, (j\,+\,1)\,(2j\,+\,1) \nonumber \\
             &\,+&\,  V_{T=1}^m (j,j)  \, 3 \, j \, (2j \,+\,1)    \,\, .
\end{eqnarray}
This expression is in agreement with the result shown in Ref. \cite{Poves1981}.



\section{An example of the shell evolution due to like-particle interaction}
\label{app:N=28 gap}

We now move on to the shell evolution 
due to like-particle ($T$=1)  interactions, by pursuing neutron shell structure with 
varying the neutron number $N$ 
[see also \citep{Sorlin2008, Sorlin_nobel}]. 

The magic number 28 is the first SO one in those of Mayer and Jensen
(see Fig.~\ref{fig:HO-LS}).
It is rather stable at least for $N=28$ isotones with $Z\ge 20$. 
Actually, the strength of the $N=28$ shell gap, i.e.,
single-particle-energy difference 
between $1f_{7/2}$ and $2p_{3/2}$, is estimated to be $4.8$~MeV 
on top of the $^{48}$Ca core 
from one-neutron separation energies of $^{48,49}$Ca. 
The $\sim 5$~MeV shell gap at $N=28$ is also supported by 
comparing the first excited levels in $^{47,48}$Ca, 
which are dominated by neutron $(2p_{3/2})^1(1f_{7/2})^{-2}$ 
and neutron $(2p_{3/2})^1(1f_{7/2})^{-1}$ configurations, respectively, 
between experiment and shell-model calculations. 
Note that the $7/2^-_1$ level at 3.13 MeV in $^{49}$Ca, 
whose spin and parity has recently given by \cite{Broda2006,
Montanari2011}, is also consistent with 
shell-model results with the $\sim 5$~MeV shell gap. 

The $N=28$ shell gap is, on the other hand, much smaller 
on top of the $^{40}$Ca core. 
Experimentally, this quantity is deduced to be 2.5 MeV 
from the distribution of the spectroscopic strengths for 
the $^{40}$Ca$(\vec{d},p)$ reaction \citep{Uozumi1994}. 

\begin{figure}[bt]     
 \begin{center}
  \includegraphics[width=6cm]{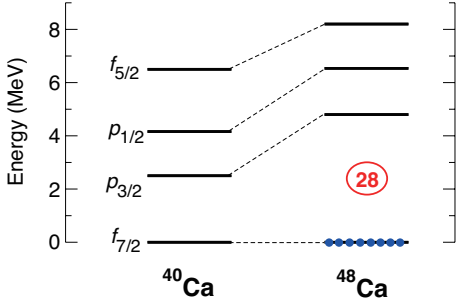} 
 \caption{Evolution of neutron $pf$ orbitals in going from $^{40}$Ca to 
$^{48}$Ca based on experimental data \citep{Uozumi1994, Uozumi1994A}. 
}
 \label{fig:ca40-48}
 \end{center}
\end{figure}

Thus, as illustrated in Fig.~\ref{fig:ca40-48}, 
$N=28$ becomes one of the classical magic numbers 
by $2$-$3$~MeV enlarging the neutron 
$1f_{7/2}$-$2p_{3/2}$ shell gap 
when the neutron $1f_{7/2}$ orbit is filled. 
Assuming the inert $^{40}$Ca core, 
the present shell evolution can be attributed to 
like-particle interaction. 
By using Eqs.~(\ref{eq:Delta_e}) and (\ref{eq:Delta_e-2}), the change of the 
$N=28$ shell gap is written as 
\begin{equation}
\begin{array}{cl}
 &  \Delta^{(f_{7/2},nn)}\epsilon_{p_{3/2}} - 
 \Delta^{(f_{7/2},nn)}\epsilon_{f_{7/2}}  \\
 = & 8\times \left\{ V^{m}_{T=1}(p_{3/2},f_{7/2}) 
  - V^{m}_{T=1}(f_{7/2},f_{7/2})
  \right\}.
\end{array}
\end{equation}

The impact of the monopole interaction on the evolution of the 
$N=28$ shell gap was first pointed out by \citet{McGrory1970}, where 
the Kuo-Brown interaction is shown to be incapable of reproducing  
low-lying energy levels in Ca isotopes in the vicinity of $N=28$.
This defect was then shown to be remedied by shifting 
$V^{m}_{T=1}(p_{3/2},f_{7/2})$ by $+0.3$ MeV, which leads to 
the additional 2.4 MeV enlargement of the $N=28$ shell gap. 
A similar monopole correction was also implemented by 
\citet{Poves1981} in their KB3 interaction, making the KB3 interaction 
frequently used for the $pf$-shell nuclei. 


\section{2-body LS force}
\label{app:LS}

We describe here the monopole component of the 2-body LS (2b-$LS$) force and its effect 
in some details but with simple terms, partly because the following discussions may not be 
found elsewhere.
       
The 2b-$LS$ force between the particles indexed as 1 and 2 contains a scalar product of 
the orbital angular momentum of the relative motion, denoted as $\vec{L}_{rel;12}$, and the total spin, 
$\vec{S}_{12}$.
Here, these operators can be written as
\begin{equation}
\vec{L}_{rel;12} \,=\, (\vec{r}_1 \,-\, \vec{r}_2) \times (\vec{p}_1 \,-\, \vec{p}_2), 
\label{eq:LS_rel}
\end{equation}
with $\vec{r}_{1,2}$ ($\vec{p}_{1,2}$) being the coordinate (momentum) 
of the particles 1 and 2, respectively.  The total spin is defined as
\begin{equation}
\vec{S}_{12} \,=\, \vec{s}_{1} \,+\, \vec{s}_{2}, 
\label{eq:LS_S12}
\end{equation}
where $\vec{s}_{1,2}$ implies the spin operator of the particles 1 and 2, respectively.

They are coupled to the 2b-$LS$ operator
\begin{equation}
W_{LS} \,=\, (\vec{L}_{rel;12} \cdot \vec{S}_{12}), 
\end{equation}
where the symbol $( \, \cdot \, )$ indicates a scalar product as usual. 

The 2-body LS force is then given by 
\begin{equation}
V_{LS} \,=\, w_{0}(r) W_{LS} \,+\, w_{1}(r) (\vec{\tau}_1 \cdot \vec{\tau}_2 ) W_{LS},
\label{eq:V_LS}
\end{equation}
where $w_{0}(r)$ and $w_{1}(r)$ are appropriate functions of the relative distance $r$.    
The functions depend on the choice of the 2b-$LS$ interaction.  We choose the 2b-$LS$ interaction 
of the M3Y interaction \citep{Bertsch1977}.  
We emphasize that the 2b-$LS$ interaction is characterized as a short-range interaction, 
and is indeed described, in this paper, by a short-range Yukawa-type $r$-dependence 
(range parameter being 0.25 and 0.4 fm), with $w_{0}(r)$ and $w_{1}(r)$ being negative.
This short range is considered to be induced by $\sigma+\rho+\omega$ meson exchanges 
\citep{Bertsch1977},
and can be compared to that of the corresponding parameter of the Millner-Kurath interaction, $\sim 0.7$ fm \citep{Millener1975}.
Because $w_{0}(r)$ and $w_{1}(r)$ are negative, if $\vec{L}_{rel;12}$ and $\vec{S}_{12}$ are oriented to
the same (opposite) direction, an attractive (repulsive) effect is produced. 
It is noted that between the total spin $S_{12}$=0 and 1 states of particles 1 and 2, only the $S_{12}=1$ state is affected by the 2b-$LS$ interaction, implying that the spins of the particles 1 and 2 must be parallel.

We here show the relations
\begin{equation}
V_{odd} (T=1) \,=\, ( w_{0}(r) + \frac{1}{4} w_{1}(r) ) \, W_{LS}
\end{equation}
and
\begin{equation}
V_{even} (T=0) \,=\, ( w_{0}(r) \,-\, \frac{3}{4} w_{1}(r) ) \, W_{LS}.
\end{equation}
If $w_{0}(r)$ and $w_{1}(r)$ are both negative as in the M3Y interaction  \citep{Bertsch1977}, 
the $T$=1 interaction becomes stronger in magnitude than the $T$=0 interaction.

\begin{figure}[t]    
 \begin{center}
\includegraphics[width=7.0cm,clip]{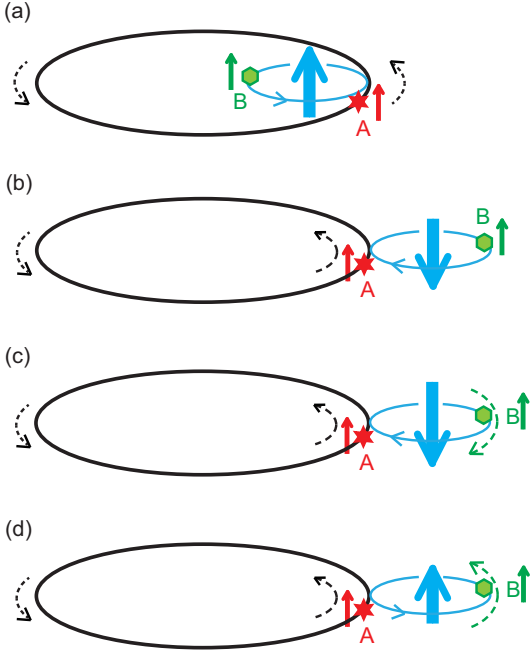}
 \caption{Schematic illustration of the monopole matrix elements of the 2-body LS (2b-$LS$) force.
 Black large circle stands for the orbit of nucleon A (red star) with spin (red short arrow).  
 The 2-body LS force works between this nucleon and another
 nucleon B (green hexagon) (a) inside or (b) outside the orbit of A.  
 The spin of the other nucleon is depicted by a green short arrow.  
 Nucleon B is assumed to be at rest in (a,b), whereas the motion of B is activated in (c) and (d) 
 as indicated by outmost green dashed arrows.  In (d), nucleon B is assumed to have an
 angular momentum larger than that of nucleon A.    
 Blue large arrows indicate the orbital angular momentum of the relative motion between A
 and B.
}
 \label{fig:2bLS}
 \end{center}
\end{figure}

Based on these basic features, we can draw an intuitive schematic picture, Fig.~\ref{fig:2bLS}, 
as to how the 2b-$LS$ force works.
This figure shows that the 2b-$LS$ force acts on the nucleon A (red star)  
circulating on an orbit (black big circle), and this force was due to the nucleon B 
(green hexagon) being either (a) inside the orbit or (b) outside the orbit.
The spin is assumed to be upward, without generality.  The spins are shown by small
red and green arrows next to the corresponding nucleons in the figure.

\begin{figure}[t]    
 \begin{center}
 \includegraphics[width=7.5cm,clip]{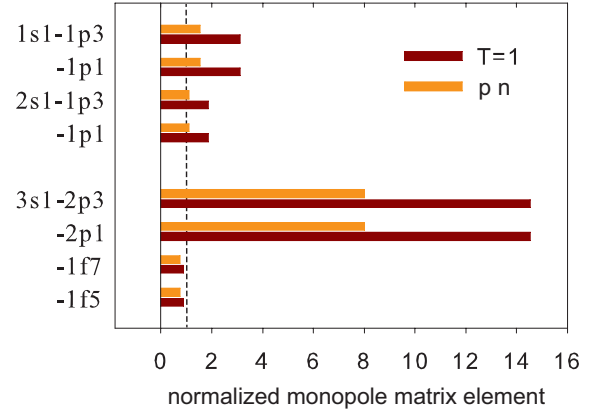}
 \caption{Monopole matrix elements obtained from the 2-body spin-orbit (2b-$LS$) force.
 The orbit on the far left indicates the orbit for nucleon B and that on the second left for nucleon A in 
 Fig.~\ref{fig:2bLS}, while the same entry for B is omitted.  
 The matrix elements are normalized by referring a standard spin-orbit splitting 
 (see the text for eq.~(\ref{eq:LS-normalized}).  The mass number $A$=16 is taken for the upper part, 
 while $A$=80 was chosen for the lower part.   
 The vertical dashed line means the unity.}
 \label{fig:LSmono_spf}
 \end{center}
\end{figure}

We assume that the nucleon A is moving in the direction shown by the 
dashed arrows in the figure.  The nucleon B is assumed to be at rest for the time being, 
while we will come back to this point.

In Fig.~\ref{fig:2bLS} (a), the orbital angular momentum of the relative motion between the
two nucleons is invoked by the motion of the nucleon A as depicted by the blue small circle, 
which is represented by the $\vec{L}_{rel;12}$  (blue middle-sized arrow in the figure).  
In this case, $\vec{L}_{rel;12}$ and $\vec{S}_{12}$ are in the same direction, which makes
the 2b-$LS$ force attractive, as described above. 
In Fig~\ref{fig:2bLS} (b), the nucleon B is located outside the orbit of the nucleon A,
and the $\vec{L}_{rel;12}$ points to the opposite direction, but the spins are the same as in Fig~\ref{fig:2bLS} (a).
Thus, the effect is opposite, {\it i.e.}, repulsive.
The net effect is obtained by combining the cases (a) and (b) over all possible locations of nucleon B.  
We emphasize here that the cases (a) and (b) cancel each other in general.  
If the density of nucleon B inside the orbit of nucleon A is higher than that outside, the case (a) survives
the cancellation, and  
the monopole matrix element becomes negative, yielding the normal spin-orbit
splitting for $j_> \,(=\ell + 1/2)$.   
If one reverses the orbital motion of nucleon A, one ends up again with the normal
spin-orbit splitting for $j_< \,(=\ell - 1/2)$.   

\begin{figure}[t]    
\begin{center}
\includegraphics[width=7.0cm,clip]{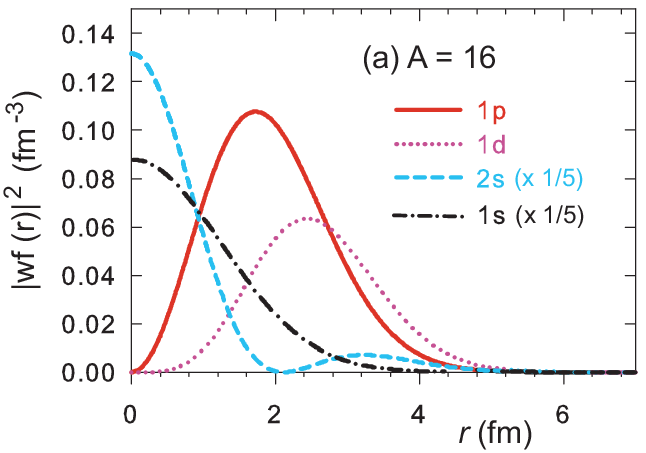}
 \vspace{0.5cm}
  \includegraphics[width=7.0cm,clip]{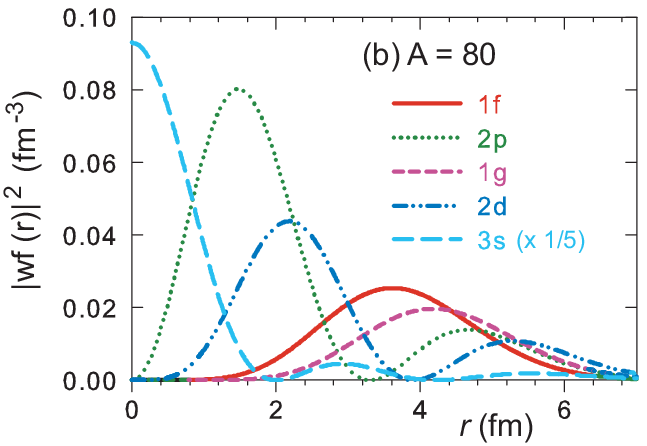}
\caption{Radial wave functions squared for (a) 1$p$, 1$d$ and 2$s$ orbits
 with $A$=16 
 and (b) 1$f$,  2$p$, 1$g$, 2$d$ and 3$s$ orbits with $A$=80.
 Harmonic Oscillator wave functions are taken. 
 }
 \label{fig:LS_radial}
 \end{center}
\end{figure}

Because the 2b-$LS$ force is of short range, the present cancellation effect may be well 
simulated by the radial derivative of the density of nucleon B at the location of nucleon A. 
This approximation becomes generally better with a large number of nucleons due to growing  
collectivity, and appears to be in accordance with the usual argument with the radial derivative of 
the density \citep{BM1}.   
The approximation is expected to be particularly adequate if nucleon B 
runs through all nucleons in a closed shell, yielding the final spin-orbit splitting.
However, there are different aspects also, as discussed below.

We shall look into actual monopole matrix elements of the 2b-$LS$ force.  
We here introduce, for the sake of a global comparison, a  
normalization factor for the monopole matrix element 
based on the usual spin-orbit splitting \citep{BM1} 
with an additional division by the mass number, $A$.  The factor is given as
\begin{equation}
- 20 \, ( \vec{\ell} \, \cdot \, \vec{s} ) \,  \, A^{-5/3} \,\,\, {\rm MeV}.
\label{eq:LS-normalized}
\end{equation}
The division by $A$ gives a better scaling for the comparison to the monopole matrix element, 
as the spin-orbit splitting is due to all other nucleons in the nucleus.
Throughout this article, unless otherwise specified, 
each monopole matrix element of the 2b-$LS$ force is divided by this factor,  
where the $\ell$ and $j$ stand for those of nucleon A.
Actual values are calculated with Harmonic Oscillator radial wave functions for a given mass number $A$.

Figure~\ref{fig:LSmono_spf} shows several cases with nucleon B in the 1, 2 or 3$s_{1/2}$-orbit.  
In Fig~\ref{fig:LSmono_spf}, the orbital entries like ``1s1-1p3'' are given to the left.
This means that nucleon A is in the 1$p_{3/2}$ orbit and nucleon B is in the 1$s_{1/2}$ orbit;
nucleon B is the source of the monopole matrix element appearing first, and nucleon A is 
its recipient appearing next.    
Each case ({\it i.e.}, orbital combination) is composed of two horizontal histograms; the upper one 
stands for the proton-neutron matrix element and the lower for the $T$=1 one.   
Both are normalized by the factor in eq.~(\ref{eq:LS-normalized}).    
The second case implies that nucleon B is the same as above but 
nucleon A is in the orbit 1$p_{1/2}$.    
The first four cases (from the top) are calculated with $A=16$, and indicate that (i) those monopole matrix elements 
show values $\approx 1-3$ in the present scaling, and (ii) $T$=1 contributions are larger than the proton-neutron ones, with the latter being about one half of the former.  
Point (i) suggests that the 2b-$LS$ force is, in these cases, one of the primary origins of the 
spin-orbit splitting.  Point (ii) implies that the dominant contribution from the present 2b-$LS$ force
is in the $T$=1 channel.  
We show the radial wave functions in Fig~\ref{fig:LS_radial} (a) as functions of the distance from the
center of the nucleus, $r$, in order to understand these monopole matrix elements.  
The squared radial wave function of the 1$s$ orbits decreases for $r$= 0-3 fm, while the 1$p$ orbit
is well outside of it.
The squared radial wave function of the 2$s$ orbits decreases for $r$= 0-2 fm.  Although it increases
for $r \approx$ 2-3 fm, the magnitude is much smaller, and does not reverse the final result.
Thus, the argument presented with Fig~\ref{fig:2bLS} (a,b) can be applied.    

The lower part of Fig.~\ref{fig:LSmono_spf} is calculated with $A=80$, and shows the monopole matrix elements for the $pf$-shell orbit due to nucleon B on the 3$s_{1/2}$ orbit.   The contributions to the 1$f_{7/2,5/2}$ orbits appear to be close to unity.

Notable are large values of the contributions to the 2$p_{3/2,1/2}$ orbits
which are much larger than 
those involving 1$f$.   They are different by a factor of 15.   
We shall discuss this intriguing problem now.  
In the coupling between the 3$s$ and 2$p$ orbits, the major components of 
the wave function of the relative motion correspond to small values of $L_{rel;{\rm AB}}$.  
On the other hand, in the coupling between the 3$s$ and 1$f$ orbits, $L_{rel;{\rm AB}}$
is shifted naturally to larger values.   Because of the short-range character of the 2b-$LS$ force, 
if relevant $L_{rel;{\rm AB}}$ values are larger, 
the effect of the 2b-$LS$ force becomes weaker.  In fact, the combination of 3$s$ and 2$p$
produces the relative motion in the $s$ or $p$ state with high probabilities, and then 
the larger monopole matrix elements are a natural consequence.   
Thus, large monopole matrix elements between $s$ and $p$ orbits are a general 
feature, and monopole matrix elements can take larger values in the combinations of other $s$ and $p$ orbits.
Without the normalization in eq.~(\ref{eq:LS-normalized}), the monopole matrix element between the 3$s_{1/2}$ and 2$p_{3/2}$ is smaller only by  
a factor of about two from that between the 2$s_{1/2}$ and 1$p_{3/2}$.   This can be
understood from a similarity between the relevant parts of the panels (a) and (b) of 
Fig~\ref{fig:LS_radial}.   On the other hand, the factor $A^{5/3}$ in eq.~(\ref{eq:LS-normalized}) 
yields about 1/15 from $A$=16 to 80, and the case with the $1f$ orbit fits well to this change.  
It is thus of much interest and significance that the $s$-$p$ coupling deviates from the trend given by 
eq.~(\ref{eq:LS-normalized}). 

\begin{figure}[t]    
 \begin{center}
  \includegraphics[width=7.5cm,clip]{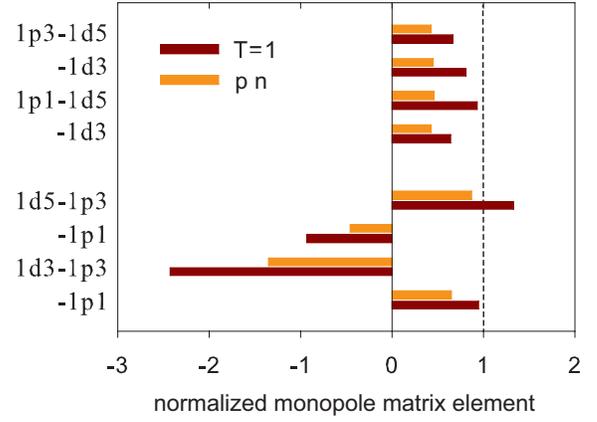} 
 \caption{Monopole matrix elements obtained from the 2-body spin-orbit (2b-$LS$) force
 between the $p$ and $sd$ shells.  See the caption of Fig.~\ref{fig:LSmono_spf}.
 }
 \label{fig:LSmono_pd}
 \end{center}
\end{figure}

Figure~\ref{fig:LSmono_pd} presents the monopole matrix element between 1$p$ and 
1$d$ orbits for $A=16$.  The upper four cases are in accordance with the usual spin-orbit splitting
shown in eq.~(\ref{eq:LS-normalized}).  
Figure~\ref{fig:LS_radial} (a) indicates that the 1$p$ orbit is inside the 1$d$ orbit for the dominant 
part of 1$d$ orbit, and then the picture explained by Fig.~\ref{fig:2bLS} works again
with nucleon A (B) in the 1$d$ (1$p$) orbit. 
   
We now exchange the orbits for nucleons A and B; A in 1$p$ and B in 1$d$.  
As the radial inner-outer relation is reversed, the sign of the monopole matrix elements are altered.   
There is, however, another factor to be considered now.    
We assumed so far that nucleon B is at rest.   We now include the motion of nucleon B, and B is 
either in the orbit $j'_{>}$ or $j'_{<}$.  Let us start with the case that nucleon A is in the $j_{>}$ orbit, 
whereas B is in the $j'_{<}$, as shown in Fig.~\ref{fig:2bLS} (c).   
The axial vector $\vec{L}_{rel;{\rm AB}}$ in Fig.~\ref{fig:2bLS} (c) increases its magnitude
keeping the direction as in Fig.~\ref{fig:2bLS} (b).  
Because $\vec{L}_{rel;{\rm AB}}$ and $\vec{S}_{12}$ are anti-parallel now, 
the monopole matrix element is positive (repulsive), and the value normalized
by the spin-orbit splitting turns out to be negative, because of the $j_{>}$ orbit for nucleon A.  
Such a case is found in the entry ``1d3-1p3'' in Fig.~\ref{fig:LSmono_pd}.   A similar case is seen in 
the entry ``1d5-1p1'' of the same figure.

We next consider, in Fig.~\ref{fig:2bLS} (d), nucleon B in $j'_{>}$ with the orbital angular momentum 
for B greater than the orbital angular momentum for A.  
The axial vector $\vec{L}_{rel;{\rm AB}}$ becomes reversed as compared to Fig.~\ref{fig:2bLS} (b).    
This occurs, for instance, if $j = p_{3/2}$ and $j' = d_{5/2}$.  In such a case,
the radial inner-outer inversion and the reversed relative rotation make sign changes twice with no net change.  This is the case with ``1d5-1p3'' and ``1d3-1p1'' in Fig.~\ref{fig:LSmono_pd}.    
The unusual cases (``1d3-1p3'' and ``1d5-1p1'' in Fig.~\ref{fig:LSmono_pd}) may not be so important 
practically but may be of certain interest;
for instance, the occupation of the 1$d_{3/2}$ orbits reduces the 1$p_{3/2}$-1$p_{1/2}$ 
splitting by $\sim$ 0.05 MeV with the $T=1$ interaction by raising the 1$p_{3/2}$ orbit more.   
    
\begin{figure}[t]    
 \begin{center}
 \includegraphics[width=8.5cm,clip]{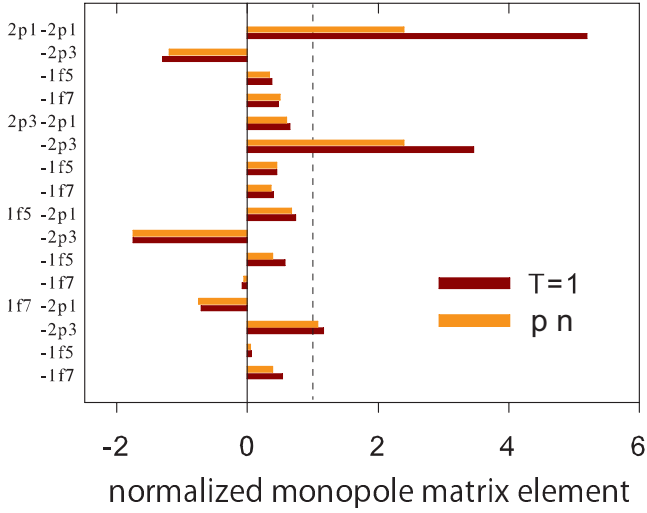}
 \caption{Monopole matrix elements obtained from the 2-body spin-orbit (2b-$LS$) force
 in the $pf$ shell.  See the caption of Fig.~\ref{fig:LSmono_spf}.
 }
 \label{fig:LSmono_pf}
 \end{center}
\end{figure}

Figure~\ref{fig:LSmono_pf} displays the monopole matrix elements 
between two nucleons in the same ($pf$) shell with normalization by eq.~(\ref{eq:LS-normalized}). 
Many cases show values less than unity in magnitude.  This is expected from 
the argument of the inside/outside cancellations discussed for Fig.~\ref{fig:2bLS}, because 
nucleons A and B are on orbits in the same shell.  There are anomalously large values in the cases 
the 2$p_{1/2}$-2$p_{1/2}$ and 2$p_{3/2}$-2$p_{3/2}$ combinations.  For instance, 
the $T$=1 monopole matrix element before the normalization 
is as large as 0.223 MeV for $A$=40.
The interpretation of this matrix element in terms of Fig.~\ref{fig:2bLS} seems to be 
inappropriate, because such interpretation is somewhat classical but the two-nucleon state in 
the 2$p_{1/2}$ orbit can be treated only quantum mechanically.   Instead, we can present a
clear quantum mechanical picture.  The $T$=1 state in the 2$p_{1/2}$ orbit is nothing but
a $J$=0 state.  This $J$=0 is coupled by the spin $\vec{S}_{12}$ in eq. (\ref{eq:LS_S12}) 
and the orbital angular momentum.   The former must fulfill $S_{12} =1$ for the 2b-$LS$ 
force, and the latter is composed of $\vec{L}_{rel;{\rm AB}}$ in eq.~(\ref{eq:LS_rel}) as well as
the corresponding center-of-mass angular momentum $\vec{L}_{cm;{\rm AB}}$.   
The component with $L_{rel;{\rm AB}}=0$ and $L_{cm;{\rm AB}}=1$ does not contribute
to the present case, whereas that with $L_{rel;{\rm AB}}=1$ and $L_{cm;{\rm AB}}=0$ 
gives a large contribution because of the short-range character of the 2b-$LS$ force. 
The $J$=0 coupling implies $(\vec{L}_{rel;{\rm AB}} \cdot \vec{S}_{12}) < 0$, leading to
the repulsive contribution.  Thus, we can expect a rather strong repulsive effect on the 
2$p_{1/2}$-2$p_{1/2}$ case. 

\begin{figure}[t]    
 \begin{center}
  \includegraphics[width=8.5cm,clip]{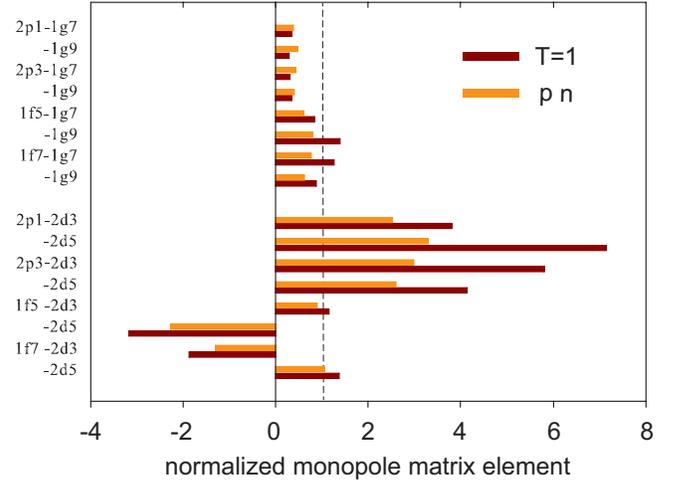} 
 \caption{Monopole matrix elements obtained from the 2-body spin-orbit (2b-$LS$) force
 between the $pf$ and $sdg$ shells.  See the caption of Fig.~\ref{fig:LSmono_spf}.
 }
 \label{fig:LSmono_pfsdg}
 \end{center}
\end{figure}

This mechanism can be applied to the 2$p_{3/2}$-2$p_{3/2}$ case for its $J$=0 component, 
but its $J$=2 component may contain $L_{rel;{\rm AB}}=1$ and $S_{12} =1$ coupled in parallel, 
leading to an opposite effect.  Thus, the net contribution becomes smaller than for the 
2$p_{1/2}$-2$p_{1/2}$ case.   
The cases containing both the $p$ and $f$ orbits can be interpreted 
already with Fig.~\ref{fig:2bLS}.  

Thus, the two nucleons in the same shell can be explained, and the exceptionally strong 
monopole matrix element between the two $p_{1/2}$ orbits is emphasized once more, as its
origin is general and robust.  
This feature produces visible effects on the appearance of the $N$=34 new magic
number (see Sec.~\ref{subsubsect:34}).  

Figure~\ref{fig:LSmono_pfsdg} depicts monopole matrix elements for the $pf$-$sdg$ shell, 
as examples for the generality of the various effects discussed so far.  
Figure~\ref{fig:LS_radial} (b) shows that the 1$g$ orbits are located in the outermost region. 
The large values for the $p$-$d$ combinations in Fig.~\ref{fig:LSmono_pfsdg} 
are due to small values of $L_{rel;{\rm AB}}$
in major components of the relative-motion wave functions.  
They increase the spin-orbit splitting also because the 2$p$ orbits are inside the 1$d$ orbit 
as can be seen from Fig.~\ref{fig:LSmono_pfsdg}. 
The negative values appear in the $f$-$d$ combination, where $j=7/2 > j'=3/2$ occurs and 
the 1$f$ orbits is outside the 2$d$ (see Fig.~\ref{fig:LSmono_pfsdg}).

\section{2$p_{3/2}$-2$p_{1/2}$ splitting in $N$=21 isotones}
\label{app:p-split}

The neutron 2$p_{3/2}$-2$p_{1/2}$ splitting has been studied recently \citep{Kay2017}, where   
the change of this splitting was interpreted in terms of the loose binding effect for 
\nuc{35}{Si}, \nuc{37}{S}, \nuc{39}{Ar} and \nuc{41}{Ca} in \citep{Kay2017}. 
We discuss here this change also in terms of the monopole effect of 2b-LS forces.  

\begin{figure}[tb]    
\includegraphics[width=8.5cm]{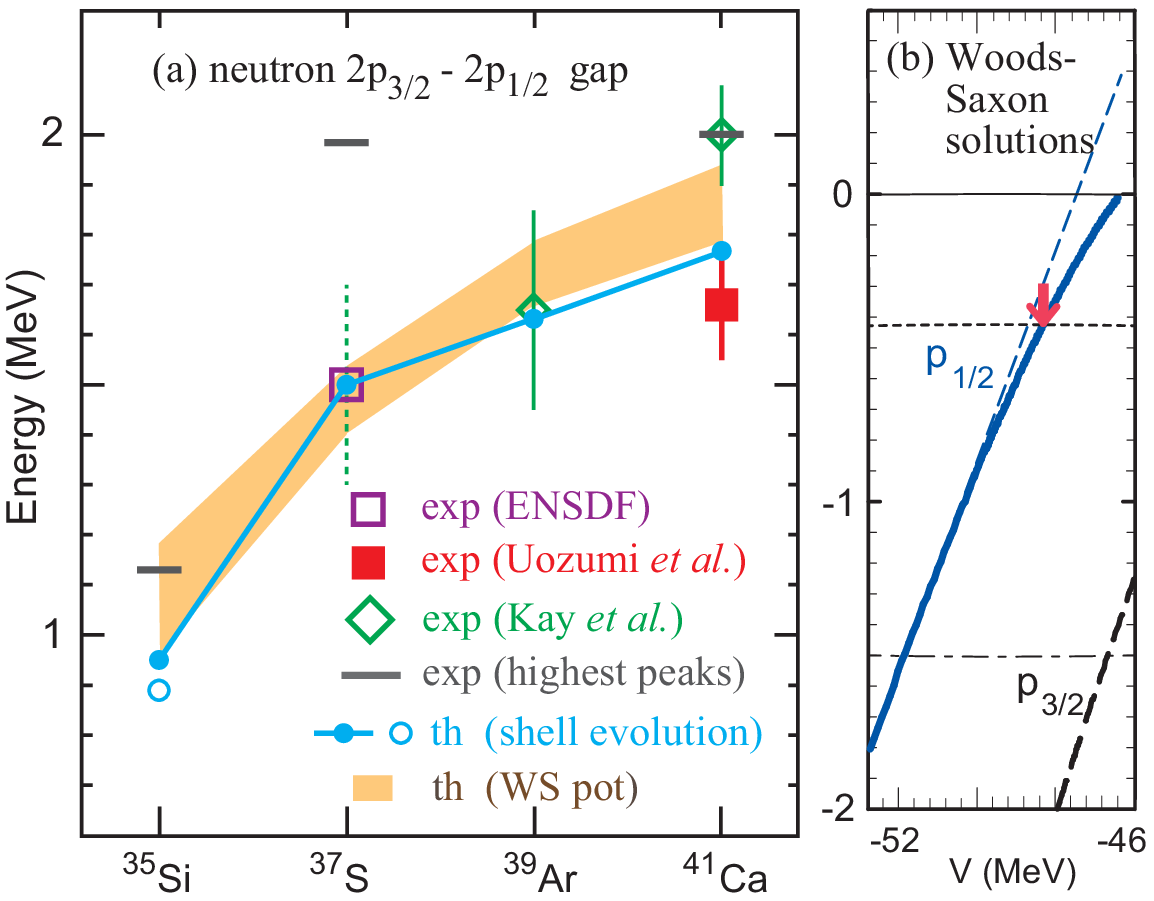} 
\caption{(a) Neutron 2$p_{3/2}$-2$p_{1/2}$ spin-orbit splitting in the $N=21$ isotonic chain. 
The symbols denote the centroids for \nuc{37}{S} \cite{ensdf}, \nuc{39}{Ar} \citep{Kay2017} 
and \nuc{41}{Ca} \citep{Uozumi1994} and \citep{Kay2017}.  The horizontal bars imply the energy differences between relevant highest peaks \citep{Burgunder2014}, \citep{Kay2017}. 
Shell evolution predictions are shown by blue closed symbols and solid line connecting them. 
The loose binding effect for $^{35}$Si is included in the open circle.
The calculation with Woods-Saxon potential with parameters adjusted are shown by the yellowish shaded area \citep{Kay2017}. 
(b) Neutron 2$p_{1/2}$ single-particle energy (blue solid line) obtained by a standard Woods-Saxon potential
\cite{BM1} as a function of the depth parameter, $V$.  The linear dependence of the deeply bound region is linearly extrapolated (blue dashed line) and is compared to the curved dependence that results from the proximity of the continuum.   The dashed line is for the 2$p_{3/2}$ orbit.   
Horizontal lines denote their single-particle energies. 
}
\label{fig:so_se} 
\end{figure}

Some relevant results obtained to date are summarized in Fig.~\ref{fig:so_se} (a), where the
centroid of the single-particle strengths presented in \cite{ensdf} for \nuc{37}{S} is used as the origin of the changes.   From experimental side, 
the centroid obtained in the case of \nuc{41}{Ca} using the $^{40}$Ca$(\vec{d},p)$ reaction \citep{Uozumi1994} is also shown in Fig.~\ref{fig:so_se}, where we estimated the error from the cross sections.   
The centroids were reported by \citep{Kay2017} for \nuc{41}{Ca} and \nuc{39}{Ar}, as well as  
an error bar for \nuc{37}{S}.  
Figure~\ref{fig:so_se} displays also the splitting taken from the energy difference between peaks of the highest strength.  This is the only quantity available for \nuc{35}{Si}.
One sees a monotonic decrease from \nuc{41}{Ca} to \nuc{35}{Si}, while the reduction from \nuc{37}{S} overwhelms the rest.  
We now turn to quantitative comparisons to theoretical approaches.  

The present splitting was described by varying Wood-Saxon potential parameters in   
\citep{Kay2017}, as shown in Fig.~\ref{fig:so_se} (a) where the yellowish area represents the uncertainty of the parameters also. 
In addition, Fig.~\ref{fig:so_se} (a) includes the monopole effects calculated 
with the filling configuration.  The decrease from \nuc{41}{Ca} to \nuc{37}{S} is then mainly due to the monopole interaction of the central and tensor forces, and their contributions are destructive with a weak net effect consistent with experiment.  
The subsequent change down to \nuc{35}{Si} is much larger similarly to the experimental trend obtained earlier \citep{Burgunder2014}, and is attributed to the effect of 2b-LS forces.  
The first 1/2$^-$ state of \nuc{35}{Si} is weakly bound ($\sim$ -0.43 MeV) \cite{ensdf}, and we take this into account: Fig.~\ref{fig:so_se} (b) shows the energy of the 2$p_{1/2}$ orbit calculated by a standard Woods-Saxon potential \cite{BM1}, as a function of the depth parameter, $V$.  
As the depth of the potential is raised, this energy
goes up almost linearly for deeply bound cases (see Fig.~\ref{fig:so_se} (b)).  
When the potential becomes shallower, a deviation from this linear dependence arises.
This deviation stands basically for the additional binding energy due to a sizable tunneling effect, {\it i.e.}, a more extended radial wave function.
When this energy is equal to the experimental energy, -0.43 MeV, of the 1/2$^-$ state, the corresponding deviation to linear trend is 130 keV
(see the red arrow in Fig.~\ref{fig:so_se} (b)).  The 2$p_{3/2}$ orbit moves 
in parallel to the 2$p_{1/2}$ orbit for deeply bound cases, with almost no change of the 2$p_{3/2}$-2$p_{1/2}$ splitting.  Similarly to the 2$p_{1/2}$ orbit, a deviation appears, being $\sim$70 keV with the same depth as for the 2$p_{1/2}$ orbit at -0.43 MeV.  The difference is 60 keV, which appears to be a reasonable estimate of the loose-binding effect on the splitting, and we indicate 
the splitting including it (open circle) in Fig.~\ref{fig:so_se} (a).  Although this is a simple estimate, it most likely conveys the order of magnitude of the effect properly.
    
Thus, the major driving force of the sudden drop of the spin-orbit splitting from \nuc{37}{S} to \nuc{35}{Si} is indicated to be the 2b-$LS$ force, with the effect an order of magnitude larger than the loose binding or, more generally the continuum.     
Although the two theoretical approaches yield somewhat similar trends, the underlying ideas are different.   We emphasize that our present result is obtained with globally used nuclear forces 
with certain microscopic origins, explaining many data simultaneously.   
The same situation may be found in 
neutron-rich C-N-O nuclei, as discussed in sec.~\ref{subsec:2-body LS} as well as in \cite{Hoffman2016}. 
In other general cases, this competition depends on the degree of the binding, 
while an extreme weak binding may be required to compete against the shell evolution effect.

\section{Examples of other relevant works on heavy nuclei}
\label{app:others}

 
The monopole interaction between proton $2p_{3/2,1/2}$ orbit and neutron $2d_{5/2}$ orbit was discussed 
for the description of Zr-Sr isotopes in \cite{Federman1984} by using the empirical interaction introduced in \cite{schiffer}.  We can now see that this is a nice visible example of the monopole property of the tensor force. 

Similarly, there have been earlier works where some properties were discussed without mentioning the tensor force, {\it e.g.} \cite{Federman1977,Federman1979a,Federman1984,Pittel1993,Goodman,Zeldes}, while those properties can be shown to be consistent with the general monopole properties discussed so far.

\begin{figure}[tb]    
 \begin{center}
 \includegraphics[width=7cm,clip]{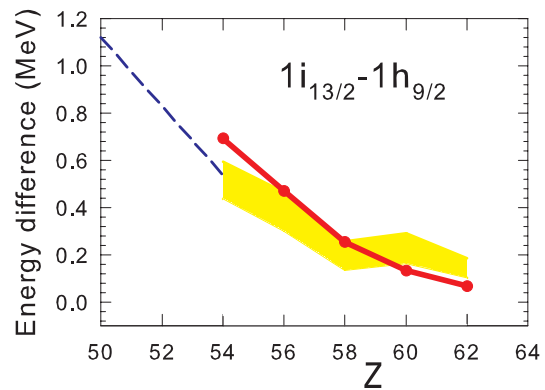}
 \caption{Experimental energy difference (filled circles) in the centroids of neutron $i_{13/2}$ and $h_{9/2}$ orbits compared to the tensor-force prediction (yellow band), where 
the band reflects the uncertainties in the proton 1$g_{7/2}$ and 2$d_{5/2}$ occupancies.
Dashed line is the tensor-force prediction for the lowest 13/2$^{+}$ and 9/2$^{-}$ states
obtained with the 85$\%$ (15$\%$) proton 1g$_{7/2}$ (2d$_{5/2}$) occupancy
extracted in \citep{Wildenthal1971}.  Based on Kay et al. (2011).
}
 \label{fig:i13h9}
 \end{center}
\end{figure}

A more visible example of the shell evolution in heavier nuclei 
is the separation between the neutron 1$i_{13/2}$ and 1$h_{9/2}$ orbits 
in $N$ =83 isotones with even $Z$ =54-62 \citep{Kay2008,Kay2011}.
The centroids of the strengths of 1$i_{13/2}$ and 1$h_{9/2}$ single-neutron states
were obtained by using spectroscopic factors measured by ($\alpha$, $^{3}$He)
and (d, p) reactions. 
As proton 1$g_{7/2}$ (2$d_{5/2}$) orbit is occupied for larger $Z$, attraction and
repulsion (repulsion and attraction) increase for neutron $i_{13/2}$ and $h_{9/2}$ orbits,
respectively, due to the tensor interaction (see  Fig.~\ref{intuition} for instance)).
Thus, the separation between the 1$i_{13/2}$ and 1$h_{9/2}$ states decreases (increases)
as the 1$g_{7/2}$ (2d$_{5/2}$) orbit is occupied by more protons. 
Figure~\ref{fig:i13h9} indicates that the observed energy difference of the centroids of 13/2$^+$ and 9/2$^-$ 
states can be compared well to the calculation with the tensor interaction, by using
proton occupation numbers of 1$g_{7/2}$ and 2d$_{5/2}$ orbits deduced from one-nucleon
transfer reactions \citep{Wildenthal1971}. 

Another example is the $Z=64$ gap, which is seen at $N=82$ \cite{Ogawa1978} 
but disappears as $N$ increases (see, {\it e.g.}, \cite{Casten1981}). 
This gap is washed away for $N > 82$, largely because neutrons in 2$f_{7/2}$ orbits
decreases the proton 2$d_{5/2}$-2$d_{3/2}$ splitting (see  Fig.~\ref{intuition} for instance)).
Many similar cases have been and will be found in other isotopic and isotonic chains, for instance, the $N=40$ gap is influenced by proton 1$f_{7/2}$ occupancy 
(as a recent example, see \cite{Santamaria2015}).  
We refer the reader to the work by \citep{Schiffer2013} for a state-of-the-art exploration of the valence nucleon population in the Ni isotopes within a consistent sum-rule formalism.


\end{appendix}


\bibliography{ref_sub_6}

\end{document}